\documentclass[apj,twocolumn]{openjournal}

\usepackage{newtxtext,newtxmath,enumitem,multirow,ctable}
\usepackage[T1]{fontenc}
\usepackage[breaklinks,colorlinks,citecolor=blue,urlcolor=blue]{hyperref}

\newcommand{\Alf}{{Alfv\'en}}

\newcommand{\gizmourl}{\href{http://www.tapir.caltech.edu/~phopkins/Site/GIZMO.html}{\url{http://www.tapir.caltech.edu/~phopkins/Site/GIZMO.html}}}

\newcommand{\paperone}{Paper {\small I}}
\newcommand{\papertwo}{Paper {\small II}}
\newcommand{\paperthree}{Paper {\small III}}
\newcommand{\paperfour}{Paper {\small IV}}

\newcommand{\orcidauthor}[3]{\author{\href{http://orcid.org/#1}{#2$^{#3}$}}}

\shorttitle{FORGE'd in FIRE II}
\shortauthors{Hopkins et al.}

\begin{document}

\title{\vspace{-0.8cm}FORGE'd in FIRE II: The Formation of Magnetically-Dominated Quasar Accretion Disks from Cosmological Initial Conditions\vspace{-1.5cm}}

\orcidauthor{0000-0003-3729-1684}{Philip F. Hopkins}{1,*}
\orcidauthor{0000-0001-8479-962X}{Jonathan Squire}{2}
\orcidauthor{0000-0003-1598-0083}{Kung-Yi Su}{3}
\orcidauthor{0000-0001-8867-5026}{Ulrich P. Steinwandel}{4}
\orcidauthor{0000-0002-4086-3180}{Kyle Kremer}{1,\dagger}
\orcidauthor{0000-0002-0087-3237}{Yanlong Shi}{1}
\orcidauthor{0000-0002-1655-5604}{Michael Y. Grudi{\'c}}{5,\dagger}
\orcidauthor{0000-0002-3977-2724}{Sarah Wellons}{6}
\orcidauthor{0000-0002-4900-6628}{Claude-Andr\'{e} Faucher-Gigu\`{e}re}{7}
\orcidauthor{0000-0001-5769-4945}{Daniel Angl{\'e}s-Alc{\'a}zar}{8,6}
\orcidauthor{0000-0002-8659-3729}{Norman Murray}{9}
\orcidauthor{0000-0001-9185-5044}{Eliot Quataert}{10}

\affiliation{$^{1}$TAPIR, Mailcode 350-17, California Institute of Technology, Pasadena, CA 91125, USA}
\affiliation{$^{2}$Physics Department, University of Otago, 730 Cumberland St., Dunedin 9016, New Zealand}
\affiliation{$^{3}$Black Hole Initiative, Harvard University, 20 Garden St, Cambridge, MA 02138, USA}
\affiliation{$^{4}$Center for Computational Astrophysics, Flatiron Institute, 162 5th Ave., New York, NY 10010 USA}
\affiliation{$^{5}$Carnegie Observatories, 813 Santa Barbara St, Pasadena, CA 91101, USA}
\affiliation{$^{6}$Department of Astronomy, Van Vleck Observatory, Wesleyan University, 96 Foss Hill Drive, Middletown, CT 06459, USA}
\affiliation{$^{7}$CIERA and Department of Physics and Astronomy, Northwestern University, 1800 Sherman Ave, Evanston, IL 60201, USA}
\affiliation{$^{8}$Department of Physics, University of Connecticut, 196 Auditorium Road, U-3046, Storrs, CT 06269, USA}
\affiliation{$^{9}$Canadian Institute for Theoretical Astrophysics, University of Toronto, Toronto, ON M5S 3H8, Canada}
\affiliation{$^{10}$Department of Astrophysical Sciences, Princeton University, Princeton, NJ 08544, USA}

\thanks{$^*$E-mail: \href{mailto:phopkins@caltech.edu}{phopkins@caltech.edu}},
\thanks{$\dagger$NASA Hubble Fellow}

\begin{abstract}
In a companion paper, we reported the formation of quasar accretion disks with inflow rates $\sim 10\,{\rm M_{\odot}\,yr^{-1}}$ down to $<300$ Schwarzschild radii from {\em cosmological} radiation-magneto-thermochemical-hydrodynamical galaxy and star formation simulations. We see the formation of a well-defined, steady-state accretion disk which is stable against star formation at sub-pc scales. The disks are optically thick, with radiative cooling balancing accretion, but with properties that are distinct from those assumed in most previous accretion disk models. The pressure is strongly dominated by (primarily toroidal) magnetic fields, with a plasma $\beta \sim 10^{-4}$ even in the disk midplane. They are qualitatively distinct from magnetically elevated or arrested disks. The disks are strongly turbulent, with trans-\Alf{ic} and highly super-sonic turbulence, and balance this via a cooling time that is short compared to the disk dynamical time, and can sustain highly super-Eddington accretion rates. Their surface and 3D densities at $\sim 10^{3}-10^{5}$ gravitational radii are much lower than in a Shakura-Sunyaev disk, with important implications for their thermo-chemistry and stability. We show how the magnetic field strengths and geometries arise from rapid advection of flux with the inflow from much weaker galaxy-scale fields in these ``flux-frozen'' disks, and how this stabilizes the disk and gives rise to efficient torques. Re-simulating without magnetic fields produces catastrophic fragmentation with a vastly smaller, lower-$\dot{M}$ Shakura-Sunyaev-like disk.
\end{abstract}


\maketitle

\section{Introduction}
\label{sec:intro}

Accretion disks are important in a wide variety of astrophysical contexts, ranging from supermassive black hole (BH) growth and evolution to star and planet and satellite formation to X-ray binaries and neutron-star mergers. Around supermassive BHs in particular, these disks are believed to be the engine that powers quasars, the most luminous sources in the Universe \citep{schmidt:1963.qso.redshift,salpeter64}, as well as less-luminous active galactic nuclei (AGN). As such, they funnel mass at enormous rates even exceeding $\gtrsim 10\,{\rm M_{\odot}\,yr^{-1}}$ to the BH, and ultimately provide most of the mass in SMBHs today \citep{soltan82}. The radiation, outflows, and jets launched from the inner regions of such disks \citep{laor:warm.absorber,crenshaw:nlr,dunn:agn.fb.from.strong.outflows,sturm:2011.ulirg.herschel.outflows,fgq2012,fgqm2012,Zakamska2016,Williams2017} -- collectively ``AGN feedback'' -- are also widely believed to explain \citep{silkrees:msigma,king:msigma.superfb.1,dimatteo:msigma,murray:momentum.winds,hopkins:lifetimes.methods,hopkins:lifetimes.obscuration,torrey:2020.agn.wind.bal.gal.fx.fire} the observed remarkable correlations between BH and host galaxy properties \citep{magorrian,FM00,Gebhardt00,hopkins:bhfp.obs,aller:mbh.esph,kormendy:2011.bh.nodisk.corr} and to dramatically influence galaxy formation and evolution \citep{croton:sam,hopkins:qso.all,hopkins:groups.ell,wellons:2022.smbh.growth,mercedes.feliz:2023.agn.feedback.positive.negative,cochrane:2023.agn.winds.galaxy.size.effects}. Understanding the nature, origins, and dynamics of quasar accretion disks, therefore, remains a crucial challenge in theoretical astrophysics.

Since the seminal work by \citet[][SS73]{shakurasunyaev73} (and others like \citealt{novikov.thorne:1973.astro.of.bhs}), much of the work on quasar accretion disks has assumed as a starting point some variation of the ``SS73 $\alpha$-disk'' model: this takes   disks to be geometrically thin (height $H \ll R$), optically thick (black-body like), sub-sonically turbulent (sonic Mach number $\mathcal{M}_{s} < 1$), slowly cooling ($t_{\rm cool} \gg t_{\rm dyn} = 1/\Omega$), thermal-pressure-dominated (plasma $\beta \equiv c_{s}^{2}/v_{A}^{2} > 1$), radiatively efficient, well-ionized, and  parameterized by an effectively constant-$\alpha$ viscosity, where $\alpha \sim (\delta v)^{2}/c_{s}^{2} < 1$ represents some Maxwell or Reynolds stresses and the kinematic viscosity scales as $\nu \equiv \alpha c_{s} ^{2}/\Omega$. Numerous variations have been introduced, including e.g.\ radiatively inefficient and/or advection-dominated, optically thin disks believed to be relevant for very low accretion rates \citep[e.g.][]{narayan.yi.95:adaf.lowmass.bhs}; radiatively inefficient but still optically-thick ``slim'' disks at super-Eddington accretion rates \citep{paczynsky.wiita:1980.slim.disk}; magnetically ``elevated'' disks with upper atmospheres (at multiple scale-heights above the midplane) or coronae with $\beta_{\rm |z|\gg H} \lesssim 1$ \citep{miller.stone:2000.magnetically.elevated.disk}; magnetically ``arrested'' disks where magnetic pressure halts accretion near the innermost stable circular orbit (ISCO) at low accretion rates \citep{tchekhovskoy:2011.mad.disk.jets}; gravitoturbulent disks relevant at low values of Toomre $Q$ \citep{gammie:2001.cooling.in.keplerian.disks}; and many others \citep[for reviews, see e.g.][]{pringle:accretion.review,frank:2002.accretion.book,abramowicz:accretion.theory.review,jafari:2019.mhd.ppd.review}. Yet for typical quasars accreting around the Eddington limit, some form of SS73-like $\alpha$-model (whether ``thin'' or ``slim'' disk in flavor) is still overwhelmingly the ``default'' model of reference. Likewise, it is usually assumed (for typical quasars) that the effective viscosity in the disk is dominated by a combination of Maxwell and Reynolds stresses produced by the weak-field ($\beta\gg1$) magneto-rotational instability (MRI; \citealt{balbus.hawley.review.1998}).

This leaves some crucial questions unresolved, however. For one, it has been known for decades that if one simply extrapolates an SS73 disk to large radii with quasar-level luminosities, then outside a few hundred gravitational radii ($\sim 10^{-4}-10^{-3}$\,pc, much smaller than typical ISM or even obscuring ``torus'' scales), it would naively become gravitationally unstable and should rapidly fragment rather than fueling the BH \citep{shlosman:inefficient.viscosities,shlosman:midscale.accretion,goodman:qso.disk.selfgrav}. Moreover, the properties of the disks in the models above (including both analytic models and traditional ``accretion disk'' simulations which can only extend to some modest number of gravitational radii from the BH), and even ``which type of disk'' one actually has, depend fundamentally on the ``outer boundary conditions'' set by larger-scale inflows into the accretion disk region. Most notably, the accretion rate $\dot{M}$ itself is simply taken as some constant input, and this has a major effect on the qualitative properties of the disks in the models above. But even for a fixed $\dot{M}$, one can imagine different distributions of angular momentum of inflowing material, which can produce qualitatively distinct phenomena (including e.g.\ warps, precessions, flips, or dynamical instabilities, if not highly coherent and close-to-circular; see \citealt{scheuer:1996.bh.acc.disk.alignment,nayakshin:2005.warped.disk.obsc,hobbs:2009.mw.nucdisk.sim}). And both the magnetic flux and geometry of the magnetic fields (e.g.\ primarily toroidal or poloidal, tangled or coherent) generally must  be assumed. Likewise, many other possible boundary condition effects are often ignored -- for example, the effects of global gravitational modes (e.g.\ coherent eccentric/lopsided disk modes) sourced by external perturbations or collective effects of stars at larger radii outside the disk \citep{hopkins:m31.disk,hopkins:zoom.sims,daa:20.hyperrefinement.bh.growth}. As a result, historical simulations of ``strongly magnetized disks'' \citep[for example][]{gaburov:2012.public.moving.mesh.code,forgan:2017.mhd.gravitoturb.sims,ju:2017.mri.acc.disks.with.spiral.modes,white:2019.mad.disk.sims,mishra:2020.elevated.disks.sims.strongly.sensitive.initial.beta,kudoh:2020.strong.b.field.agn.acc.disk.sims.compare}, 
while crucial for understanding the internal evolution, structure, dynamics, and variability of such disks, must adopt critical parameters like the magnetic flux ad-hoc in their initial conditions and so cannot answer the question of whether or not such disks {\em should} or even {\em could} arise in real quasars and AGN. Doing so would require  a self-consistent predictive model that follows the gas flows and magnetic field dynamics, star formation and feedback on much larger (galactic and intergalactic) scales, all the way down to the BH accretion disk scales.

\begin{table*}
\begin{center}
\begin{footnotesize}
\caption{Some common variables used throughout this manuscript (others are defined throughout where relevant).\label{tbl:variables}}
\begin{tabular}{cl}
\hline\hline
$R$, $\phi$, $z$ & Cylindrical radial, azimuthal, and vertical coordinates (centered on the SMBH, aligned with the inner disk) \\
$r$ & Spherical (3D) distance $r \equiv |{\bf r}|$ from the SMBH \\
${\bf B}$, $B_{i}$ & Magnetic field ${\bf B}$ and components $B_{i}$ (e.g.\ radial, toroidal, poloidal components $B_{R}$, $B_{\phi}$, $B_{z}$) \\
${\bf v}$, $v_{i}$ & Gas velocity ${\bf v}$ and components $v_{i}$ (e.g.\ radial, azimuthal, vertical components $v_{R}$, $v_{\phi}$, $v_{z}$) \\
\hline
$\delta {\bf B}$, $\delta {\bf v}$ & Fluctuating part of ${\bf B}$ or ${\bf v}$ (value relative to mean in some annulus/region, e.g.\ $\delta {\bf B} \equiv {\bf B} - \langle {\bf B} \rangle$) \\
$\rho$, $n_{\rm gas}$, $\Sigma_{\rm gas}$ & Gas 3D density $\rho$ or number density $n_{\rm gas}$ (in ${\rm particles\,cm^{-3}}$), and projected surface density $\Sigma_{\rm gas}$ \\
$T$, $c_{s}$ & Gas temperature $T$ ($T_{\rm rad}$, $T_{\rm dust}$ denote radiation and dust temperatures) and thermal sound speed $c_{s} \equiv \sqrt{k_{B}\,T/\mu\,m_{p}}$ \\
$v_{A}$, $v_{A,\,i}$ & \Alf\ speed $v_{A} \equiv |{\bf B}|/\sqrt{4\pi\rho}$, and component-wise $v_{A,\,i} \equiv B_{i}/\sqrt{4\pi\rho}$ \\
\hline
$\beta$, $\mathcal{M}_{s}$, $\mathcal{M}_{A}$ & Plasma $\beta \equiv c_{s}^{2}/v_{A}^{2}$ parameter, sonic $\mathcal{M}_{s} \equiv |\delta {\bf v}|/c_{s}$ and \Alf\ $\mathcal{M}_{A} \equiv |\delta {\bf v}|/v_{A}$ Mach numbers \\
$H$ & Gas disk vertical scale-height $H$ (defined within a given annulus $R$) \\
${\bf j}$, $\boldsymbol{\tau}$ & Specific angular momentum vector ${\bf j} \equiv {\bf r} \times {\bf v}$ and specific torque vector $\boldsymbol{\tau} \equiv {\bf r} \times {\bf a}$ \\
$V_{c}$, $v_{\rm K}$ & Total circular velocity $V_{c}^{2} \equiv G\,M_{\rm enclosed}(<r)/r$ (including all mass), Keplerian speed $v_{\rm K} \equiv G\,M_{\rm BH}/r$ (so $V_{c} \rightarrow v_{\rm K}$ as $r\rightarrow 0$) \\
\hline
$M_{\rm BH}$, $\dot{M}_{i}$ & SMBH mass $M_{\rm BH}$ and inflow/outflow/SF rates $\dot{M}_{i} = $ $\dot{M}_{\rm in}$, $\dot{M}_{\rm out}$, $\dot{M}_{\ast}$, respectively \\
$\boldsymbol{\Pi}_{ij}^{a}$ & Stress tensor $\boldsymbol{\Pi}$: tensor component $ij$, $a$ denotes different physical contributions (e.g.\ magnetic, kinetic, thermal $\boldsymbol{\Pi}^{\rm mag}$, $\boldsymbol{\Pi}^{\rm kin}$, $\boldsymbol{\Pi}^{\rm therm}$) \\
$Q_{i}$ & Effective Toomre $Q$ parameter $Q_{i} \equiv \sigma_{i}\,\kappa_{\rm ep}/\pi\,G\,\Sigma_{\rm gas}$ where $\sigma_{i} = c_{s},\,v_{A},\,|\delta v_{R}|,\,(c_{s}^{2} + v_{A}^{2} + \delta v_{R}^{2})^{1/2}$ appear in \\
\, & \ \ \ \ \ \ \ \ \ \ the thermal $Q_{\rm therm}$, magnetic $Q_{\rm mag}$, turbulent $Q_{\rm turb}$, and ``total'' $Q_{\rm eff}$ parameters, respectively (and $\kappa_{\rm ep}^{2} \equiv (2\Omega/R)\,\partial_{R}(R^{2}\Omega)$) \\
\hline
$\Omega$, $t_{\rm dyn}$, $t_{\rm cool}$ & Orbital frequency $\Omega \equiv V_{c}/r$, dynamical time $t_{\rm dyn} \equiv \Omega^{-1}$, and gas cooling time $t_{\rm cool}$ \\
$|a_{m}|$ & Mode amplitude for non-axisymmetric modes with azimuthal integer wavenumber $m$ (e.g.\ eccentricity $|a_{1}|$) \\
$R_{g}$, $R_{\rm schw}$, $R_{\rm BHROI}$ & BH gravitational radius $R_{g} \equiv G\,M_{\rm BH}/c^{2}$, Schwarzschild radius $R_{\rm schw}\equiv 2\,R_{g}$, and BH radius of influence \\
\, & \ \ \ \ \ \ \ \ \ \ $R_{\rm BHROI} \equiv G\,M_{\rm BH}/\sigma_{\rm gal}^{2}$ (where $\sigma_{\rm gal}$ represents the ``parent'' galactic velocity dispersion) \\
\hline\hline
\end{tabular}
\end{footnotesize}
\end{center}
\end{table*}

Motivated by this, in \citet{hopkins:superzoom.overview} (henceforth \paperone) we presented the first simulations to follow all of these physical and dynamical effects in a single simulation around a SMBH, from cosmological simulations (using a super-Lagrangian hyper-refinement technique) down to scales of $\sim 80\,$au (less than $\sim 300\,R_{\rm schw}$). We observed the formation of a true ($Q\gg 1$) ``accretion disk.'' In \paperone, we focused on the hierarchy of processes driving angular momentum loss and gas inflow from scales as large as $\gtrsim$\,Mpc onto the galaxy, through the galactic nucleus, the BH radius of influence (BHROI), torus-like regions, all the way down to the accretion-disk scales. We also studied how turbulence, magnetic fields, and radiation-hydrodynamics produced star formation on large scales and lead to the suppression of star formation on small scales ($\lesssim 0.1-1\,$pc) around the SMBH, allowing for the conditions to transition naturally from ``galactic-type'' or ``ISM-like'' conditions at $\gtrsim$\,pc scales to ``accretion-disk-like'' at $\ll $\,pc. 

Here, in \papertwo\ of the series, we focus on the emergent properties of the accretion disks in these simulations, and the physics that give rise to their key behaviors. Specifically, we show that the simulations  naturally produce disks that are {\em strongly} magnetically dominated ($\beta \ll 1$, with values much smaller in the midplane than usually assumed in historical models), specifically dominated by a toroidal magnetic field (but with substantial ``turbulent,'' radial, and poloidal field components), with vigorous trans-\Alf{ic}, highly super-sonic turbulence, large coherent eccentricities and coherent global modes, as well as gravito-turbulence and spiral arm-like structures. All together these produce rapid radiatively efficient and potentially super-Eddington accretion. We show that the fields are amplified by simple flux-freezing -- or similarly, that the toroidal field dynamo is ``closed'' by rapid advection of new magnetic flux -- with completely ``normal'' ISM magnetic field strengths (themselves built up from extremely small trace cosmological fields). We therefore refer to them as ``flux-frozen disks'' for simplicity. In a companion paper (\citealt{hopkins:superzoom.analytic}; hereafter \paperthree), we further demonstrate that these ideas are supported by a simple analytic accretion disk model. As such, while this is just one simulation, we might expect these behaviors to be quite common. Moreover, we show that the local turbulence may not, in fact, be dominated by the traditional weak-field MRI, but perhaps by distinct instabilities or variants of the MRI that arise when the magnetic fields are extremely strong. 

In \S~\ref{sec:methods} we summarize the numerical methods, and Table~\ref{tbl:variables} defines some useful variables we will refer to throughout. In \S~\ref{sec:basic} we summarize the basic conditions and properties of the ISM predicted on sub-pc scales to set the context here, including the connection to large radii in \S~\ref{sec:basic:morph}, basic gas properties in \S~\ref{sec:basic:profile}, and (lack of) star formation in \S~\ref{sec:sf}. In \S~\ref{sec:b} we examine the magnetic structure of the disks in more detail, discussing both the strength and detailed structure (\S~\ref{sec:b:ov}) and physical origins (\S~\ref{sec:b:origin}) of the strong magnetic fields. In \S~\ref{sec:v} we consider the same for the velocity field structure in the midplane (\S~\ref{sec:v:ov}) and out-of-plane (\S~\ref{sec:v:vertical}), its relation to global coherent eccentric/lopsided disk modes (\S~\ref{sec:v:ecc}) and the details of the turbulent structure (\S~\ref{sec:v:turb}), its physical origins/driving (\S~\ref{sec:v:drive}), and the (relatively weak) role of turbulent resistivity (\S~\ref{sec:why.flux.freezing}). In \S~\ref{sec:vertical} we discuss the vertical structure and profiles of various thermo-chemical and magnetic disk properties (\S~\ref{sec:vertical:profiles}) and their (weak) stratification (\S~\ref{sec:vertical:nostrat}) as well as the physics behind this (\S~\ref{sec:vertical:physics}). In \S~\ref{sec:torque}, we explore the physical torques and angular momentum exchange processes in the disk (\S~\ref{sec:torque:torque}) and their relation to different stresses including traditional Maxwell and Reynolds stresses (\S~\ref{sec:torque:stress}). With this in mind, \S~\ref{sec:lit.mag.disks} compares to a variety of previous literature models of magnetized accretion disks, including magnetically arrested (\S~\ref{sec:mad}), magnetically elevated (\S~\ref{sec:elevated}), galactic/star-forming (\S~\ref{sec:galactic.vertical.mag.disks}), toroidal-field dominated (\S~\ref{sec:other.models}), and decaying (\S~\ref{sec:no.decay}) disks. We briefly describe how the disks are likely to be mis-aligned with the pre-existing BH spin in \S~\ref{sec:angmom.alignment}. In \S~\ref{sec:no.mhd}, we contrast a simulation without magnetic fields and describe how this leads to runaway nuclear star formation (\S~\ref{sec:no.mhd:sf}), orders-of-magnitude lower accretion rates (\S~\ref{sec:no.mhd:mdot}), and a razor-thin, much smaller and lower-mass gravitoturbulent $\alpha$ disk (\S~\ref{sec:no.mhd:disk}). We summarize and conclude in \S~\ref{sec:conclusions}.

\begin{figure*}
	\centering\includegraphics[width=0.9\textwidth]{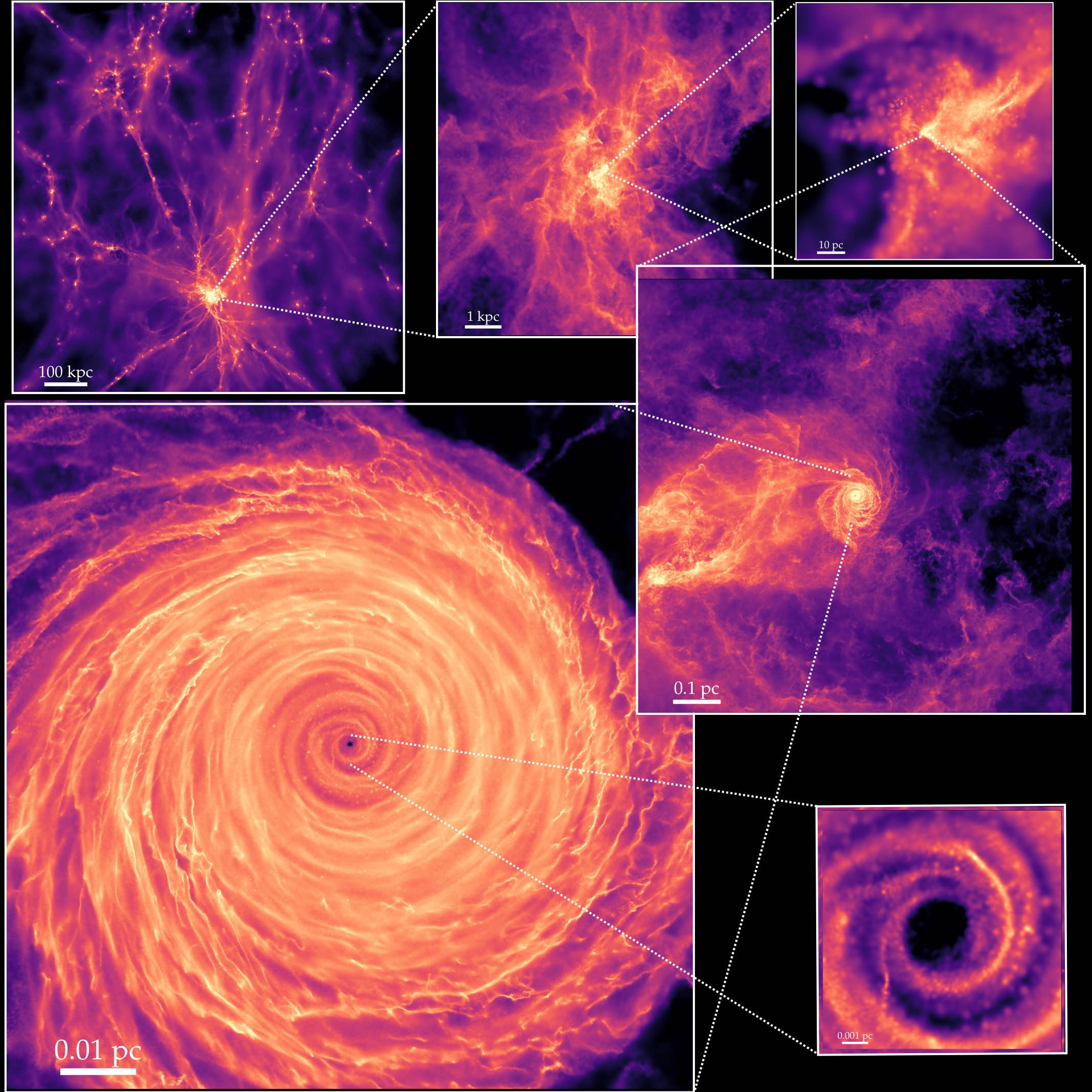}
	\caption{Image of the gas surface density in our fiducial cosmological simulation. We show projected gas density on a logarithmic scale (increasing dark-to-bright, dynamic range rescaled in each panel from a median $N_{H} \sim 10^{19}\,{\rm cm^{-2}}$ at the largest scale to $\sim 10^{8}$ times larger at the smallest scale). Multiple scales are shown to illustrate the dynamic range of the simulation, which zooms in down to $\approx 80\,$au scales around a $\sim 10^{7}\,M_{\odot}$ SMBH in the center of a massive galaxy at redshift $z\approx 4.4$ in a $\sim (100\,{\rm Mpc})^{3}$ cosmological box. The simulations include explicit multi-band radiation-magnetohydrodynamics, detailed thermochemistry/cooling, self-gravity with resolved individual (proto)star formation, accretion, evolution, and feedback, and many other physical processes (\S~\ref{sec:methods}). Tidally captured gas streams from an encounter with a massive star-forming cloud complex triggered via gravitational torques in a galaxy-scale merger fall into the BH radius of influence (BHROI) at a few $\sim$\,pc and circularize at $\sim 0.1-1\,$pc to form an accretion disk which we follow down to $\sim 300$ BH Schwarzschild radii. 
	\label{fig:image.zoom}}
\end{figure*}
\begin{figure*}
	\centering\includegraphics[width=0.9\textwidth]{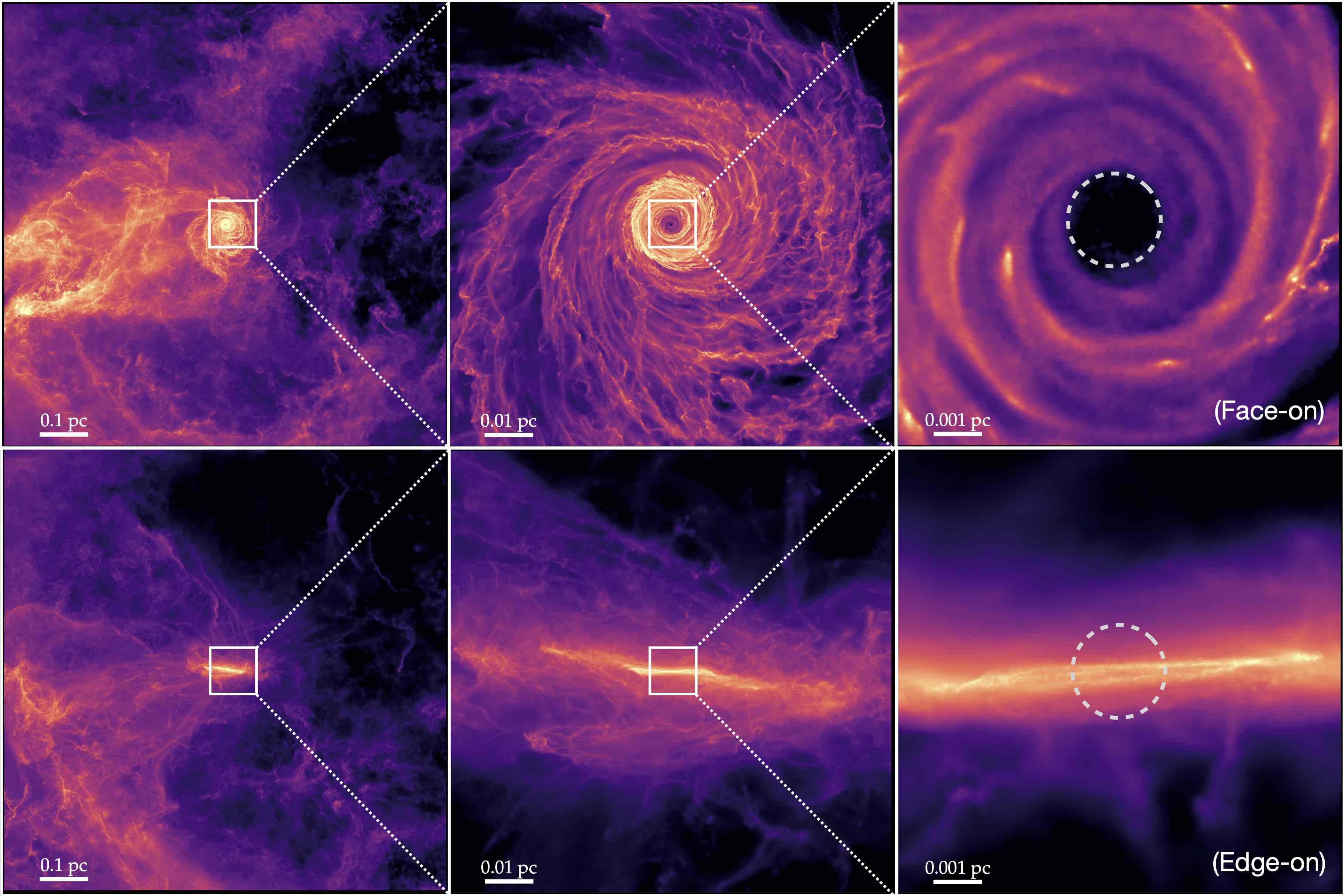}
	\caption{As Fig.~\ref{fig:image.zoom}, showing face-on ({\em top}) and edge-on ({\em bottom}) projections (relative to the nuclear disk) with order-of-magnitude different spatial scales (dynamic range a factor of $\sim 100$ in each panel, rescaled as Fig.~\ref{fig:image.zoom}). We focus on sub-pc scales; for discussion of the dynamics on larger scales driving these flows see \paperone. The dashed circle at {\em right} denotes the inner accretion boundary at $r < 80\,$au.
	We see the disk circularization radius from the inflowing filament. The disk is thin but has complex structure with spiral arms and multiple warps and some large-scale arms at different angles tracing new inflow with slightly different impact parameter. 
	\label{fig:image.faceonedgeon.inner}}
\end{figure*}

\section{Methods}
\label{sec:methods}

\subsection{Physics \&\ Resolution}

The simulations studied here are presented and extensively described in \paperone. Briefly, we begin from a $\sim (100\,{\rm cMpc})^{3}$ cosmological periodic box at redshift $z\sim 100$ with a primordial trace magnetic field, and follow it as a cosmological galaxy formation simulation following the full combined Feedback In Realistic Environments \citep[FIRE;][]{hopkins:fire2.methods,hopkins:fire3.methods} and STARFORGE \citep{grudic:starforge.methods,guszejnov:2020.starforge.jets} physics in the code {\small GIZMO}\footnote{A public version of {\small GIZMO} is available at \gizmourl} \citep{hopkins:gizmo}. We evolve the simulation with a modest refinement (target mass resolution $\sim 4000\,{\rm M_{\odot}}$ in the galaxy, or spatial resolution $\sim 10\,$pc in the galaxy nucleus) until a redshift $z\sim 4.4$ when a period of violent merging and starburst activity induces large inflows into the central $\sim\,$kpc of the galaxy. We then initiate an additional hyper-refinement layer (as in e.g.\ \citet{daa:20.hyperrefinement.bh.growth}) to go to higher and higher resolution following the gas inflows, to reach sufficient resolution to resolve {\em individual} (proto)star formation, accretion and evolution and protostellar disk structure in the central $\lesssim 10-100\,$pc of the galaxy. We continue to refine to a target resolution of $\Delta m <0.01\,{\rm M_{\odot}}$ in the central $\lesssim 10\,$pc, to follow gas inflows and disk formation down to $<300\,$ Schwarzschild radii around the super-massive black hole of mass $\sim 1.3\times10^{7}\,{\rm M_{\odot}}$. Figs.~\ref{fig:image.zoom}-\ref{fig:image.faceonedgeon.inner} show some illustrative images of the circum-BH disk which forms in the fiducial simulation.

\begin{figure}
	\centering\includegraphics[width=0.99\columnwidth]{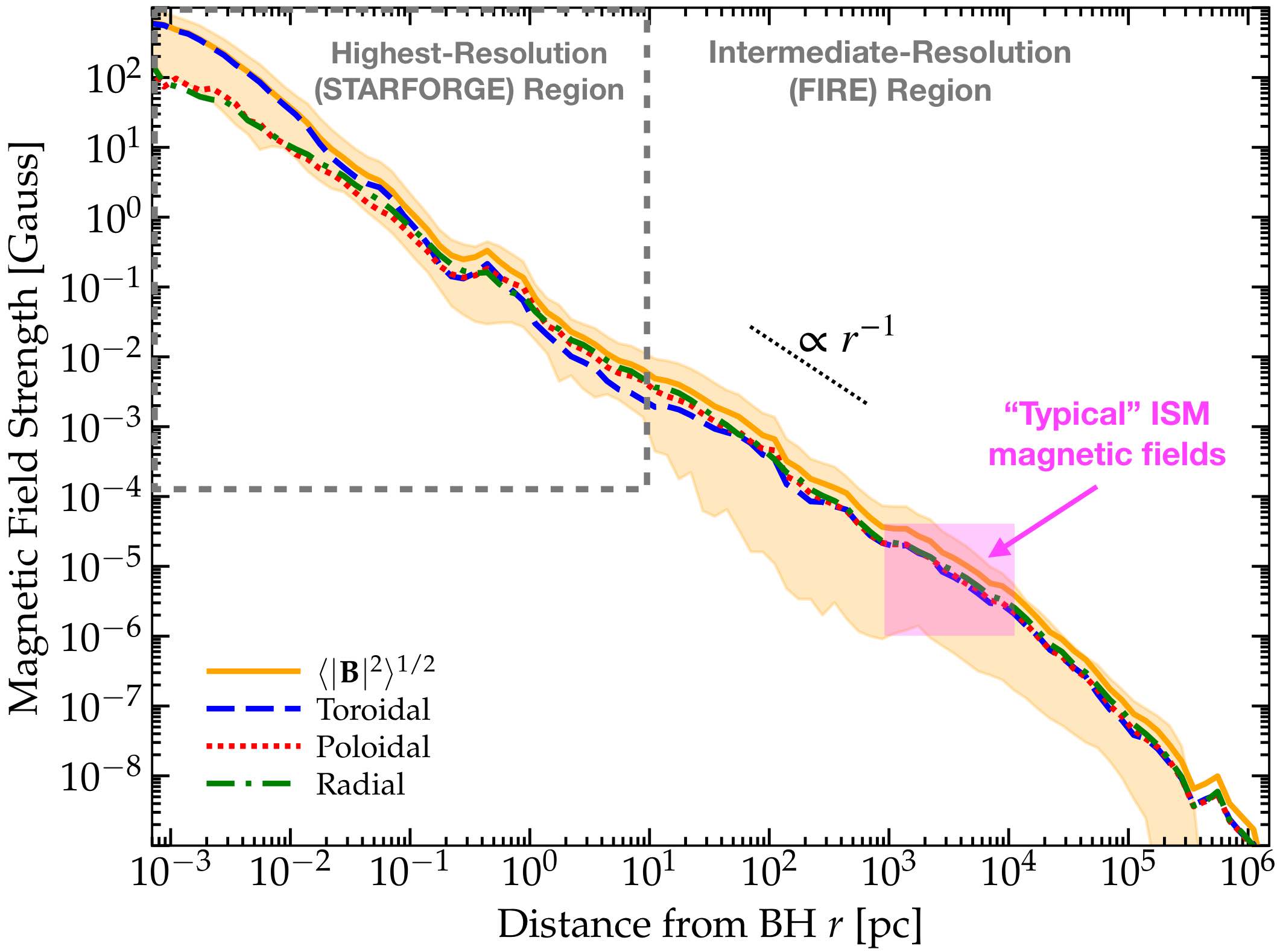}
	\caption{Magnetic field strengths as a function of BH-centric radius $r$. We measure $\langle |{\bf B}|^{2} \rangle^{1/2}$ in spherical annuli from our inner boundary at $\sim 80\,$au to $>$\,Mpc scales, and show the corresponding $\langle B_{i}^{2} \rangle^{1/2}$ for the radial, toroidal, and poloidal components of the field (with directions defined relative to the angular momentum vector of the BH accretion disk at $r<0.1\,$pc). Inside $\ll 10\,$pc, the resolution is uniformly $<0.01\,M_{\odot}$ and star formation follows the {\small STARFORGE} individual-resolved-stars physics. Outside of this radius the resolution is lower and the {\small FIRE} prescriptions form stellar population particles representing multiple stars sampling an assumed IMF (see \S~\ref{sec:methods}). Intergalactic sub-nG magnetic fields are amplified by the turbulent and galactic dynamo to ``typical'' ISM magnetic field strengths of $\sim 1-15\,{\rm \mu G}$ at $\sim 1-10\,$kpc (labeled), but the fields are amplified and approach kG at $r \sim 100\,$au. At most radii the fields are roughly isotropic (representing the dis-ordered ISM and CGM without a preferred direction), with a mild radial bias in the infall region, before becoming predominantly toroidal where the clear rotating thin BH accretion disk forms (see Fig.~\ref{fig:image.zoom}).
	\label{fig:bfield.demo}}
\end{figure}

The simulations include a wide range of physics including magnetic fields (see Fig.~\ref{fig:bfield.demo}), using the high-order constrained-gradient method from \citealt{hopkins:mhd.gizmo,hopkins:cg.mhd.gizmo}, with kinetic (anisotropic Braginskii viscosity and conduction) and non-ideal (Ohmic, ambipolar, Hall) effects \citep{su:2016.weak.mhd.cond.visc.turbdiff.fx,hopkins:gizmo.diffusion}; cosmic ray transport and coupling to gas dynamics \citep{hopkins:m1.cr.closure,hopkins:cr.spectra.accurate.integration,hopkins:cr.multibin.mw.comparison,hopkins:2021.sc.et.models.incompatible.obs}; self-gravity with adaptive softenings scaling with the resolution and high-order Hermite integrators capable of accurately integrating $\gtrsim 10^{5}$ orbits in hard binaries \citep{grudic:2020.tidal.timestep.criterion,grudic:starforge.methods,grudic:2021.accelerating.hydro.with.adaptive.force.updates,hopkins:tidal.softening}; metal enrichment and dust destruction/sublimation \citep{ma:2016.disk.structure,gandhi:2022.sne.1a.comparisons,choban:2022.fire.dust.growth.destruction.chemistry}; super-massive black hole seed formation and growth via gravitational capture of gas \citep{hopkins:qso.stellar.fb.together,shi:2022.hyper.eddington.no.bhfb,wellons:2022.smbh.growth}; (proto)star formation and accretion and explicit feedback from stars in the form of protostellar jets, main-sequence stellar mass-loss, radiation, and supernovae \citep{grudic:2022.sf.fullstarforge.imf,guszejnov:2022.starforge.cluster.assembly,guszejnov:environment.feedback.starforge.imf,guszejnov:starforge.environment.multiplicity}. They include explicit multi-band M1 radiation-hydrodynamics with adaptive-wavelength bands \citep{hopkins:radiation.methods,hopkins:2019.grudic.photon.momentum.rad.pressure.coupling,grudic:starforge.methods} coupled explicitly to all the thermo-chemical processes, radiative cooling and thermo-chemistry incorporating cosmic backgrounds, radiation from local stars, re-radiated cooling radiation, dust, molecular, atomic, metal-line, and ionized species opacities and processes, cosmic rays, and other processes allowing us to robustly model the thermochemistry and opacities in gas with densities from $n \sim 10^{-8} - 10^{16}\,{\rm cm^{-3}}$ and temperatures $\sim 1-10^{10}\,$K in a range of radiation and cosmic ray environments. A pure inflow/accretion boundary is enforced at $\approx 80\,$au from the central SMBH -- we do not model any flux from e.g.\ jets or radiation emerging from the inner region, as these should depend on the accretion disk properties themselves. Fig.~\ref{fig:image.faceonedgeon.inner} illustrates some of the complex phase structure that emerges even in just the nuclear regions.

We stress that the entire simulation uses the identical, full physics -- there is no discontinuous change in the equations integrated in space nor time. Instead, as described in \paperone, we evolve the full self-gravitating radiation-magneto-thermochemical-hydrodynamics for all gas cells, and simply allow the code to form two distinct types of star particles: (1) FIRE ``stellar population'' particles which form from star-forming gas in the low-resolution cells (resolution $\gg 1\,{\rm M}_{\odot}$), and therefore sample an assumed stellar initial mass function (IMF) and calculate IMF-integrated rates for stellar feedback; and (2) STARFORGE ``individual star'' particles which form in the high-resolution cells ($\ll 1\,{\rm M}_{\odot}$)  and therefore evolve along individual (proto)stellar+main sequence evolutionary tracks.

Some of these physics are not important on the scales we will study here, although they may play a crucial role in determining the boundary conditions via their role on larger scales. On all scales we study in detail in this paper, the refinement has reached the target resolution of $<0.01\,{\rm M_{\odot}}$ (we briefly re-ran with refinement a factor $\sim 8$ higher, and see no difference in our results), and in the densest regions just outside our inner boundary condition we reach local spatial resolution $\sim 10-20\,{\rm au}$ and time resolution as small as $\sim$\,days. The most relevant physics on these scales are gravity, (ideal) MHD, and explicit radiation-thermodynamics. 

\subsection{Analysis and Definitions}
\label{sec:methods:analysis}

Table~\ref{tbl:variables} defines a number of variables we use throughout. In this manuscript, we will often refer to cylindrical radial/azimuthal/vertical coordinates $R$, $\phi$, $z$, defined with respect to the angular momentum vector of the inner accretion disk (e.g.\ gas at $r < 0.01\,$pc) and centered on the SMBH, so $\hat{z}$ points along the angular momentum vector and $\hat{\phi}$ points in the rotation direction. We distinguish the spherical radius/distance $r$ (with spherical radial/polar/azimuthal angles $r$, $\theta$, $\tilde{\phi}$ defined in the same way so that $z=r\,\cos{\theta}$) from the cylindrical radius $R$. Given our focus, we will generally use the terms toroidal and azimuthal interchangeably (and likewise for poloidal and vertical). 

The instantaneous values of fluctuating quantities like $\delta {\bf B}({\bf x},\,t) \equiv {\bf B}({\bf x},\,t) - \langle {\bf B}({\bf x},\,t) \rangle$ are defined by respect to their appropriately weighted averages $\langle {\bf B} \rangle \equiv \langle {\bf B}({\bf x},\,t)\rangle_{\bf x} \equiv (\int w\,{\bf B}({\bf x},\,t)\,d^{3}{\bf x})/(\int w\,d^{3}{\bf x}) = (\sum_{i} {\bf B}_{i}\,w_{i}\,{\rm Vol}_{i}) / (\sum_{i} w_{i}\,{\rm Vol}_{i})$ at a given time (with $w$ the chosen weight, and the summation over all cells $i$). For example, unless otherwise specified we will define volume-weighted averages in radial annuli, i.e.\ $w=1$ for $r$ or $R$ (whichever is plotted) within some narrow logarithmic radial annulus of width $\sim 0.1\,$dex, and $w=0$ outside the annulus (though we have checked that the exact choice of bin widths makes no appreciable differences to any plot here). We will sometimes compare mass-weighted ($w=\rho$) or other explicitly weighted distributions, where stated. We also define the corresponding weighted $90\%$ ($5-95\%$) or $\pm1\,\sigma$ ($16-84\%$) inclusion intervals, as e.g.\ the values of ${\bf B}$ above/below which correspond to the appropriate fraction of the total weight ($\int w\,d^{3}{\bf x}$). We note below that the stress tensor $\boldsymbol{\Pi}$ can be written in terms of a mean and fluctuating component, e.g.\ $\langle \boldsymbol{\Pi}_{R\phi}^{\rm kin} \rangle = \langle \rho\,v_{R}\,v_{\phi}\rangle + \langle \rho\,\delta v_{R}\,\delta v_{\phi} \rangle$, so will specify throughout whether we refer to the ``total'' or ``fluctuating'' components. 

As described below, we consider a few different definitions of the (gas mass) scale height $H$ and show they give nearly identical results, including measuring the mass-weighted median $|z|$ in annuli, the mass-weighted rms $\langle z^{2}\rangle^{1/2}$, or fitting a vertical Gaussian or ${\rm sech}^{2}$ profile to the density. Projected quantities like $\Sigma_{\rm gas}$ are defined as the sum within cylindrical annuli. For vector/tensor quantities like ${\bf B}$, ${\bf v}$, $\boldsymbol{\Pi}$, we follow usual convention and define them in terms of cylindrical components. However, we have remade all salient plots of these quantities, instead (1) defining in terms of the spherical components (e.g.\ $\langle B_{r,\,\theta,\,\tilde{\phi}} \rangle$ instead of $B_{R,\,z,\,\phi}$); (2) plotting versus spherical radius $r$ instead of cylindrical $R$; (3) allowing the $\hat{z}$ axis to vary in annuli (defining it with respect to the angular momentum axis {\em in each annulus} instead of a global, fixed coordinate system); and (4) re-defining them in eccentric annuli or similarly subtracting the mean $m=1$ non-axisymmetric component (defined and plotted below) from the fluctuating components. Unless we explicitly state otherwise for specific quantities in the text below, these choices do not qualitatively change any of our conclusions or comparisons.

Because we are interested in various non-equilibrium properties and dynamics and their time evolution, we will focus on specific representative times chosen after the simulation reaches steady-state, to represent the range of instantaneous behaviors. However we have surveyed hundreds of snapshots of the simulations and confirm that quantities studied here such as the accretion rates and mass profiles are representative of the range over all times after the simulation reaches its highest refinement level. We explicitly show this for several properties (showing their time evolution) below. For radial profiles of positive-definite quantities like $\rho$ or $\langle |{\bf B}|^{2} \rangle$, we show in \paperone\ and \paperthree\ that their time-averaged behaviors over the entire simulation lie well within the scatter we plot at a given time and radius. For many signed quantities, like the toroidal and radial magnetic fields, time-averaging the profiles would artificially obscure the interesting behaviors.

\begin{figure*}
	\centering\includegraphics[width=0.9\textwidth]{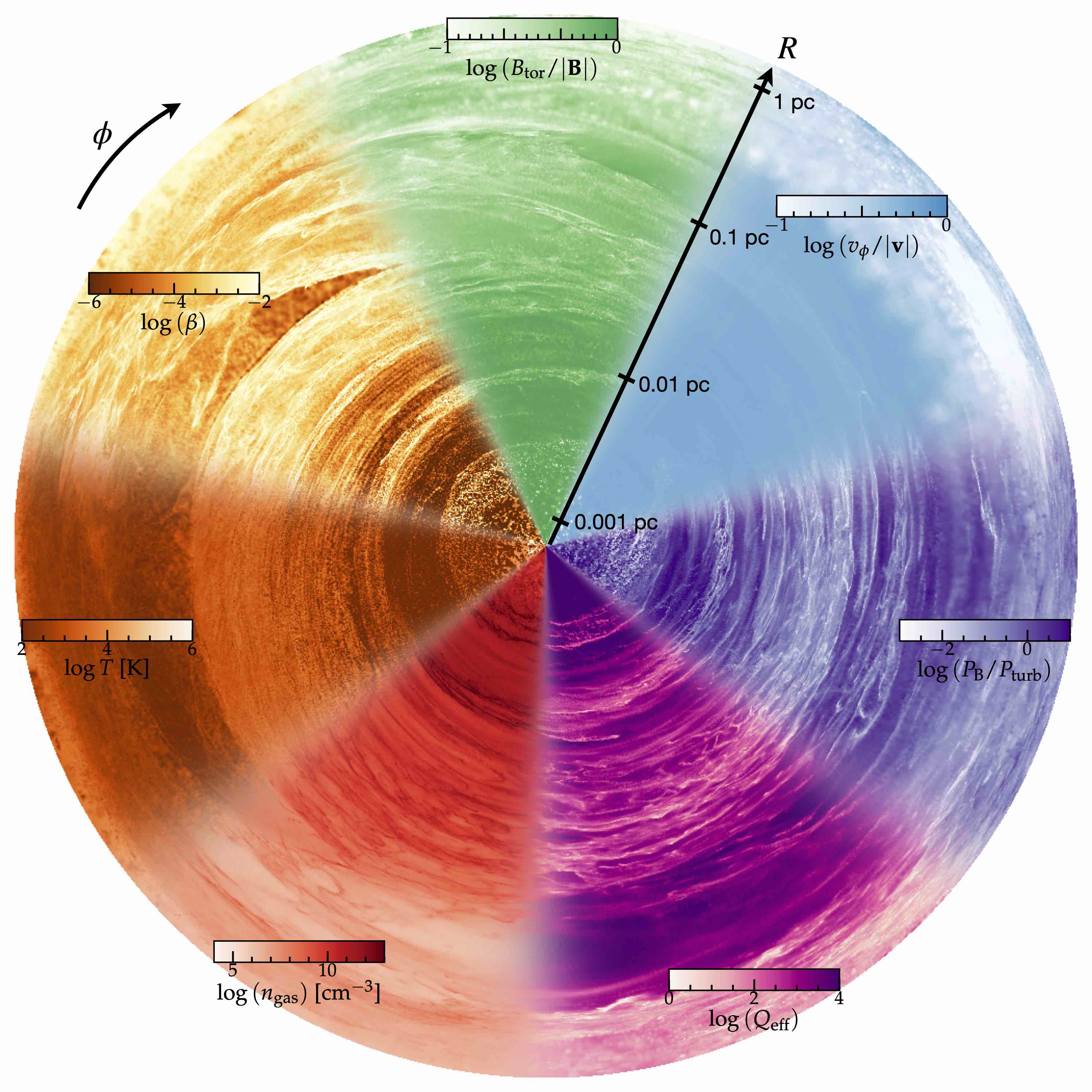}
	\caption{Wedge plot showing various disk properties (colormaps labeled) including: 3D gas density $n_{\rm gas}$, gas temperature $T$, plasma $\beta\equiv c_{s}^{2}/v_{A}^{2}$, ratio of toroidal to total field strength $|B_{\rm tor}|/|{\bf B}|$, ratio of azimuthal to total velocity $|v_{\phi}|/|{\bf v}|$, ratio of magnetic pressure $P_{\rm B}$ to total non-rotational kinetic/turbulent ram pressure $P_{\rm turb}$, and ``effective'' total (including thermal+magnetic+turbulent support) local Toomre $Q_{\rm eff}$ parameter. The plot shows cylindrical coordinates $R,\,\phi$, but with the cylindrical radius $R$ stretched on a log scale from $\sim 80\,$au to $\sim 2\,$pc (labeled), to show the behavior over a large range of scales. All quantities are mass-weighted averages within each image pixel, in a slice through the disk midplane ($|z|/R<0.1$). Densities increase inwards; temperatures vary but tend to be mostly cold-to-warm at these scales; $\beta\ll1$  and decreases in the inner disk $R\lesssim 0.01\,$pc; the toroidal field and azimuthal (rotational) velocities dominate inside $\lesssim 0.1\,$pc; the magnetic pressure is crudely of order turbulent pressure but varies locally from $\sim 0.01-10$ times its value; and the effective stability parameter $Q_{\rm eff} \gg 1$ on all these scales. 
	\label{fig:image.wedgeplot}}
\end{figure*}

\begin{figure*}
	\centering\includegraphics[width=0.95\textwidth]{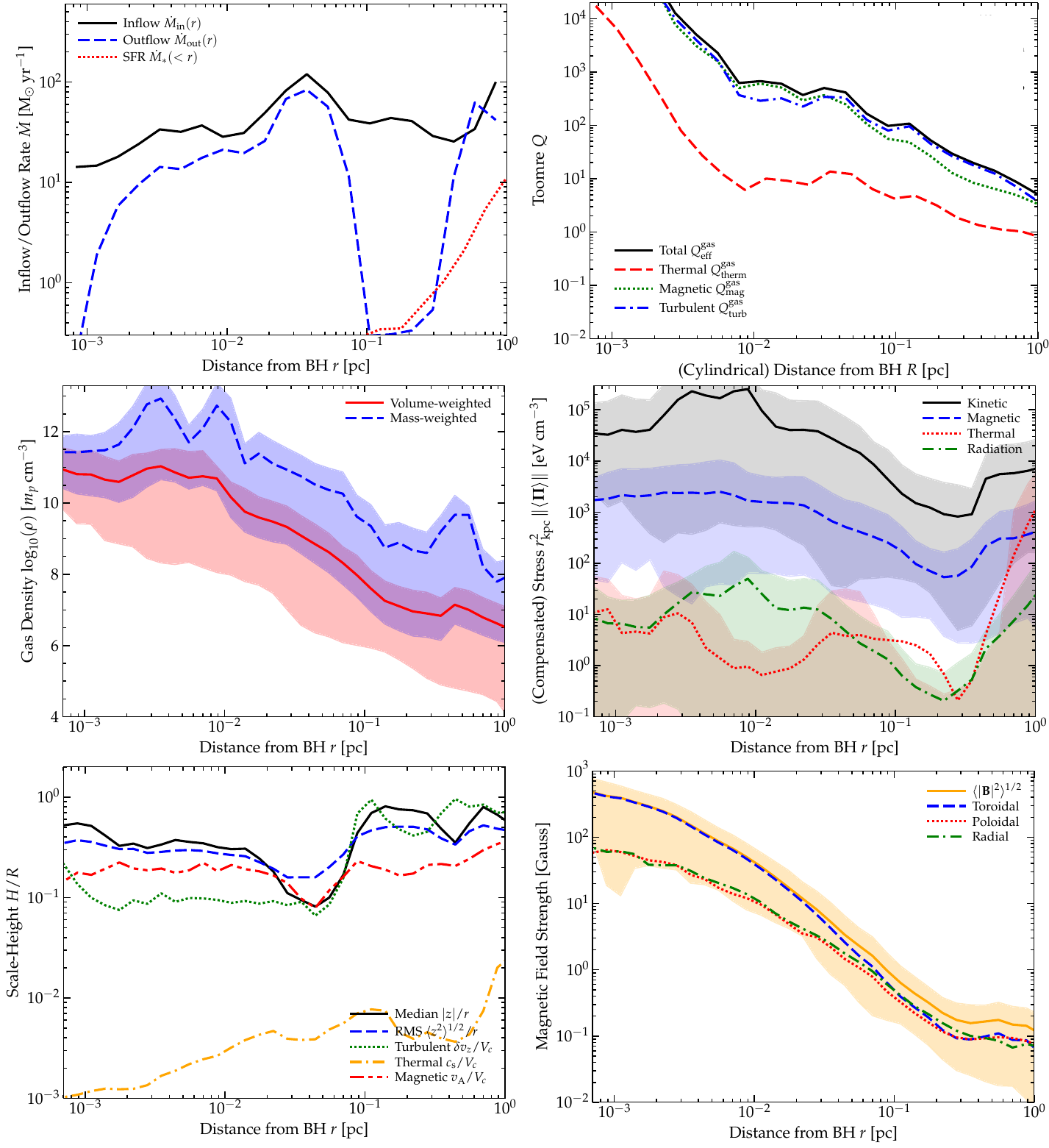}
	\caption{Radial profiles  of properties versus spherical BH-centric distance $r$, from $<10^{-3}$\,pc to $\sim 1\,$pc   (properties are mass-averaged in concentric shells unless otherwise stated; for discussion of properties on larger (extra)galactic scales, see \paperone). 
	{\em Top left:} Inflow rate (summing all gas with $v_{r} < 0$), outflow rate ($v_{r} > 0$), and star formation rate (stellar mass formed/accreted within the last dynamical time, in each shell). We see consistent inflow at up to $\sim 20\,{\rm M_{\odot}\,yr^{-1}}$. 
	{\em Top right:} Toomre $Q$ parameter for the gas alone, separating thermal support ($c_{s}$), magnetic ($v_{A}$), turbulent ($v_{\rm turb}$) and total ($\delta v_{r,\,{\rm eff}}^{2} \equiv c_{s}^{2} + v_{A}^{2} + v_{\rm turb}^{2}$). The system is stable on all these scales, but the relatively modest thermal-only $Q$ produces some of the large-scale gravitoturbulence and some (very slow/weak) star formation.
	{\em Middle left:} Gas density (lines show mean, shaded range $\sim 90\%$ inclusion interval) weighted by mass (spikes correspond to denser arms/clumps) or volume. 
	{\em Middle right:} Total contribution to the stress/pressure tensor (Frobenius norm of each tensor, lines show volume-weighted mean, shaded range $\sim 90\%$ interval), for kinetic ($\rho\,{\bf v}\,{\bf v}$), magnetic ($|{\bf B}|^{2}\,\mathbb{I}/8\pi -{\bf B}{\bf B}/4\pi$), thermal ($n\,k\,T\,\mathbb{I}$), and radiation ($\int {\rm d}\nu\,e_{\nu}\,\mathbb{D}_{\nu}$) energies. The thermal and radiation terms are in rough equilibrium as expected (the disk is optically thick) but are strongly dominated by magnetic and kinetic terms. Note the kinetic energy is primarily rotation of the disk, the turbulent kinetic energy is order-of-magnitude comparable to magnetic (turbulence is broadly trans-\Alf{ic}).
	{\em Bottom left:} Scale height $H/R$ directly measured (median or rms $z$, as labeled), and expected ($\approx \sigma/V_{c}$) from turbulent/magnetic/thermal support. Given the above, we see turbulent+magnetic support clearly dominate the disks' vertical structure.
	{\em Bottom right:} Magnetic field strength (rms volume-weighted value in solid, shaded shows $90\%$ range for total field strength), for total and by toroidal/poloidal/radial component. The fields rise to very large values at small $r$, and are toroidal-dominated in the disk but with non-negligible poloidal+radial terms. In a companion paper (\paperthree) we note that all of these scalings can be reasonably approximated with a simple analytic similarity model. 
	\label{fig:radial.profile.general}}
\end{figure*}

\section{Basic Properties \&\ Conditions on Sub-pc Scales}
\label{sec:basic}

\subsection{Morphology and Connection to Larger Scales}
\label{sec:basic:morph}

Figs.~\ref{fig:image.zoom}-\ref{fig:image.faceonedgeon.inner} illustrate the nuclear disk, which forms from cosmological initial conditions. We see the disk forms within a chaotic, clumpy, massive high-redshift ($z\approx 4.5$) starburst (galaxy-integrated SFR $> 100\,{\rm M_{\odot}\,yr^{-1}}$) galaxy, where a massive star-forming cloud complex has a close passage to the $\sim 10^{7}\,{\rm M_{\odot}}$ SMBH in the galaxy nucleus. This outer galaxy is studied in \paperone. Some of the material from the cloud is tidally stripped by the SMBH around its radius of influence (BHROI, a few pc, interior to which the BH dominates the potential) and falls in initially in a radial stream (a tidal tail) and we see it circularize at $\sim 0.1-1\,$pc. As expected given the large cloud size and inhomogeneous structure, not all the inflowing gas has an identical impact parameter, so some falls in at slightly different angles (giving rise to warps and eccentricity in the outer disk). 

Fig.~\ref{fig:bfield.demo} illustrates the scaling of the magnetic fields with BH-centric radius, showing the full dynamic range of the simulation. We will study scales within the disk below but, because we will argue below that the magnetic flux carried into the disk is important, we wish to highlight that the fields grow ultimately from sub-nG intergalactic magnetic fields, with ISM magnetic fields on scales $\sim 1-10\,$kpc, which have completely ``typical'' values of $\sim 1-15\,{\rm \mu G}$ \citep{beck:2015.b.field.review}, similar to those observed in the local ISM of the Milky Way (despite this being a massive, high-redshift galaxy). The magnetic fields (and other properties like gas densities) grow relatively smoothly down to disk scales, without a sharp discontinuity at some particular radius. We also identify the range of scales where our simulation reaches maximum target resolution ($r\lesssim 10\,$pc), to make it clear that the entire dynamic range we study here uniformly has mass resolution $<0.01\,{\rm M}_{\odot}$.

Fig.~\ref{fig:image.wedgeplot} visualizes several different properties in a projected 2D image over a dynamic range of a factor $\gtrsim 1000$ in radius, now focusing just on the innermost regions of our simulation from $10^{-3}-1$\,pc where the disk forms. These include: 3D gas density $n_{\rm gas}$, temperature $T$, plasma $\beta\equiv c_{s}^{2}/v_{A}^{2}$, ratio of toroidal to total field strength $|B_{\rm tor}|/|{\bf B}|$, ratio of azimuthal to total velocity $|v_{\phi}|/|{\bf v}|$, ratio of magnetic pressure $P_{\rm B}$ to total non-rotational kinetic/turbulent ram pressure $P_{\rm turb}$, and ``effective'' total local Toomre $Q_{\rm eff}$ parameter (including thermal+magnetic+turbulent support). Fig.~\ref{fig:radial.profile.general} complements this, showing the 1D (averaged in concentric radial shells) radial profiles of different properties out to $\sim$\,pc scales.

\subsection{Radial Trends and Basic Scalings in the Disk}
\label{sec:basic:profile}

On sub-pc scales, Figs.~\ref{fig:image.wedgeplot}-\ref{fig:radial.profile.general} show: 
\begin{enumerate}

\item Inflow is systematically larger than outflow, and star formation is largely negligible (note the three can co-exist, as the medium is clearly not spherically symmetric, nor in strict long-term equilibrium), with an order-of-magnitude constant $\dot{M}_{\rm in} \sim 20-30\,{\rm M_{\odot}\,yr^{-1}}$ sustained into the central $\sim 80$\,au around the SMBH. 

\item The gas densities increase towards the center, with a slightly more shallow profile than a singular isothermal sphere ($\rho \propto r^{-2}$), with clumpiness evident in the visual projection and in the mass-weighted spherical profile. 

\item The radiation and thermal energy densities are in rough equilibrium (with the radiation, dust, and gas kinetic temperatures coming into increasing equilibrium at $r \ll \,$pc, and the dust mostly sublimated at $\lesssim$\,pc as shown in \paperone), but with both of their energy densities well below the magnetic energy density in the disk, with $\beta$ ranging from $10^{-6}-10^{-2}$ depending on the local phase. The temperature is dominated by cold and warm atomic media (with some warm ionized gas), and rises weakly towards the center.

\item The kinetic energy of the gas is uniformly larger than magnetic, but most of that on scales $\ll\,$pc is the disk rotation, i.e. $|{\bf v}|\sim v_{\phi} \sim V_{c}(r)$, and the disk is rotation-dominated within $\lesssim 0.1\,$pc. The remaining non-rotational or ``turbulent'' kinetic energy density is more comparable to magnetic energy densities, with $P_{\rm B}/P_{\rm turb}$ ranging from $\sim 0.01-10$ depending on the local phase sub-structure and density of the gas. In other words, the {\em turbulence is broadly trans-\Alf{ic}}.

\item The magnetic fields are stronger in the center, with $|{\bf B}|$ increasing slightly steeper than $r^{-1}$, and the toroidal component of the field dominating inside the radii where the disk is ordered and rotation-dominated. 

\item A combination of turbulence and magnetic pressure support the measured disk scale height (i.e.\ the disk is in approximate vertical equilibrium) with $H/R \sim 0.1$ in the central regions (and quasi-spherical structure at $\sim$\,pc, outside the rotation-dominated region). At all radii $\ll $\,pc, the ``effective' Toomre $Q$ parameter including thermal+magnetic+turbulent support  is large. But even the pure-thermal Toomre $Q_{\rm thermal} \sim 10$ from $r\sim10^{-2}-10^{-1}$\,pc inside the disk and rises rapidly to $Q_{\rm thermal} \gg 10$ at $r< 0.01\,$pc.

\end{enumerate}

Given that the accretion rates at $\sim 80\,$au (our inner boundary) correspond to super-Eddington accretion if they remained constant down to horizon scales with a fixed radiative efficiency of $\epsilon_{r}=0.1$, predicting in detail the quasar luminosities requires radiation-GRMHD simulations which can extend the disk simulations here to those scales \citep[e.g.][]{jiang:2019.supereddington.sims.low.outflow.efficiency.low.rad.eff}. If we assume that the accretion becomes radiatively inefficient (or strong outflows suppress $\dot{M}$ on near-horizon scales) so that the luminosity remains limited to Eddington \citep{abramowicz:1988.slim.disks}, we would predict $L_{\rm bol} \sim L_{\rm Edd} \sim 10^{45}\,{\rm erg\,s^{-1}}$, if on the other hand the radiative efficiency remains high ($\epsilon_{r} \sim 0.1$) and $\dot{M}$ remains constant to the horizon (essentially an upper limit), we would predict $L_{\rm bol} \sim 10^{47}\,{\rm erg\,s^{-1}}$. At the redshift $z\approx 4.5$ here, this range brackets the ``knee'' of the observed bolometric quasar luminosity function \citep{hopkins:bol.qlf,shen:bolometric.qlf.update}, so in terms of luminosities, this should correspond to a ``typical'' quasar at the redshifts simulated.

Because we will refer to it below, we note that \paperthree\ compares these profiles to those predicted for a \citet{shakurasunyaev73} or SS73 $\alpha$-disk with the same accretion rate $\dot{M}$, to show they differ by orders of magnitude. Briefly, SS73 and other ``weakly-magnetized'' models {\em assume} $\beta \gtrsim 1$, so the vertical support in the outer disk comes only from thermal pressure ($H/R \sim c_{s}/V_{c}$), with an effective viscosity ($\nu \propto \alpha\,c_{s}\,H$) provided by some Reynolds/Maxwell stresses with $\alpha \sim \delta v_{\rm turb}^{2}/c_{s}^{2} \sim v_{A}^{2}/c_{s}^{2} \sim 1/\beta \ll 1$ leading to inflow ($\dot{M} \propto \nu\,\Sigma_{\rm gas}$). So $H/R$, $v_{A}/V_{c}$, and $\delta v_{\rm turb}/V_{c}$ are predicted by SS73 to be smaller by factors of $\sim 100-1000$ than the values here, while (for the same\footnote{We compare at fixed $\dot{M}$ because this is the key boundary condition which determines the SS73 solution properties for quantities like $\Sigma_{\rm gas}$, and is set by our cosmological/galaxy ISM-scale inflows.} $\dot{M}$) $\Sigma_{\rm gas}$ would be larger in SS73 by factors $\sim 10^{4}-10^{6}$. Correspondingly, the midplane density ($\rho \propto \Sigma_{\rm gas}/H$) is larger and effective Toomre $Q$ ($\propto \delta v_{\rm eff}\,\Omega/G\,\Sigma_{\rm gas} \propto H\,\Omega^{2}/\Sigma_{\rm gas}$) is smaller by factors $\sim 10^{6}-10^{8}$ in SS73 compared to these simulations.

\subsection{Stability and (Lack of) Star Formation}
\label{sec:sf}

As noted above, the SFR interior to $\lesssim$\,pc is much smaller than inflow rates. The physics of this suppression is discussed in \paperone, but as we will see below in our tests without MHD, magnetic fields play an important role (raising $Q$, preventing local collapse perpendicular to the mean toroidal field, and promoting faster torques and more rapid accretion). For our purposes here, we show therein and in \citet{hopkins:superzoom.imf} (where the properties of the few stars that do form in the outer accretion disk are studied) that the density or total mass of stars on these scales is very small compared to the gas mass, and that the stars contribute negligibly to the dynamics or stresses (momentum/energy flux or turbulence driving) or heating in the disk either via their gravitational influence or via their stellar ``feedback'' effects (jets, winds, radiation). Indeed, if we re-start the simulations from a snapshot and simply delete all stars at $<1$\,pc (and disable new star formation at these radii) and run for $\sim 10-20$ dynamical times, we see no appreciable difference in any of the properties we study on these scales.

So we are justified in neglecting stars and stellar feedback in our discussion of the accretion disk structure, on these scales. This is distinct from larger, more ``ISM-like'' scales, where stars dominate the mass and dynamics \citep[see e.g.][]{hopkins:inflow.analytics,daa:20.hyperrefinement.bh.growth}.

\begin{figure*}
	\centering\includegraphics[width=0.83\textwidth]{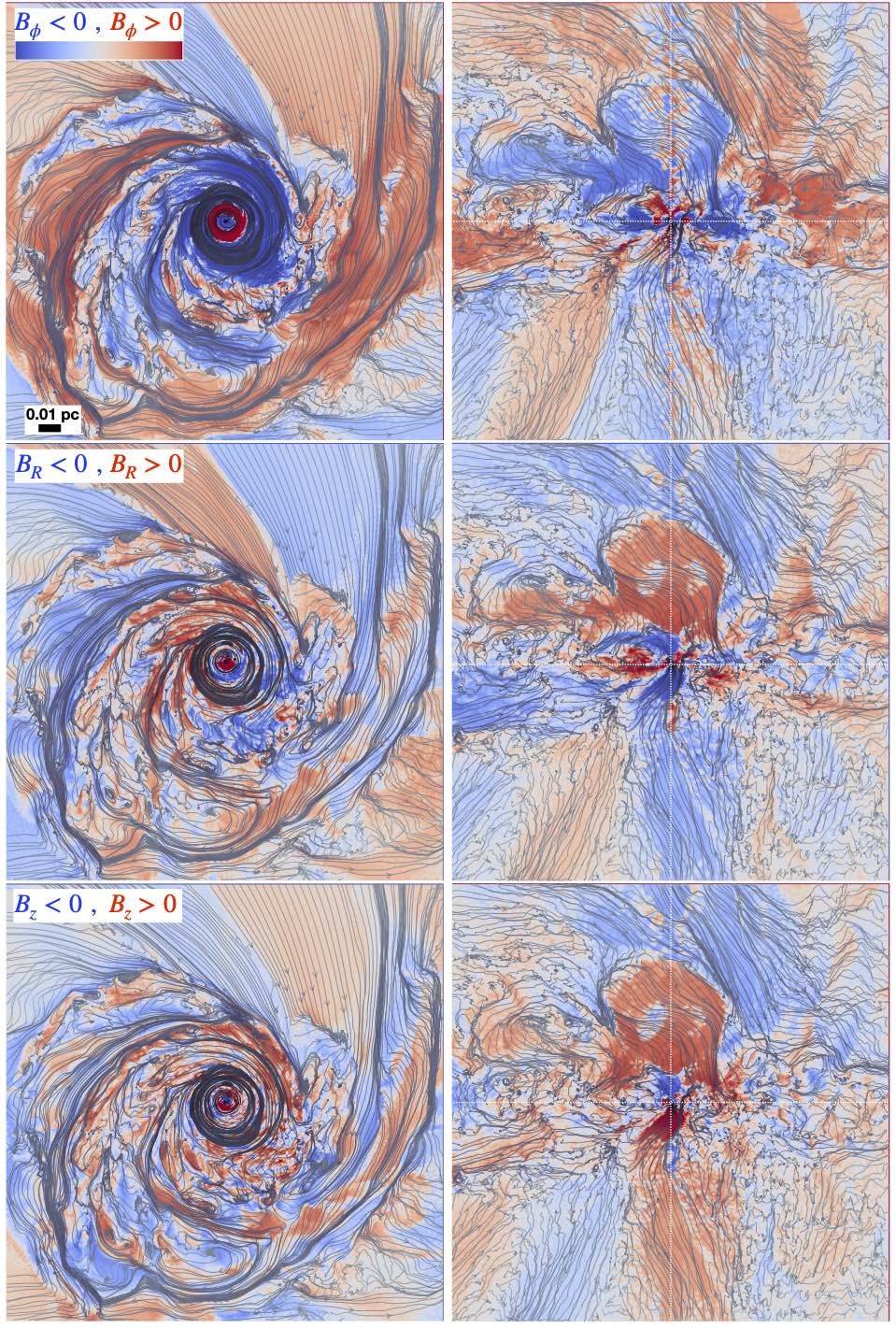}
	\caption{The nuclear disk in a $\pm 0.1\,$pc box, showing the magnetic field lines ({\em grey}) in a face-on slice through the midplane ({\em left}; lines show the in-plane $B_{R}-B_{\phi}$ field, in gas with $|z/R|<0.1$) or edge-on slice through the disk in cylindrical $R$-$z$ coordinates ({\em right}; lines show the in-plane $B_{R}-B_{z}$ field, in gas with $|\sin{\phi}|<0.1$). For each, colors denote the sign of toroidal/azimuthal ($B_{\phi}$; {\em top}), radial ($B_{R}$; {\em middle}), or poloidal/vertical ($B_{z}$; {\em bottom}) field. Our sign convention is that $\hat{z}$ and $\hat{\phi}$ point in the direction of the angular momentum vector and direction of co-rotation of the inner gas disk (so $B_{\phi}>0$ is prograde), and $\hat{R}$ points away from the BH. We see large-scale ordered structure tracing the radial inflows from scales outside the BHROI being wrapped around the center as the disk circularizes at $\sim 0.1\,$pc, with sign changes on large scales determined by the large-scale inflow structure (and reversals or occasional loops forming at the sign change boundaries). There is also an obvious anti-correlation in the sign of $B_{\phi}$ and $B_{R}$.
	\label{fig:bfields.faceon.edgeon}}
\end{figure*}

\begin{figure*}
	\centering\includegraphics[width=0.97\textwidth]{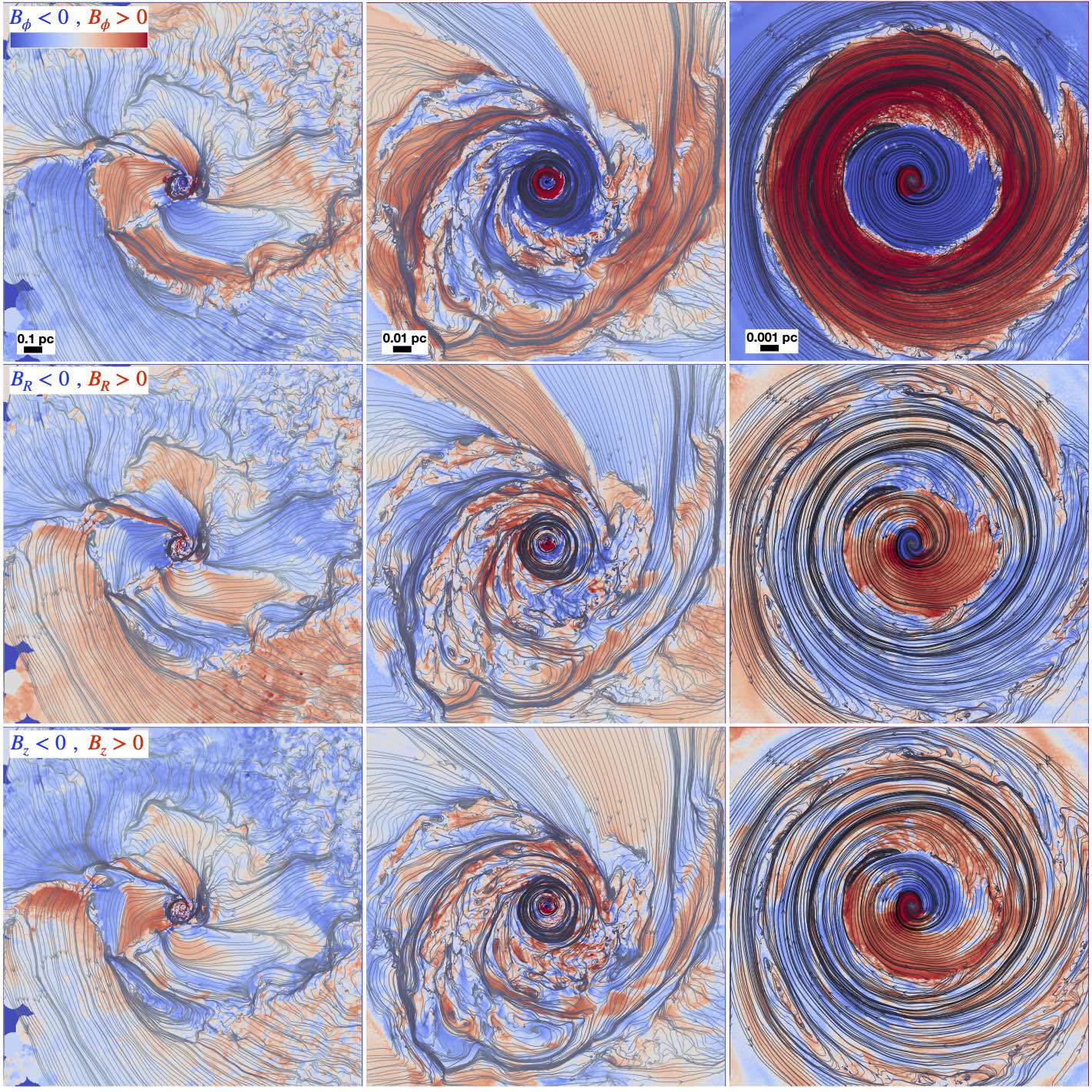}
	\caption{Field lines as Fig.~\ref{fig:bfields.faceon.edgeon}, in face-on projection of the midplane on three different spatial scales: $\pm 1\,$pc ({\em left}), $\pm 0.1\,$pc ({\em middle}), $\pm 0.01\,$pc ({\em right}). We more clearly see the field lines transition from a somewhat less coherent and radially-biased flow with the filamentary inflow on the largest $\sim 1\,$pc scales to the more ordered and more clearly azimuthal field geometry on $\ll 0.01\,$pc scales. We also see the expected $B_{\phi}-B_{R}$ anti-correlation on all scales, with a somewhat less correlated/more turbulent $B_{z}$.
	\label{fig:bfields.faceon.zoom}}
\end{figure*}

\begin{figure*}
	\centering\includegraphics[width=0.9\textwidth]{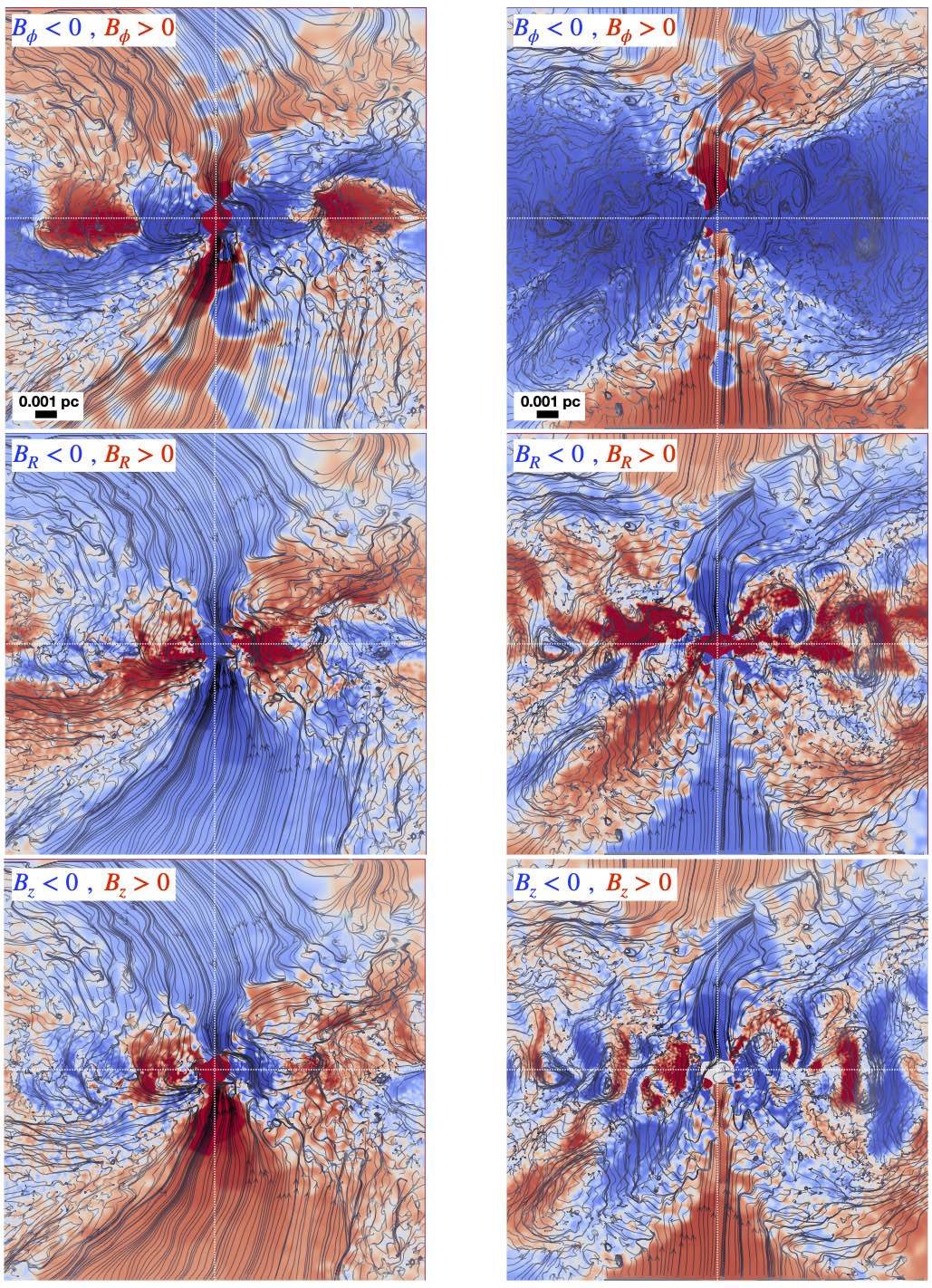}
	\caption{Field lines as Fig.~\ref{fig:bfields.faceon.edgeon}, in edge-on projection on smaller $\pm 0.01\,$pc scales, at two different times (earlier at {\em left}, later at {\em right}), separated by $\sim 2\times10^{4}$ dynamical times at the innermost boundary radius. The times are chosen to represent times before ({\em left}) and after ({\em right}) the sign-flip in $B_{\phi}$ on small scales accretes through our inner boundary at $<80\,$au, leaving a somewhat larger coherent toroidal field at the late time (but still with successive sign-flips on larger scales). In both cases we see some $B_{\phi}-B_{R}$ anti-correlation but at the later time it is less obvious  (but would become more obvious if we subtract the coherent mean-field $B_{\phi}$ and plot the fluctuations $\delta B_{\phi} \equiv B_{\phi} - \langle B_{\phi}(R) \rangle $ and $\delta B_{R}$). Coherent modes with wavenumber $k_{z} \sim k_{R} \sim 1/H$ are obvious and especially prominent in $B_{z}$ at the later time. We denote the $z=0$ and $R=0$ axes with the thin white dotted line for clarity: we see no evidence for a sign switch in $B_{\phi}$ at $z=0$ as would be predicted if the toroidal $B_{\phi}$ were ultimately sourced by a dominant mean-field $B_{z}$. 
	\label{fig:bfields.edgeon.smallscale}}
\end{figure*}

\begin{figure*}
	\centering\includegraphics[width=0.97\textwidth]{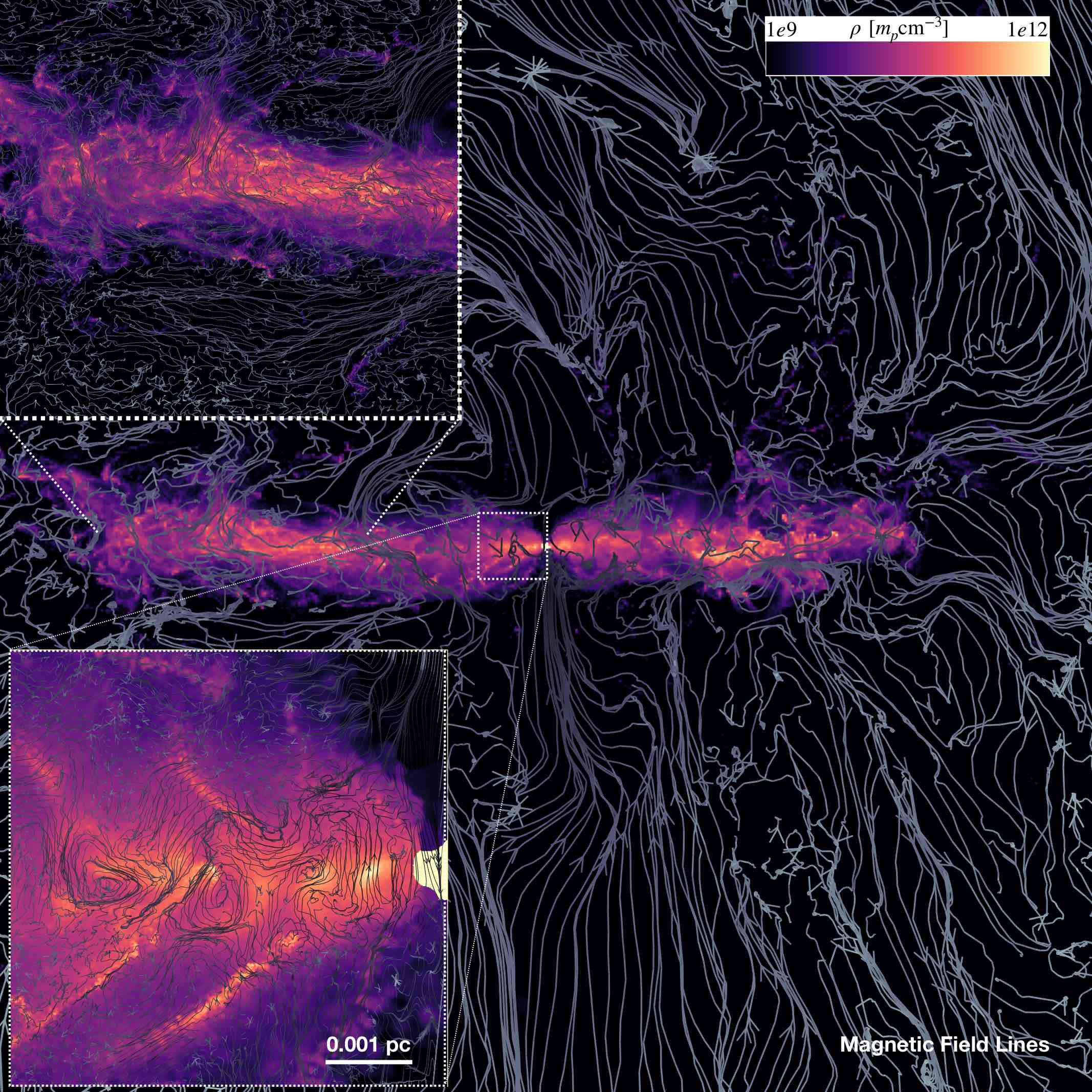}
	\caption{Field lines (as Fig.~\ref{fig:bfields.edgeon.smallscale}; {\em grey}) superimposed on an edge-on gas density projection (colors, $\rho$ as labeled) in $R$-$z$ coordinates, in a narrow wedge ($\sin{\phi}<0.1$). There is a large-scale asymmetry in the disk density, reflecting the coherent large-scale eccentric disk (Figs.~\ref{fig:image.zoom}-\ref{fig:image.faceonedgeon.inner}). Insets show higher-resolution examples of a couple of regions ({\em top-left} inset zooms in on the midplane gas just below in the image). There is considerable density substructure, consistent with highly super-sonic and trans-\Alf{ic} turbulence in the disk, and tangled internal disk fields. We see clear radial/vertical field structure and cells/loops with scale length $\sim H$. Approaching $R\rightarrow 0$, the disk becomes relatively thick, while being thinner at intermediate radii.
	\label{fig:bfields.edgeon.oplotrho}}
\end{figure*}

\begin{figure*}
	\centering\includegraphics[width=0.97\textwidth]{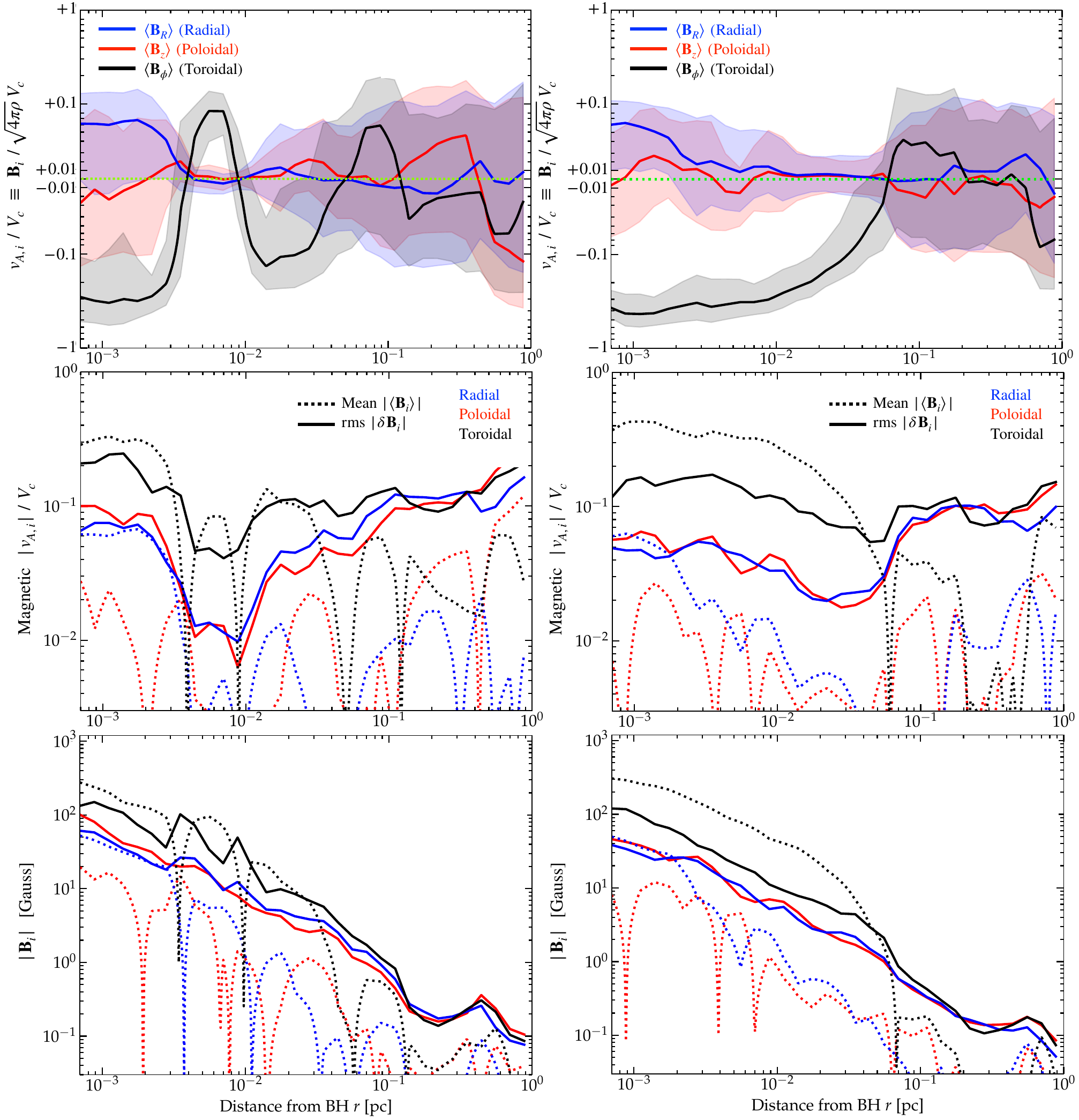}
	\caption{Magnetic field strength and \Alf\ speed (defined by $v_{A,\,i} \equiv B_{i}/\sqrt{4\pi\,\rho}$) relative to circular velocity $V_{c} \equiv \sqrt{G\,M_{\rm enc}(<r)/r}$ in spherical annuli at BH-centric radii $r$ out to $\sim 1\,$pc. We compare two different well-separated times ({\em left} and {\em right}) corresponding to the two times shown in Fig.~\ref{fig:bfields.edgeon.smallscale}. 
	{\em Top:} $v_{A,\,i}/V_{c}$, showing the (mass-weighted) mean $\langle B_{i} \rangle$ ({\em line}) and $\pm1\sigma$ ($16-84\%$) range ({\em shaded}) in each annulus. Dotted green line shows $B=0$ for reference. We clearly see that the toroidal field is dominated by the mean component in the ordered disk region ($r \lesssim 0.1\,$pc), with coherent sign flips, while the radial and poloidal fields are dominated by the fluctuating component (that vary in time). The toroidal field also becomes non-negligible compared to gravity at the smallest radii.
	{\em Middle:} $|v_{A,\,i}|/V_{c}$ on a logarithmic scale, showing the mean $|\langle B_{i} \rangle |$ ({\em dotted}) and fluctuating rms component $\delta B_{i} \equiv B_{i,\,84\%}-B_{i,\,16\%} \sim \langle |B_{i} - \langle B_{i}\rangle|^{2} \rangle^{1/2}$ ({\em solid}; we use a percentile-based definition to avoid biasing the component from a few dense star-forming cells with much-larger $|{\bf B}|$). We see only the toroidal field is dominated by the mean component, and even then always has a non-negligible fluctuating component; the fluctuating radial and poloidal fields are comparable to each other but smaller than $\delta B_{\phi}$ by a factor of a few in the inner disk (outside the disk region, $|\delta B| \gg |\langle B\rangle|$ and all components are comparable, indicating a turbulence-dominated system). $B_{R}$ has mean comparable to $\delta B_{R}$ in the inner regions, while $|\langle B_{z} \rangle | \ll |\delta B_{z}|$.
	{\em Bottom:} As {\em middle}, but plotting the components $B_{i}$ in Gauss. The field strength rises crudely as $\propto r^{-1}$. 
	\label{fig:b.profile}}
\end{figure*}

\begin{figure}
	\centering\includegraphics[width=0.99\columnwidth]{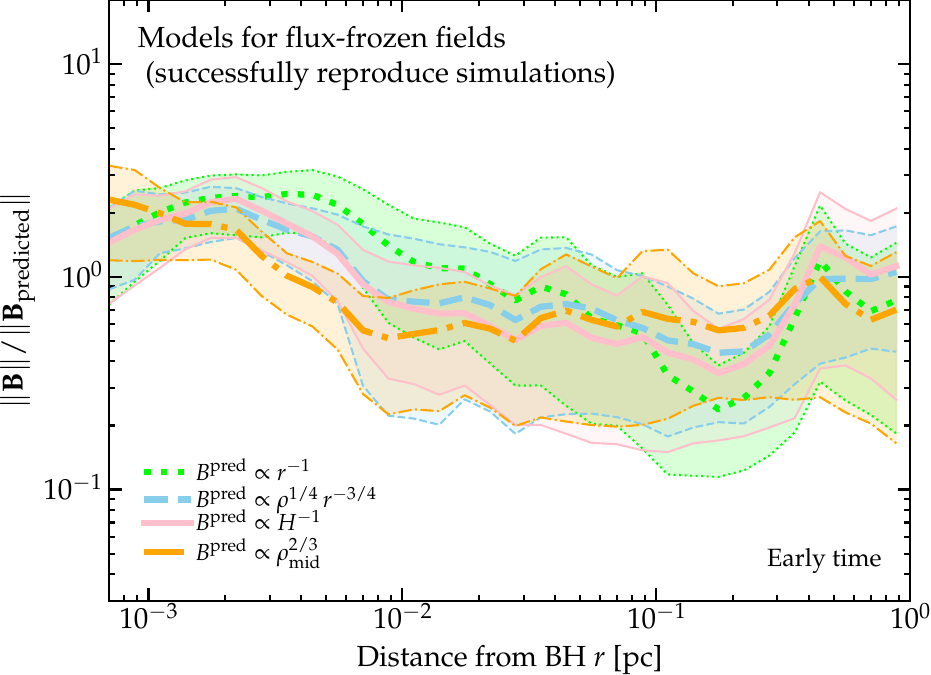}
	\centering\includegraphics[width=0.99\columnwidth]{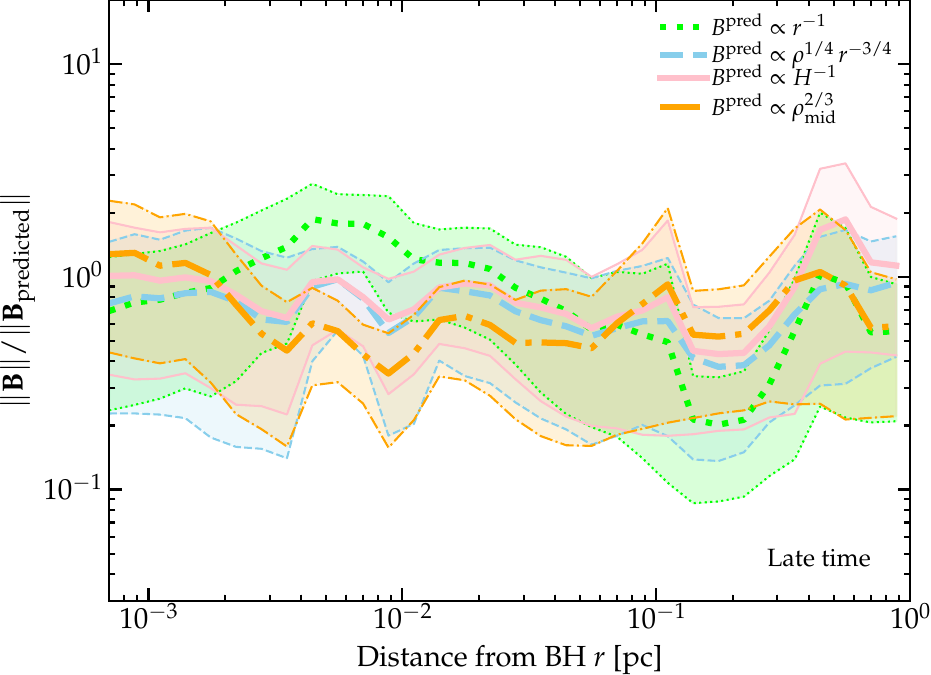}
	\caption{Comparison of the analytic predictions for the amplification of ${\bf B}$ versus radius for the set of models motivated by simple flux-freezing considerations for the toroidal and radial flux, as described in the text (\S~\ref{sec:b:origin:ov}; Eqs.~\ref{eqn:bpred.fluxfreezing.r}-\ref{eqn:bpred.fluxfreezing.rhor}). These successfully reproduce the key behaviors in the simulations. We compare the mass-weighted simulation mean ({\em thick line}) value and $90\%$ inclusion range ({\em shaded}) of $|{\bf B}|$ in spherical annuli to the value $|{\bf B}_{\rm predicted}|$ predicted by the models (labeled), versus radius $r$. We show one early time ({\em top}) and one late time ({\em bottom}), matching the times in Figs.~\ref{fig:bfields.edgeon.smallscale} \&\ \ref{fig:b.profile}. These different scalings, which make slightly different assumptions about how the gas is compressed as it accretes, all predict $|{\bf B}|$ within factors $\sim 2$ without a large systematic residual or offset.
	\label{fig:b.tests.goodmodels}}
\end{figure}

\begin{figure}
	\centering\includegraphics[width=0.99\columnwidth]{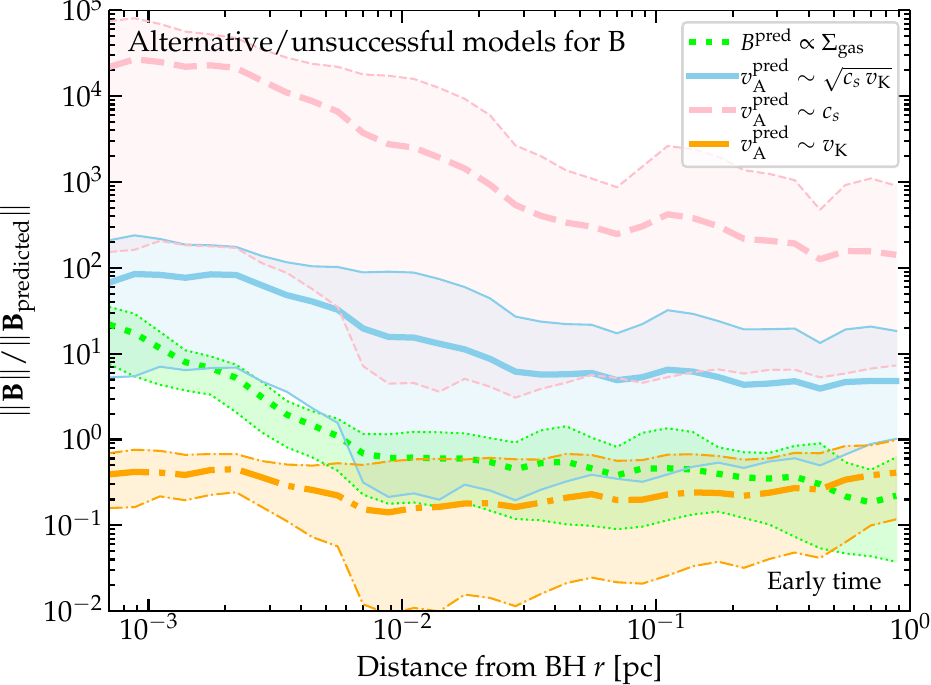}
	\centering\includegraphics[width=0.99\columnwidth]{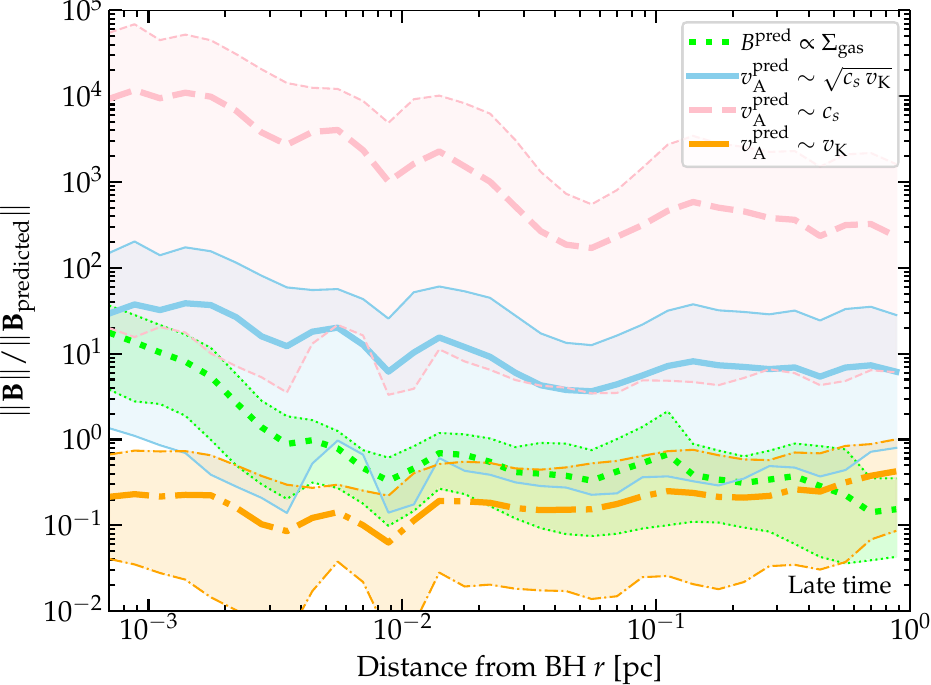}
	\caption{As in Fig.~\ref{fig:b.tests.goodmodels}, but comparing alternative previously proposed analytic models for amplification of $|{\bf B}|$, which are {\em not} based on advection/freezing of toroidal/radial magnetic flux (see \S~\ref{sec:b:origin:tests}, scalings (2), (3), (5), and (6)). We see that these do not accurately describe the simulations, especially noting the much larger vertical axis range here compared to Fig.~\ref{fig:b.tests.goodmodels}. The model used for most previous discussions of toroidal magnetically dominated disks is the predicted saturation value of the linear (local, unstratified) MRI at $v_{A}^{\rm pred} \sim \sqrt{c_{s}\,v_{\rm K}}$ from \citet{pessah.psaltis:2005.mri.extensions.stronger.fields}, but we see that the mean fields here are already well above this value at {\em all} nuclear radii, the scatter is very large, and there is a systematic trend in the median with radius (so no simple re-normalization leads to agreement). The same is true for constant-$\beta$ models or models with $|{\bf B}| \propto \Sigma_{\rm gas}$ sometimes invoked in the literature. A model where $v_{A}$ saturates at $\sim v_{\rm K}$ fares somewhat better but still shows a systemic deviation with radius and a normalization offset from the simulations.
	\label{fig:b.tests.badmodels}}
\end{figure}

\begin{figure}
	\centering\includegraphics[width=0.92\columnwidth]{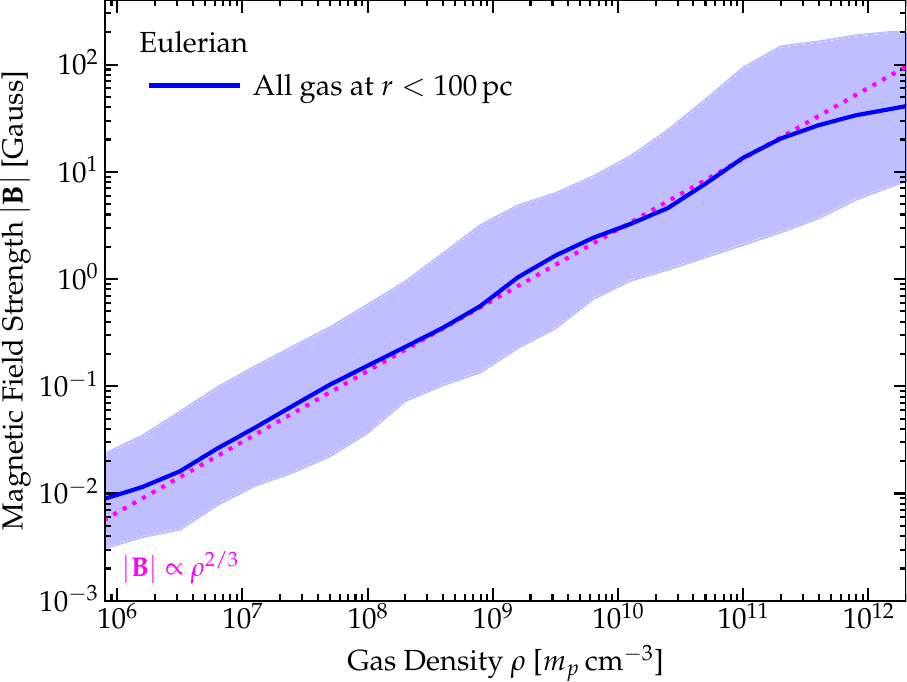}
	\centering\includegraphics[width=0.92\columnwidth]{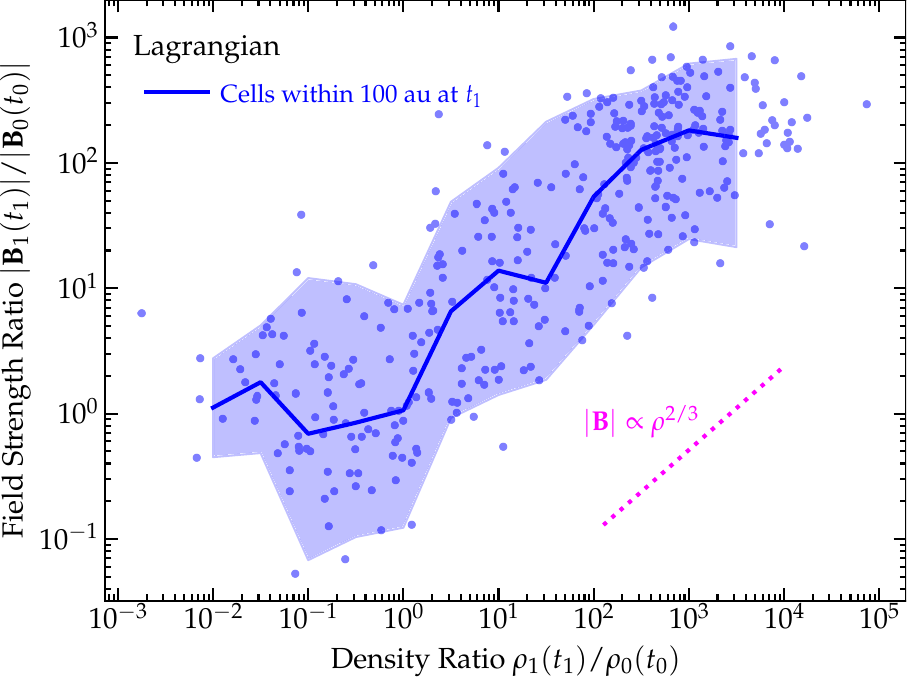}
	\centering\includegraphics[width=0.92\columnwidth]{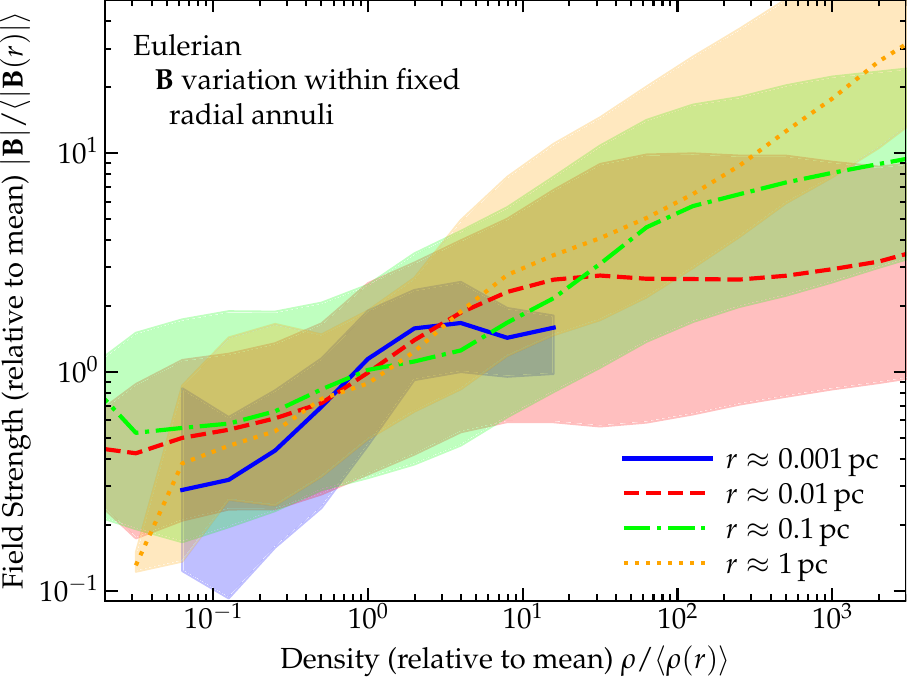}
	\caption{Magnetic field strength $|{\bf B}|$ versus gas density $\rho$. 
	{\em Top:} All gas cells within $r<100\,$pc at a fixed time (Eulerian), showing mass-weighted median ({\em solid line}) and $90\%$ interval ({\em shaded}). We compare an analytic $|{\bf B}| \propto \rho^{2/3}$ scaling motivated by flux-freezing of the mean midplane toroidal/radial fields as they accrete (\S~\ref{sec:b:origin:ov}-\ref{sec:b:origin:tests}). 
	{\em Middle:} Evolution of $|{\bf B}|$ and $\rho$ over time for fixed Lagrangian gas elements (points; the line shows running median). We select random cells that are close to being accreted (within $r<100\,$au) in the final simulation time $t_{1}$, and trace each back to an earlier time $t_{0}$ ($\sim 5000\,\Omega_{\rm inner}^{-1}$ dynamical times earlier), to plot $|{\bf B}(t_{1})|/|{\bf B}(t_{0})|$ versus $\rho(t_{1})/\rho(t_{0})$. As expected for a mean-field-dominated disk in steady-state with most of the mass in midplane inflow, this traces a similar power-law $|{\bf B}|\propto \rho^{2/3}$.
	{\em Bottom:} Variation in $|{\bf B}|$ with $\rho$ {\em at a given radius} (narrow radial annuli at the different $r$ shown). To compare we normalize $|{\bf B}|$ and $\rho$ to their mean within each annulus. Here variation in $|{\bf B}|$ in an annulus is not dominated by global compression of the mean field, so is {\em not} expected to obey the same power-law, and instead varies weakly ($\propto \rho^{0.2-0.3}$), as expected if fluctuations are generally modest relative to the mean field. The tail to high $\rho / \langle \rho(r) \rangle$ is dominated by rare sites of star-formation, so this reflects the physics of local collapse (along field lines).
	\label{fig:Bcompare.eulerian.lagrangian}}
\end{figure}

\begin{figure}
	\centering\includegraphics[width=0.99\columnwidth]{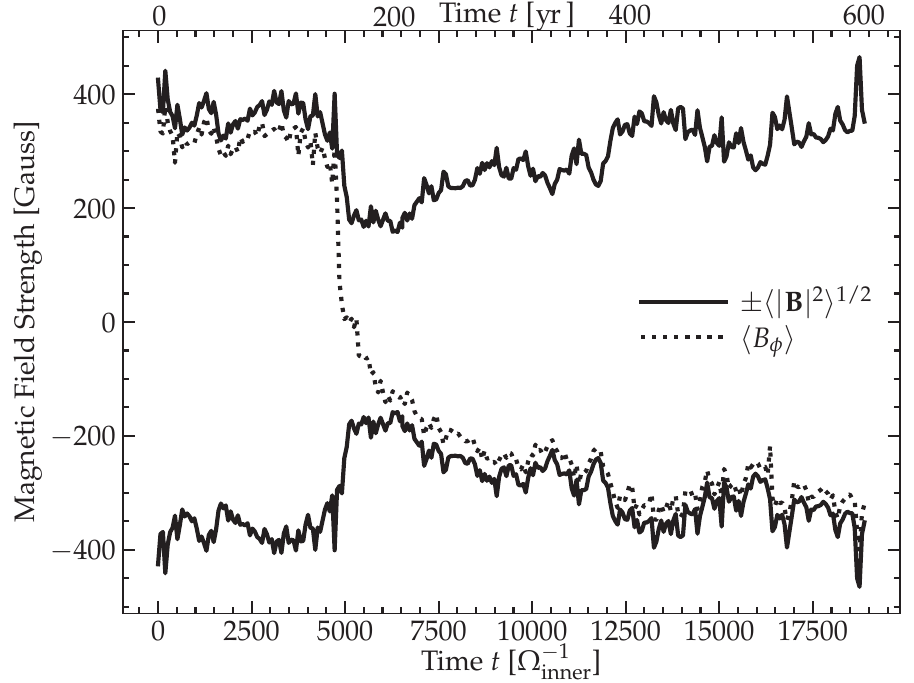}\\
	\centering\hspace{0.5cm}\includegraphics[width=0.93\columnwidth]{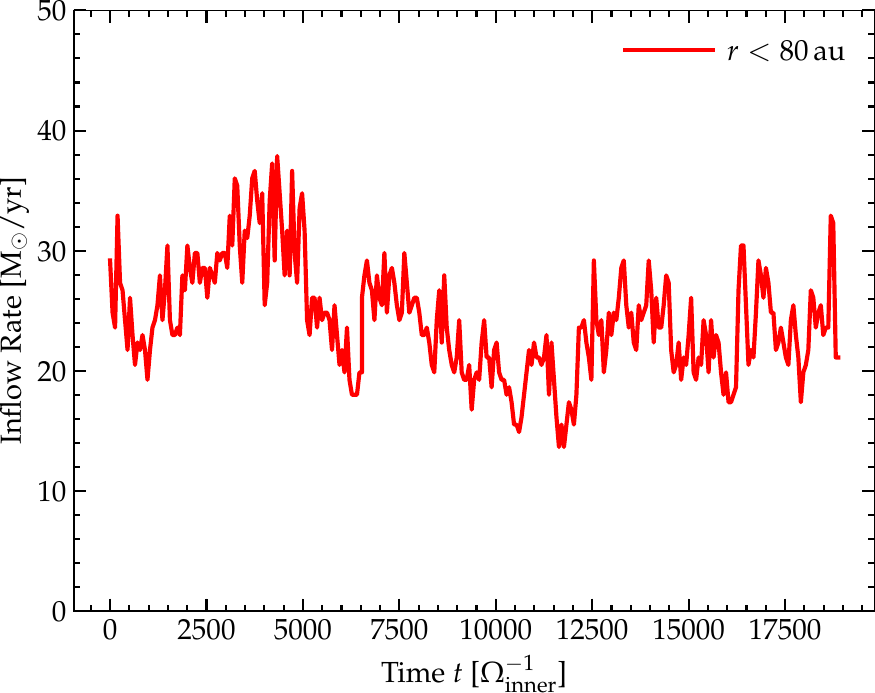}
	\caption{{\em Top:} Gas mass-weighted magnetic field strength $\langle |{\bf B}|^{2} \rangle^{1/2}$ and mean toroidal field $\langle B_{\phi} \rangle$ in the innermost region of the simulation (radii $80\,{\rm au} < r < 120\,{\rm au}$), as a function of time. The time is in units of the dynamical time $1/\Omega_{\rm inner} \equiv \sqrt{r_{\rm inner}^{3} / G\,M_{\rm enc}(<r_{\rm inner})} \approx \sqrt{r_{\rm inner}^{3}/G\,M_{\rm BH}}$, with $r_{\rm inner} \equiv 80\,$au (the time zero-point is arbitrary). We zoom in to show times a few thousand dynamical times before and after a sign flip event where the innermost gas with $\langle B_{\phi} \rangle > 0$, as evident in e.g.\ Fig.~\ref{fig:bfields.faceon.edgeon}, is accreted, with gas from larger radii with $\langle B_{\phi} \rangle < 0$ moving in towards these smaller radii. We see that the magnitude of both $\langle |{\bf B}|^{2} \rangle^{1/2}$ and $|\langle B_{\phi} \rangle|$ recover quickly after sign-flip events and remain stable for $\gtrsim 10^{4}$ dynamical times.
	{\em Bottom:} Accretion rate into the central $<80\,{\rm au}$, averaged in $\Delta t \sim 2\,$yr ($\sim 50\,\Omega_{\rm inner}^{-1}$) increments.
	\label{fig:b.time.flip}}
\end{figure}

\begin{figure}
	\centering\includegraphics[width=0.99\columnwidth]{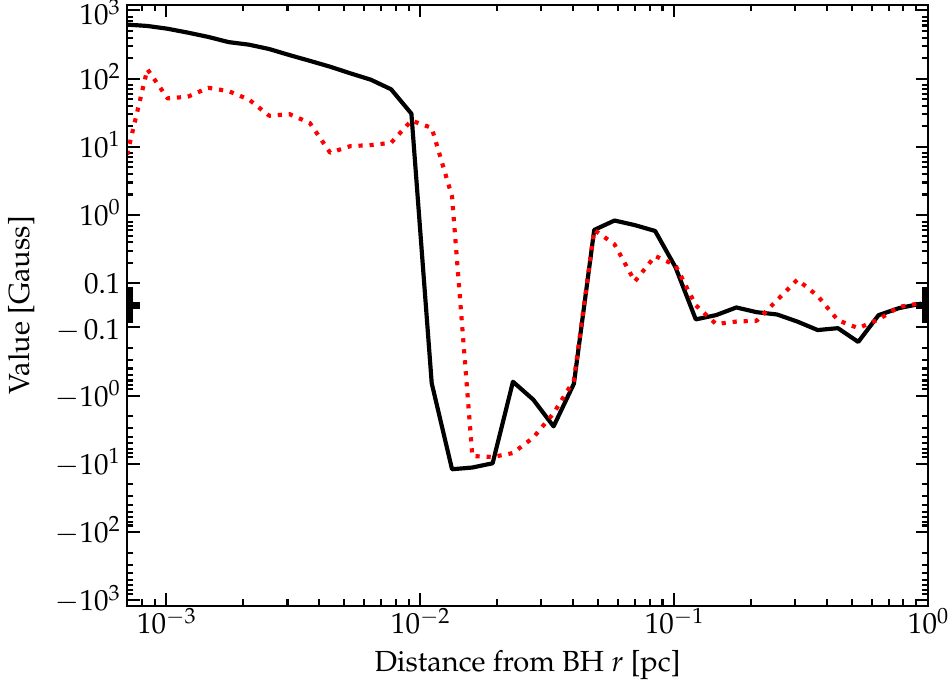}
	\centering\includegraphics[width=0.99\columnwidth]{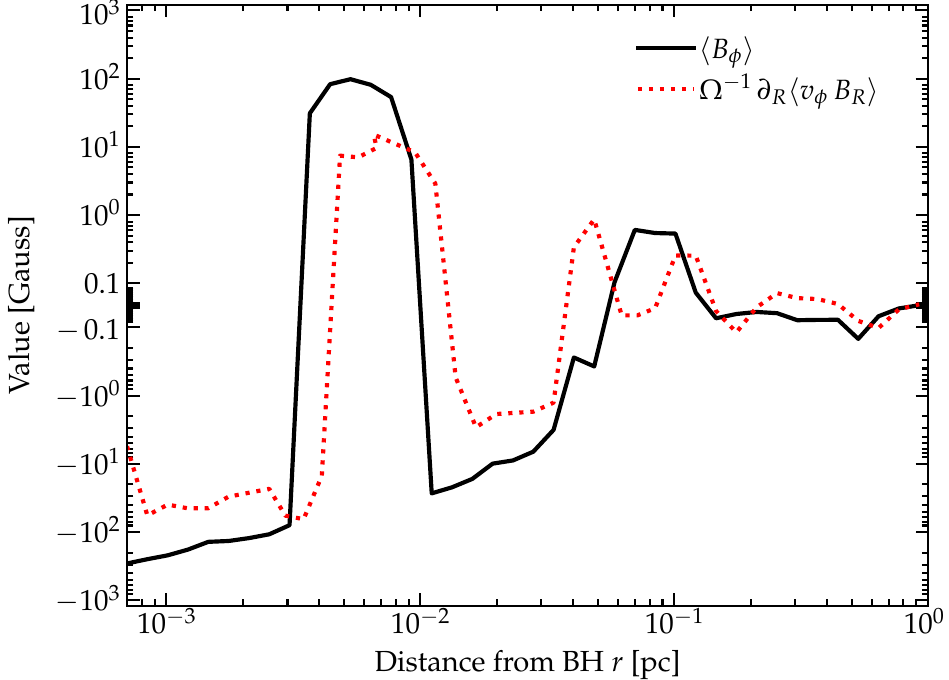}
	\caption{Mean toroidal $\langle B_{\phi} \rangle$ in azimuthal annuli (as Fig.~\ref{fig:b.profile}), versus the expected leading-order term from the induction equation for amplification of the toroidal field via advection/stretching of radial magnetic flux $B_{R}$ ($\partial \langle B_{\phi} \rangle/\partial t \approx \partial( v_{\phi}\,B_{R})/\partial R + ...$), multiplied by the dynamical time $\Omega^{-1}$ at each radius. We show two times ({\em top} and {\em bottom}), before and after the sign flip in Fig.~\ref{fig:b.time.flip}. We see a more quantitative version of the $B_{\phi}-B_{R}$ anti-correlation and that this leading-order average roughly predicts the correct mean toroidal field. Together with the previous figures, this suggests that the strong toroidal field arises primarily from flux conservation (advection of radial flux ``closes the dynamo loop'').
	\label{fig:b.induction.eqn}}
\end{figure}

\section{Structure and Origins of the Magnetic Fields}
\label{sec:b}

\subsection{Overview of Properties \&\ Transition to Toroidal}
\label{sec:b:ov}

Now we turn to study of the magnetic fields in the simulations. Fig.~\ref{fig:bfield.demo} and its ``zoomed-in'' version, Fig.~\ref{fig:radial.profile.general}, show that the total magnetic field strength rises to values approaching $\sim\,$kG at $\lesssim 10^{-3}\,$pc, somewhat steeper than $|{\bf B}|\propto r^{-1}$, from $\sim\,$pc scales. We see directly in Fig.~\ref{fig:bfield.demo} that something like $\langle |{\bf B}|\rangle \propto r^{-1}$ also describes the field strength (very crudely) at $>$\,pc scales out to $\gtrsim$\,kpc. Defining the poloidal, toroidal, and radial components of the field,\footnote{Throughout this paper, we consider a fixed angular momentum axis at any given time defined by the net angular momentum of gas interior to $r < 0.01\,$pc, though our results are not particularly sensitive to exactly where we define this cutoff radius so long as it is within the visually-obvious disk. The poloidal and toroidal and radial field components and azimuthal, vertical, radial velocity components are defined with respect to this. Our sign convention is such that for a toroidal/azimuthal field, a positive value indicates prograde fields (aligned with the gas rotation).} we see in Figs.~\ref{fig:bfield.demo}, \ref{fig:image.wedgeplot}, \&\ \ref{fig:radial.profile.general} a clear transition from an isotropic or slightly radially-biased field at $\gtrsim 0.1\,$pc (which Fig.~\ref{fig:bfield.demo} shows is true at much larger radii as well) to a toroidal-dominated field at smaller radii, coinciding with the visually well-ordered disk in Fig.~\ref{fig:image.zoom}. 

Fig.~\ref{fig:bfields.faceon.edgeon} plots the projected structure of the field lines in the disk plane (taking a slice through the midplane, so in cylindrical $R$-$\phi$ coordinates considering a wedge of some width in $|z|/R$) and edge-on (in cylindrical $R$-$z$ coordinates considering a wedge of some width in $\phi$), in one example time snapshot within the ordered disk ($R \lesssim 0.1\,$pc). Fig.~\ref{fig:bfields.faceon.zoom} considers the face-on field structure at three different scales $R\lesssim\,(0.01,\,0.1,\,1)$\,pc. Fig.~\ref{fig:bfields.edgeon.smallscale} shows the edge-on field-line structure on the smaller scales $\lesssim 0.01\,$pc at two different times, to illustrate how they can change, and Fig.~\ref{fig:bfields.edgeon.oplotrho} shows the field lines overplotted on an edge-on density map to see how illustrate the relation to the density substructure within the disk. 

In the face-on projections, we see the fields are fairly well ordered, with a clear transition from more radial field lines pointing along the direction of gas infall onto the nuclear disk, circularizing where the disk forms, to become toroidal, with increasing order in the toroidal field at smaller radii (by eye, it becomes closer to purely azimuthal). We also see clear repeating sign flips in the toroidal field joined by field reversals (occasionally breaking off into large-scale loop-type structures), with some intermediate turbulent zones. However, the inflow continues to be traced even as the toroidal field strengthens as we see a spiral-type structure (i.e.\ non-negligible coherent radial field components pointing inwards). We also plainly visually see an anti-correlation in the signs of $B_{\phi}$ and $B_{R}$, as expected if the toroidal field is sourced by radial flux.

In the edge-on projections we see less large-scale coherence. In the midplane there is a (much weaker) coherent radial/vertical field component in the $R$-$z$ projection, and there is some vaguely ``jet like'' vertical bipolar field in $R$-$z$ (the vertical/conical fields at small $R$ with coherent $B_{\phi}$ and $B_{R}$ but oppositely-signed $B_{z}$ above/below the disk). Interestingly, looking at the sign of the toroidal field in the edge-on projection, we see that there can be sign flips of $B_{\rm tor}$ at different vertical heights, as well as at different radial intervals; but, these are generally at $|z| \gtrsim H$ -- i.e.\ outside the body of the disk (with coherence lengths $\gtrsim H$). It is worth noting that there is no sign flip of $B_{\phi}$ across the midplane, as is often seen if $B_{\phi}$ were sourced by a much stronger mean poloidal field (although there can be exceptions to this). We see clear evidence for some mode structure in $B_{R}$ and $B_{z}$ interior to the disk, with wavelength $\sim H$.

In Fig.~\ref{fig:b.profile} we examine the magnetic field profile on these scales somewhat more quantitatively. Here, we consider a 1D profile (averaged in spherical shells), plotting both the mass-weighted mean-field values of the cylindrical $R$, $\phi$, $z$ components, as well as their $\sim1\sigma$ range.\footnote{To be robust against outliers from e.g.\ small-scale structure in small sub-volumes containing nascent protostellar disks, for example, we define the $\sim 1\sigma$ value of the value of each magnetic component as $1/2$ of the $16-84\%$ mass-weighted inclusion interval. If we simply consider the usual RMS $\langle B_{i}^{2} \rangle^{1/2}$ we see similar trends, but slightly systematically larger values, indicating that the ``tails'' of the distribution are somewhat fatter than Gaussian, as well as slightly more noise owing to some sub-structure. The exact quantitative values also vary somewhat if we mass or volume or magnetic-energy weight the results, but this does not change any of our relative comparisons or conclusions.} We plot these both in absolute units, as well as in units of the \Alf\ speed $v_{A,\,i} \equiv B_{i}/\sqrt{4\pi\,\rho}$, relative to the circular velocity $V_{\rm c} \equiv \sqrt{G\,M_{\rm enc}(<r)/r}$ at each radius. We see the same rise in the field values as Fig.~\ref{fig:radial.profile.general}, and the increasing preference for toroidal fields within the disk as above, but now more quantitatively see the transition from a primarily turbulent field (the rms/dispersion component larger than mean) to more coherent (mean similar to or even larger than dispersion, for $B_{\phi}$). The mean $\langle B_{\phi} \rangle$ dominates at the inner disk, but the ``turbulent'' or rms components are not vastly smaller. There appears to be a robust sort of ``hierarchy'' of the different field components in the inner disk, with $|\langle B_{\phi} \rangle| \gtrsim |\delta B_{\phi} | \gtrsim |\delta B_{R}| \sim |\delta B_{z}| \gtrsim |\langle B_{R} \rangle| \gtrsim |\langle B_{z} \rangle |$, i.e.\ the mean toroidal and vertical fields are the strongest and weakest, respectively, with the fluctuating toroidal second-strongest, followed by the fluctuating radial and vertical fields. This is independent of whether we define these components in a global cylindrical coordinate system (shown), or a spherical coordinate system, or a spatially-variable coordinate system where we rotate each annulus independently to correspond to the angular momentum axis of gas just in that annulus, and/or whether we subtract the mean $m=1$ component in each annulus (to control for a coherent eccentric mode). We also see that the sign flips in the disk clearly evolve, as mass is accreted through the disk, but the magnitude of the total field stays broadly consistent over tens of thousands of disk dynamical times.

As discussed above (\S~\ref{sec:basic:profile}), the magnetic pressure dominates the vertical support of the disk but with $\mathcal{O}(1)$  contribution from trans-\Alf{ic} turbulence: this is reflected both in direct comparison of the turbulent velocity components to the various $v_{A,\,i}$ below; or from the kinetic energy densities, or relative contribution to the effective stability parameter $Q$, or comparison of the disk $H/R$ to $v_{A}/V_{c}$ in Fig.~\ref{fig:radial.profile.general}. In other words, $\rho\,v_{\rm turb}^{2} \sim \langle |{\bf B}|^{2} \rangle$ broadly speaking, as we quantify in more detail below. Of course, both $v_{A}$ and $v_{\rm turb}$ are much larger in the simulations than in an SS73 disk which assumes $\beta \gg 1$.

\begin{figure*}
	\centering\includegraphics[width=0.94\textwidth]{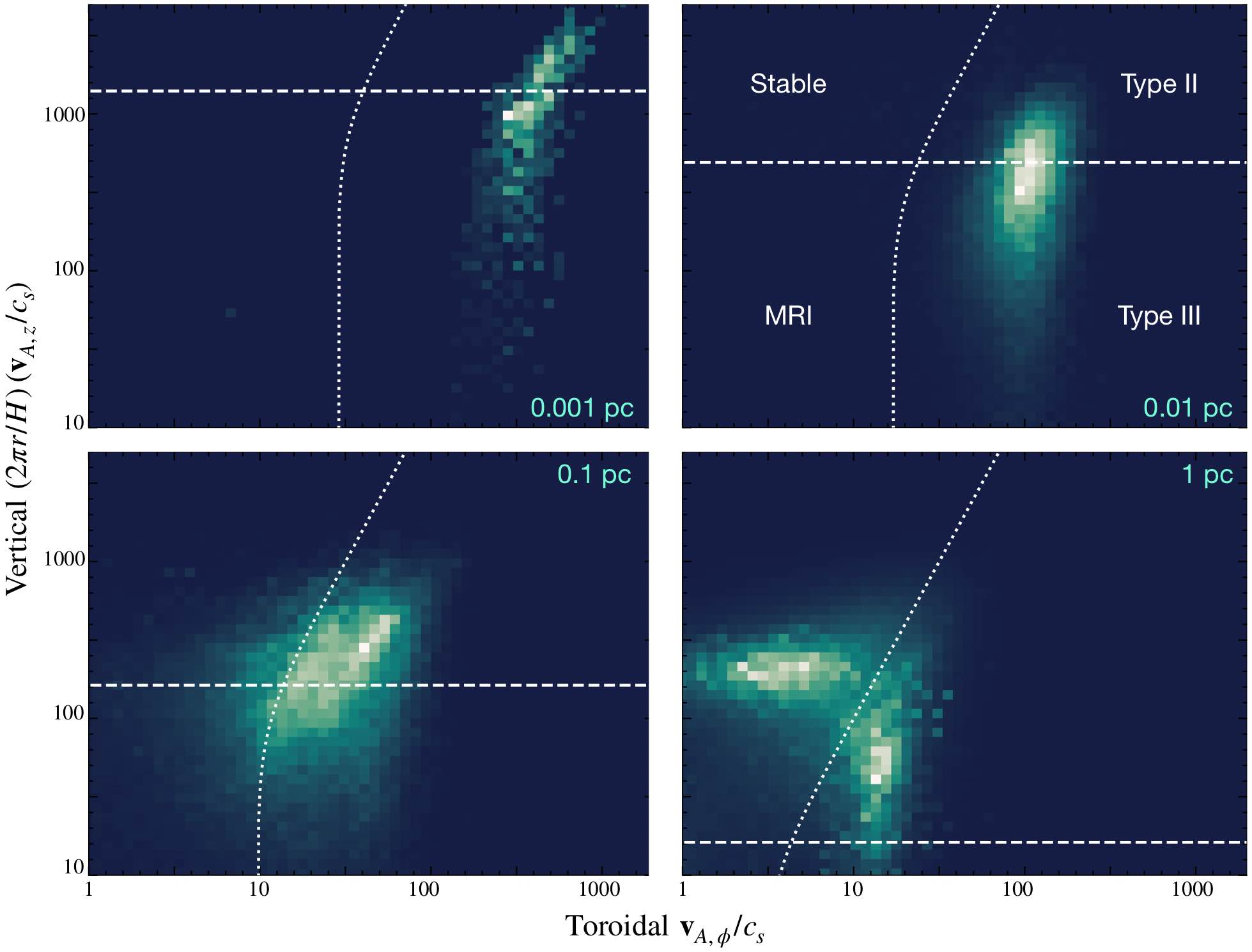}
	\caption{Instability parameter space map, following e.g.\ Fig.~3 of \citet{pessah.psaltis:2005.mri.extensions.stronger.fields}. We select gas in the warm, volume-filling phases (excluding gas with $T<1000\,$K) in a radial annulus (factor of $\sim 2$ range of $R$ centered on $R=(0.001,\,0.01,\,0.1,\,1)$\,pc, corresponding to the different panels, as labeled), within the disk midplane ($|z| < H$ at each $R$), and plot the volume-weighted histogram (with a linear scale, increasing from $0$ in black to $1$ in white) of the gas in the space of the toroidal $v_{A,\,\phi}/c_{s} \equiv B_{\phi}/(c_{s}\,\sqrt{4\pi\,\rho})$ versus vertical ``effective minimum wavenumber'' $k^{\rm eff}_{z}\,r\,v_{A,\,z}/c_{s} \sim (2\pi\,r/H)\,(B_{z}/(c_{s}\,\sqrt{4\pi\,\rho}))$. {\em Dashed} and {\em dotted} white lines show the critical wavenumbers $k_{z}^{c1}$ and $k_{z}^{c2}$ defined in \citet{pessah.psaltis:2005.mri.extensions.stronger.fields} which separate the different regimes of instability labeled (for an analytic, linear, laminar, unstratified system). The bottom-left quadrant corresponds to the classical (linear, unstratified, weak-field) MRI, top-left to a stable regime, while the top-right and bottom-right to the ``Type II'' and ``Type III'' (bouyancy-related) instabilities defined therein (or the suprathermal slow mode instability [SSMI] and suprathermal hybrid mode instability [SHMI], respectively, in \citealt{das:2018.pessah.psaltis.limit.mri}). We see that the volume-filling midplane phases in the simulations reside well into the Type II/III (SSMI/SHMI) regime, with the largest wavelengths ($\sim H$) broadly similar to the dividing line between those modes, i.e.\ the toroidal field is stronger than the maximum amplitude at which the linear MRI grows under the usual laminar, unstratified disk assumptions ($v_{A,\,\phi}^{\rm max,\,MRI} \sim \sqrt{c_{s}\,v_{\rm K}}$, as shown in Fig.~\ref{fig:b.tests.badmodels} above). Colder-phase gas (not shown here) shifts diagonally to the upper right, further into the Type II/SSMI region.
	\label{fig:instability.map}}
\end{figure*}

\subsection{Physical Origins of the Field Strength \&\ Structure}
\label{sec:b:origin}

\subsubsection{Origins of the Mean Field in Flux-Freezing/Advection of Flux with Accreting Gas}
\label{sec:b:origin:ov}

We find that the qualitative behavior of the dominant mean toroidal field $\langle B_{\phi} \rangle$ can be understood primarily from simple flux-freezing considerations. Below, we test and validate this in more detail, but first let us describe the qualitative scenario and key behaviors in the simulations. To begin, recall that the gas forming the disk is tidally captured from a close passage by a molecular cloud complex to the BHROI (Figs.~\ref{fig:image.zoom}-\ref{fig:image.faceonedgeon.inner}), so behaves as an initially ``cold'' (weakly pressurized) tidal filament/stream, akin to satellite galaxy encounters on large scales \citep{hernquist.mihos:minor.mergers,bullock.johnston:2005:stellar.halos,younger:minor.mergers,moster:2010.thin.disk.heating.vs.gas} and similarly analogous to some of the behaviors seen in simulations of magnetic fields in stellar tidal disruption events (TDEs) \citep{bonnerot:2017.bfield.in.tdes,2017ApJ...834L..19G}. 

As shown in Fig.~\ref{fig:bfield.demo}, the gas at large radii (e.g.\ in the sub-kpc galactic nucleus) has broadly isotropic turbulent fields, with magnetic energy density a few percent of the kinetic energy density (see also Fig.~9 in \paperone), as expected for the supersonic dynamo in both idealized \citep{federrath:supersonic.turb.dynamo,rieder.teyssier:2017.turb.dynamo.saturation.few.percent.fraction.b} and multi-phase galaxy formation simulations \citep{martin.alvarez:2022.cosmological.turb.dynamo,guszejnov:fire.gmc.props.vs.z,seta.federrath:2022.turb.dynamo.twophase.medium}. When some portion of this is captured and falls in (with tangential velocity below $V_{\rm c}$), it is tidally stretched into a radial stream of length $\ell$ and width $w$ (as we see occurring in Figs.~\ref{fig:bfields.faceon.edgeon}-\ref{fig:bfields.faceon.zoom}). For $B_{z}$ and $B_{\phi}$, the perpendicular areas --  $(\Delta r)\,(r\,\Delta \phi) \sim \ell\,w$ and $(\Delta r)\,(\Delta z) \sim \ell\,w$, respectively -- are increased by the stretching\footnote{For pure radial infall in a Keplerian potential, the perpendicular extent $w$ is tidally compressed with a transverse acceleration, $\Delta a_{\bot} = w\,\partial a_{\rm grav,\,\bot}/\partial w = -\Omega^{2}\,w$, while the radial extent $\ell$ is stretched with a parallel acceleration, $\Delta a_{\|} = \ell\,\partial a_{\rm grav,\,\|}/\partial \ell = 2\,\Omega^{2}\,\ell$. If one begins from small coherent shear/expansion velocities, this means the perpendicular areas for $B_{z}$ and $B_{\phi}$ ($\sim \ell\,w$) will increase while the perpendicular area for $B_{R}$ ($\sim w^{2}$) decreases.} so these components are weakened; in contrast, $B_{R}$ is amplified, owing to the perpendicular compression (perpendicular area $(r\,\Delta \phi)\,(\Delta z) \sim w^{2}$). Equivalently we can think of the field lines as being ``stretched'' in the radial direction as they are dragged. The expected amplification of $B_{R}$ ranges between $r^{-0}$ (for e.g.\ an infalling clump with $w\ll r$, or no radial motion) to $r^{-2}$ (maximal case where $w\sim r$ and pure-radial inflow), depending on the efficiency of the perpendicular compression. This agrees with the behavior seen in Figs.~\ref{fig:radial.profile.general} \&\ \ref{fig:b.profile}.

The infalling gas has non-zero impact parameter $b$ and circularizes at $\sim 0.1\,$pc (plainly visible in Figs.~\ref{fig:image.faceonedgeon.inner} \&\ \ref{fig:image.wedgeplot}). The ``initially'' radial field therefore follows the gas flow and wraps/winds up to become toroidal (Fig.~\ref{fig:b.profile}). Equivalently the compression ratios rotate as shear now means that the elongation direction is azimuthal, so the mean azimuthal field $\langle B_{\phi} \rangle $ is amplified strongly while the mean radial and vertical fields grow less rapidly (remaining sub-dominant to their fluctuating field components). Since the field was initially tangled and isotropic (with $|\delta B_{i} | \gtrsim |\langle B_{i} \rangle|$ for all components of ${\bf B}$) at radii outside of the BHROI, there are sign flips in the ``initial'' field which is now stretched into some mean $\langle B_{\phi} \rangle$ as it was radially accreted -- these become the successive sign flips in the radial direction. Indeed, following the fluid over time in Fig.~\ref{fig:bfields.edgeon.smallscale}, we note below that the sign flips in $B_{\phi}$ simply reflect these frozen-in trends and are advected inwards with the fluid as it accretes. In steady-state, even if there is some damping of the coherent toroidal field owing to turbulent resistivity or buoyant escape, $\langle B_{\phi} \rangle$ is constantly replenished by the steady supply of radial magnetic flux into and through the disk.

For flux-freezing, the compression in the $z$ direction suggests $B \sim B_{\phi} \propto 1/H$ \citep{machida:2006.mag.supported.disks.from.cooling.around.adafs}. Meanwhile direct analysis of the simulations or simple analytic considerations give weak compression in the $R$ direction.\footnote{This is expected if the solution is wind-like (constant $v_{r}$),  free-fall-like,  spherical ($H\sim r$, akin to the Solar wind), or simply circularizing while conserving specific angular momentum from a broad/flat initial distribution (for some uniform initial bulk velocity dispersion of the captured cloud, $dr \propto db$).} As shown below, a yet simpler isotropic-flux-freezing expectation $B \propto \rho^{2/3}$ works equally well in explaining the evolution of the field strengths both in time (following a Lagrangian parcel) or space (fitting the radial profile). Checking the normalization of $|{\bf B}|$, if we fit $B \propto \rho^{2/3}$ to the simulations using the midplane values of $\langle |{\bf B}|^{2} \rangle^{1/2}$ and $\langle \rho_{\rm midplane} \rangle$ (Figs.~\ref{fig:bfield.demo} \&\ \ref{fig:radial.profile.general}), we obtain $|{\bf B}| \sim (2-8)\,{\rm \mu G}\,(n/{\rm cm^{-3}})^{2/3}$ (depending on how we weight it; or $|{\bf B}| \sim (2-6)\,{\rm \mu G}\,(r/10\,{\rm kpc})^{-1}$) -- i.e.\ at Galactic radii this extrapolates to typical mundane values of $|{\bf B}|$, consistent with our direct estimates from the simulations in Fig.~\ref{fig:bfield.demo}. Moreover, as discussed below and in \paperthree, the absolute field strengths here (in Gauss) are actually {\em smaller} than in an SS73-like $\alpha \sim 0.1$ disk with the same $\dot{M}$ (even though $\beta \ll 1$ is much smaller). Thus no extreme fields at large radii are needed to sustain these strong fields in the disk. 

This is consistent with the global field geometry, and at least qualitatively explains the sign flips in the toroidal field in successive radial annuli (as this reflects field reversals from the turbulent fields at much larger radii, before amplification). The idea also explains the lack of systematic sign flips in $\langle B_{\phi} \rangle$ at $z=0$ (i.e. reversals of $\langle B_{\phi} \rangle$ as one vertically crosses the midplane), as well as the broad anti-correlation between $B_{\phi}$ and $B_{R}$. The relative amplification versus suppression above also explains why we see a dominant mean $\langle B_{\phi} \rangle$ component with much weaker mean $\langle B_{R} \rangle$ and $\langle B_{z} \rangle$ components (Fig.~\ref{fig:b.profile}; note also that $\langle B_{z} \rangle$ seems to grow somewhat more slowly than $\langle B_{R} \rangle$). Instead, the radial and poloidal fields are  more dominated by their turbulent/fluctuating components, whose amplitudes are of order $|\delta B_{R,\,z}| \sim | \delta (v_{R,\,z}  / v_{A})\,\langle B_{\phi} \rangle | $ -- i.e.\ consistent with field lines being stretched, distorted, and perhaps modestly amplified by turbulence within the disk (as we appear to see occurring in Figs.~\ref{fig:bfields.edgeon.smallscale}-\ref{fig:bfields.edgeon.oplotrho}). 

We stress that, as discussed in the numerical tests in \paperone, microphysical resistivity is not expected to play a significant role here owing to the fact that the gas is still relatively diffuse and, in the nuclear regions of importance, still highly ionized with ionization fraction $\gtrsim 0.01$ (so ideal MHD remains a good approximation). This owes to a combination of high temperatures ($\gtrsim 1000\,$K), dust destruction, high cosmic-ray energy densities and ionization rates (even in the Milky Way center, these exceed typical Solar circle values by several orders of magnitude; see \citealt{indriolo:2015.cr.ionization.rate.vs.galactic.radius}), and high interstellar radiation field densities due to the concentrated intense star formation ($\gtrsim 100\,{\rm M_{\odot}\,yr^{-1}}$ in the central $\sim 100\,$pc; see \paperone). This is unlike accretion of magnetic fields onto a protostar via a protostellar disk, where ionized fractions are expected to be in the range $\lesssim 10^{-17}-10^{-15}$ in much of the disk.\footnote{Calculating the generalized Elsasser number $N_{\rm E} \equiv B^{2}/(\rho\,\eta\,\Omega) \sim v_{A}^{2}\,t_{\rm orbit} / \eta$, where $\eta$ is the largest of e.g.\ ambipolar, Hall, or Ohmic resistivities, and plugging in numbers for typical values in the simulations at $\lesssim$\,pc scales gives $N_{\rm E} \gtrsim 10^{13}$ through most of the inner disk, while $N_{\rm E} \lesssim 1$ is required for resistivity to have a large global effect (or for e.g.\ ambipolar diffusion to generate substantial ``drift'' between ions and neutrals on the timescales of interest). That would generally require an ion fraction $x_{i} \sim n_{i}/n_{\rm neutral} \ll 10^{-15}$ in the accretion disk.} In \S~\ref{sec:why.flux.freezing}, we discuss the role of turbulent resistivity in detail and show that it is also sub-dominant: damping of the dominant radial and toroidal fields via turbulent resistivity is generically slower than their growth via flux-freezing and advection, though such damping may be important for the less-coherent poloidal field. Thus our discussion in this section does not assume weak turbulence and indeed we have assumed trans-\Alf{ic} or even modestly super-\Alf{ic} (and highly super-sonic) turbulence could be present throughout (\S~\ref{sec:why.flux.freezing}), so long as the turbulence is sub-virial ($v_{\rm turb} \lesssim V_{c}$), so that the turbulent coherence length is smaller than the characteristic radial distance $\gtrsim r$ over which the field lines are being stretched as part of the tidal inflow/stream.

\subsubsection{Validating that the Mean Field is Indeed Driven by Flux-Freezing and Advected Flux}
\label{sec:b:origin:tests}

We now consider various quantitative tests of the picture described above in \S~\ref{sec:b:origin:ov}, to validate more rigorously that it is a reasonable description of the simulations. Moreover, one might imagine several alternative scenarios/models that attempt to explain or predict strong magnetic fields in an accretion disk, but we find most of these do not reproduce the behaviors seen in our simulation. Consider the following possibilities:
\begin{enumerate}[labelindent=0pt,labelwidth=10pt,labelsep*=0pt,leftmargin=!,align=parleft]
\item{Flux-freezing of the radial/azimuthal magnetic flux}, the scenario from \S~\ref{sec:b:origin:ov}. Here $\langle B_{\phi} \rangle$ is sourced by flux-frozen accreted fields, and the mean toroidal field originates from advection of radial flux with the radially-infalling gas from outside the BHROI. Its evolution following the description in \S~\ref{sec:b:origin:ov} (Eqs.~\ref{eqn:bpred.fluxfreezing.r}-\ref{eqn:bpred.fluxfreezing.rhor} below).

\item{Flux-freezing of a dominant mean dipolar/poloidal field}. One could instead assume that the magnetic field was dominated by a mean poloidal/vertical/dipolar field $\langle B_{z} \rangle$, which was flux-frozen as mass advected inwards in a laminar disk (e.g.\ torqued by magnetic braking). Then the fixed flux-to-mass ratio would predict $|{\bf B}^{\rm pred}| \sim |\langle B_{z} \rangle | \propto \Sigma_{\rm gas}$ in the disk.

\item{Traditional MRI-Driven Fields (following \citealt{begelman.pringle:2007.acc.disks.strong.toroidal.fields})}. If the fields (including the mean field) were primarily driven by the traditional weak-field MRI (sourced as usually assumed by some initial poloidal field), then this would predict that the maximum saturation value of the magnetic field strength should be given by the value above which the MRI ceases to grow efficiently. For the analytic models of magnetically-dominated disks in \citet{begelman.pringle:2007.acc.disks.strong.toroidal.fields}, this is taken to be $v_{A}^{\rm pred} \approx \sqrt{c_{s}\,v_{\rm K}}$, or $B^{\rm pred}\sim \sqrt{4\pi\,\rho\,c_{s}\,v_{\rm K}}$ (the value derived for linear MRI growth in a laminar, unstratified, local analysis in \citealt{pessah.psaltis:2005.mri.extensions.stronger.fields}).\footnote{Technically, following the more general dispersion relation in \citet{pessah.psaltis:2005.mri.extensions.stronger.fields}, this should be modified to $v_{A}^{\rm pred} \approx \sqrt{c_{s}\,v_{\rm K}/|1 + \partial \ln{B}_{\phi}/\partial \ln{R}|}$ in the presence of radial magnetic stratification \citep[see discussion in][]{begelman:2023.mri.saturation.estimates}. This correction generally makes a small (tens of percent) difference, but even in the annulus where the correction is maximized (near $\sim 0.002$\,pc where $\partial \ln{B}_{\phi}/\partial \ln{R}$ is close to $-1$) the correction never exceeds a factor of $\sim 2.5$, so does not change any of our conclusions.}

\item{The Small-Scale Super-Sonic, Rapidly-Cooling Turbulent Dynamo}. In the standard super-sonic turbulent dynamo, the fields saturate with magnetic energy a few percent the turbulent kinetic energy (i.e.\ \Alf\ Mach numbers $\sim 3-10$), with isotropic, tangled fields, and fluctuating components much larger than mean ($|\delta {\bf B}^{\rm pred}| \gg |{\bf B}|$). 

\item{The Small-Scale Sub-Sonic, Gravitational/Protostellar Dynamo}. Various models for the local dynamo in more slowly cooling dense gas in molecular clouds/clumps/cores collapsing to proto-stellar disks have argued for saturation at fixed $\beta \sim 1$, giving $v_{A}^{\rm pred} \sim c_{s}$ \citep{2017ApJ...838...40M},  again with isotropically tangled fields ($|\delta {\bf B}^{\rm pred}| \gg |{\bf B}|$). 

\item{``Arrested'' Fields}. If the field simply saturated at a value where it would dynamically arrest further inflow (as in e.g.\ ``magnetically arrested disks'', discussed further below), we might expect $v_{A}^{\rm pred} \sim V_{c}$

\end{enumerate}

First, we note that the measurements of the different mean and fluctuating components (e.g.\ which components are dominant where, and the ratio of mean-to-fluctuating component amplitudes), as well as the presence/absence of different sign flips, shown in Figs.~\ref{fig:bfields.faceon.edgeon}-\ref{fig:b.profile}, are all consistent with our favored flux-freezing picture {\em (i)}, as described in \S~\ref{sec:b:origin:ov}. The field geometry we see is immediately inconsistent with model {\em (ii)} (which assumes the field is dominated by a mean poloidal/vertical field) and models {\em (iii)}, {\em (iv)} and {\em (v)} (which all predict that the mean toroidal field should be much smaller than the fluctuating field). Moreover, phenomena such as the sign flips are either not predicted or predicted to have qualitatively different behaviors in models {\em (ii)}-{\em (vi)}.

Second, in Figs.~\ref{fig:b.tests.goodmodels}-\ref{fig:b.tests.badmodels}, we specifically compare the measured $|{\bf B}|(r)$ in the simulation  to the value predicted by the different simple assumptions/models above (at two different times). To begin, in Fig.~\ref{fig:b.tests.goodmodels} we compare a group of models for $|{\bf B}| \sim |\langle B_{\phi} \rangle|$ motivated by the simple flux-freezing considerations of model {\em (i)} as described in  \S~\ref{sec:b:origin:ov}. We see that these reasonably reproduce the absolute magnitude and radial trend of $|{\bf B}|$ with relatively little scatter. If we compare 
\begin{align}
\label{eqn:bpred.fluxfreezing.r} B^{\rm pred} &\propto r^{-1}
\end{align}
or 
\begin{align}
\label{eqn:bpred.fluxfreezing.H} B^{\rm pred} &\propto H^{-1}
\end{align}
(both noted in \S~\ref{sec:b:origin:ov}), the $B^{\rm pred} \propto H^{-1}$ scaling clearly exhibits even smaller scatter and more accurate prediction of the mean $|{\bf B}|$ compared to $B^{\rm pred}\propto r^{-1}$. We obtain almost as good a fit with the  simpler expression 
\begin{align}
\label{eqn:bpred.fluxfreezing.rho} B^{\rm pred} &\propto \langle\rho_{\rm mid}\rangle^{2/3}
\end{align}
(shown explicitly in Fig.~\ref{fig:Bcompare.eulerian.lagrangian}).\footnote{Note this does more accurately predict $|{\bf B}|$ than assuming $|{\bf B}|\propto \rho^{2/3}$ for all $\rho$ within a given disk annulus, as some of the dense, cold and/or hot, diffuse gas phases actually have more similar $|{\bf B}|$ if they are at the same radial annulus near the midplane, owing to the fact that their $|{\bf B}|$ is dominated by the mean $\langle B_{\phi} \rangle$ component and collapse/expansion can occur along these field lines. So {\em locally} on small scales within the disk isotropic flux-freezing is not always a good approximation, even if it is not a bad approximation for the global behavior of the mass in the disk on large scales.} We can also assume $\langle |B_{\phi}|\rangle  \propto H_{A}^{-1}$ where $H_{A} \approx \langle v_{A,\,\phi} \rangle/ \Omega \approx \langle |B_{\phi}|/(\Omega\,\sqrt{4\pi\,\rho})$ is the scale height set by the magnetic pressure (specifically focusing on the toroidal component providing the vertical support). This gives 
\begin{align}
\label{eqn:bpred.fluxfreezing.rhor} B^{\rm pred} \propto \rho^{1/4}\,r^{-3/4} \ , 
\end{align} 
which also provides a good fit. 

To compare, Fig.~\ref{fig:b.tests.badmodels} plots the predicted value of $|{\bf B}^{\rm pred}|$ from models {\em (ii)}, {\em (iii)}, {\em (v}), and {\em (vi)}. We see that all of these models fail to correctly predict the magnitude of the typical fields and their dependence on radius within the disk (not just the field geometries and more detailed structure). Models {\em (ii)}, {\em (iii)}, and {\em (v)} fare especially poorly, predicting huge variations in the local value of $|{\bf B}|$ at a given annulus that we do not see in the simulations, predicting the incorrect radial trend (there is a systematic trend in the mean offset from the actual simulation $|{\bf B}|$), and predicting the incorrect normalization of $|{\bf B}|$. Model {\em (vi)} fares somewhat better but is still notably offset from the actual field strengths (and the fact that the disk is actually accreting appears to immediately contradict model {\em (vi)}). Model {\em (iv)} is not shown here because we have not explicitly separated the turbulent velocity fields/kinetic energies, but we show below (\S~\ref{sec:v:turb}) that the typical \Alf\ Mach numbers in the disk at $\lesssim 0.1\,$pc are $\mathcal{M}_{A} \sim 0.3-1$ -- i.e.\ the saturation magnetic field strength in the simulation is an order-of-magnitude larger than model {\em (iv)} would predict (and, again, the field geometry is completely different). Nonetheless, it is worth noting model {\em (iv)} appears perfectly reasonable as a description of both the field geometry and strength at much larger radii $R \gg $\,pc, far outside the disk (in the ISM). Thus we see that the model {\em (i)} variants clearly provide a much better fit to the simulated values of $|{\bf B}|$, compared to other hypotheses {\em (ii)}-{\em (vi)} above.

Third, in Fig.~\ref{fig:Bcompare.eulerian.lagrangian} we have followed the Lagrangian time evolution of individual fluid elements (since this is a Lagrangian code, this is numerically trivial), and verified that they obey the approximate model-{\em (i)} scalings and behaviors  {\em in time}, as well as in space. This is expected if the disk is in steady state with steady inwards accretion, and $\langle \rho \rangle$ or $H$ are monotonic functions of $R$, but it is important to validate. It also allows us to  confirm that the evolution occurs continuously as the material advects, and not, for example, only after it reaches some critical radius. We can also immediately confirm that the mean toroidal field is not amplified from some trace/turbulent seed field as models {\em (iii)}, {\em (iv)}, and ({\em vi}) predict.

Fourth, by examining different snapshots, we have verified that the sign flips in $B_{\phi}$ move with Lagrangian fluid elements over time exactly as expected in model {\em (i)}, as opposed to oscillating as they would if they arose from e.g.\ instabilities like the MRI (model {\em (iii)}) or the turbulent dynamo (models {\em (iv)} and {\em (v)}). We see this directly in Figs.~\ref{fig:bfields.edgeon.smallscale} and Fig.~\ref{fig:b.profile}, as well as via the fact that when following Lagrangian parcels we do not see sign flips. We illustrate this physics more explicitly in Fig.~\ref{fig:b.time.flip}, where we follow the gas at the inner radii of the simulation just before it is accreted. This Figure illustrates a number of important properties: (1) that the magnetic field strength, toroidal field prominence, and accretion rate into $<80\,$au are stable (to within a factor of a couple) over tens of thousands of dynamical times at our inner radii; (2) that sign flips occur, on a timescale comparable to the accretion timescale ($\sim \dot{M}_{\rm in} / M_{\rm gas}(<r)$), as new gas moves from larger radii into the annulus of interest; (3) that the field strength at a given radius is robust to these flips and restores quickly ``through'' the flip. We also see this in Fig.~\ref{fig:b.induction.eqn}, where at two different times more closely separated than the times in Fig.~\ref{fig:b.profile} we more plainly see the sign reversals in $\langle B_{\phi} \rangle$ systematically propagating inwards with the gas. As noted above, the form of the sign flips is also qualitatively inconsistent with model {\em (ii)}. 

Fifth, Fig.~\ref{fig:b.induction.eqn} compares the mean toroidal field $\langle B_{\phi} \rangle$ with its expected growth  if it was ultimately sourced primarily from advection of radial flux  -- as predicted by model {\em (i)} -- in a close-to-Keplerian disk. In particular, we approximate the induction equation for $\langle B_{\phi}\rangle$, $\partial_{t} \langle B_{\phi}\rangle = \langle\partial_{z}(B_{z} v_{\phi} - B_{\phi} v_{z} )\rangle + \langle\partial_{R}(B_{R} v_{\phi} - B_{\phi} v_{R})\rangle$, to include only the term that accounts for the stretching of the radial field by the toroidal flow $\partial_{t} \langle B_{\phi} \rangle \approx \partial_{R}\langle v_{\phi}\,B_{R} \rangle$. Note the pure radial $B_{\phi}$ transport term $\partial_{R} \langle B_{\phi}\,v_{R} \rangle$ is usually smaller, but not always negligible, while the vertical flux divergence $B_{z}\,v_{\phi}$ vanishes and the vertical inflow term $B_{\phi}\,v_{z}$ is small. In any case we see this approximation works well at describing quantitatively the sign flips and trends in $\langle B_{\phi}\rangle$. This behavior is generally distinct from the predictions of the alternative models {\em (ii)}-{\em (vi)} above. Moreover, if we multiply  $\partial_{R} \langle B_{\phi}\,v_{R} \rangle$  by the characteristic dynamical time $t_{\rm dyn} = \Omega^{-1}$, this appears to provide a remarkably good order-of-magnitude estimate of the saturation $\langle B_{\phi} \rangle$ -- this is expected if either (i) the accretion is dynamical (so the amplification time is limited to some multiple of $\Omega^{-1}$) or (ii) if the disk is trans-\Alf{ically} turbulent (so  $H \sim v_{\rm turb}/\Omega$ by definition in vertical equilibrium, and the turbulent magnetic dissipation time is $\sim \Omega^{-1}$) or (iii) if the midplane toroidal flux is lost via buoyancy on the vertical buoyancy timescale (a few to tens of times $\Omega^{-1}$). In any of these cases, provided the disk support is dominated by a mean toroidal field and Maxwell stresses with trans-\Alf{ic} turbulence (as we have here), the dimensional expectation for the rate at which flux is ``lost'' through a fixed Eulerian annulus is similar, and we can think of it as being ``replenished'' by advection of new radial+toroidal flux with the inflow from larger radii. So we can say -- effectively equivalently to our description above -- that the dynamo is ``closed'' by advection of mean radial+toroidal magnetic flux from the inflowing gas. 

Together, all of these comparisons quantitatively (and qualitatively) support our argument from \S~\ref{sec:b:origin:ov} that the mean field is ultimately sourced via flux-freezing (model {\em (i)} here), rather than via some other scenario (e.g.\ models {\em (ii)}-{\em (vi)}).

\begin{figure*}
	\centering\includegraphics[width=0.97\textwidth]{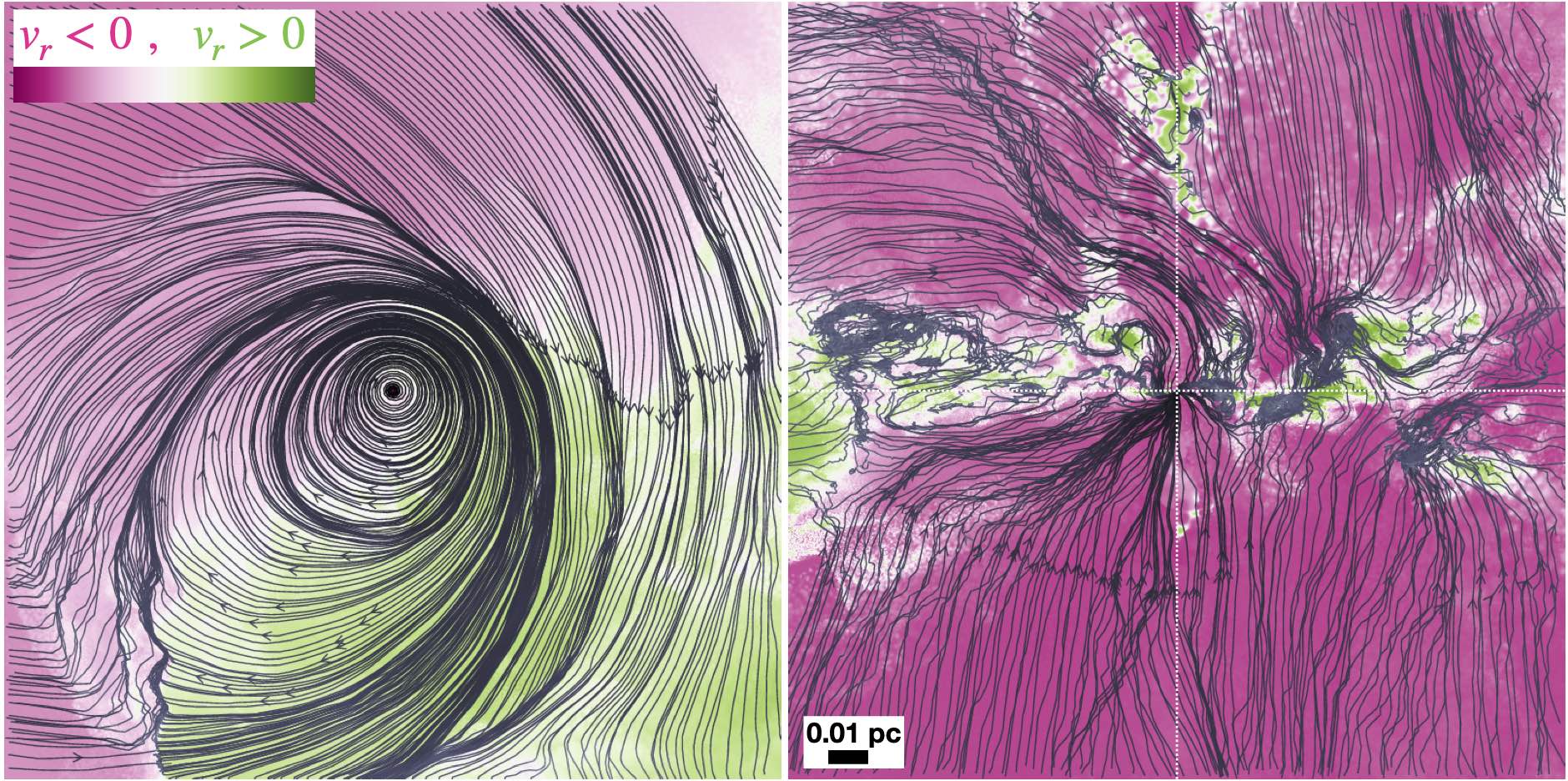}
	\caption{Velocity streamlines in a cylindrical wedge inside $<0.1\,$pc, as in  Fig.~\ref{fig:bfields.faceon.edgeon}, face-on in the midplane ({\em left}, showing $v_{R}$-$v_{\phi}$) and edge-on ({\em right}, showing $v_{R}$-$v_{z}$), colored by the sign of the (spherical) $v_{r} \equiv {\bf v} \cdot \hat{\bf r}$ (where the rest frame of the system is defined as the velocity of the SMBH). Compared to the magnetic field, the velocity fields are much more clearly dominated by coherent rotation in the disk (there are no sign flips), and vertical infall onto the disk. Gas pileups at spiral shocks forming the arms in Figs.~\ref{fig:image.zoom}-\ref{fig:image.faceonedgeon.inner} are obvious, with increasing order inside $<0.01\,$pc. The next-order inflow/outflow motion in the midplane is clearly dominated by a coherent, large-scale eccentric disk ($m=1$) ``slow mode'' which propagates inwards from large $r$ where the disk is self-gravitating, but the shocks and turbulence break the exact cancellation of this motion and lead to net angular momentum loss. In the edge-on midplane, we more clearly see the turbulent internal disk dynamics. The eccentric disk slowly precesses around the SMBH lagging the path of the infalling cloud from which it is forming, as expected.
	\label{fig:vfields.lines}}
\end{figure*}

\begin{figure}
	\centering\includegraphics[width=1.01\columnwidth]{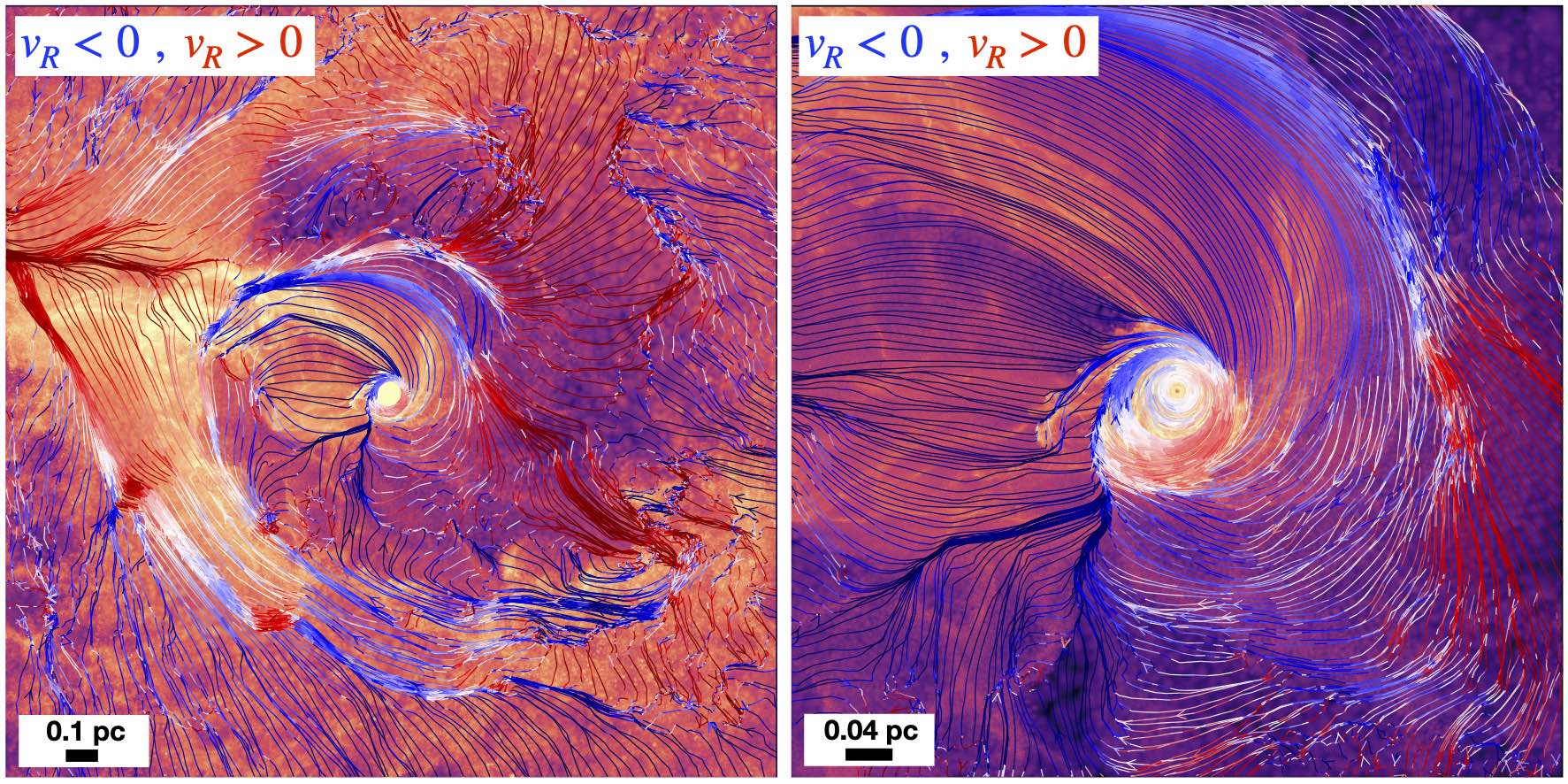}
	\caption{Velocity streamlines (as Fig.~\ref{fig:vfields.lines}), but now with the lines color-coded by the sign of the radial velocity (inflow/outflow) and super-imposed on a face-on projected density map of the gas in the disk midplane (as Figs.~\ref{fig:image.zoom}-\ref{fig:image.faceonedgeon.inner}) and on somewhat larger scales. This highlights the flows that form the disk from captured gas initially beyond the BHROI in a massive star-forming complex. As generically expected in tidal capture/disruption events, a large-scale stream falls in primarily radially, shocking on self-intersection, and circularizes to form the disk, with an order-unity fraction of the complex mass not bound and instead in a large unbound tidal stream or outflow ``fan.'' The inflow, being tidally compressed, forms the denser filamentary structure while the outflow becomes lower-density gas. Within the inflowing filament in this ``free fall'' zone before it circularizes, the flow is not highly turbulent: there is both outflow and inflow (so the variance in $v_{R}$ can be large, at a given $R$), but these clearly reflect large-scale, geometrically separate structures rather than turbulent local motions.
	\label{fig:vfields.inflow.outflow.largescales}}
\end{figure}

\subsubsection{Discriminating Between Models for the ``Turbulent'' Field}
\label{sec:b:origin:turb}

Now we consider the sub-dominant, but still not negligible, fluctuating or ``turbulent'' field components $\delta {\bf B}$. Given the observed strength of the turbulent velocity fields, which are trans-\Alf{ic} and crudely isotropic, the typical magnitude of the fluctuating field components $\delta B_{i}$ in Fig.~\ref{fig:b.profile} are consistent with the usual trans/sub-\Alf{ic} relation $|\delta B| / |{\bf B}| \sim |\delta v|/ v_{A}$, as expected. This implies that the origins of the turbulent/fluctuating magnetic-field components are likely related to the origins of the turbulence in the disk, which we will examine in detail in the next section.

However, one important question is whether we are in the regime of the ``traditional'' MRI. As shown in Fig.~\ref{fig:b.tests.badmodels}, the typical magnetic-field strength in the disk (both $|\langle B_{\phi} \rangle|$ and $|{\bf B}|$) is an order-of-magnitude or more larger than the characteristic strength $v_{A} \sim \sqrt{c_{s}\,v_{\rm K}}$ at which the linear growth rate of the MRI is usually assumed to vanish following the analytic analysis in e.g.\ \citet{pessah:2006.mri.signature.ratio.maxwell.reynolds}. And Figs.~\ref{fig:bfields.faceon.edgeon}-\ref{fig:bfields.edgeon.oplotrho} do not show obvious magnetic morphological signatures of the MRI (e.g.\ channel modes). We further show below that the ratio of Maxwell to Reynolds stress differs significantly from commonly-quoted saturated weak-seed-field MRI simulation results through much of the disk  \citep{brandenburg:mri.predicted.high.ratio.maxwell.to.reynolds}.

But it is well known that for fields stronger than the commonly-quoted MRI limit ($v_{A} \gtrsim \sqrt{c_{s}\,v_{\rm K}}$), magnetized disks are still unstable -- the instabilities can simply change in character \citep{pitts.tayler:1985.toroidal.buoyancy.modes.rotating.stars,terquem:1996.toroidal.field.bouyancy.instability,kim.ostriker:2000.mhd.instab.shearing.cold.winds,pessah.psaltis:2005.mri.extensions.stronger.fields,hirabayashi:2016.toroidal.field.bouyancy.modes,das:2018.pessah.psaltis.limit.mri}. We therefore follow \citet{pessah.psaltis:2005.mri.extensions.stronger.fields} and plot the simulations on the ``mode diagram'' shown in e.g.\ Fig.~3 therein, in our Fig.~\ref{fig:instability.map}. Specifically, those authors show that the parameter space of different characteristic unstable modes -- at least in a simplified analytic linear stability analysis of a laminar, unstratified disk -- can be, to leading order, represented in a two-dimensional plot of dimensionless vertical wavenumber $\tilde{k}_{z} \equiv k_{z}\,r\,v_{A,\,z}/c_{s}$ (with $v_{A,\,i} \equiv |B_{i}|/\sqrt{4\pi\,\rho}$) versus a measure of the dimensionless toroidal magnetic field strength $v_{A,\,\phi}/c_{s}$. The parameter space is then divided into four characteristic regimes based on the intersection of two critical wavenumbers: $\tilde{k}_{z}^{c1} \equiv k_{z}^{c1}\,r\,v_{A,\,z}/c_{s} \approx (V_{c}/c_{s})\,\sqrt{-2\,{\rm d}\ln{\Omega}/{\rm d}\ln{r}}\approx \sqrt{3}\,v_{\rm K}/c_{s}$, and $\tilde{k}_{z}^{c2} \equiv k_{z}^{c2}\,r\,v_{A,\,z}/c_{s} \approx (V_{c}/c_{s})\,\left[v_{A,\,\phi}^{4}/c_{s}^{2}\,V_{c}^{2} - \kappa^{2}/\Omega^{2}\right]^{1/2} \approx \sqrt{(v_{A,\,\phi}/c_{s})^{4} - (v_{\rm K}/c_{s})^{2}}$. Modes with $\tilde{k}_{z} > {\rm MAX}(\tilde{k}_{z}^{c1}\, , \, \tilde{k}_{z}^{c2})$ are stable; those with $\tilde{k}_{z}^{c2} < \tilde{k}_{z} < \tilde{k}_{z}^{c1}$ are unstable to the traditional MRI (with the near-vertical part of $\tilde{k}_{z}^{c2}$ defining the upper limit $v_{A,\,\phi} < \sqrt{c_{s}\,v_{\rm K}}$ above which, for the conditions considered in \citealt{pessah.psaltis:2005.mri.extensions.stronger.fields}, the MRI growth rate vanishes);\footnote{If we retain the radial stratification term $\hat{B}_{\phi} \equiv \partial \ln{B_{\phi}}/\partial \ln{R}$, then $\tilde{k}_{z}^{c2}$ is modified to $\sqrt{(1+\hat{B}_{\phi})^{2}\,(v_{A,\,\phi}/c_{s})^{4}-(v_{\rm K}/c_{s})^{2}}$, so the boundary for MRI-like behavior shifts to $v_{A,\,\phi} < \sqrt{c_{s}\,v_{\rm K}/|1+\hat{B}_{\phi}|}$. If we include this $|1+\hat{B}_{\phi}|^{-1/2}$ correction in Fig.~\ref{fig:instability.map}, the effect is small here: the $\tilde{k}_{z}^{c2}$ boundary between MRI and ``Type III'' shifts upwards by a factor $\sim 1.1-1.4$ in each panel, which has no effect on our conclusions. Although $\hat{B}_{\phi}$ is broadly similar to $-1$ on average in Fig.~\ref{fig:bfield.demo}, evaluating this correction term in factor $\sim 2$ radial intervals from $\sim 80$\,au to $\sim 10\,$pc, we find that $|1+\hat{B}_{\phi}|^{-1/2}$ never exceeds a factor of $\sim 3$.} those with $\tilde{k}_{z}^{c1} < \tilde{k}_{z} < \tilde{k}_{z}^{c2}$ are unstable to a second or ``Type II'' instability; and those with $\tilde{k}_{z} < {\rm MIN}(\tilde{k}_{z}^{c1}\, , \, \tilde{k}_{z}^{c2})$ are unstable to a third or ``Type III'' instability.\footnote{Technically, in \citet{pessah.psaltis:2005.mri.extensions.stronger.fields} there is a small stable ``strip'' just interior to the boundaries of the ``Type III'' parameter space. However on the (large) dynamic range plotted in Fig.~\ref{fig:instability.map}, this occupies a negligibly small fraction of the parameter space and is unimportant for our comparison. Moreover, \citet{das:2018.pessah.psaltis.limit.mri} argue this strip vanishes (the full Type-III regime is unstable) in a global mode analysis.} Note that the ``Type II'' and ``Type III'' instabilities in \citet{pessah.psaltis:2005.mri.extensions.stronger.fields} are also called axisymmetric toroidal buoyancy (ATB) mode[s] in \citet{kim.ostriker:2000.mhd.instab.shearing.cold.winds}, or superthermal slow mode instability (SSMI) and suprathermal hybrid mode instability (SHMI) in \citet{das:2018.pessah.psaltis.limit.mri} -- while there are subtle but important differences in these analyses, the order-of-magnitude dividing criteria between the different instability regimes and key behaviors are, for our purposes, identical. We plot the simulation gas in the disk, at each of several radii, on this diagram, assuming that the characteristic minimum wavenumber of interest (and wavenumber containing most of the power) is $k_{z} \sim 2\pi/H$. We see that the simulations lie solidly in the ``Type II/III range,'' even if we focus only on the warmer (higher-$\beta$) volume-filling phases of the gas in the disk where it is multi-phase (in the colder, denser gas, $\beta$ is even smaller and the simulations lie even further from the traditional MRI regime). This confirms our intuition and quantitative statement that $v_{A,\,\phi} \gg \sqrt{c_{s}\,v_{\rm K}}$ above.

These specific modes are fundamentally related to radial magnetic buoyancy (see references above), operating near the midplane, but as noted in \citet{pessah.psaltis:2005.mri.extensions.stronger.fields} they generically involve comparable in-plane and vertical displacements. Moreover, given that the disk is vertically stratified at some level (see \S~\ref{sec:v:vertical}), additional buoyancy modes in a manner potentially similar to that discussed in idealized simulation studies such as \citet{johansen.levin:2008.high.mdot.magnetized.disks}. While the analytic models of such modes are somewhat less clear in the regime here (non-constant $\beta \ll 1$, non-isothermal ${\rm d}T/{\rm d}|z| > 0$, strongly differentially rotating), dimensional considerations and simpler versions of said instabilities suggest that the fastest-possible growth timescales (a few times the vertical \Alf\ crossing time, $\sim {\rm a\ few}\,\Omega^{-1}$) at the characteristic scales $\sim H$ should be order-of-magnitude similar to the Type II/III modes above \citep[see e.g.][and references therein]{foglizzo:1994.parker.instab.with.differential.rotation,vishniac:1995.mhd.turb.acc.disks,rodrigues:2016.parker.instab.disk.galaxies,salveson:2016.sims.mri.dominated.bfield.disks}. Together this would naturally explain why we see broadly similar turbulent vertical and radial components $|\delta B_{z}| \sim |\delta B_{R}|$ (Fig.~\ref{fig:b.profile}) with similar-scale structures (Fig.~\ref{fig:bfields.faceon.edgeon}). 
The Type III/SHMI instability in particular is also robust to the different vertical and radial density/pressure/magnetic stratification terms and range of mode propagation angles considered in \citet{pessah.psaltis:2005.mri.extensions.stronger.fields,das:2018.pessah.psaltis.limit.mri}. It depends on differential rotation in a similar manner to the traditional MRI and its linear eigenvectors feature a broadly similar structure: notably the linear Type III/SHMI instability, like the linear MRI, always produces both Maxwell and Reynolds stresses which transport angular momentum outwards (dominated by the $R\phi$ component), although the linear mode Reynolds-to-Maxwell ratio can be higher or lower than the traditional MRI over the parameter space spanned by the simulations. And its growth rate peaks at relatively long wavelengths. In fact, if we insert values of the simulation parameters (including the radial gradients in $\rho$ and $B_{\phi}$) from Fig.~\ref{fig:instability.map} into the equations from \citet{pessah.psaltis:2005.mri.extensions.stronger.fields}, we find the simulations can often be in a parameter space which produces an even faster-growing, longer-wavelength version of their Type III instability (Maxwell-stress dominated, with peak growth rate of $\sim (|v_{A,\,z}|/v_{\rm K})^{1/3}\,\Omega$ at wavenumber $k \sim 1/H \sim \Omega/v_{A,\,\phi}$). Altogether, this suggests these modes could play an important role in driving turbulence and angular momentum transport, but clearly further non-linear simulation studies are needed.\footnote{In contrast, the Type II/SSMI instabilities operate on much shorter wavelengths, do not depend on differential rotation, and their linear eigenmodes feature weak Maxwell and Reynolds stresses with anti-aligned/opposing angular momentum transport, so in this sense they are more akin to local convective instabilities.}

Briefly, the fact that the simulation modes with wavelength $\sim H$ reside order-of-magnitude around the Type II/III (or SHMI/SSMI) dividing line ($\tilde{k}_{z}^{c1}$) is not actually surprising: when $(2\pi\,r/H)\,(v_{A,\,z}/c_{s}) \sim \tilde{k}_{z}^{c1} \sim \sqrt{3}\,v_{\rm K}/c_{s}$ in a disk dominated by toroidal magnetic pressure with $H \sim v_{A,\,\phi}/\Omega$, the Type II/III dividing line is equivalent to $v_{A,\,\phi} \sim 4\,v_{A,\,z}$. So residing broadly ``near'' this line is simply a statement that the typical $|B_{\phi}|$ is not orders-of-magnitude larger than the typical $|B_{z}|$ in a cell. Note that we use the value of $|B_{\phi}|$ and $|B_{z}|$ in each cell for this histogram (to define $v_{A,\,\phi}$ and $v_{A,\,z}$) -- if we instead replaced these values with the mean field $|\langle B_{\phi}\rangle|$, $|\langle B_{z} \rangle|$ (closer to the assumption in \citealt{pessah.psaltis:2005.mri.extensions.stronger.fields}) then $v_{A,\phi}$ and the horizontal position of the simulation changes very little (since this is mean-field dominated per Fig.~\ref{fig:b.profile}), but $v_{A,\,z}$ is reduced by a factor of $\sim 3-10$. This places modes with wavelength $\sim H$ more firmly in the lower-wavelength or ``Type III/SHMI'' regime (these modes more strongly depend on, and interact with, the differential rotation). But if we considered a somewhat larger wavenumber $|k_{z}| > 1/H$, then the vertical position of the simulations would instead shift upwards by a corresponding factor towards the more local-mode ``Type II/SSMI'' regime.

Briefly, some other instabilities may be less likely to drive the magnetic fluctuations we see. As discussed in a number of previous studies \citet{machida:2006.mag.supported.disks.from.cooling.around.adafs,begelman.pringle:2007.acc.disks.strong.toroidal.fields,oda:2009.analytic.mag.disk.structure.models,sadowski:2016.mag.elevated.disk.sims.radpressure.midplane.thermal.instability.suppressed,habibi:2019.thermal.instabilities.magnetically.dominated.disks}, magnetically-dominated disks such as these are generically stable against the usual (linear) viscous and thermal instabilities. The disks here are also (linearly) stable against the short-wavelength (${\bf k} \perp {\bf B}$) Parker-like magnetic convective/Rayleigh-Taylor/interchange modes, as these require $d\ln{(|B|/\rho)}/dz < 0$ (for $z>0$ and $\beta \ll 1$; \citealt{tayler:1973.magnetic.instabilities.interchange,terquem:1996.toroidal.field.bouyancy.instability,kim:2002.mhd.disk.instabilities}), i.e.\ that the magnetic scale-height is smaller than the density scale-height, which we plainly  show in \S~\ref{sec:v:vertical} is not satisfied. And although the long-wavelength Parker instability requires only a vertically-decreasing $|{\bf B}|$ ($d\ln{B^{2}}/dz < 0$), the characteristic wavelengths $\lambda \gtrsim \lambda_{\rm crit} \sim (5-10) \times H$ \citep{parker:1966.magnetic.instability.disks,kim:1997.parker.instab.linear.gravity.relevant.for.acc.disks,lee:2007.parker.instab.analysis.char.modes.wavelengths} are extremely large ($> R$, given the relatively large $H$ here), and the trans-\Alf{ic} magnetic fluctuations are much larger than commonly-quoted thresholds above which the instability may be strongly suppressed \citep[see][and references therein]{kim:2001.turb.strongly.suppresses.parker.instability}, so it is not clear if these modes can actually exist (at a minimum, a global analytic treatment is required).

Of course, all of these analytic ``dividing lines'' are predicated on analytic linear stability analysis, with a number of simplifying assumptions (e.g.\ that the disk is azimuthally symmetric, laminar, and adiabatic, and that various vertical and radial stratification terms can be neglected). It is not obvious, therefore, how much can be applied to simulations like ours with complicated stratification, fully-developed strong turbulence, cooling, and highly non-linear modes. And other instabilities or variants of those discussed above may be present as well. For example, as discussed in \citet{begelman:2023.mri.saturation.estimates}, the ``traditional'' MRI can persist in a supra-thermal form if $\partial \ln{B_{\phi}}/\partial \ln{R}$ is very close to $-1$ (interestingly similar to the ``average'' slope we see in Fig.~\ref{fig:bfield.demo}), although from the linear analysis in \citet{pessah.psaltis:2005.mri.extensions.stronger.fields} this would require $|1+\partial \ln{B_{\phi}}/\partial \ln{R}| < \beta\,(v_{\rm K}/v_{A,\,\phi})^{2} \sim \beta\,(H/R)^{-2} \ll 0.01-0.1$ for the disk parameters here, so if it is occurring here it may be a transient phenomenon in space and/or time. It is difficult to speculate further -- our intent here is to motivate more exploration in both analytic studies and non-linear but idealized numerical simulations of magnetic instabilities in the strong toroidal field parameter space of interest here, and to highlight that the field strengths here are not in fact restricted to the sometimes-quoted value of $v_{A} \lesssim \sqrt{c_{s}\,v_{\rm K}}$ (as assumed in e.g.\  \citealt{begelman.pringle:2007.acc.disks.strong.toroidal.fields}).

\subsubsection{Comparison to Field Strengths in ``Traditional'' $\alpha$ Disks}
\label{sec:magnetic.compare.SS73}

As shown in \paperthree, if we take the scalings for the outer accretion disk from \citet{shakurasunyaev73} for an SS73-like ``weakly-magnetized'' ($\beta \ll 1$) $\alpha$-disk, then building up a sufficient Maxwell stress to produce the canonical $\alpha \sim v_{A}^{2}/c_{s}^{2} \sim 0.1$ in such a disk would actually require magnetic fields whose absolute strength (in Gauss) is approximately a factor of $\sim 10$ larger than those seen in our simulations  (e.g.\ from SS73 Eq.~2.19 therein, $B_{\rm SS73} \sim 5000\,{\rm G}\,(R/0.001\,{\rm pc})^{-1.3}$ for $\dot{M} \sim 20-30\,{\rm M_{\odot}\,yr^{-1}}$ and $M_{\rm BH} \sim 10^{7}\,{\rm M_{\odot}}$). Essentially, ``weakly-magnetized'' models such as SS73, as well as the vast majority of historical accretion disk simulations, make the implicit assumption that the disk ``initially'' formed with negligible vertical magnetic support (i.e.\ vanishingly small or strictly vertical magnetic fields). This implies the disk would collapse to much smaller scale-heights $H/R \sim c_{s}/V_{c}$, with midplane densities $\rho$ a factor of $\sim 10^{6}-10^{8}$ larger than those seen here (\S~\ref{sec:basic:profile}). This in turn would require some process like the MRI to very efficiently amplify $|{\bf B}|$ up to quite large absolute values to produce even a modest \Alf\ speed $v_{A} \sim |{\bf B}|/\sqrt{4\pi\,\rho}$, as needed to produce a Maxwell stress that is any appreciable fraction of $\rho\,V_{c}^{2}$. 

This of course has important consequences for the observational properties of disks like those simulated here: even though $v_{A}/c_{s}$ and $v_{A}/V_{c}$ are much larger here than in a traditional SS73-like disk, $|{\bf B}|$ can actually be significantly {\em smaller}. It also re-emphasizes that the absolute field strengths in our simulations are not particularly extreme or implausible (\S~\ref{sec:b:ov}). And it further suggests that if one ``initially'' forms the disk from gas with more realistic ISM magnetic field strengths (with non-trivial toroidal and/or radial fields), it will likely reach a magnetically-dominated state akin to the simulations here, well before it could actually collapse to extreme densities like those assumed for ``weakly-magnetized'' SS73-like disks.

\begin{figure*}
	\centering\includegraphics[width=0.97\textwidth]{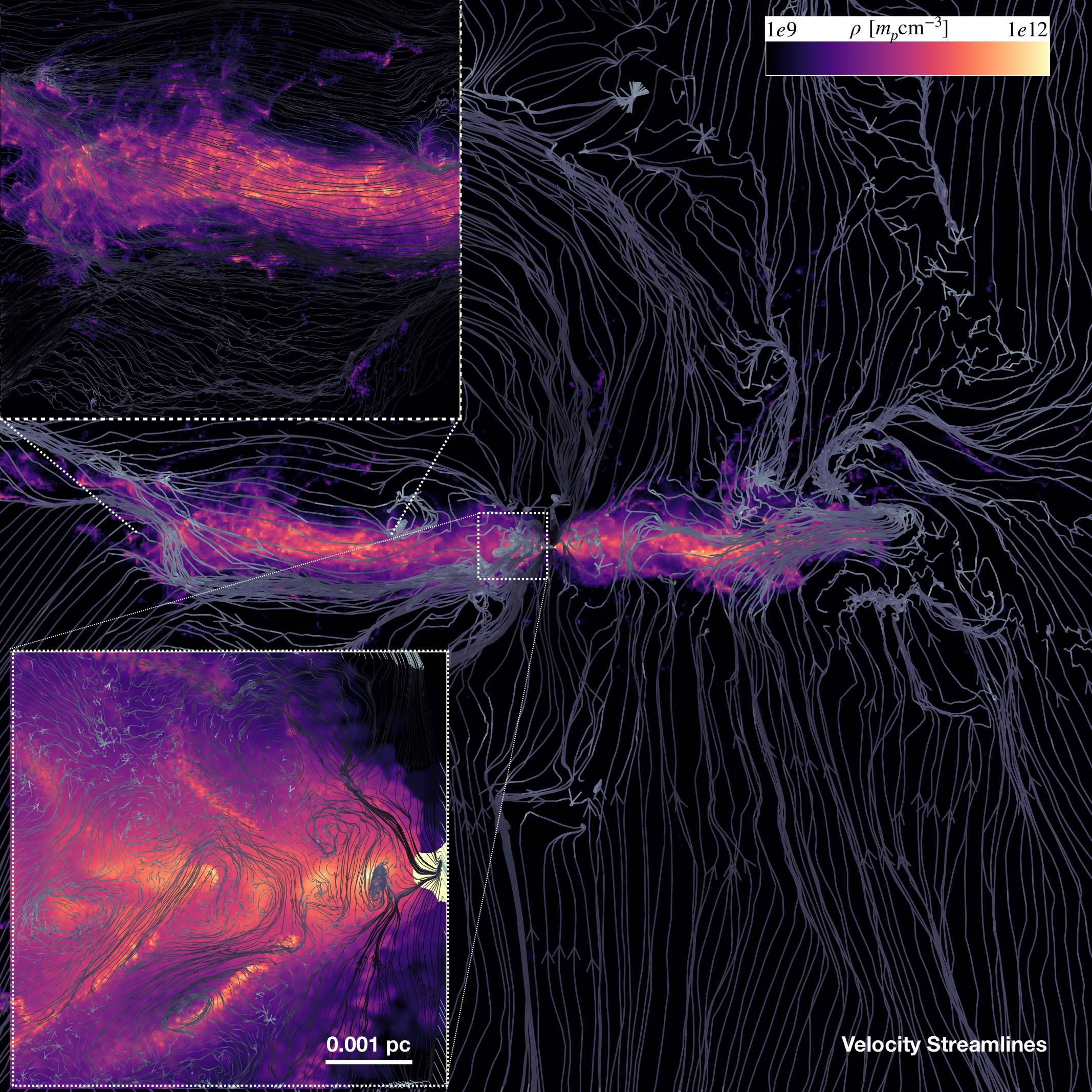}
	\caption{Velocity field line structure on top of a gas density projection, as Fig.~\ref{fig:bfields.edgeon.oplotrho}. We see clear radial motion through the midplane of the disk carrying most of the mass (more obviously showing mean velocity in this direction, as compared to the magnetic field line structure in Fig.~\ref{fig:bfields.edgeon.oplotrho}), and vertical inflow onto the disk from the tenuous atmosphere. Interior to the disk we see turbulent cells of size $\sim H$.
	\label{fig:vfields.lines.rho}}
\end{figure*}

\section{Structure of the Velocity Fields}
\label{sec:v}

\subsection{Overview}
\label{sec:v:ov}

\begin{figure*}
	\centering\includegraphics[width=0.97\textwidth]{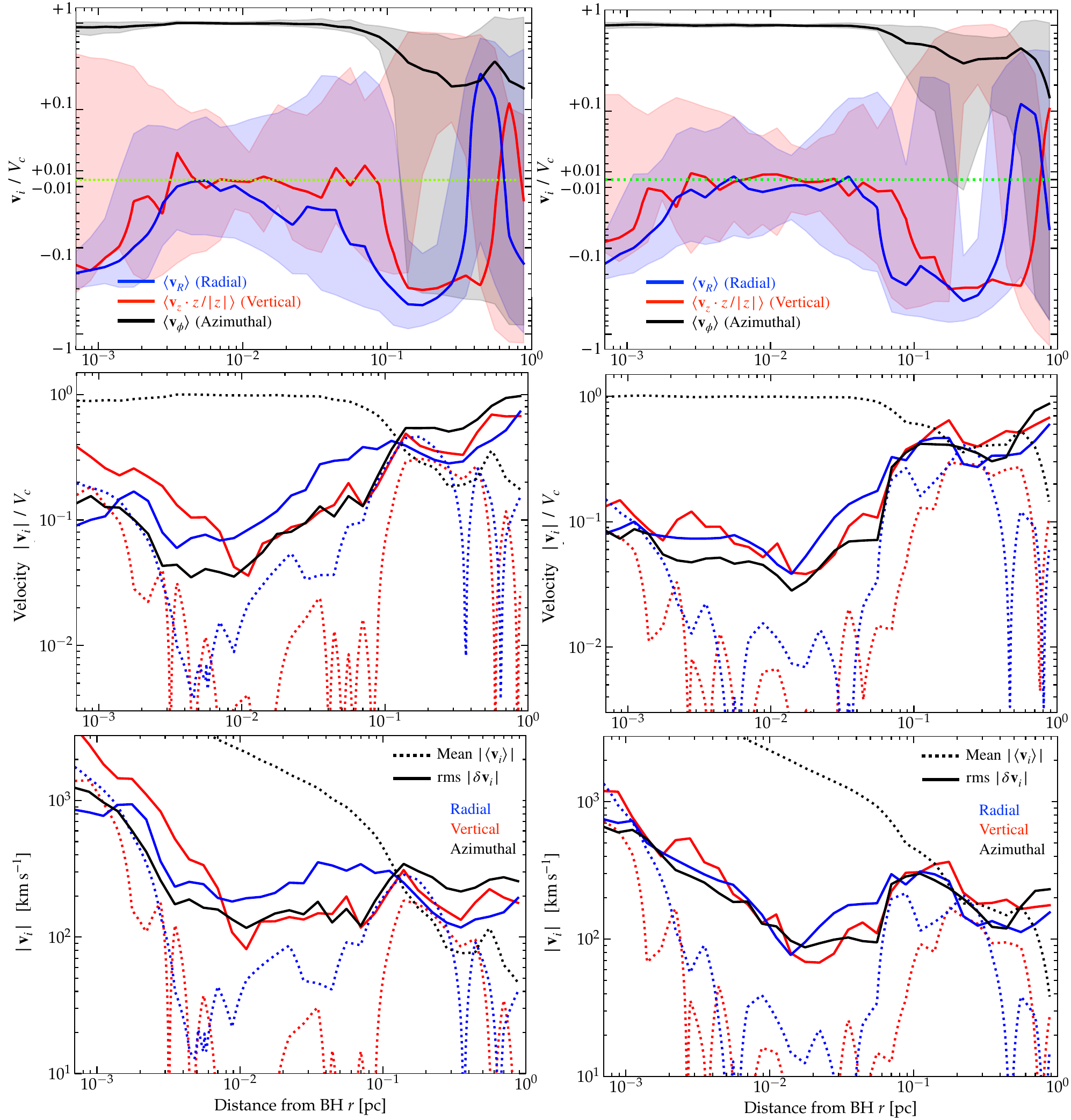}
	\caption{Mean velocity with its dispersion ({\em top}), and separation of the mean $\langle {\bf v}_{i} \rangle$ and fluctuating $\delta {\bf v}_{i}$ velocity components in both units of $V_{c}$ ({\em middle}) and ${\rm km\,s^{-1}}$ ({\em bottom}), in the same style and showing the same two times ({\em left} and {\em right}) as Fig.~\ref{fig:b.profile}. We see quantitatively that the velocity fields are more ordered than the magnetic fields. Inside the disk ($\lesssim 0.1$\,pc) the velocity is clearly dominated by rotational motion, with the turbulent/fluctuating components broadly all similar to each other in magnitude at $\delta {\bf v}_{i} \sim 0.1\,V_{c}$. Almost always we see $\langle v_{R} \rangle < 0$, indicating inflow, as expected, with modest amplitude. For $v_{z}$ we show $\langle v_{z} \cdot z/|z| \rangle$, so that a value of $>0$ or $<0$ indicates vertical outflow or inflow respectively (regardless of whether the gas parcel happens to be ``above'' or ``below'' the midplane), showing weak vertical inflows as well.
	\label{fig:v.profile}}
\end{figure*}

In Figs.~\ref{fig:vfields.lines}-\ref{fig:vfields.lines.rho} we plot the structure of the velocity fields (face-on and edge-on). We clearly see radial infall at large $r$, circularizing and forming the disk (Figs.~\ref{fig:vfields.lines}-\ref{fig:vfields.inflow.outflow.largescales}). Face-on, the disk appears highly ordered, but with an obvious coherent eccentric disk mode present in Fig.~\ref{fig:vfields.lines}. Edge-on, we see vertical inflow onto the disk, with a thick turbulent midplane layer evident in Fig.~\ref{fig:vfields.lines.rho}. Note that edge-on, since we show below $|\delta v_{R}| \gtrsim |v_{R}|$ and $|\delta v_{z}| \gtrsim |v_{z}|$, this is essentially the same as a plot of the velocity fluctuations. This shows, as we saw above in the morphology and ${\bf B}$, that the midplane is not a uniform, perfectly rigid layer but features a complex internal density structure with many warps and even streams with somewhat different orientations at large radii. Fig.~\ref{fig:v.profile} plots the radial profile of the different velocity components (in absolute units and relative to $V_{\rm c}(r)$), showing both the mean and fluctuating velocities. 

As discussed in \paperone, at radii $\gg$\,pc, outside the BHROI, the velocity structure in the nucleus is primarily turbulent with quasi-isotropic velocity fields -- we see this already at $\gtrsim 0.3\,$pc in Figs.~\ref{fig:vfields.lines}-\ref{fig:v.profile}, where the incoherent or ``turbulent'' velocity components dominate over the mean velocities, and are comparable both to one-another and to the circular velocity $\delta v_{i} = \langle |{\bf v}_{i} - \langle {\bf v}_{i} \rangle|^{2} \rangle^{1/2} \sim V_{\rm c} \gg \langle {\bf v}_{i} \rangle$. At $\lesssim 0.1\,$pc, we see the disk clearly in kinematic space, with $\langle v_{\phi} \rangle \approx V_{\rm c}$ much larger than other components, which have a dispersion $\sim 0.1\,v_{\phi}$. The kinematics of the disk are much more coherent compared to ${\bf B}$ above: there is a global smooth rotation-dominated flow with no sign flips (i.e.\ all the material is rotating the same direction, as expected).

\subsection{Vertical Structure}
\label{sec:v:vertical}

Vertically we see infall from above/below onto the disk, at most radii; but we stress this does not dominate the total inflow rate, most of which occurs through the thicker and denser midplane. This may change if we modeled emergent ``feedback'' (e.g.\ jets or high-velocity outflows) from the un-resolved portions of the accretion disk at $<80\,$au. 

We discuss stratification in greater detail below, but we do not see any evidence for coherent stratification of the velocity structure within the disk (at $-H \lesssim z \lesssim H$).

\subsection{The Eccentric Disk}
\label{sec:v:ecc}

\begin{figure}
	\centering\includegraphics[width=0.95\columnwidth]{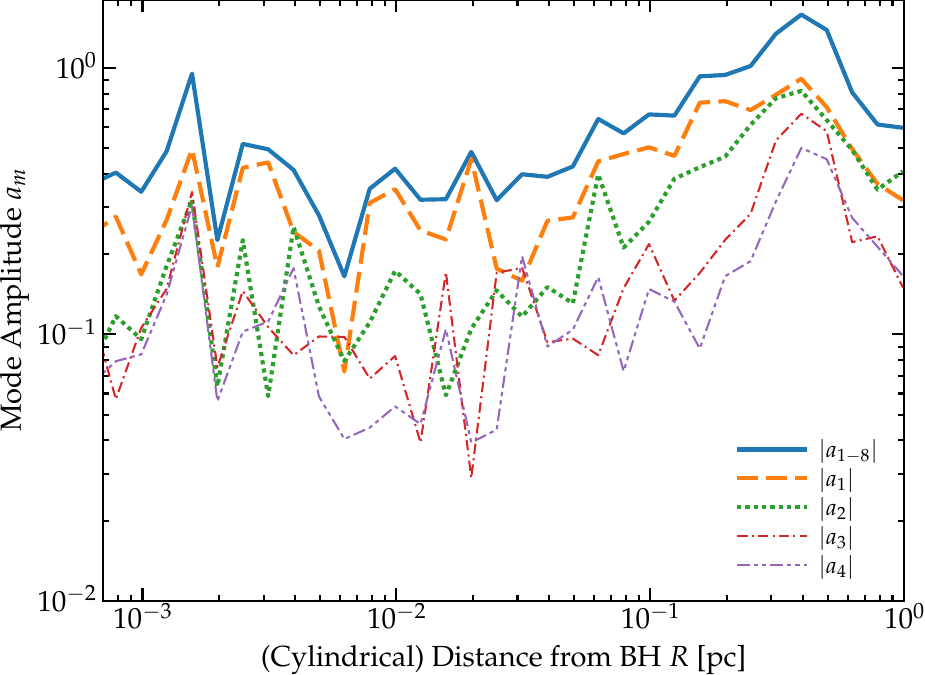}
	\caption{Amplitude of the asymmetric modes in the gas density distribution: fitting the (face-on) projected surface gas density to $\Sigma_{\rm gas}(R) = \Sigma_{0}(R)\,[ 1 + \sum_{m=1}^{\infty}\,a_{m}(r)\,\cos{(m\,\phi + \phi_{0,\,m}[R])}]$ and plotting $|a_{m}|$. Here $|a_{1-8}|$ sums the first 8 modes in quadrature, showing this is dominated by the global $m=1$ mode at all the radii shown. The relatively large values of $a_{m}$ clearly indicate that the disk has strong spiral and coherent eccentric disk structure that can drive shocks and inflow, as clearly evident in the velocity field lines in Fig.~\ref{fig:vfields.lines}.
	\label{fig:eccentricity.profile}}
\end{figure}

Following the system in time, we see the initial inflows as the gas captured first falls into the center and circularizes to form an eccentric, lopsided structure with a clear shock/compression region as orbits self-intersect (Fig.~\ref{fig:vfields.inflow.outflow.largescales}). This ``settles'' over some tens of dynamical times into a smoother structure with more coherent velocity structure visible in Fig.~\ref{fig:vfields.lines}. The eccentricity does not disappear even over many thousands of dynamical times. Rather, the system settles into a coherent $m=1$ eccentric disk mode, where we clearly see the eccentricities of individual gas orbits align, and the entire eccentric structure precesses with a slow pattern speed (pattern speed $\Omega_{p} \sim \Omega(r_{\rm max})$ where $r_{\rm max} \sim $\,pc, so at smaller radii $\Omega_{p} \ll \Omega(r \ll r_{\rm max})$). The amplitude of the eccentricity declines weakly as $r\rightarrow 0$. Fig.~\ref{fig:eccentricity.profile} plots the amplitude of the eccentric ($m=1$) and higher-$m$ modes in the gas surface density in the face-on surface density projection of the disk, as a function of radius, which demonstrates this explicitly.

These are exactly the predicted structures for ``slow'' $m=1$ modes in nearly-Keplerian potentials, which are well-studied and (at least in linear theory) can persist indefinitely \citep{tremaine:slow.keplerian.modes,bacon:m31.disk,hopkins:slow.modes}. It is important to note that these are unique among global/large-scale modes in the orbit structure: other modes, e.g.\ $m=2$ bars or higher $m>2$, are damped as $r\rightarrow 0$, but for slow $m=1$ modes there is essentially zero energetic cost of the mode propagating inwards as it just involves coherent alignment of already-eccentric orbits. The pattern speed is set by the ``driving'' of the eccentric mode: namely, the motion of the parent gas complex which is being tidally stripped by the SMBH to fuel the accretion event and form the disk in the first place. That complex both torques the disk directly (as its mass is larger than that of the SMBH and most of the complex lies outside the BHROI), and provides the newly-infalling gas which follows the trajectory of its parent cloud (itself on an effectively hyperbolic orbit) as it passes through some impact parameter or pericenter $b_{\rm complex}$. The cloud complex therefore drives a characteristic (lagging) pattern frequency $\Omega_{\rm eff} \sim |\delta v| / b_{\rm complex} \sim \sigma_{\rm galaxy} / b_{\rm complex} \sim 200\,{\rm km\,s^{-1}} / {\rm a\ few\ pc} \sim 10^{-12}\,{\rm s^{-1}}$, i.e.\ precession on a $\sim 10^{4}\,$yr timescale.

This means that in an {\em instantaneous} sense, for a given gas parcel in the outer disk, its radial (inflow/outflow) velocity is dominated by where it is instantaneously in its eccentric orbit. This is clear in the face-on projections of the velocity streamlines in Figs.~\ref{fig:vfields.lines}-\ref{fig:vfields.inflow.outflow.largescales}. But  to leading order, these orbits are of course closed and their radial flow cancels  over the course of the full orbit, so we stress that this should not be conflated with the systematic or {\em net} inflows feeding accretion or outflows ejecting material from the nucleus. Instead, as expected, the combination of some shocks/dissipation (which break the exact symmetries of the eccentric orbits), plus non-zero local turbulent/Reynolds/Maxwell stresses, means that there is a non-zero torque on the gas which causes a systematic {\em net} inflow/accretion of gas. But it is worth noting that the leading-order description of the outermost disk is not a perturbed circular disk, but a perturbed eccentric disk. However, the fact that there is also clearly less-coherent/global more ``turbulence''-like cells or eddies in the inner disk ($R\lesssim 0.01\,$pc) with coherence length $\sim H$ in the midplane and $|\delta {\bf v}_{z}| \sim |\delta {\bf v}_{R} |$ is evident in the edge-on projections in Figs.~\ref{fig:vfields.lines} \&\ \ref{fig:vfields.lines.rho}.  

We can attempt to separate the coherent eccentric motion in our definition of $\delta{\bf v}$ (see \S~\ref{sec:methods:analysis}), by subtracting the best fit $m=1$ component from $\delta v_{i}(R,\,\phi)$. We have done so (fitting this independently in each radial annulus, which we caution may over-estimate the coherent component), and find that it has a modest effect most notably on $\delta v_{R}$ in the outer disk ($0.01\,{\rm pc} \lesssim R \lesssim 0.1\,{\rm pc}$), reducing it by a factor of $\sim 2$ (and a much smaller effect at smaller or larger radii, as expected). Interestingly after doing so, the residual velocity fluctuations in the outer disk become closer to isotropic, suggesting the more traditionally ``turbulent'' component is indeed close-to-isotropic. We show below (\S~\ref{sec:no.mhd}) that this is quite different from the case where we re-run without magnetic fields, where the disk is much thinner and exhibits much more extreme anisotropic structure and higher eccentricities, with much weaker Reynolds stresses and lower inflow rates. Of course, in either case, the eccentric motions have no measureable effect on the vertical $\delta v_{z}$.

\subsection{Turbulent/Velocity Fluctuation Structure}
\label{sec:v:turb}

From Fig.~\ref{fig:v.profile} we see the turbulence or more general velocity fluctuations, while sub-dominant to rotation, are still vigorous, with $\langle \delta v^{2} \rangle^{1/2} \sim 0.1\,V_{\rm c} \sim v_{A}$. Comparison of Figs.~\ref{fig:b.profile} \&\ \ref{fig:v.profile} immediately shows turbulent ram pressure is comparable to magnetic pressure and, from Figs.~\ref{fig:image.wedgeplot}-\ref{fig:radial.profile.general}, we see both are much larger than thermal or radiation pressure within the disk (i.e.\ the turbulence is broadly trans-\Alf{ic}, but highly super-sonic). Fig.~\ref{fig:mach} shows this more explicitly, plotting the typical sonic and \Alf{ic} Mach numbers of the velocity fluctuations in different radial annuli.

Figs.~\ref{fig:v.profile} \&\ \ref{fig:mach} also show that the typical velocity fluctuation $\delta v$ is generally larger than the mean velocities in $z$ or $R$ directions, and not strongly anisotropic ($|\delta v_{\phi}| \sim |\delta v_{R}| \sim |\delta v_{z}|$, to within a factor of $\sim 2-3$ or so, though note that the apparent transient dominance of e.g.\ $\delta v_{z}$ at small $r$ in Fig.~\ref{fig:v.profile} owes as much to the presence of coherent warps/bends in the disk as to actual ``local'' small-scale vertical turbulence). This promotes strong mixing and turbulent structure within $-H < z < H$, contributing to the complex edge-on structure (as compared to well ordered face-on structure), as well as the Reynolds stresses  (analyzed below).

Of course, the velocity fluctuations here are much stronger than in a typical SS73-like $\alpha$-disk, where (by assumption) the sonic Mach number $\mathcal{M}_{s} \sim |\delta v| / c_{s} \sim \alpha^{1/2} < 1$ (i.e.\ the non-circular motions are always sub-sonic).

\subsection{What Drives the Turbulence (Or sets its Amplitude)?}
\label{sec:v:drive}

\subsubsection{Does Anything Actually Need to Drive The Turbulence Interior to the Disk?}
\label{sec:v:drive:needed}

Given these vigorous super-sonic and quasi-isotropic velocity dispersions, it is natural to ask what their ``driving mechanism'' may be. However, it is not necessarily obvious that a driving mechanism for the velocity fluctuations -- in the usual specific sense of some local instability constantly amplifying turbulent modes on some driving scale $\lesssim H$ -- is strictly necessary. If the turbulence is \Alf{ic}, simple arguments \citep{voelk.aplers:1973.alfven.packet.eos} show that the effective ``adiabatic index'' of a group of passively-advected\footnote{In the configuration we consider, wave packets propagating radially inwards faster than the bulk accretion flow will have $|\delta{\bf B}|$ decay relative to $|{\bf B}|$, but those moving along the mean field or otherwise transverse, or with the accretion flow, will be amplified.} \Alf\ wave packets is $3/2$ (i.e.\ we expect $|\delta {\bf B}| \propto \rho^{3/4}$ moving with a Lagrangian parcel, or equivalently ${\rm d}_{t} \langle \delta{\bf B}^{2} \rangle \sim -(3/2)\, \langle \delta{\bf B}^{2} \rangle \, (\nabla \cdot \langle {\bf v} \rangle)$). Similarly, if we instead consider purely hydrodynamic turbulence, many authors have shown the Euler equations in globally compressed turbulence (if dissipation is slow) give rise to a Lagrangian quasi-adiabatic behavior with $|\delta {\bf v}| \propto \rho^{1/3}$ \citep[i.e.\ ${\rm d}_{t} \langle \rho\,\delta {\bf v}^{2} \rangle \sim -(5/3)\, \langle \rho\,\delta {\bf v}^{2} \rangle \, (\nabla \cdot \langle {\bf v} \rangle)$; see e.g.][]{vazquez-semadeni:1995.turb.jeans.instab,robertson.goldreich:2012.adiabatic.turb.heating}. So given that we saw approximately $|{\bf B}| \propto \rho^{2/3}$ for the mean-field from flux-freezing above (${\rm d}_{t} \langle {\bf B}\rangle^{2}  \sim -(4/3)\, \langle {\bf B}\rangle^{2}\,( \nabla \cdot \langle {\bf v} \rangle)$), we would have $|\delta {\bf B}|/|{\bf B}| \propto |\delta {\bf v}|/ v_{A} \propto \rho^{1/12-1/6}$ increasing with a (weak) positive power of $\rho$ as gas accretes inwards and $\rho$ increases. This suggests that advection alone could sustain $\mathcal{M}_{A} \sim 1$ in principle, even with some damping/dissipation.

The challenge here is the turbulent dissipation time: if the turbulence dissipates on a timescale $\sim t_{\rm turb,\,diss}$, then the expressions above should be modified to ${\rm d}_{t} \langle \rho\,\delta{\bf v}^{2} \rangle \sim -C\,\langle \rho\,\delta{\bf v}^{2} \rangle \, (\nabla \cdot \langle {\bf v} \rangle) - \langle \rho\,\delta{\bf v}^{2} \rangle/t_{\rm turb,\,diss}$ (where $C$ can be in the range from $3/2$ to $5/3$ depending on the regime, as described above). Noting that $|\nabla \cdot \langle {\bf v} \rangle| \sim r/|\langle v_{r} \rangle| \sim (v_{\rm K}/|\langle v_{r} \rangle |)\,\Omega$ is the inverse of the accretion timescale, this becomes ${\rm d}_{t}\ln{\langle \rho\,\delta {\bf v}^{2}\rangle} \sim C\, (v_{\rm K}/|\langle v_{r} \rangle |)\,\Omega - t_{\rm turb,\,diss}^{-1}$, so it is clear that the scalings outlined in the previous paragraph require that the accretion/advection timescale is comparable to, or faster than, the turbulent damping timescale at each radius. For hydrodynamic, supersonic turbulence with $t_{\rm cool} \ll t_{\rm dyn}$, the expected dissipation time is of order a couple of turnover times at the driving scale $t_{\rm turb,\,diss} \sim t_{\rm turnover} \sim L_{\rm drive}/v_{\rm drive} \sim H/|\delta {\bf v}_{\rm turb}| \sim \Omega^{-1} \sim t_{\rm dyn}$ \citep{thompson:rad.pressure,pan:2010.turbulent.mixing.times,hopkins:frag.theory}; the expectation is similar for magnetized trans-\Alf{ic} but subsonic/incompressible turbulence \citep{GS95.turbulence}. But the inflow time $\sim M_{\rm gas}/\dot{M}$ is $\sim( v_{\rm K}/|\langle v_{r} \rangle|)\,\Omega^{-1} \sim (v_{t}/v_{\rm K})^{-2}\,t_{\rm dyn}$. While the accretion here is very fast compared to more traditional $\alpha$-disk models (as we discuss below), inserting typical values from e.g.\ Fig.~\ref{fig:v.profile} into the above still gives an inflow time $\sim 30-100\,t_{\rm dyn}$ or $\sim 5-15\,t_{\rm orbit}$. So we would naively expect the gas to sit too long at a given $r$, meaning the fluctuations will damp faster 
than they are amplified via compression.

However, we stress that we use the term ``turbulence'' rather loosely here, encompassing any non-zero rms fluctuating motions/fields $\delta {\bf v} \equiv {\bf v} - \langle {\bf v} \rangle$. So some of the power in these fluctuations could be in structures which are not strictly ``turbulent'' in the classical sense (i.e.\ not part of some cascade) and therefore potentially dissipate on slower timescales. For example, in addition to the eccentric motions discussed above, recent idealized studies \citep{skalidis:2021.sub.alfvenic.turb.cdf.method.revision,skalidis:2022.lagrangian.models.sub.alfvenic.supersonic.turb,beattie:2020.sims.field.fluctuations.supersonic.subalfvenic.turb,beattie:2022.energy.equipartition.structure.supersonic.subalfvenic.turbulence} of randomly-forced super-sonic but sub/trans-\Alf{ic} motions (akin to what we see here; Fig.~\ref{fig:mach}) have argued that most of the kinetic energy at the driving scale ends up neither going into shocks (as it would with weaker fields) nor a classical MHD cascade (as in the sub-sonic case) but into structures such as compressive transverse non-linear magnetosonic modes (e.g.\ ram pressure of flows transverse to ${\bf B}$ with $|{\bf B} \cdot \delta {\bf B}| \sim \rho\,\delta v^{2}$), which do not dissipate rapidly. So the damping time for these types of motions could be much longer and potentially remain in balance with amplification via inflow, obviating the need for an explicit ``driving'' mechanism. A more detailed analysis of the dissipation and power spectrum/structure functions of the velocity field is therefore an interesting subject for future work.

\subsubsection{Viable Driving Mechanisms}
\label{sec:v:drive:mech}

That said, there are multiple viable driving mechanisms operating here in principle. As expected given the lack of e.g.\ strong star formation and therefore stellar feedback from within the disk, the ultimate source of energy for the turbulence on these scales is the gravitational energy of the gas. We see this reflected in the fact that the inflows obey the usual Reynolds/Maxwell energy balance condition $\Sigma_{\rm gas}\,v_{t}^{2}\,\Omega \sim \dot{M}\,\Omega^{2}$ (at least at the order-of-magnitude level; this is discussed in detail below). But this does not immediately provide a local mechanism. 

One obvious possibility is the strong global eccentric mode discussed above (quantified in Fig.~\ref{fig:eccentricity.profile}). Per \citet{noguchi:merger.induced.bars.gas.forcing,barneshernquist96,hopkins:inflow.analytics}, we would expect this to excite non-circular motions of the order $v_{t}/V_{\rm c} \sim |a_{m=1}| \sim 0.1$, which is not far from what we see. Given that the mode is a slow mode (pattern speed small compared to $V_{\rm c}$), gas will intersect the compressive orbit crossing/pileup locations in Fig.~\ref{fig:vfields.lines} once per orbit, so the turbulence can be rapidly ``rejuvenated'' on a timescale by definition comparable to its decay/damping time, even in the fast-decay ($t_{\rm decay} \sim 1/\Omega$) limit.

Another possibility is local gravito-turbulence. This is distinct from the global-eccentric-mode driving discussed above in that it depends on the local self-gravity of the gas collapsing on itself and driving high-$m$ ($k \gtrsim 1/H$) modes in Fig.~\ref{fig:eccentricity.profile}, while the global mode is driven by an external perturbation (in this case, the self-sustaining asymmetric distribution at large-$r$). This would naturally give rise to turbulence of the form $\Sigma_{\rm gas}\,v_{t}^{2}\,\Omega \sim \dot{M}\,\Omega^{2}$ \citep{gammie:2001.cooling.in.keplerian.disks}.

\begin{figure}
	\centering\includegraphics[width=0.95\columnwidth]{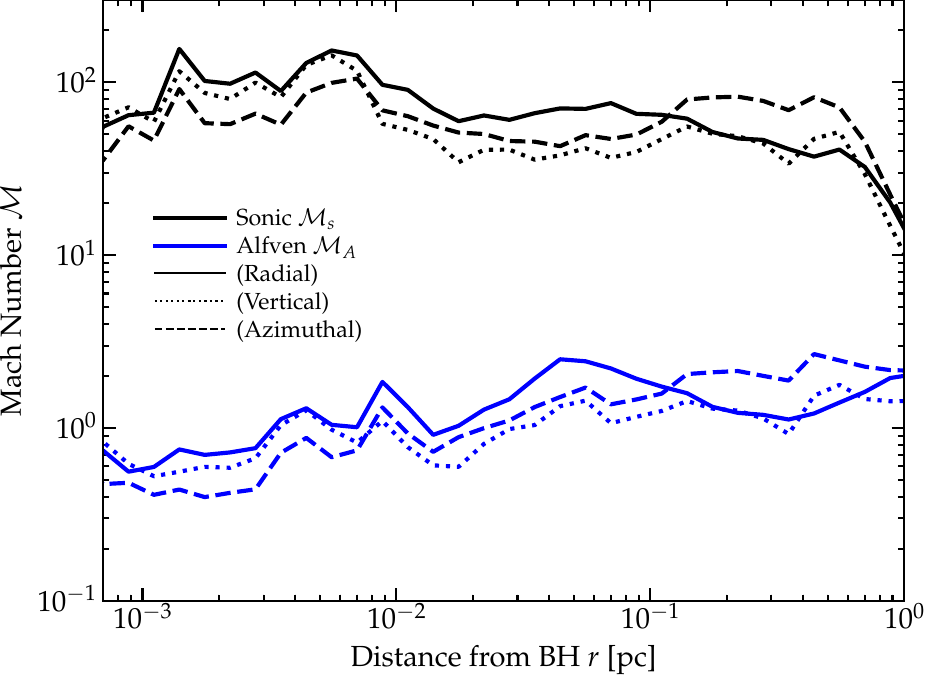}
	\centering\includegraphics[width=0.95\columnwidth]{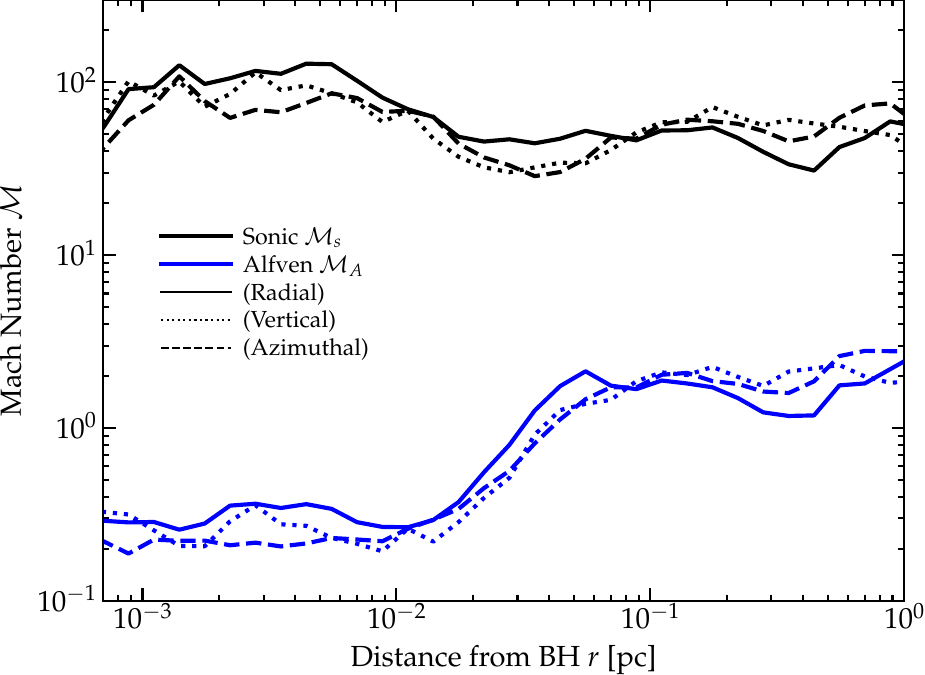}
	\caption{Turbulent Mach numbers of the gas in radial annuli, at two times as Figs.~\ref{fig:b.profile} \&\ \ref{fig:v.profile}. We take the velocity dispersion $\delta v_{i}$ estimated as $1/2$ the $16-84\%$ range (to avoid being pulled by outliers) of $v_{i}$, weighted by gas mass, in radial annuli restricting to gas within $|z|/r<0.3$ of the midplane, for each component of $v$ in cylindrical coordinates. We then define the sonic Mach number $\mathcal{M}_{s,\,i} \equiv \delta v_{i} / \langle c_{s} \rangle$ and \Alf\ $\mathcal{M}_{A,\,i} \equiv \delta v_{i} / \langle v_{A} \rangle$. The turbulence is broadly trans-\Alf{ic} and highly super-sonic (reflecting $\beta \ll 1$ in the disk). In the inner disk, the turbulence becomes mildly sub-\Alf{ic} ($\delta v_{R,\,\phi,\,z} \sim 0.3\,v_{A}$, so $\delta v_{\rm total} \sim 0.5\,v_{A}$) -- this closely maps to the radii where the mean toroidal field $\langle B_{\phi} \rangle$ becomes larger than the fluctuating field in Fig.~\ref{fig:b.profile}, as expected. The radius of this transition can fluctuate by a factor of a few as new magnetic flux advects into and through the disk (Figs.~\ref{fig:b.time.flip}-\ref{fig:b.induction.eqn}).
	\label{fig:mach}}
\end{figure}

Both of these are likely playing some role, especially at larger radii ($\gg 0.01\,$pc), where Reynolds stresses dominate over Maxwell stresses and the turbulence is modestly super-\Alf{ic}. This is further reinforced by our comparisons in \S~\ref{sec:no.mhd} below to simulations without magnetic fields. However, at smaller radii in the disk ($\lesssim 0.03\,$pc), it seems unlikely that these dominate the turbulence at saturation, for several reasons. The growth/driving rates of these modes are independent of ${\bf B}$ and neither therefore provides an obvious reason the turbulence should saturate at trans-\Alf{ic} values in Fig.~\ref{fig:mach}. Moreover we show below that the Reynolds stresses at these radii are much weaker (and often opposite in sign) if we re-run without magnetic fields, whereas the gravitoturbulence and disk asymmetry are stronger. For the eccentric mode, examining the spatial pattern of the turbulent torques, we see a hint at the largest radii that these may trace the eccentric shock but comparing Fig.~\ref{fig:vfields.lines} it is clear that the torques in the inner disk are dominated by less coherent (non-global) structures. Moreover, such a mechanism would not efficiently drive out-of-plane turbulence (leading to a large anisotropy, unlike what we see; see e.g.\ \citealt{hopkins:fb.ism.prop} and discussion below). It is possible the vertical turbulence is instead driven by the weak vertical inflows but again this would not explain why it is roughly isotropic with the radial/azimuthal turbulence, and moreover it is challenging to explain how the turbulent velocity would be much larger than the mean inflow velocity as seen at most radii in Fig.~\ref{fig:v.profile}. For gravitoturbulence, as discussed in \paperone, at small radii $Q_{\rm thermal}$ is too large, and the magnetic field strength {\em much} too large, to actually sustain vigorous gravitoturbulence \citep[see][]{kim.ostriker:2001.gravitoturb.galactic.disks.mhd.conditions,lizano:2010.stab.grav.instab.ppds.via.bfields,riols:2016.mhd.ppd.gravitoturb,forgan:2017.mhd.gravitoturb.sims}. Further, such vigorous gravitoturbulence would also be inconsistent with the low star formation rates we actually see at these radii.

As discussed in detail in \S~\ref{sec:b:origin:turb}, the ``traditional'' weak-field MRI -- at least in the simplest linear form usually invoked -- does not appear viable as the primary turbulent driving mechanism. Specifically, as shown above (there and \S~\ref{sec:b:origin:tests}), the toroidal $B_{\phi}$ is sufficiently large that the system should be stable against the standard MRI in the usual idealized context where it has been studied (i.e.\ for the mean-field parameters here, the standard weak seed-field MRI modes should generally be quenched and linearly stable, at least for a laminar, azimuthally-symmetric, unstratified system; see \citealt{terquem:1996.toroidal.field.bouyancy.instability,kim.ostriker:2000.mhd.instab.shearing.cold.winds,pessah.psaltis:2005.mri.extensions.stronger.fields,lin:2014.linear.stability.magnetized.selfgrav.ppds,hirabayashi:2016.toroidal.field.bouyancy.modes,das:2018.pessah.psaltis.limit.mri}). Also when we analyze the structure of the torques in more detail below, neither the mode morphology nor ratio of Reynolds to Maxwell stresses agrees quantitatively with the predictions of some previous saturated MRI simulations in similarly simplified/idealized shearing boxes \citep[compare e.g.][]{brandenburg:mri.predicted.high.ratio.maxwell.to.reynolds,pessah:2006.mri.signature.ratio.maxwell.reynolds}. But as reviewed in \S~\ref{sec:b:origin:turb}, while this particular form of the MRI is stabilized under such conditions at sufficiently high $B_{\phi}$, this just means that different linear instabilities appear. Specifically, even under those conditions, at $B_{\phi}$ above the threshold where the MRI growth rate vanishes, in a $\beta \ll 1$ disk, the Type II/III \citep{pessah.psaltis:2005.mri.extensions.stronger.fields} or ATB \citep{kim.ostriker:2000.mhd.instab.shearing.cold.winds} or SSMI/SHMI \citep{das:2018.pessah.psaltis.limit.mri} modes appear (see Fig.~\ref{fig:instability.map}). As shown in those studies, these modes are essentially radial buoyancy modes driven by the competition between dominant magnetic pressure support of the gas and the Keplerian gravity/differential rotation around the BH. For the conditions here, these have rapid growth rates ($\sim (v_{A}/v_{\rm K})\,\Omega$ for the largest-wavelength unstable modes with $k\sim 1/H$), are at least plausibly consistent with the turbulent morphologies and Reynolds stresses below, and -- given their dependence on the mean-field direction/structure -- would plausibly saturate at trans-\Alf{ic} turbulence. Likewise vertical buoyancy modes, if unstable, should have broadly similar growth rates for modes with $k\sim 1/H$, potentially explaining (at least at the order-of-magnitude level) the broadly isotropic turbulent velocity and magnetic field structure. And modified (supra-thermal) versions of the MRI could exist where $\partial \ln{B}_{\phi}/\partial \ln{R}$ is very close to $-1$ \citep{begelman:2023.mri.saturation.estimates}. Of course, it is also plausible that other, distinct instabilities appear, or even that some alternative modified versions of the MRI persist, under the more complicated conditions present in our simulations (e.g.\ allowing for eccentricity, stratification, different equations-of-state, turbulence, non-linear perturbations, etc.). It therefore seems likely that some unstable modes are at least present, and could play an important role in the accretion dynamics.

Unfortunately, identifying a ``unique'' driver in fully non-linear, saturated, multi-physics and multi-scale simulations like these (as opposed to more idealized controlled numerical experiments) is challenging, and will require additional work with different simulations motivated by the discussion above to isolate the distinct mechanisms discussed above in more controlled circumstances. Importantly, however, this has allowed us to identify a relatively small list of likely processes which are occurring, in order to guide said simulations. We also wish to stress that there is no smooth or laminar ``initial'' phase from which the turbulence and/or magnetic field strengths in the disk amplify -- in idealized simulations, such a phase is an artifact of artificially smooth initial conditions. Instead, the system here {\em begins} from a highly out-of-equilibrium gas configuration as the disk forms from infall, with strong turbulence and magnetic flux already {\em well above} the commonly-quoted MRI saturation threshold and inhomogeneity in the velocity, density, temperature, and magnetic fields, before relaxing into a steady-state strongly-turbulent disk.

In potential future, more idealized/controlled numerical experiments with initial and boundary conditions motivated by these simulations, our discussion above also demonstrates that global simulations may be necessary to properly capture the magnetic curvature effects and low-$k$ ``Type III'' as well as eccentric modes, as well as the inflow of gas and magnetic flux. But our discussion above also shows that such studies could be made more simplified in various ways -- e.g.\ by replacing the explicit radiation-hydrodynamics and thermochemistry with more idealized models, or neglecting star formation and stellar dynamics -- and still be used to explore e.g.\ the effects of self-gravity and different instabilities. It would be particularly interesting to study these in the context of angular momentum transfer in the disk. While this has been studied more extensively (under more idealized conditions) for the weak-field MRI, gravito-turbulence, and global gravity (eccentric/spiral) modes (all of which are known to produce efficient angular momentum exchange), there has been much less study of the non-linear buoyancy modes we discuss above (let alone other instabilities that may be present here). 

At least in linear theory, the longer-wavelength ``Type III/SHMI'' instabilities (active at wavenumbers $|k_{z}| \lesssim k_{z}^{c1}$) appear to depend on differential rotation in a similar manner to the MRI \citep{pessah.psaltis:2005.mri.extensions.stronger.fields}, and the linear mode eigenvectors of the Type III/SHMI instabilities are similar to the (compressible) MRI both in the relative magnitude of the density/velocity/magnetic field perturbation components and in the fact that both the resulting Maxwell and Reynolds stresses always transport angular momentum outwards (although the Maxwell-to-Reynolds ratio can vary with respect to the MRI). This suggests that their non-linear behavior and ability to produce efficient angular momentum transport may also be similar to the MRI. On the other hand, the short-wavelength ``Type II/SSMI'' modes are more local and persist without differential rotation (and their linear eigenvectors involve weak Maxwell/Reynolds stresses with angular momentum transport of opposing signs), so may be more akin to e.g.\ convective instabilities in their non-linear behavior. But further study will be needed for more rigorous conclusions. 

\subsection{What Is the ``Turbulent Resistivity'' and Why Does It Not Destroy Flux-Freezing?}
\label{sec:why.flux.freezing}

Since we have shown that the magnetic fields in the accretion disk are largely ``sourced'' by simple flux-freezing/advection (\S~\ref{sec:b}) and that microphysical resistivity is negligible (\S~\ref{sec:b:origin:ov}), it is worth briefly discussing the role of the effective turbulent resistivity $\eta_{\rm turb}$, and whether this could destroy flux-freezing in the disk \citep[see e.g.][for a detailed discussion of this for pure-poloidal fields, although this is the opposite of the regime of interest here]{lubow:1994.bfield.advection.acc.disks.vs.resistivity}. 

Consider first the radial field component $B_{R}$, as we argue this sources $B_{\phi}$, following a Lagrangian parcel of gas. The turbulent resistivity should damp the field as ${\rm d}^{\rm turb}_{t} B_{R} \sim -\eta_{\rm turb}\,k_{B_{R}}^{2}\,B_{R}$, where $k_{B_{R}}$ gives some inverse gradient scale length of $B_{R}$ and $\eta_{\rm turb} \sim v_{\rm turb}/k_{\rm turb}$. Making the standard assumption that genuinely turbulent eddies with crossing time larger than $\sim \Omega^{-1}$ will be sheared out (which appears to be valid in the simulations) we have\footnote{We will ignore $\mathcal{O}(1)$ prefactors here but note that $\eta_{\rm turb}$ could be much smaller by a factor $\sim 10-100$ than this estimate (see e.g.\ \citealt{salveson:2016.sims.mri.dominated.bfield.disks}, who find a prefactor $\sim 0.3/{\rm Pr} \sim 0.01-0.1$ in terms of the turbulent Prandtl number ${\rm Pr}$). So we are effectively considering the maximum possible effect of turbulent resistivity.} $k_{\rm turb}\,v_{\rm turb} \sim \Omega$ so $\eta_{\rm turb} \sim v_{\rm turb}^{2}/\Omega$. Meanwhile, flux-freezing will grow $B_{R}$ on the inflow/compression timescale $\sim r/\langle v_{r} \rangle$, so\footnote{Again we ignore order-unity coefficients but these can be derived from the induction equation for specific geometries, e.g.\ we show in \S~\ref{sec:b:origin} that for a tidally captured radially infalling stream, the compression in the $\hat{z}$ and $\hat{\phi}$ directions ($R^{-1}\,\partial_{\phi}\,\langle B_{R}\,v_{\phi} \rangle$ and $\partial_{z} \langle B_{R}\,v_{z} \rangle$) give a coefficient $=2$. So, we are again being conservative and assuming a somewhat weaker growth via flux-freezing by neglecting such order-unity factors.} ${\rm d}^{\rm adv}_{t} B_{R} \sim -(\langle v_{r} \rangle /r)\,B_{R}$. For turbulence to completely destroy flux-freezing, clearly we require $|{\rm d}^{\rm turb}_{t} B_{R}| \gg |{\rm d}^{\rm adv}_{t} B_{R}|$, or $v_{\rm turb}^{2} \gg |\langle v_{r} \rangle|\,k_{B_{R}}^{-2}\,\Omega/r$. 

With this in mind, is it helpful to divide the simulation into three regimes or ``zones'' based on distance to the SMBH, which exhibit three qualitatively different behaviors: 
\begin{enumerate}[labelindent=0pt,labelwidth=10pt,labelsep*=0pt,leftmargin=!,align=parleft]
\item{The outer/ISM/turbulent ``zone''} well outside the BHROI and accretion disk ($R \gg $\,pc). Here the physics are ``ISM-like'' (per above): there is no disk or any coherent BH ``accretion flow'' to speak of (the BH by definition does not dominate the potential). Instead we see self-gravitating, collapsing, rapidly star-forming GMC-like gas complexes ``stirred'' by stellar feedback, with super-\Alf{ic} internal turbulence (these can globally lose angular momentum and get closer to the galactic nucleus, but are not systematically ``accreting''). In this regime (see e.g.\ Fig.~\ref{sec:basic} and \paperone) the magnetic fields are tangled and quasi-isotropic, mean fields are much smaller than fluctuating fields, and ``global'' compression (with coherent infall towards $R\rightarrow 0$) is small compared to compression of gas internal to the cloud via shocks and self-gravity. So inserting values we naturally find $|{\rm d}^{\rm turb}_{t} B_{R}| \gg |{\rm d}^{\rm adv}_{t} B_{R}|$: as discussed in \paperone, the clouds internally reach the expected saturation for the small-scale dynamo with turbulent amplification balancing turbulent resistivity (magnetic energy a few percent of turbulent energy).\footnote{It is worth noting that the bulk velocity of individual clouds/complexes is larger than their internal turbulence, as expected for any clouds that are smaller than the galaxy in which they are embedded. This is important insofar as it explains why on a close passage to the SMBH, complexes will be tidally disrupted and an $\mathcal{O}(1)$ fraction of the mass will be strongly bound and begin to radially free-fall with relatively small impact parameter.}

\item{The intermediate/capture/free-fall zone} inside the BHROI ($R \sim 0.1-$a few pc). Here we see a complex is tidally disrupted on close passage to the SMBH, leading to a tidally elongated, radially free-falling stream of bound gas towards the SMBH (see \S~\ref{sec:basic}). We stress that while the rms $\langle \delta v_{R} \rangle^{2}$ might appear large in e.g.\ Fig.~\ref{fig:v.profile} at these radii, a cursory examination of the visual morphology of the gas stream (Fig.~\ref{fig:image.zoom}) or velocity fields (Figs.~\ref{fig:vfields.lines}-\ref{fig:vfields.inflow.outflow.largescales}) or magnetic field lines (Fig.~\ref{fig:bfields.faceon.zoom}), or a more rigorous quantitative diagnostic of the local, genuinely turbulent components of ${\bf v}$, immediately makes it obvious that the flow is locally coherent and strongly dominated by radial motion stretching the radial field lines, as opposed to turbulence (\S~\ref{sec:b:ov}). The large rms $\langle \delta v_{R} \rangle^{2}$ does not reflect local turbulence, but the fact that just like in any tidal disruption event, of order half the mass is unbound, so there is a free-falling stream inwards and a large unbound stream/fan being ejected. Within the accreting filament (which dominates the gas supply so is what actually matters here), the gas is effectively free-falling radially onto the SMBH, so $v_{r}/r \sim v_{\rm freefall}/r \gtrsim \Omega$ reaches its maximum possible value, while $v_{\rm turb}$ is relatively weak ($\ll v_{\rm freefall}$), hence the net effect is stretching $k_{B_{R}} \rightarrow 1/r$. So in this regime, $|{\rm d}^{\rm turb}_{t} B_{R}| \ll  |{\rm d}^{\rm adv}_{t} B_{R}|$ is easily ensured even if the gas supply at larger radii is turbulent.

\item{The accretion disk zone} at radii $\lesssim 0.1\,$pc, where the gas circularizes and an ordered accretion disk forms. This is the regime of interest and unlike the previous zones, the behavior is not trivial. Here, the relations noted above should hold for $|{\rm d}^{\rm turb}_{t} B_{R}|$ and $|{\rm d}^{\rm adv}_{t} B_{R}|$ so flux-freezing is destroyed if $v_{\rm turb}^{2} \gg |\langle v_{r} \rangle|\,k_{B_{R}}^{-2}\,\Omega/r$. In our simulation (Figs.~\ref{fig:bfields.faceon.edgeon}-\ref{fig:b.profile} \&\ \ref{fig:b.induction.eqn}) we clearly have $k_{B_{R}} \sim 1/R$ in this regime, which should be expected analytically from the boundary conditions forming the disk (from zone {\em (ii)} above), and -- if flux-freezing is valid -- will be maintained by the disk\footnote{If the disk begins from an initial condition with $k_{B_{R}} \sim 1/R$ with flux-freezing valid, then we can approximate the evolution of $k_{B_{R}}$ through the disk analytically via the stretching between two gas elements in the radial direction. The radial distance between two elements $\Delta R$ evolves in a Lagrangian manner as they migrate radially (in the midplane of a steady azimuthally-symmetric accretion flow) as ${\rm d}_{R} \Delta R = v_{R}(R+\Delta R) - v_{R}(R) \approx \Delta R \partial_{R} v_{R}$, and noting in this limit we can replace ${\rm d}_{t}$ with $v_{R}\,\partial_{R}$, this becomes $\partial_{R}(\ln{\Delta R}) \approx \partial_{R}(\ln{[-v_{R}]})$. So if we begin from $k_{B_{R}} \sim 1/R$, $k_{B_{R}} \lesssim 1/R$ will be preserved so long as $|v_{R}| \propto R^{\alpha}$ with $\alpha \le 1$. This is easily satisfied in the simulation in a time-averaged sense (ignoring obvious transient features; see Fig.~\ref{fig:v.profile}), and in the analytic model presented in \paperthree\ this is automatically ensured not just for our default trans-sonic flow assumption ($\zeta \sim 1/3$ in the variables defined therein) but for any $\zeta >= -1/4$ (where any physical solution requires $\zeta > 0$, so this condition is always met).} as $R \rightarrow 0$. This means that the resistivity condition becomes $v_{\rm turb}^{2} \gg |\langle v_{r} \rangle|\,v_{\rm K}$. But for a system dominated by Maxwell and Reynolds stresses we expect and measure in the simulation $|\langle v_{r} \rangle | \sim (\Pi_{\rm Reynolds} + \Pi_{\rm Maxwell})/(\rho\,R\,\Omega) \sim (v_{\rm turb}^{2} + v_{A}^{2})/v_{\rm K}$ (see \S~\ref{sec:torque} and Figures therein below, as well as \paperthree), so this is impossible to satisfy for $v^{2}_{A} > 0$, and hence flux-freezing/advection should not be strongly modified by turbulent resistivity (though the resistivity is likely to be a non-negligible correction).

\end{enumerate}

Note that for the accretion disk zone {\em (iii)}, if we had made the more common accretion disk assumption -- {\em opposite} the behavior we actually see in the simulations -- that the disk ``begins'' from negligible mean $B_{R}$ (e.g.\ a strictly poloidal field with all of $B_{R}$ sourced from the MRI), then $k_{B_{R}}$ would necessarily be $\sim k_{\rm turb} \sim \Omega/v_{\rm turb}$. In this case, the condition  $|{\rm d}^{\rm turb}_{t} B_{R}| \gg |{\rm d}^{\rm adv}_{t} B_{R}|$ or $v_{\rm turb}^{2} \gg |\langle v_{r} \rangle|\,k_{B_{R}}^{-2}\,\Omega/r$ would simply reduce to $|\langle v_{r} \rangle| \ll v_{\rm K}$, which is always true in the disk. Thus there are two qualitatively distinct but each internally self-consistent regimes: (i) the ``traditional'' weakly-magnetized assumption where $B_{R}$ is sourced primarily by a poloidal field and turbulent resistivity invalidates flux-freezing, and (ii) the flux-frozen and therefore hyper-magnetized regime where there is some coherent $B_{R}$ being ``fed'' to the disk, ensuring that flux-freezing remains a valid assumption throughout.\footnote{This has some features in common with the scenario discussed in \citet{johansen.levin:2008.high.mdot.magnetized.disks}, though we stress that the source and coherence of $B_{R}$, and resulting relative magnitude of the terms here, are qualitatively and physically distinct.} The regime of interest clearly depends on the disk outer boundary condition: thus even though the dynamic range of the ``free-fall'' zone in radius $R$ is quite small compared to the ``ISM zone'' or ``accretion disk zone,'' it plays a crucial role in establishing these conditions. 

We can repeat these arguments for the toroidal $B_{\phi}$ and reach similar conclusions in the disk zone where $B_{\phi}$ is already dominant, but it is more interesting to consider the ``sourcing'' of $B_{\phi}$ from $B_{R}$ which we demonstrated in \S~\ref{sec:b:origin:ov}. Motivated by Fig.~\ref{fig:b.induction.eqn}, if we compare turbulent resistivity ${\rm d}_{t} B_{\phi} \sim -\eta_{\rm turb}\,k_{B_{\phi}}^{2}\,B_{\phi}$ to the source term ${\rm d}_{t} B_{\phi} \sim \partial_{R}\langle v_{\phi}\,B_{R} \rangle \sim -\Omega\,B_{R}$, then (given that it is even more obvious that $k_{B_{\phi}} \sim 1/R$ is coherent in the disk zone) turbulent resistivity is a small effect so long as $|\langle B_{\phi} \rangle | \lesssim (v_{\rm K}/v_{\rm turb})^{2}\,|\langle B_{R} \rangle | \sim 100\, |\langle B_{R} \rangle |$ (inserting typical $v_{\rm turb}$ from Fig.~\ref{fig:v.profile}). This is easily satisfied in the disk (Fig.~\ref{fig:b.profile}); indeed, $|\langle B_{\phi} \rangle|$ never really reaches this upper limit because other terms dominate (e.g.\ the gas and field line advection/accretion time is short enough that this is balancing the induction term, rather than resistivity).

Distinct from the usual assumption for magnetized accretion disks, the relevant interactions here are between $B_{R}$ and $B_{\phi}$: the poloidal field plays a sub-dominant dynamical role in either (though it does provide some additional stability and source for $B_{R}$ and $B_{\phi}$, see \citealt{salveson:2016.decaying.fields.poloidal.flux}). This is consistent with the relatively weak, more turbulence-dominated poloidal fields we see in the midplane (Fig.~\ref{fig:b.profile}). However we caution that this does not mean the behaviors of interest here could be captured in a strictly 2D simulation: phenomena such as turbulence and vertical compression of the disk ($H\rightarrow 0$ as $R\rightarrow 0$, contributing to the flux-freezing amplification of both $B_{\phi}$ and $B_{R}$) still depend on 3D structure.

\begin{figure*}
	\centering\includegraphics[width=0.45\textwidth]{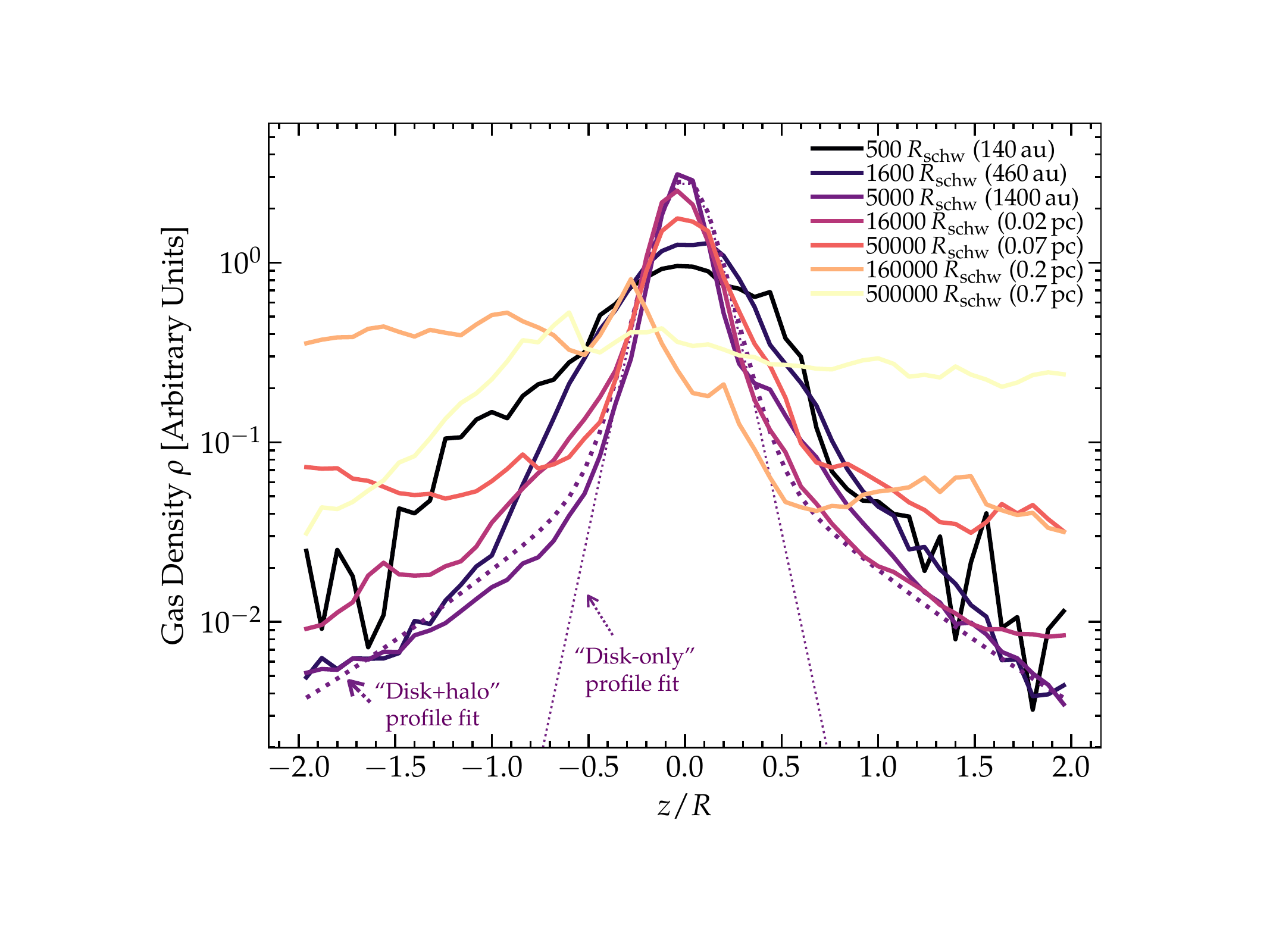}
	\centering\includegraphics[width=0.45\textwidth]{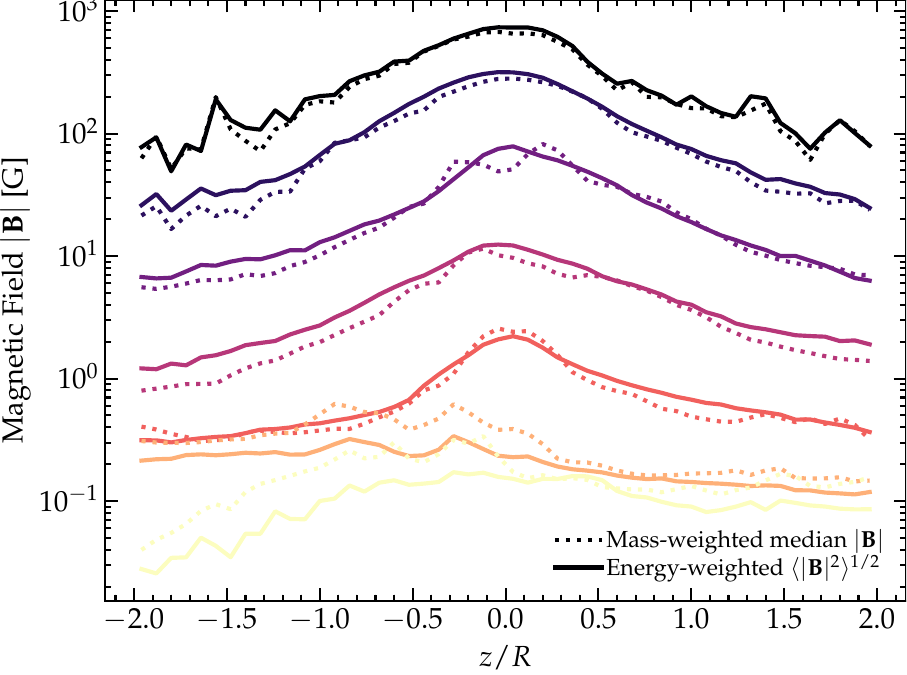}
	\centering\includegraphics[width=0.45\textwidth]{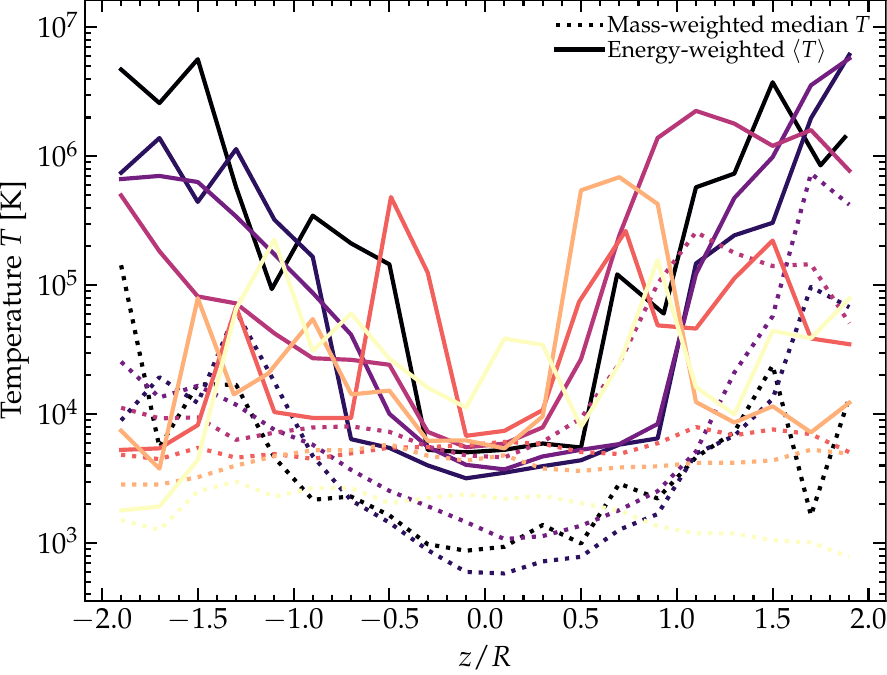}
	\centering\includegraphics[width=0.45\textwidth]{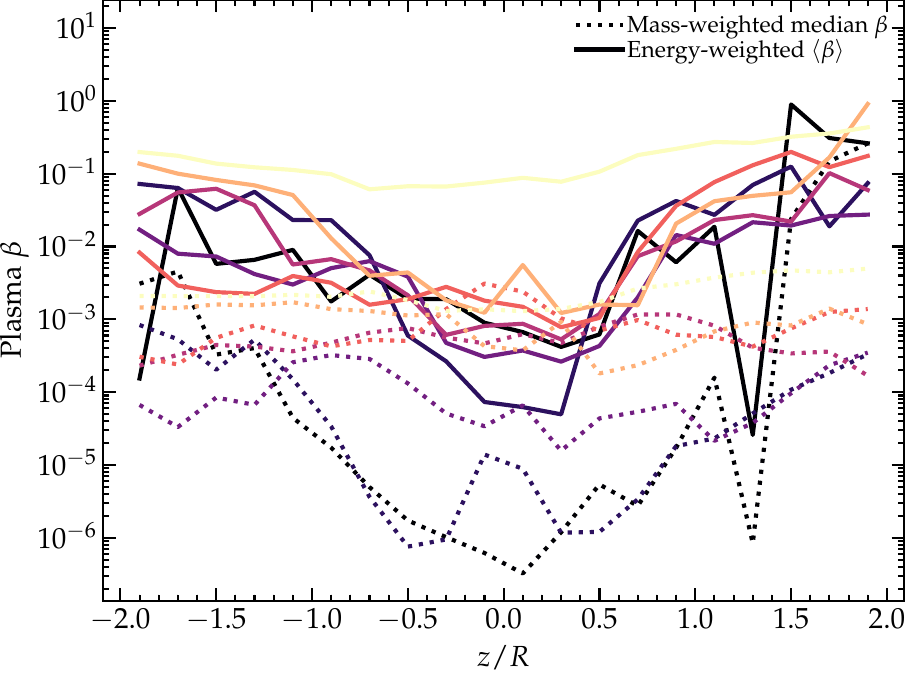}
	\caption{Vertical profile of disk properties (\S~\ref{sec:vertical}): gas density $\rho$ ({\em top left}), magnetic field strength $|{\bf B}|$ ({\em top right}), temperature $T$ ({\em bottom left}), and plasma $\beta$ ({\em bottom right}). For each, we mass or energy-average in cylindrical annuli at radius $R$ ({\em line colors}, labeled) in factor $\sim 3$ intervals from $\sim 100\,$au to $\sim 1\,$pc. The density is stratified as expected for $H/R \sim 0.1-0.3$ at this range of radii; magnetic field strengths peak in the midplane but are more weakly stratified. The temperatures are coolest in the midplane and rise in the diffuse, more optically-thin corona, leading to the plasma $\beta$ also rising (weakly) with scale height. Energy-weighted thermal averages appear more ``noisy'' owing to physical multi-phase structure (the average can be dominated by e.g.\ hot gas in shocks or SNe at large radii). In the density plot, we illustrate for one radius ($\sim 5000\,R_{\rm schw}$) a decomposition into a two-component ``disk'' (Gaussian/exponential or $\rho \propto {\rm sech}^{2}(z/\alpha_{\rm H}\,H)$ here; {\em thin dotted}) plus power-law ``halo/coronal'' ($\rho \propto (R^{2} + z^{2})^{-\alpha_{\rm halo}}$ with $\alpha_{\rm halo} \sim 3.5$ here; {\em thick dotted}) profile. In this decomposition, the disk component with $H/R \sim 0.1$ contains most of the mass, but there is clearly an extended vertical gas distribution beyond the extrapolation of the vertical profile from smaller $z$. The vertical thermal structure is expected for a more tenuous corona/atmosphere above the disk, but the weak stratification within the disk of $T$ and $|{\bf B}|$ owes to a combination of optical depth effects and strong super-sonic turbulence driving efficient vertical mixing (\S~\ref{sec:vertical:physics}). 
 	\label{fig:vertical}}
\end{figure*}

\section{Vertical Structure \&\ Thermo-Chemical Properties of the Disk}
\label{sec:vertical}

Briefly, we examine the vertical and thermo-chemical structure of the disk. The details of the thermo-chemistry as a function of scale are discussed in much greater detail in \paperone\ to which we refer interested readers, because it is of great potential importance to the cessation of star formation in the disk. But, because $\beta \ll 1$, the thermal properties of the disk actually play relatively little role in the accretion dynamics. We thus only briefly discuss them here insofar as they are useful to inform our understanding of the disk structural properties and to help distinguish the behaviors we observe from some other possible models (as with e.g.\ the amplitude of $|{\bf B}|$ as discussed above).

\subsection{General Structure \&\ Vertical Density Profiles}
\label{sec:vertical:profiles}

Recall, Fig.~\ref{fig:image.wedgeplot} shows various thermo-chemical properties such as the temperature and plasma $\beta$ in face-on projection. Edge-on, Fig.~\ref{fig:vertical} shows more detailed quantitative vertical profiles at a variety of BH-centric radii $R$ (we use cylindrical $R$, $\phi$, $z$ coordinates, and azimuthally average over $\phi$ in cylindrical rings). We see a broad radial trend in Fig.~\ref{fig:image.wedgeplot} where the gas becomes less dramatically multi-phase at smaller radii, and starts to become somewhat warmer in denser regions towards small $R$, as expected in an optically thick disk (though much of the disk at the radii resolved here is still atomic with modest ionization fraction $\sim 1-10\%$; see \paperone). In Fig.~\ref{fig:vertical}, the obvious take-away is that the thermal and magnetic properties of the disk are quite weakly stratified.

In detail, in Fig.~\ref{fig:vertical}, we see that the vertical {\em density} profile follows a very typical ``disk+halo/corona'' profile. Specifically we can fit $\rho(R,\,z)$ fairly well with the sum of a ``disk'' and ``halo'' component. The disk is  Gaussian or ${\rm sech}^{2}$,  $\rho_{\rm disk} \approx \rho_{\rm mid} \,{\rm sech}^{2}(z/\alpha_{\rm H}\,H)$, as is standard in the resolved/galactic disk literature (\citealp{vanderkruit:1988.vertical.light.profile.fitting.functions}; the constant $\alpha_{\rm H} \sim 1$ depends on whether we define $H$ as an rms, median, half-mass height, etc.). The halo follows a  power-law  $\rho_{\rm halo} \propto r^{-\alpha_{\rm halo}}$ with lower normalization, $\alpha_{\rm halo}$ between $2-4$, and $r^{2} = R^{2}+z^{2}$. This is quite generic to disks with extended gaseous halos, coronae, or outflows, and very similar to the profiles typically used to fit e.g.\ edge-on gas densities in galaxies or protostellar disks. The disk is clearly evident here, as it should be given its morphological presence in e.g.\ Fig.~\ref{fig:image.faceonedgeon.inner}. There is significant skewness/asymmetry especially at large radii owing to the asymmetric inflow, but following Lagrangian parcels in time we typically see the vertical profile of ``new'' material settle into this profile over just a couple of dynamical times once it reaches its circularization radius (inside e.g.\ $\lesssim 0.1\,$pc).

\subsection{Weak or Inverse Stratification of $|{\bf B}|$, $T$, and $\beta$}
\label{sec:vertical:nostrat}

Comparing the magnetic field strength versus height $z$ in Fig.~\ref{fig:vertical}, we see notably weaker vertical stratification compared to the gas density. At various radii we see that a $\sim 3$\,dex decrease in $\rho$ from midplane to large heights $|z| \gtrsim 2\,R$ corresponds to a $<1$\,dex change in $|{\bf B}|$. This means that, as noted in \S~\ref{sec:b:origin:tests}, although we can approximate the midplane scaling of $\langle |{\bf B}| \rangle_{\rm midplane}  \propto \langle \rho \rangle_{\rm midplane}^{2/3}$, we cannot assume $|{\bf B}| \propto \rho^{2/3}$ everywhere outside the disk. Interestingly, we can actually model the vertical $|{\bf B}|$ profiles in  Fig.~\ref{fig:vertical} fairly well if we assume $|{\bf B}|(z)$ scales with the ``halo'' or ``coronal'' component of $\rho$ as defined in \S~\ref{sec:vertical:profiles}, as $|{\bf B}| \propto \rho_{\rm halo}^{2/3}$ -- i.e.\ if the shape of $|{\bf B}|(z)$ only follows the diffuse/slowly varying ``corona'' component, with a normalization scaling as $\langle |{\bf B}| \rangle \propto \langle \rho_{\rm mid} \rangle^{2/3}$, such that the ``disk'' component and stratification effectively disappears when we consider $|{\bf B}|$. As noted below, this is naturally connected to the strong turbulence in the disk, itself potentially related to strong magnetic buoyancy, convective, and other instabilities producing effective vertical motions.

Turning to the gas temperature $T$, we again do not see any strong stratification {\em within} the disk in Fig.~\ref{fig:vertical} -- i.e.\ at $|z|/R \lesssim 0.3$ or so where we see clear structure in $\rho(z)$. And the (weaker) stratification between the entire disk ``zone'' and extended halo at $|z| > R$ is inverted -- i.e.\ we see higher temperatures at large $|z|/R$. The absolute values of the temperatures we obtain depends on how we weight the average in each annulus (a generic expectation if there is any multi-phase structure, since most of the thermal energy often resides in phases that do not dominate the total gas mass), and the trend is more clear if we weight by thermal energy as compared to e.g.\ gas mass.\footnote{We caution that care is required in interpreting the exact thermal properties of some of the gas at very low temperatures in the midplane, where $\beta$ can locally be extremely small. For extremely-low $\beta$ the thermal energy of the gas (being just a tiny fraction of the total internal energy) can be affected significantly by numerical integration error, or even machine roundoff errors, in the energy solutions to the Riemann problem.}

The inverse stratification we see in Fig.~\ref{fig:vertical} is generic to almost all diffuse coronae/halo type-systems, and is seen in e.g.\ AGN accretion disk coronae, proto-planetary disks, stellar atmospheres and their coronae, and galactic disk-halo (or CGM) interface regions. Indeed it has also been seen in other idealized simulations of magnetized disks not unlike those here \citep{kudoh:2020.strong.b.field.agn.acc.disk.sims.compare}. Like in all these other systems, in the more diffuse gas at $|z| \gtrsim R$, the absorption optical depth is low and the gas is far from LTE. The diffuse gas cools relatively inefficiently and is heated by a wide range of non-equilibrium and/or non-local processes, including shocks and turbulent dissipation, magnetic reconnection, radiation from the disk, external radiation from starlight in the galaxy, the galactic cosmic ray background, and stellar feedback (e.g.\ jets, winds, supernovae) from the stars further out in the disk.

Combining these trends, we see that $\beta$ is both $\ll 1$ everywhere, and inversely or weakly stratified (with $\beta$ lower near the midplane). Unsurprisingly the absolute value of the mean $\beta$ and precise degree of inverse stratification depend systematically on the averaging method. For example, weighting by thermal energy  will give the highest $\langle \beta \rangle$,  magnetic energy is biased towards the lowest $\langle \beta \rangle$,  gas mass gives results in between,  or one could also use $\langle \beta \rangle \equiv \langle c_{s}^{2} \rangle / \langle v_{A}^{2} \rangle$. But the qualitative results  are the same for all weightings.

\begin{figure*}
	\centering\includegraphics[width=0.97\textwidth]{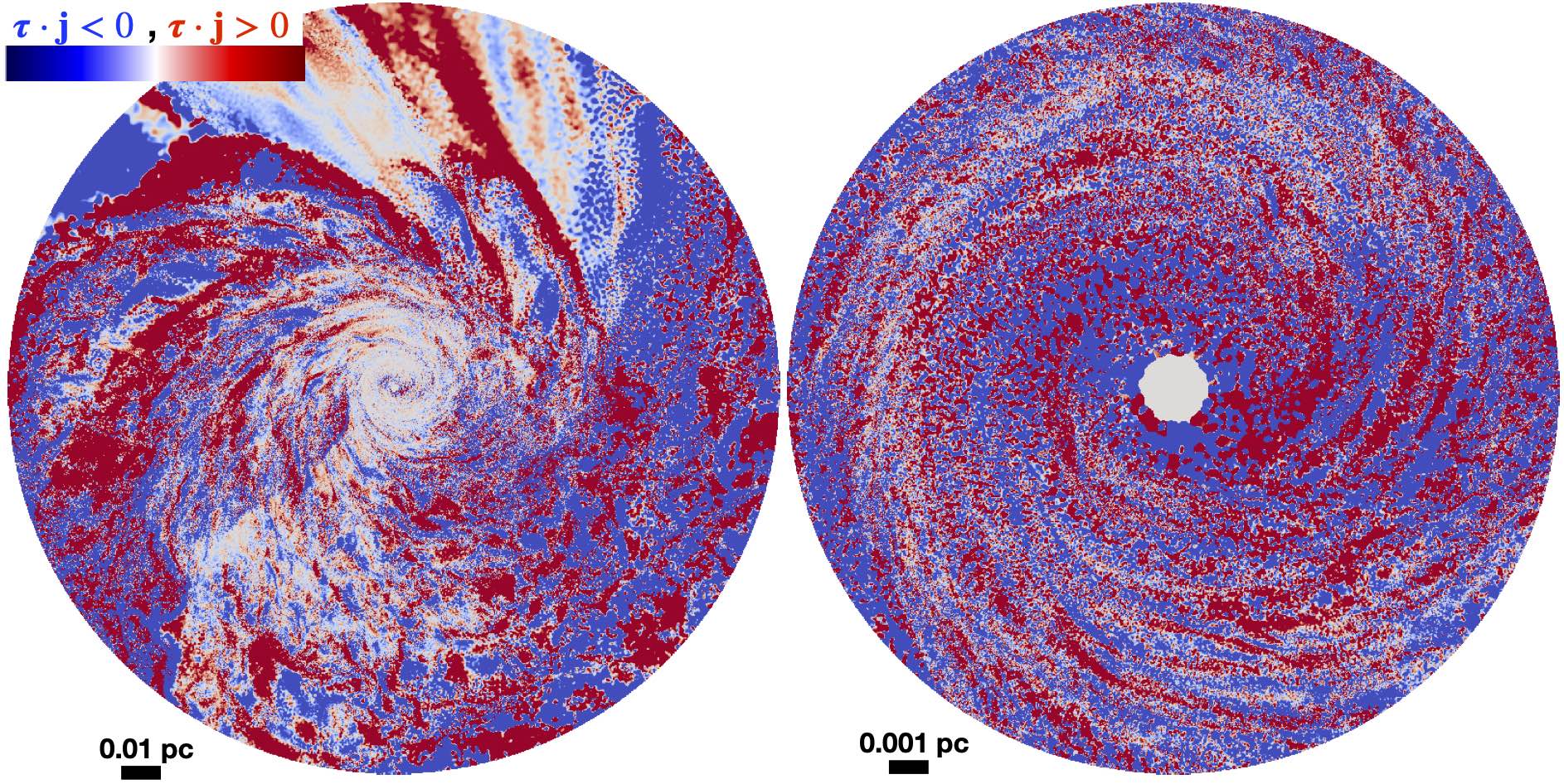}
	\caption{Map of the net torque $\boldsymbol{\tau} \equiv {\bf r} \times {\bf a}$ (taken directly from the simulation) on each gas element $i$ (inside $r<0.1\,$pc {\em left}, or $<0.01\,$pc {\em right}), in the direction of the local angular momentum vector ${\bf j} \equiv {\bf r} \times {\bf v}$: $|{\boldsymbol{\tau}}_{i} \cdot \hat{\bf j}| / V_{c}^{2}$ (from $-1$ in {\em blue} to $+1$ in {\em red}). The image is a face-on projection of the disk midplane, as in Fig.~\ref{fig:bfields.faceon.edgeon}-\ref{fig:bfields.faceon.zoom}. The torques are clearly dominated by a fluctuating component, involving non-axisymmetric/tightly-wound modes with wavenumbers $k\,H \sim $ a few. This is broadly expected if a combination of Reynolds \&\ Maxwell stresses dominate the torques, potentially originating  from the instabilities considered in Fig.~\ref{fig:instability.map}.
	\label{fig:torque.map}}
\end{figure*}

\subsection{Physics of the Weak Disk Stratification}
\label{sec:vertical:physics}

The lack of strong vertical stratification in $|{\bf B}|$ and $T$ within the disk is a prediction distinct from e.g.\ a classic weakly-magnetized, weakly-turbulent \citet{shakurasunyaev73} $\alpha$-disk, but should be expected for the disks here, given the vigorous turbulence present. As noted above (Fig.~\ref{fig:mach}), the disk is exhibits super-sonic and trans-\Alf{ic}, quasi-isotropic  turbulence. This means, essentially by definition, that an order-unity fraction of the turbulent power will be in vertical flows with coherence length $\sim H$, crossing/turnover time $\sim H/v_{\rm t} \sim \Omega^{-1} \sim t_{\rm dyn}$, and ram pressure comparable to or greater than the total (thermal+radiation+magnetic) pressure in the disk. 

As discussed in more detail in \paperone\ and above, we also should not expect LTE to apply in the diffuse gas above the disk nor in the outermost regions of the disk. In the inner disk midplane, the behavior does begin to resemble more blackbody-like LTE behavior. As we showed in \paperone, the radiation temperature and gas temperature begin to converge in the dense gas with optical depth $\gg 1$, and the gas radiation flux roughly balances the change in gas energy from accretion -- but this really only applies to the most dense, nuclear midplane gas. And even that gas features rapid cooling ($t_{\rm cool} \ll t_{\rm dyn}$), and is primarily atomic with abundances of important species such as H$^{-}$ and free electrons far from the naive Saha equation estimate, owing to the large deviations from LTE and large contribution to ionization and reactions from terms like external irradiation, shocks, cosmic rays, and other processes described above.

\begin{figure}
	\centering\includegraphics[width=0.95\columnwidth]{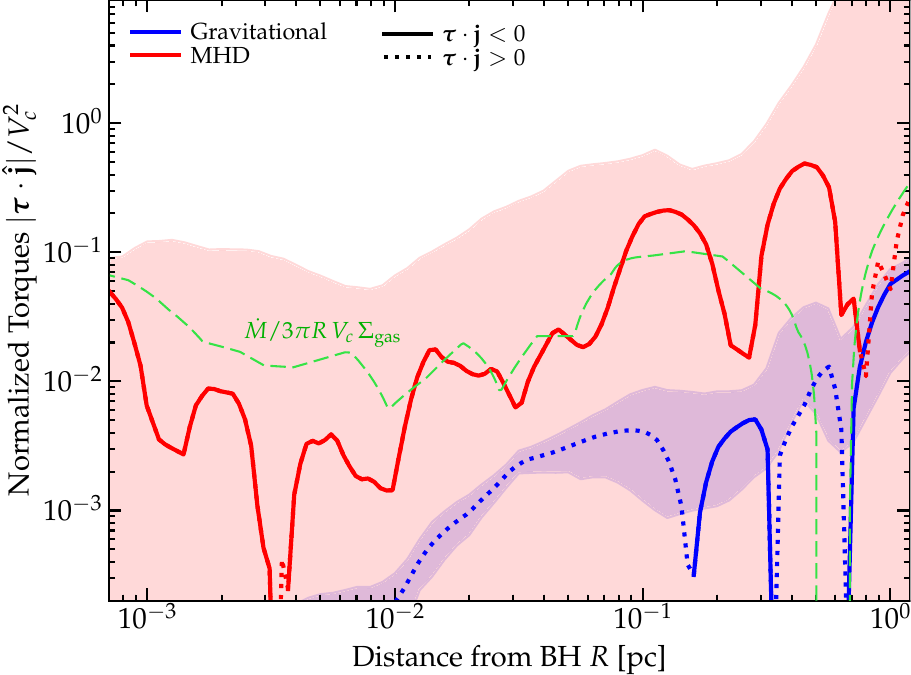}
	\caption{Quantitative properties of the torques from Fig.~\ref{fig:torque.map}. 
	We plot the instantaneous value of the torques ($\boldsymbol{\tau}  \equiv {\bf r} \times {\bf a}$) in cylindrical annuli (restricting to gas with $|z|<0.3\,R$), divided into two components (separating the acceleration ${\bf a}$ into said components as they are operator-split in-code): the ``MHD'' torque (all terms solved in the Riemann problem including magnetic+kinetic+thermal stresses), and the ``gravitational'' torque (from ${\bf a}_{\rm grav} = {\bf g}$). Other terms (\S~\ref{sec:torque:torque}) such as the torque from radiation pressure, cosmic rays, or microphysical viscosity fall below the plotted range and are negligible here.
	As in Fig.~\ref{fig:torque.map} we take the component of the torque in the angular momentum [AM] direction $\boldsymbol{\tau}  \cdot \hat{\bf j}$ (so values $<0$ [{\em solid}] indicate AM loss, while values $>0$ [{\em dotted}] indicate AM gain) , and normalize to $V_{c}^{2}(r)$ so that a value $\sim 1$ indicates an order-unity change in AM in a dynamical time. 
	We show the mean ({\em line}) in each annulus and $90\%$ range of individual cell values ({\em shaded}). 
	While \paperone\ showed gravitational torques dominate at $\gtrsim\,$pc scales, MHD torques clearly dominate here on smaller scales within the disk, and generate net angular momentum loss within the disk. There is large cell-to-cell scatter in the instantaneous torque, consistent with the fluctuating picture in Fig.~\ref{fig:torque.map}. We compare ({\em green dashed}) the instantaneous net inflow rate $\dot{M}/3\pi R V_{c}\Sigma_{\rm gas}$ measured (Fig.~\ref{fig:radial.profile.general}): this should be $\sim |\tau|/V_{c}^{2}$ for a thin, homogeneous, steady-state, axisymmetric disk on slowly-decaying circular orbits with constant $\boldsymbol{\tau}$. Given the large deviations from these assumptions the two agree fairly well.
	\label{fig:torque.profile}}
\end{figure}

\section{Structure of the Torques and Stresses Within The Disk}
\label{sec:torque}

We now explore the origins and physics of the torques/stresses/angular momentum transfer processes in these disks in more detail.

\subsection{The Torques and Net Angular Momentum Change of the Gas}
\label{sec:torque:torque}

First, consider the torques themselves. We directly compute and record the ``true'' in-code instantaneous specific torque $\boldsymbol{\tau} \equiv {\bf r} \times {\bf a}$ on every Lagrangian gas element in the simulations (where ${\bf r}$ is the vector distance from the center and ${\bf a}$ is the acceleration of the parcel).\footnote{Note that we define our coordinate system centered on the BH, not the center-of-mass of the material inside $|{\bf r}|$. At small radii where the BH dominates the potential this produces negligible effects, but at $\sim$\,pc scales the offset can start to become significant. However re-computing all our measures for either choice produces identical qualitative conclusions.} We compare this to the in-code specific angular momentum ${\bf j} \equiv {\bf r} \times {\bf v}$, to calculate the change in the scalar specific angular momentum $\boldsymbol{\tau}\cdot \hat{\bf j}$. It is convenient to express this in units of $R\,V_{\rm c}\,\Omega = V_{\rm c}^{2}$, with $\boldsymbol{\tau}\cdot \hat{\bf j}/V_{\rm c}^{2}$ representing the fractional angular momentum loss of a quasi-circular orbit per dynamical time. These are shown in a two-dimensional projection in the disk in Fig.~\ref{fig:torque.map}. We clearly see that the torques follow a complicated structure in space, with turbulent/fluctuating and non-axisymmetric modes dominating (with wavenumbers $k\,H \sim$\,a few and $k_{\phi} \gtrsim k_{R}$, $k_{z} \gtrsim k_{R}$). In a local, instantaneous sense, there is a broadly comparable volume with $\boldsymbol{\tau}\cdot\hat{\bf j}>0$ and $\boldsymbol{\tau}\cdot\hat{\bf j}<0$, though of course the mean torque causes angular momentum loss. This must be the case given the steady inflow and inwards radial flow seen above, and is consistent with the mean Reynolds and Maxwell-type stresses as we show below.

Fig.~\ref{fig:torque.profile} plots the azimuthally-and-vertically-averaged torques as a function of cylindrical radius $R$, now separating by components. We again in-code separate different contributions to the torques, so this is decomposing exactly the contributions to the torque seen by the gas cells. Specifically, we divide the acceleration (and therefore resulting torque) into several terms: the gravitational term (from all gravitational forces), the ``MHD'' term (from all MHD forces), the radiative term (from radiation pressure/photon momentum), the cosmic ray term (from cosmic ray pressure forces/scattering) and the viscous term (from physical  molecular, atomic, and Spitzer-Braginskii viscosity). Since the computation of these terms is operator-split in {\small GIZMO}, their decomposition is straightforward here.\footnote{One caveat is that, given our hierarchical timestepping scheme, we must be careful to synchronize/drift the forces to the same instant in time in the outputs, rather than simply using the value from the last ``kick'' update for each cell.} The dominant component on-average at $r \ll $\,pc is clearly the ``MHD torque,'' which includes all the terms from the Riemann problem solved in-code (essentially, the sum of torques from magnetic stresses, thermal pressure gradients, kinetic/Reynolds stresses or winds/outflows, etc). This shows a very large variance and (per Fig.~\ref{fig:torque.map}) often local sign changes, but the {\em net} torque, azimuthally averaged, is negative (i.e.\ causing a net loss of angular momentum). The variation of the instantaneous torques in time (whether we consider a given location ${\bf x}$ or Lagrangian parcel) is comparable to their variation in space shown in Figs.~\ref{fig:torque.map}-\ref{fig:torque.profile}, and the variation of the spatially-averaged torque at a given annulus $R$ is a factor of a few, as the various ``bumps'' and ``dips'' seen in Fig.~\ref{fig:torque.profile} appear and disappear. 

Given these broad fluctuations, the range of torques are consistent with the measured instantaneous inflow rates at a given time and radius, to within a factor of a few. But since, as noted above, neither is in exact steady state and both show inhomogeneity in space and time, we do not expect perfect agreement. The agreement improves of course, as it must, if we average over time as well as space.

The ``gravitational torque'' (torques arising from the gravitational forces themselves directly, in a non-axisymmetric potential) are much weaker but the next-strongest on average. At large radii $\gtrsim$\,pc, these torques actually dominate, and they are extensively discussed in \paperone\ as a result (see also \citealt{daa:20.hyperrefinement.bh.growth}). However, a dominant gravitational torque relies on (a) the non-BH contributions to the potential being non-negligible (i.e.\ terms that could conceivably be non-axisymmetric), and (b) the presence of a dominant collisionless (in this case stellar) disk, with mass much larger than the gas disk at a given radius, which can efficiently torque the gas disk (thus giving rise to torques that are stronger than would appear with a gas disk alone). The cessation of efficient star formation at scales $\lesssim\,$pc therefore produces the transition to MHD torques dominating. In addition to being relatively weak, the gravitational torque is quite smooth (being dominated by the external $m=1$ mode of the stellar disk at $\gtrsim\,$pc scales) and flips sign (so actually spins up the disk) at sub-pc scales (as predicted and discussed in \paperone). So it cannot be the dominant source of inflow.

The radiative torques from radiation pressure forces are much smaller at all radii plotted. Also not shown because they fall entirely off the plot, but computed nonetheless in our simulations, are the torques from the micro-physical Spitzer-Braginskii and atomic/molecular viscosities (many orders-of-magnitude smaller than those here) and anisotropic cosmic ray forces. 

It is also easy to verify -- and indeed, is expected due to the very low values of $\beta \ll 1$ in the disk -- that the thermal pressure gradient contribution to the MHD torque, is negligible. So the net torque inside the disk is dominated by some combination of the magnetic and kinetic stresses.

\subsection{Contributions to the Magnetic and Kinetic Stresses}
\label{sec:torque:stress}

\begin{figure*}
	\centering\includegraphics[width=0.48\textwidth]{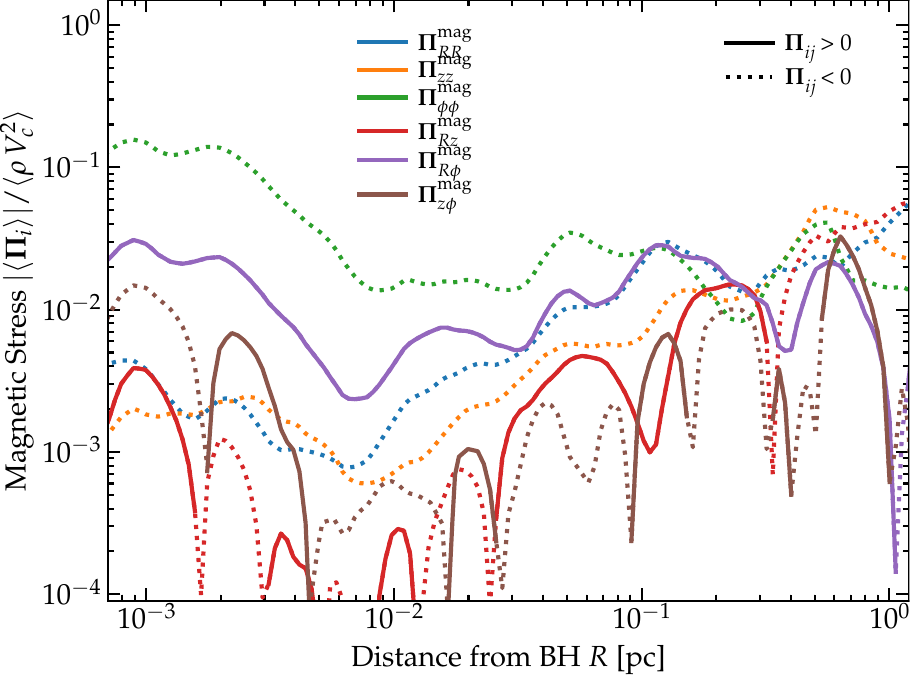}
	\centering\includegraphics[width=0.48\textwidth]{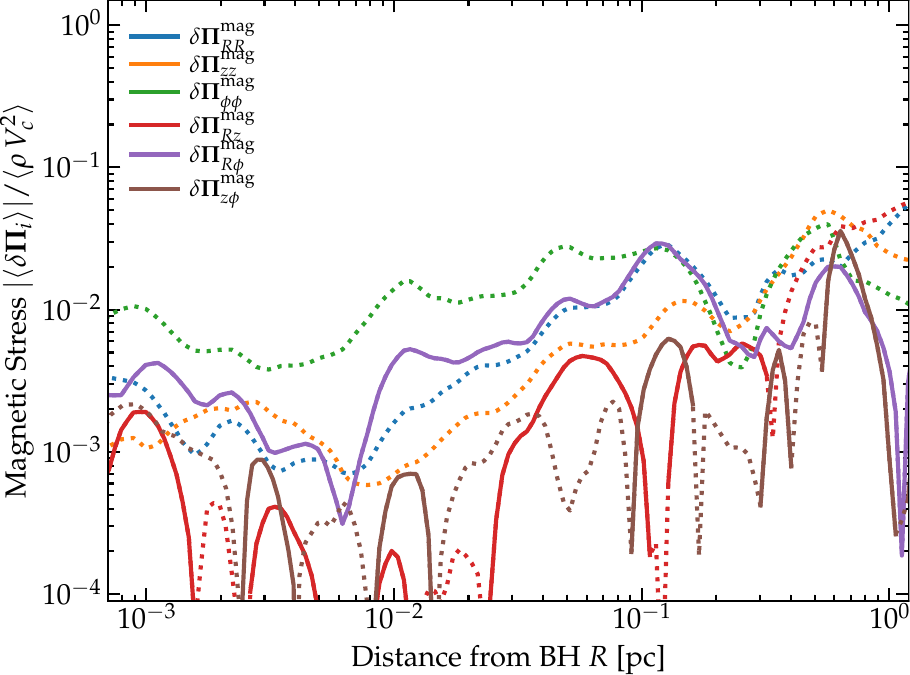}
	\centering\includegraphics[width=0.48\textwidth]{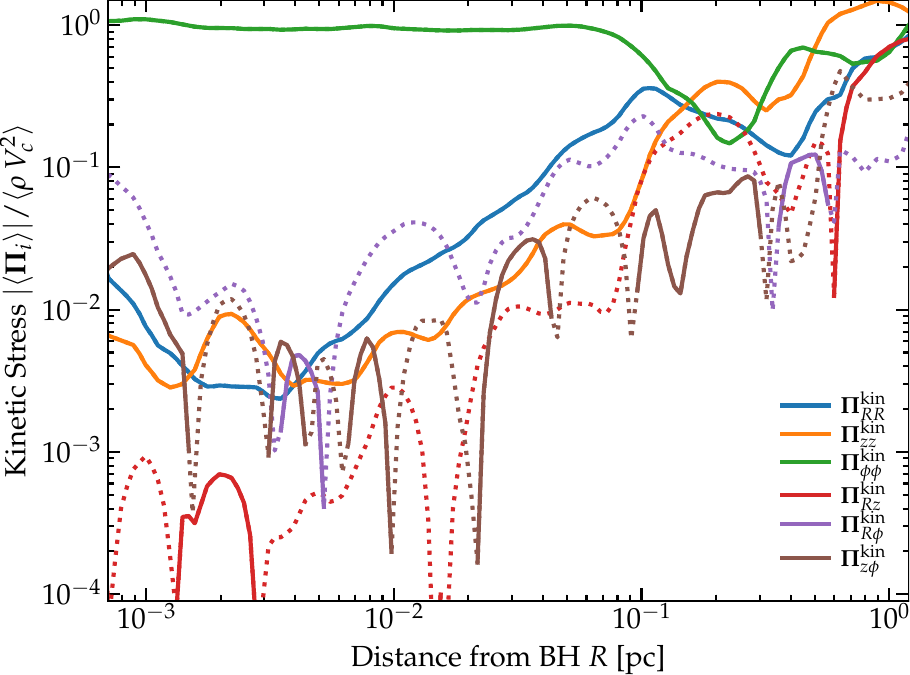}
	\centering\includegraphics[width=0.48\textwidth]{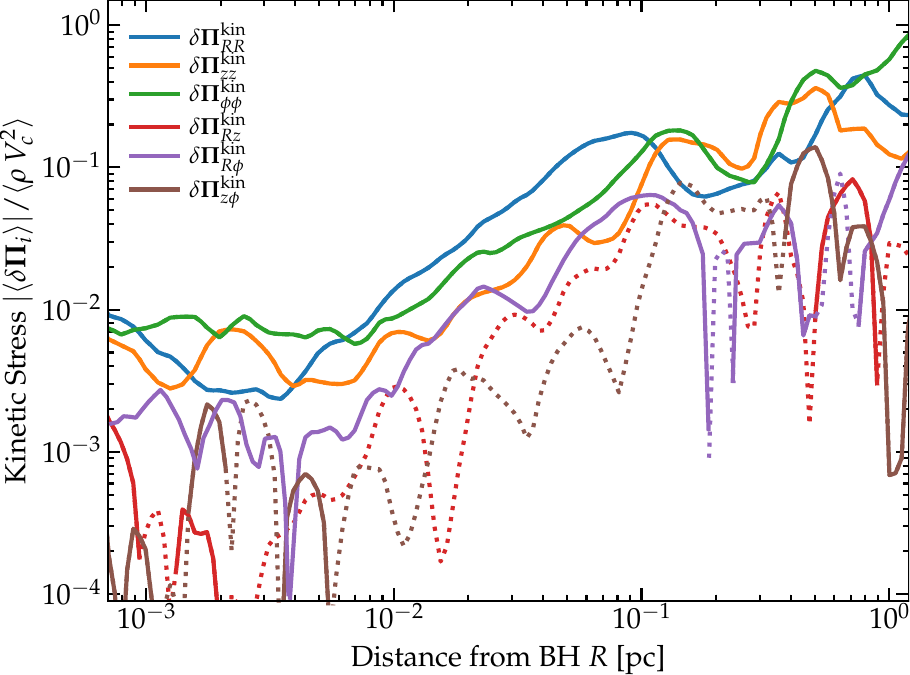}
s	
	\caption{Magnetic ($\boldsymbol{\Pi}_{\rm mag} \equiv (|{\bf B}|^{2} {\bf I}/2 - {\bf B}{\bf B})/4\pi$; {\em top}) and kinetic ($\boldsymbol{\Pi}_{\rm kin} \equiv  \rho {\bf v} {\bf v}$; {\em bottom}) contributions to the total stress tensor $\boldsymbol{\Pi}_{\rm tot}$. 
	{\em Left:} We plot the mean $\langle \boldsymbol{\Pi} \rangle$ (within the disk, excluding dense star-forming clumps where present) of each component of $\boldsymbol{\Pi}$ in radial annuli, as a function of distance to the SMBH, normalized to $\langle \rho\,V_{c}^{2} \rangle$. 
	{\em Right:} Same, but plotting just the fluctuating terms: $\delta \boldsymbol{\Pi}_{\rm mag} \equiv (|\delta {\bf B}|^{2} {\bf I}/2 - \delta{ \bf B}\delta{ \bf B})/4\pi$, $\delta \boldsymbol{\Pi}_{\rm kin} \equiv  \rho \delta {\bf v} \delta {\bf v}$ (where $\delta {\bf B} \equiv {\bf B} - \langle {\bf B} \rangle$, $\delta {\bf v} \equiv {\bf v} - \langle {\bf v} \rangle$ in each annulus).
	Positive (negative) terms are shown as {\em solid} ({\em dotted}) lines. 
	In the disk the rotation/centrifugal term $\Pi^{\rm kin}_{\phi\phi}$ is the most important as expected, followed by the toroidal magnetic term $\Pi^{\rm mag}_{\phi\phi}$, which provides most of the pressure. 
	The $RR$, $zz$, $R\phi$, and $z\phi$ terms are broadly order-of-magnitude comparable, with $Rz$ generally smaller. 
	Of the terms that can drive angular momentum transfer ($\boldsymbol{\Pi}^{\rm mag}_{i\phi}$ and $\delta \boldsymbol{\Pi}^{\rm kin}_{i\phi}$), the $R\phi$ (traditional Maxwell/Reynolds) terms not only appear most important, but consistently $>0$ (indicating angular momentum transfers outwards and material accretes) -- the $z\phi$ component for example is generally smaller but also much more clearly oscillating in sign (often driving weak angular momentum gain via vertical accretion onto the disk). 
	\label{fig:stress.terms}}
\end{figure*}

\subsubsection{Which Components of the Kinetic/Magnetic Stress Regulate Angular Momentum Transfer?}
\label{sec:torque:stress:which}

Given that the torques are dominated by a combination of magnetic and kinetic stresses, we now turn to examine those stresses directly. Recall, the gas momentum equation can be written $\partial(\rho {\bf v})/\partial t + \nabla \cdot \boldsymbol{\Pi}_{\rm internal} = {\bf S} + \rho\,{\bf g}$, with ${\bf S}$ being a source term representing e.g.\ non-hyperbolic terms from CR and radiation partial-coupling (which are small in the disk region of interest) and ${\bf g} = -\nabla \Phi$ the gravitational acceleration. $\boldsymbol{\Pi}_{\rm internal} \equiv \boldsymbol{\Pi}_{\rm kin} + \boldsymbol{\Pi}_{\rm mag} + \boldsymbol{\Pi}_{\rm therm} + \boldsymbol{\Pi}_{\rm visc} + \boldsymbol{\Pi}_{\rm cr} + \boldsymbol{\Pi}_{\rm rad}$ represents the usual pressure tensor decomposed into kinetic ($\boldsymbol{\Pi}_{\rm kin} \equiv  \rho {\bf v} {\bf v} $), magnetic ($\boldsymbol{\Pi}_{\rm mag} \equiv (|{\bf B}|^{2} {\bf I}/2 - {\bf B}{\bf B})/4\pi$), thermal/``hydrodynamic'' ($\boldsymbol{\Pi}_{\rm therm} = n\,k_{\rm B}\,T\,{\bf I}$), viscous ($\boldsymbol{\Pi}_{\rm visc} \equiv ({\nu_{\rm visc}}/{3})\,( 3 \hat{\bf B} \hat{\bf B} - {\bf I} ) (3 \hat{\bf B} \hat{\bf B} - \bf{I} ) : (\nabla {\bf v} )$), cosmic-ray ($\boldsymbol{\Pi}_{\rm cr} \equiv \int {\bf p}_{\rm cr}\,{\bf v}_{\rm cr}({\bf p}_{\rm cr}) f_{\rm cr}({\bf p}_{\rm cr}) \, d^{3}{\bf p}_{\rm cr}$) and radiation ($\boldsymbol{\Pi}_{\rm rad} \equiv \int \frac{e_{{\rm rad},\,\nu}}{3}\,\mathbb{D}_{\nu}\,{\rm d}\nu$) components respectively. As shown in Figs.~\ref{fig:image.wedgeplot}, \ref{fig:radial.profile.general}, \ref{fig:torque.profile}, and discussed further in \S~\ref{sec:torque:torque}, the thermal/hydrodynamic, viscous, cosmic-ray, and radiation stresses, as well as the source terms ${\bf S}$, are small compared to the leading-order magnetic $\boldsymbol{\Pi}_{\rm mag}$ and kinetic $\boldsymbol{\Pi}_{\rm kin}$ terms (and of course gravity). Fig.~\ref{fig:stress.terms} therefore plots the average value of each component of $\boldsymbol{\Pi}_{\rm kin} \equiv  \rho {\bf v} {\bf v} $ and $\boldsymbol{\Pi}_{\rm mag} \equiv (|{\bf B}|^{2} {\bf I}/2 - {\bf B}{\bf B})/4\pi$ in radial annuli around the BH, where again we directly extract the stress tensor from the simulation at each time. 

As expected (per Fig.~\ref{fig:v.profile}), within the disk ($\lesssim 0.1\,$pc), the rotational/centrifugal/angular momentum term $\langle \boldsymbol{\Pi}^{\rm kin}_{\phi\phi} \rangle \equiv \langle \rho\,v_{\phi}^{2} \rangle$ dominates, and provides the dominant support versus the radial gravitational force from the SMBH. The next-most-prominent term, at least in the inner disk, is the azimuthal/toroidal magnetic term $\langle \boldsymbol{\Pi}^{\rm mag}_{\phi\phi} \rangle \equiv -\langle B_{\phi}^{2} \rangle/8\pi$, which, as we showed above, dominates the internal disk pressure and provides most of the vertical support (Figs.~\ref{fig:radial.profile.general} \&\ \ref{fig:b.profile}). Then there are a group of broadly order-of-magnitude comparable terms (as anticipated from Figs.~\ref{fig:b.profile} \&\ \ref{fig:v.profile}) including the $RR$, $zz$, $R\phi$, and $z\phi$ terms. These reflect the not-extremely-anisotropic fluctuating velocity and magnetic field terms seen in Figs.~\ref{fig:b.profile} \&\ \ref{fig:v.profile}, and of course the $RR$ and $zz$ terms are just dominated by the normal radial and vertical dispersions. The $Rz$ term is often significantly weaker in the disk.  

Of course, these are the {\em total} stresses, so for example $\langle \boldsymbol{\Pi}_{R\phi}^{\rm kin} \rangle = \langle \rho\,{ v}_{R}{ v}_{\phi} \rangle = \langle \rho \rangle \langle v_{R} \rangle \langle v_{\phi} \rangle + \langle \rho  \delta v_{R} \delta v_{\phi} \rangle$ includes both the ``mean field'' term $\langle \rho \rangle \langle v_{R} \rangle \langle v_{\phi} \rangle$ and the ``fluctuating'' or traditional Reynolds stress term $\langle \delta\boldsymbol{\Pi}^{\rm kin}_{R\phi} \rangle \equiv \langle \rho \delta v_{R} \delta v_{\phi} \rangle$ (where $\delta v_{i} \equiv v_{i} - \langle v_{i} \rangle$). $\langle \boldsymbol{\Pi}_{R\phi}^{\rm kin} \rangle$ is negative therefore, as expected in any system with net inflow, because it is dominated by the mean term with $\langle v_{\phi} \rangle > 0$ (the rotational motion) and $\langle v_{R} \rangle < 0$ (inflow). We therefore also plot the separation of each component of $\boldsymbol{\Pi}$ into fluctuating components, which allows us to more clearly see (1) the crudely isotropic turbulent fluctuations,\footnote{As discussed in \S~\ref{sec:v:ecc}, if we subtract the best fit $m=1$ component from each $\delta v_{i}(R,\,\phi)$ to attempt to remove the effects of coherent eccentric motion, this has at most a modest (order-unity) effect reducing the kinetic $\langle \delta\boldsymbol{\Pi}^{\rm kin}_{RR} \rangle$ and $\langle \delta\boldsymbol{\Pi}^{\rm kin}_{\phi\phi} \rangle$, and almost no effect on the most relevant Reynolds stress $\langle \delta\boldsymbol{\Pi}^{\rm kin}_{R\phi} \rangle$.} (2) the fact that the Reynolds stress $\langle \delta v_{R} \delta v_{\phi} \rangle$ and both the total and fluctuating Maxwell stresses are almost always positive (indicating angular momentum loss, given our sign convention), and (3) that in the rotationally-dominated disk at $\ll 0.1\,$pc, the dominant torque should come from the traditional Maxwell stress $\langle \boldsymbol{\Pi}^{\rm mag}_{R\phi} \rangle$. As discussed below (\S~\ref{sec:torque:stress:ratios}), in the inner disk the Maxwell stress is itself dominated by the mean component, but with non-negligible contribution from the fluctuating ($\langle -\delta B_{\phi} \delta B_{R} \rangle/4\pi$) component. The magnitude of the fluctuating components in both cases is consistent with the properties of the turbulence discussed in \S~\ref{sec:v:turb} (e.g.\ $\langle \rho \delta v_{\phi} \delta v_{R}\rangle \sim \langle \rho\rangle\,\langle \delta v_{\rm turb}^{2} \rangle$). 

Briefly, it is worth noting that the sign and efficiency of the Maxwell stresses are expected here, given  the strong anti-correlation between $B_{\phi}$ and $B_{R}$, as  demonstrated in \S~\ref{sec:b}. Recall that this results from the simple fact that the toroidal field is supplied by radial fields in the disk plane (given the induction equation; see Fig.~\ref{fig:b.induction.eqn} and \S~\ref{sec:b:origin}), with simple flux-freezing/advection and trans-\Alf{ic} turbulence explaining the mean and fluctuating field strengths. We stress that this is distinct from some historical models wherein $B_{\phi}$ and $B_{R}$ are sourced from some strong mean poloidal field $\langle B_{z}\rangle$, in which case such an anti-correlation (ensuring inflow) is non-trivial. For the Reynolds stress in the outermost disk where it dominates, it is less obvious what exactly regulates the detailed quantitative properties of the stress, related to our discussion in \S~\ref{sec:v:turb} regarding the uncertainty in what exactly drives the turbulence. If instabilities related to self-gravity like the global $m=1$ modes and/or gravito-turbulence drive the turbulence, these would naturally produce the kind of in-plane motion with the correct sign (on average) of $\langle \rho \delta v_{R} \delta v_{\phi} \rangle$, but other instabilities that could be present, such as the low-$\beta$ extensions of the MRI (see \S~\ref{sec:b:origin:turb}), have not yet been well-studied in the non-linear regime.

For completeness, we note if we define the {\em comoving} MHD stress tensor ${\bf T} \equiv \boldsymbol{\Pi}^{\rm mag} + \delta\boldsymbol{\Pi}^{\rm kin}$, there are three terms in Fig.~\ref{fig:stress.terms} that can in principle give rise to torques in the disk plane.\footnote{The equation for the co-moving evolution of the specific angular momentum $\ell$ of some gas parcel becomes $\langle \rho \rangle \, {\rm d} \ell / {\rm d} t = -\nabla \cdot (R\,{\bf T}_{i\phi})$.} First, the usual Maxwell/Reynolds stress (${\bf T}_{R\phi}$), which we discussed above and will discuss further below. Second, the non-axisymmetric azimuthal term which can produce a local torque $\propto \partial {\bf T}_{\phi\phi} / \partial \phi$. And third, the wind or convective term $\propto \partial {\bf T}_{z\phi}/\partial z$. The azimuthal $\partial {\bf T}_{\phi\phi}/\partial \phi$ term is usually neglected (even when $|{\bf T}_{\phi\phi}|$ itself is large) because by definition the {\em net} torque integrated over $\phi$ must vanish at leading order for orbits that are approximately closed and/or circular (assuming small deviations from axisymmetry in the potential) in a disk which is not evolving on a timescale fast compared to the orbital time \citep{kalnajs:1971}. However, \citet{hopkins:inflow.analytics} showed that this term can be leading order in the net torque when the ``gravitational torques'' discussed above are important (or when non-axisymmetric terms in the potential become non-linearly large), because one can break the periodicity in $\phi$ that causes the integral of  $\partial {\bf T}_{\phi\phi}/\partial \phi$ to vanish if the external potential induces orbit crossings and shocks or pileups in the gas. So we note it briefly here because at radii $\gtrsim 0.1\,$pc, where the gravitational torques can be important (and where we see in e.g.\ Fig.~\ref{fig:vfields.lines} that the strong eccentric pattern produces a clear discontinuity in the velocity streamlines), this can actually play a leading-order role. However, in the primarily rotationally-supported inner disk at $\ll 0.1\,$pc, it becomes less important.

The vertical/convective term in $\partial {\bf T}_{z\phi}/\partial z$ is usually considered when there is a strong outflow/wind removing angular momentum from the disk. But recall here (Figs.~\ref{fig:vfields.lines}-\ref{fig:v.profile}) that the vertical velocity is primarily (weak) inflow. This means that in order to significantly lower the specific angular momentum of disk material and promote accretion, the inflowing gas would have to (1) carry a considerable fraction of the total accretion rate (which it does not, because we showed above the density structure means most of the inflow is through the gas in the midplane with $|z|\lesssim H$), and (2) be significantly sub-Keplerian (which it is usually not, since by definition it tends to join the disk at its circularization radius). 

So it seems clear that, on average, the most important MHD torques in the inner disk indeed arise predominantly from the usual Maxwell+Reynolds stress.

\begin{figure}
	\centering\includegraphics[width=0.95\columnwidth]{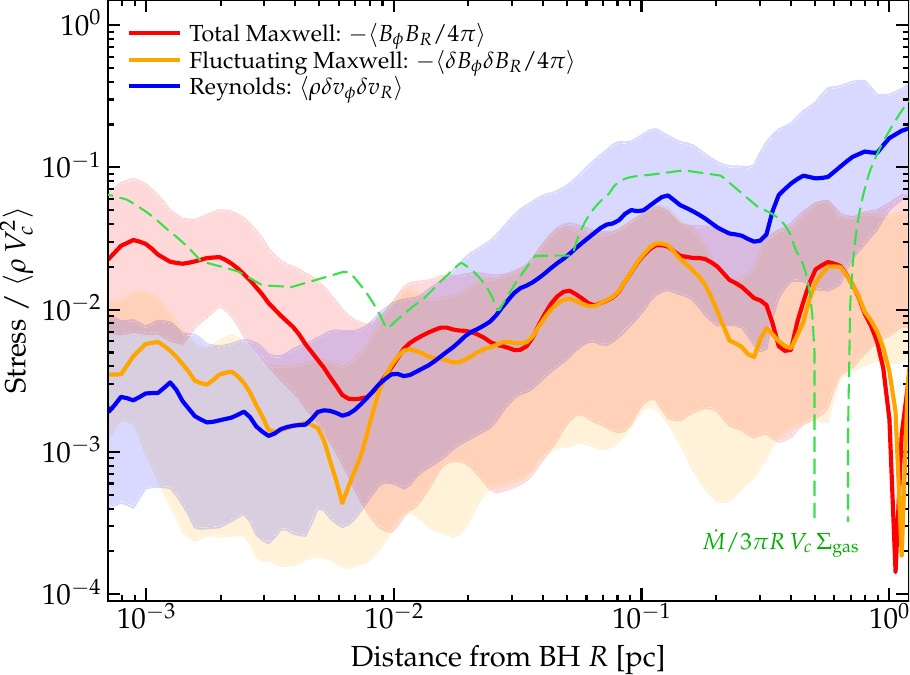}
	\centering\includegraphics[width=0.95\columnwidth]{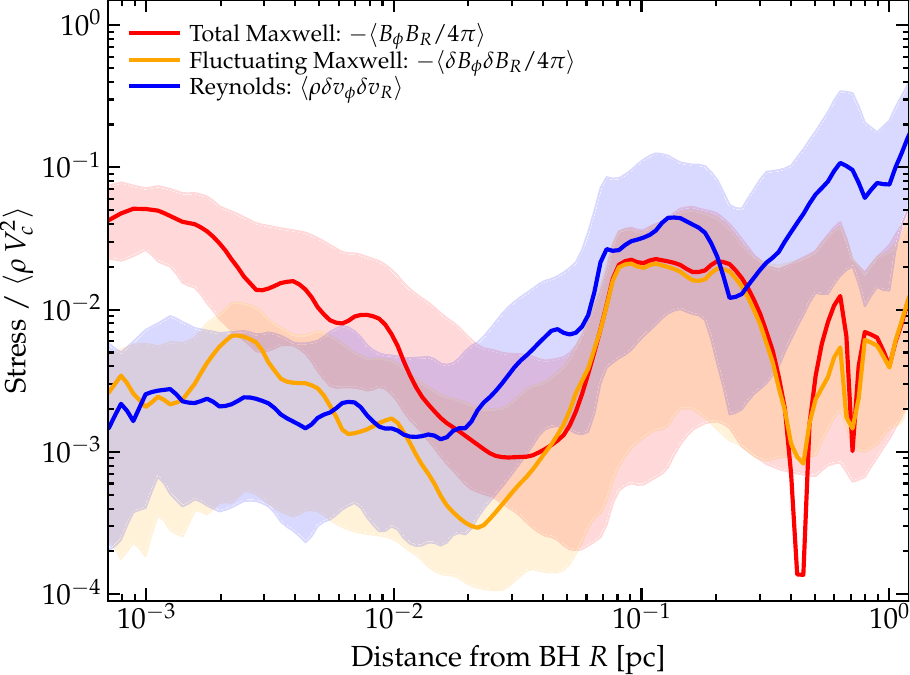}
	\caption{Quantitative properties of the Maxwell \&\ Reynolds stresses from Fig.~\ref{fig:stress.terms} in more detail. We plot two times as in Figs.~\ref{fig:b.profile} \&\ \ref{fig:v.profile} (panels), showing the $R\phi$ component of the Maxwell stress (both the total $-\langle B_{\phi} B_{R} \rangle$ and fluctuating $-\langle \delta B_{\phi} \delta B_{R} \rangle$ component, as labeled) and the Reynolds stress ($\langle \rho \delta v_{\phi} \delta v_{R} \rangle$). Note that the total Maxwell stress, $-\langle B_{\phi} B_{R}/4\pi\rangle$ (i.e.\ not just the fluctuating component), is what appears in the angular momentum (AM) equation, while $\langle \rho v_{\phi} v_{R} \rangle$ includes additional pure advection terms that do not influence the AM (i.e.\ only $\langle \rho \delta v_{\phi} \delta v_{R} \rangle$ gives rise to AM exchange). Each quantity is averaged over several snapshots around the time of interest and in annuli within the disk, excluding any dense star-forming clumps. The shaded range shows the $\pm1\sigma$ range of $|-B_{\phi} B_{R}/4\pi|$, etc., in all cells in each annulus. 
	Reynolds stresses tend to be larger at radii $\gtrsim 0.01-0.1\,$pc and especially outside the ordered disk, where the turbulence is weakly super-\Alf{ic} (Fig.~\ref{fig:mach}). Within the inner disk, Maxwell stresses dominate by a factor of $\sim 10$. The total/mean Maxwell stress is larger than the fluctuating component by factors of several, consistent with the strong mean fields in e.g.\ Fig.~\ref{fig:b.induction.eqn} but distinct from predictions of e.g.\ the weak-seed-field MRI. We also ({\em top}) plot the instantaneous inflow rate $\dot{M}/3\pi R V_{c}\Sigma_{\rm gas}$ from Fig.~\ref{fig:torque.profile}. This agrees well with the sum of total Maxwell+Reynolds stresses, as expected if they dominate the angular momentum transport.
	\label{fig:maxwell.reynolds.turb}}
\end{figure}

\subsubsection{Relative Importance of the Maxwell vs.\ Reynolds Stresses, and Fluctuating vs.\ Mean-Field Components}
\label{sec:torque:stress:ratios}

In Fig.~\ref{fig:maxwell.reynolds.turb}, we therefore plot the azimuthally-and-vertically averaged Reynolds stress $R_{R\phi} \equiv \langle \delta \boldsymbol{\Pi}^{\rm kin}_{R\phi}  \rangle \equiv \langle \rho\,({v}_{R}-\langle {v}_{R}\rangle)\,(v_{\phi}-\langle { v_{\phi}}\rangle) \rangle \equiv \langle \rho\, \delta v_{R} \delta v_{\phi} \rangle$ and Maxwell stress $-M_{R\phi} \equiv\langle \boldsymbol{\Pi}^{\rm mag}_{R\phi} \rangle \equiv - \langle { B}_{R}\,{ B}_{\phi} \rangle/4\pi$, as a function of radius from the BH, with their range and sign. Because the Maxwell term $\propto \langle B_{R}B_{\phi} \rangle$ includes both a mean-field $\langle B_{R}\rangle \langle B_{\phi}\rangle$ and fluctuating field $\langle \delta B_{R}  \, \delta B_{\phi}\rangle$ component, we also plot both the total Maxwell and fluctuating component alone. We can immediately compare the normalization of these terms to the torques $\tau$ in Fig.~\ref{fig:torque.profile} noting that, if dominant, the sum of these terms in the units given ($\rho\,V_{c}^{2}$) should approximately equal the torque in units of $V_{c}^{2}$ (up to an order one constant that depends on the averaging weighting and slope of the different terms). We see reasonably good agreement, with a factor of $<2$ difference owing to how the different weighting of the means. And of course the magnitude of the stresses are much larger than in a SS73-like $\alpha$ disk where (by assumption) the \Alf\ and turbulent velocities are much smaller than the thermal sound speed.

Next, we compare and see that at large radii $\gg 0.01\,$pc, the Reynolds stress is dominant by a factor of a couple to a few, while at smaller radii the two are comparable or the Maxwell stress is dominant by a similar factor. To understand this better we recall Fig.~\ref{fig:mach}, which plots the Mach number $\mathcal{M}$ of the turbulence versus radius. We see that it is quasi-isotropic, trans-\Alf{ic}, and highly super-sonic (consistent with all our previous analysis of $\beta$ and velocity fields directly). We note here though that although the trend of \Alf\ Mach number $\mathcal{M}_{A}$ with radius is weak, we see the turbulence transition from mildly sub-\Alf{ic} at the smallest radii, to modestly super-\Alf{ic} at large radii. This is broadly consistent with the trend in the ratio of Reynolds to Maxwell stresses. 

We also see that the although mean field Maxwell stress is usually dominant, the fluctuating field stress is non-negligible, especially at some times at very small radii. It will occasionally occur that an excess of toroidal magnetic flux will build up in the center, sometimes briefly suppressing accretion but then producing a large mean-field Maxwell stress and leading to a rapid accretion event. This leads to the fields being accreted inwards, changing the ratio of mean-to-fluctuating stresses briefly. But in general the behavior plotted holds, showing that the fluctuations $\delta B_{\phi}$ and $\delta B_{R}$ are sufficiently strongly anti-correlated that they can be comparable to the mean-field stress, even when $\langle |\delta B_{\phi}| \rangle \lesssim |\langle B_{\phi} \rangle |$. 

The ``origins'' of the angular momentum transport and stresses, therefore, are directly tied to (a) the origins of the strong magnetic fields, and (b) the origin of the turbulence within the disk, each of which was discussed above.

\section{Comparison To Previously-Studied ``Strongly-Magnetized'' Disks}
\label{sec:lit.mag.disks}

While the majority of the literature on quasar accretion disks has focused on disks with ``weak'' magnetic fields (e.g.\ magnetic pressure less than radiation or thermal pressure), there has been some discussion of disks in the strong-field limit. These can generally be divided into a couple of different categories, some of which are indeed closely related to the behaviors we report here in our ``flux-frozen'' (and flux-fed) disks, and some of which are not. We therefore find it helpful to briefly review these past models and distinguish the results from their predictions. Note that a more quantitative comparison to SS73-like ``weakly-magnetized'' ($\beta \gg 1$) disks is given in \paperthree\ (see also \S~\ref{sec:basic:profile} \&\ \ref{sec:magnetic.compare.SS73}).

\subsection{Magnetically Arrested (MAD) Disks}
\label{sec:mad}

Magnetically arrested (MAD) disks, in which accretion is halted by strong magnetic fields near the BH, are usually characterized by extremely strong poloidal magnetic flux \citep{bisnovatyi.kogan:1976.mad.disk,narayan:2003.mad.disk}. We clearly see that in many (though not all) respects, the behaviors here are opposite those predicted in the MAD regime and seen in idealized simulations of MAD disks \citep[e.g.][]{tchekhovskoy:bh.spin.vs.radio.pwr,tchekhovskoy:2011.mad.disk.jets,white:2019.mad.disk.sims,fuguo:2019.mad.disk.sims.rad.properties}. Most importantly: (1) accretion proceeds rapidly here, (2) the accretion is in fact aided by magnetic fields, (3) the fields are primarily toroidal, which has qualitatively different consequences at these scales, and (4) the disks are vigorously turbulent and cool efficiently (which also helps promote strong inflows, compared to the more well-studied MAD limit). 

This is not surprising, as it is trivial to verify that, everywhere we resolve, the magnetic field strengths are below the MAD limit of $\langle B_{z} \rangle_{\rm mad}^{2}/8\pi \sim G\,M_{\rm BH}\,\Sigma_{\rm gas}/4\,R^{2}$ (this limit is effectively equivalent to $v_{A,\,z}^{\rm mad} \gtrsim v_{\rm K}$, for the poloidal field, but we see relatively weak poloidal fields and even including toroidal components we see $v_{A} < v_{\rm K}$ everywhere). This is true both in the simulations here, and the analytic models in presented in \paperthree\ which can be extrapolated to ISCO scales. As argued therein, if one starts from an outer disk boundary like that in our simulations, then even if we ignore the important geometric constraint that one cannot generate a dominant mean vertical field via flux freezing from field configurations like those in our simulations, it would require much stronger amplification of the mean field compared to what we see ($P_{B} \propto \langle | {\bf B} |^{2} \rangle \propto \rho^{\gamma}$ with $\gamma>2$, as opposed to $\gamma \sim 4/3$ which we measure in the simulations here). In addition, the turbulence would have to be strongly suppressed to become  become highly sub-\Alf{ic} ($v_{\rm turb} \ll v_{A}$), in order to satisfy a MAD-like criterion $B^{2}/8\pi \gtrsim G\,M_{\rm BH}\,\Sigma_{\rm gas}/4\,R^{2}$ as $r\rightarrow 0$. 

However, the argument that a disk should eventually enter the MAD limit somewhere outside the ISCO or event horizon is often phrased in terms of an incoming magnetic flux from much larger scales, as e.g.\ $\Phi_{\rm mag} \sim \pi\,R^{2}\,B \gtrsim \Phi_{\rm crit} \sim {\rm \mu G\,pc^{2}}\,(M_{\rm bh}/10^{7}\,{\rm M_{\odot}})^{3/2}\,(\dot{M}_{\rm BH}/\dot{M}_{\rm Edd})^{1/2}$ using the scalings from \citet{fuguo:2019.mad.disk.sims.rad.properties}. At a glance (comparing Fig.~\ref{fig:b.profile}) it would naively seem that our simulations (and most observed AGN) exceed this limit at ISM scales. But we caution that this flux-based extrapolation makes several key assumptions which are {\em not} valid in the simulations here. Most importantly, it assumes (1) that the ``seed'' field is dominated by a coherent poloidal/vertical/dipole field (uniform ${\bf B} \approx \langle B_{z}\rangle\,\hat{z}$ with coherence length $\sim R$), which then (2) is amplified much more rapidly than other field components, assuming strictly homologous, laminar accretion with ${\bf B} = \langle B_{z} \rangle\,\hat{z}$, so that $\langle B_{z} \rangle \propto 1/R^{2}$, with (3) a thermal-pressure dominated \citet{shakurasunyaev73} $\alpha$ disk assumed to estimate $\Sigma_{\rm gas}$ and the disk thickness $H/R$ and pressure support level ``needed'' to arrest the disk, together with (4) assumed radial/toroidal field components $|B_{R}|/|B_{z}| \sim |B_{\phi}|/|B_{z}| \sim H/R \ll 1$ that are sourced only via turbulence from the dominant poloidal field. But none of these conditions are satisfied in our simulations. As shown above, at large radii the ``seed'' flux is isotropic and turbulent (with $|\delta B_{z}| \gg \langle B_{z} \rangle$; Figs.~\ref{fig:bfield.demo}, \ref{fig:radial.profile.general}, \ref{fig:bfields.faceon.edgeon} \&\ \ref{fig:b.profile}). Moreover in the ``free-fall'' region where gas is tidally captured by the SMBH and then circularizes to form the disk (see \S~\ref{sec:b:ov}), the structure of the tidal compression/expansion means that any mean poloidal $\langle B_{z} \rangle$ is preferentially suppressed relative to the dominant $\langle B_{R} \rangle$ or $\langle B_{\phi} \rangle$ components, ensuring that $|{\bf B}| \gg |\delta B_{z}| \gg \langle B_{z} \rangle$ in the disk (e.g.\ Fig.~\ref{fig:b.profile}). Per \S~\ref{sec:why.flux.freezing} and discussion in \citet{lubow:1994.bfield.advection.acc.disks.vs.resistivity}, this means that the vertical field $B_{z}$ will be ``locked'' in a regime where turbulent flux transport and resistivity suppress the amplification of $\langle B_{z} \rangle$ (qualitatively unlike $B_{R}$ and $B_{\phi}$, which grow via flux-freezing/advection rather than being sourced from $\langle B_{z} \rangle$), explaining why the mean $\langle B_{z} \rangle$ grows much more slowly than $\langle B_{R} \rangle $ and $\langle B_{\phi} \rangle$ as $r\rightarrow 0$ (opposite the MAD assumption). And we have shown that assuming a thermal-pressure-dominated ($\beta \gg 1$) \citet{shakurasunyaev73} disk would predict orders-of-magnitude different disk properties from those in our simulations (with e.g.\ $H/R$ and $\Sigma_{\rm gas}$ incorrect by factors of $\sim 300$ and $\sim 10^{4}$, respectively; see Fig.~\ref{fig:radial.profile.general}). 

Thus it is obvious that our disks are not magnetically arrested in practice, nor should they be given the physical conditions; however, it is certainly possible that if the densities of the inflows (hence accretion rates) dropped sufficiently at some later time, the system might transition to a MAD-like state (depending on whether the magnetic fields also declined, and whether or not the turbulence became less vigorous as the densities and accretion rates declined). It is also conceivable that very close to the BH (around the ISCO) the behavior of these disks could become more ``MAD-like'' (or otherwise truncated or inefficient at low accretion rates, see \citealt{hogg:2018.truncated.disks.sims,datta:2022.am.transport.acc.disk.weak.bfields}). But again, the simple analytic models we develop in \paperthree\ suggest that this would require some qualitative change in the fundamental scalings for the turbulence and magnetic field strengths, compared to the resolved behaviors in our simulations.

\subsection{Magnetically Elevated or Levitated Disks}
\label{sec:elevated} 

Magnetically levitated and/or elevated disks are disks in which magnetic fields are relatively weak in the midplane (with $\beta \gg 1$), but become fractionally more important with $\beta \lesssim 1$ in the tenuous gas at a few scale heights above the midplane ($|z| \gg H$). This can produce a variety of interesting behaviors with e.g.\ inflow/outflow along current sheets or angular momentum transport via MHD winds, but the bulk of the mass is still effectively a ``classical'' thermal-pressure-dominated $\alpha$ disk \citep{johansen.levin:2008.high.mdot.magnetized.disks,gaburov:2012.public.moving.mesh.code,sadowski:2016.mag.elevated.disk.sims.radpressure.midplane.thermal.instability.suppressed,mishra:2020.elevated.disks.sims.strongly.sensitive.initial.beta}. Again our simulations are in an opposite regime: (1) most importantly we see $\beta \ll 1$ in the midplane and in a gas-mass- and gas-density-weighted mean sense; (2) the dependence of $\beta$ on both gas density ($\beta$ is lower in denser gas) and scale height ($\beta$ is weakly stratified and often increases in the midplane) is the opposite of that in a magnetically elevated disk; (3) the thermal structure of the disk is also opposite the elevated disk model (it is warmer above the midplane); (4) the velocity structure above the disk is also distinct.

\subsection{Vertically Magnetized Star-Forming/Galactic Disks}
\label{sec:galactic.vertical.mag.disks} 

Recently, \citet{begelman.silk:2023.magnetically.boosted.accretion} suggested that strongly-magnetized galactic disks could drive BH accretion in proto-galaxies on scales much larger than the traditional accretion disk. They focus primarily on galactic radii $\gtrsim$\,pc, where they assume a magnetic field that is dominated by a mean global vertical field (with smaller, turbulence-resistivity-dominated radial and toroidal components sourced from $\langle B_{z} \rangle$), turbulent $Q \sim 1$ maintained by stellar feedback, and magnetic torques that dominate on these scales with turbulence that is sub-\Alf{ic} and/or sub-sonic relative to the dominant mean vertical field (specifically they require $\langle v_{A,\,z} \rangle = |\langle B_{z}\rangle|/\sqrt{4\pi\,\rho} \gtrsim \delta v_{\rm turb}/\sqrt{1+1/\mathcal{M}_{s}^{2}}$). While these larger scales are primarily discussed in \paperone, we note here that we do not see these conditions in our simulations. At star-forming galactic radii $\gg {\rm pc}$, \paperone\ showed that (1) magnetic forces are sub-dominant to gravity and turbulence/bulk motions; removing magnetic fields has almost no effect on the torques/inflow rates at $R\gg {\rm pc}$, which are dominated on these scales by a combination of gravitational torques and stellar-feedback-induced shocks; (2) there is no dominant coherent/uniform mean vertical field (consistent with observed galactic magnetic fields; \citealt{mao:2010.no.coherent.vertical.Bfield.through.solar.circle,beck:2015.b.field.review,jaffe:2019.magnetic.field.modeling.review,mao:galactic.bfields.revisions,krause:2020.spiral.galaxy.halo.magnetic.geometries.and.coherence}); and (3) the turbulence at $R\gg {\rm pc}$ is super-\Alf{ic}. In the accretion disk studied here (radii $\ll\,$pc), we also see a very different situation to that assumed in  \citet{begelman.silk:2023.magnetically.boosted.accretion}, with fields that are primarily toroidal and set by flux freezing (\S~\ref{sec:why.flux.freezing}), $B_{\phi}$ sourced by $B_{R}$ (\S~\ref{sec:b:origin}), weak and incoherent vertical fields in the disk (\S~\ref{sec:b}), $\langle v_{A,\,z} \rangle \ll \delta v_{\rm turb}$ (by factors $\sim 10-100$; Figs.~\ref{fig:b.profile} \&\ \ref{fig:v.profile}), thermal+magnetic $Q \gg1$ and minimal star formation (\S~\ref{sec:basic}), and qualitatively different scalings with radius.

\subsection{Previous Models of Toroidal-Field-Dominated Disks}
\label{sec:other.models}

 In previous literature, the models that come closest to capturing the properties we observe in our simulations are those of \citet{begelman.pringle:2007.acc.disks.strong.toroidal.fields,oda:2009.analytic.mag.disk.structure.models}.  These have been studied numerically in idealized simulations, which consider a relatively small section of the disk with simplified physics and fixed initial and boundary conditions, in e.g.\ \citet{salveson:2016.sims.mri.dominated.bfield.disks,kudoh:2020.strong.b.field.agn.acc.disk.sims.compare} (see also \citealt{johansen.levin:2008.high.mdot.magnetized.disks} and \citealt{gaburov:2012.public.moving.mesh.code}, though their assumptions, with $\beta \sim 1$, may be more similar to magnetically elevated disks). The main similarity is that these models posit $\beta \ll 1$ from a primarily toroidal magnetic field, which dominates over the midplane gas+radiation pressure. As we discuss in more detail in \paperthree\ with a simple analytic model motivated by our simulations, it turns out that this similarity alone is sufficient to capture most of the crucial properties of the simulations here. 

Nonetheless, there do still exist {\em qualitative} differences between the behaviors seen in our simulations and those models. Most importantly, those models implicitly assume the field is amplified from initially small values via the MRI and produces a toroidal $B_{\phi}$ which  {\bf (a)} is dominated by its fluctuating components, and so {\bf (b)} rapidly changes sign, even following a given Lagrangian parcel, {\bf (c)} is dominant over the poloidal $B_{z}$ (which sources it in the first place) but by a relatively small factor, and {\bf (d)} saturates at a value of $v_{A}\sim \sqrt{c_{s}\,v_{\rm K}}$, above which the linear growth of the MRI is assumed to be suppressed following the analytic analysis in \citet{pessah.psaltis:2005.mri.extensions.stronger.fields}. The final condition leads {\bf (e)} to relatively modest $\beta \sim 0.1$ at the radii of interest here (much higher than the values we see). This also leads {\bf (f)} to the prediction that the maximum accretion rate that can be maintained is only just about the Eddington limit around supermassive black holes, so these studies focused on much lower-density, lower-accretion rate regimes and did not consider highly super-Eddington accretion. And {\bf (g)} this means that the disks simulated here are orders-of-magnitude more gravitationally stable (e.g.\ retain $Q\gg 1$ out to orders-of-magnitude larger radii from the BH) compared to the predictions in these studies.

There are other differences as well that could be important: the models in \citet{begelman.pringle:2007.acc.disks.strong.toroidal.fields} (as well as \citealt{oda:2009.analytic.mag.disk.structure.models}) made very different assumptions for the temperature and opacity structure and predict a stratified disk with a hotter midplane; however the actual simulations in \citet{kudoh:2020.strong.b.field.agn.acc.disk.sims.compare}, which include dynamical cooling and heating (albeit with a simplified prescription compared to the detailed network here) predict an opacity and thermal structure much closer to what we see (inversely stratified, with a mostly-atomic cool midplane at these radii). Moreover these previous studies neglect the fact that these models, almost by necessity, predict highly super-sonic ($\mathcal{M}_{s} \gg 1$) accretion-disk turbulence at high $\dot{M}$, which in turn relates to very efficient/rapid cooling ($t_{\rm cool}/t_{\rm dyn} \sim \mathcal{M}_{s}^{-2} \ll 1$). This in turn leads to some star-formation but avoids catastrophic gravito-turbulent fragmentation via magnetic support. Indeed, \citet{begelman.pringle:2007.acc.disks.strong.toroidal.fields} essentially make the \citet{shakurasunyaev73}-style assumption of a laminar disk with turbulent velocity $v_{\rm t} \sim \alpha^{1/2} v_{A} \ll v_{A} \ll V_{\rm c}$. And these analytic/idealized models (including ours, in \paperthree) all assume quasi-circular disks, neglecting the potentially important role of the coherent eccentric, large-amplitude $m=1$ modes we see here.

Fundamentally, the key physical difference is that our simulations do not need to amplify ${\bf B}$ from some weak/trace ``seed'' field via the MRI in order to achieve their ``magnetically dominated'' state. Rather, they {\em begin} from this state, as the accreted gas carries in sufficiently large magnetic flux, with the consequence that the gas {\em initially} in the disk already has a magnetic field well {\em above} the specific (analytic) saturation threshold for the MRI assumed in \citet{begelman.pringle:2007.acc.disks.strong.toroidal.fields,oda:2009.analytic.mag.disk.structure.models}, namely $v_{A,\,{\rm flux-freezing}} \gg (c_{s}\,v_{\rm K})^{1/2}$. This, in turn, produces a variety of other consequences as described above, as well as different instabilities operating in the disk.

That said, some key conclusions are robust and confirmed here: the fact that $\beta \ll 1$ can be supported down to the ISCO, in principle; that the fields do not decay; that these can produce accretion rates far larger than a thermal \citet{shakurasunyaev73} disk; that they can sustain super-Eddington accretion; that the disks are much more stable at large radii than an equivalent \citet{shakurasunyaev73} disk; and that the outer disk is primarily atomic and ``cool'' and thermally weakly-stratified or even inversely stratified. These are all at least qualitatively robust conclusions comparing to the simulations in \citet{kudoh:2020.strong.b.field.agn.acc.disk.sims.compare} and analytic arguments in \citet{begelman.pringle:2007.acc.disks.strong.toroidal.fields}, even if the details and origins of the magnetic fields differ in some important respects.

\subsection{Why Does the Strong Field Not Decay (or Buoyantly Escape)?}
\label{sec:no.decay}

Related to the discussion above, there have also been some previous claims in the literature \citep{salveson:2016.sims.mri.dominated.bfield.disks,fragile.sadowski:2017.strong.magnetized.acc.disk.bfield.decay} that a strong toroidal-field dominated disk can, under the right conditions, rapidly (in tens of orbits) evolve to $\beta > 1$, due to a combination of adiabatic expansion with outflows and buoyant escape.
Clearly we do not see this ``decay,'' even having run our simulations for $\gg 10^{4}$ orbits at their innermost radii (see e.g.\ Fig.~\ref{fig:b.time.flip}). It is also worth noting that the simulations of \citet{kudoh:2020.strong.b.field.agn.acc.disk.sims.compare} discussed above also did not see any such decay/escape. Moreover as discussed extensively above, we have checked not just that the field is maintained at a given Eulerian position, but following a Lagrangian parcel over time, we see it amplified (not decaying) according to simple theoretical expectations.

There are a several important and straightforward reasons why the arguments for the specific situation considered in e.g.\ \citet{salveson:2016.sims.mri.dominated.bfield.disks,fragile.sadowski:2017.strong.magnetized.acc.disk.bfield.decay} should not hold here. 
Importantly, those authors considered a qualitatively different parameter space, initial conditions, and boundary conditions from those here. 
In those papers, the initial ``disk'' is strictly hydrostatic, with no net accretion or torques, and rapidly expands/puffs up, producing outflows almost everywhere (even in the disk midplane) and weakening the magnetic fields primarily via adiabatic expansion. This is almost the exact opposite of the behavior here, where we see $|{\bf B}|$ increase because gas flows {\em in} becoming more dense (Figs.~\ref{fig:radial.profile.general}-\ref{fig:b.induction.eqn}). Moreover, the density and temperature scales are orders-of-magnitude different, and those idealized simulation models were strictly adiabatic -- i.e.\ had no cooling -- while we see very efficient cooling $t_{\rm cool} \ll t_{\rm dyn}$. This can both maintain low-$\beta$ and dissipate the thermal energy, which in the idealized models could cause the disk to puff up or drive strong buoyancy instabilities (see also \paperthree). And they also only considered quite modest initial $\beta > 0.1$, so already close to $\beta \sim 1$, while our disks essentially ``begin'' at orders-of-magnitude lower $\beta$ (Fig.~\ref{fig:image.wedgeplot}-\ref{fig:radial.profile.general}) via the transport of magnetic flux from a weakly magnetized ISM (Fig.~\ref{fig:b.tests.goodmodels}-\ref{fig:b.induction.eqn}). This also qualitatively changes which instabilities can operate in the disk (as shown in Fig.~\ref{fig:instability.map}): for example as discussed in \S~\ref{sec:b:origin:turb}, the specific Parker-like vertical buoyancy modes discussed in \citet{johansen.levin:2008.high.mdot.magnetized.disks,salveson:2016.sims.mri.dominated.bfield.disks} may operate on such large wavelengths ($\sim 10\,H \gtrsim R$, given the large scale heights in Fig.~\ref{fig:radial.profile.general}) that they cannot fit within the disk, or they may be suppressed by the combination of strong trans-\Alf{ic} turbulence, non-negligible radial and vertical fields, and low-$\beta$ \citep{horiuchi:1988.parker.growth.rate.vs.beta,kim:2001.turb.strongly.suppresses.parker.instability}, but other radial buoyancy modes may appear which have qualitative different effects on the field (\S~\ref{sec:b:origin:turb}).

Moreover, both \citet{salveson:2016.sims.mri.dominated.bfield.disks} and \citet{fragile.sadowski:2017.strong.magnetized.acc.disk.bfield.decay} considered {\em exclusively}  azimuthal fields, and noted that adding some poloidal or radial component would prevent the field decay, as was subsequently shown explicitly in \citet{salveson:2016.decaying.fields.poloidal.flux}. Here, we do see dominant toroidal fields, but with substantial (order-unity) projections into both the poloidal and radial directions (Fig.~\ref{fig:b.profile}). Indeed, this is inevitable in our simulations, given the origin of the ``seed'' fields for the accretion disk from a quasi-isotropically-tangled field in the ISM. Thus our simulations actually reside ``safely'' in the regime where \citet{salveson:2016.decaying.fields.poloidal.flux} argued they should {\em not} experience rapid decay/loss/escape.

It is also crucial to note that there is no {\em source} of magnetic flux in the idealized simulations of \citet{salveson:2016.sims.mri.dominated.bfield.disks,fragile.sadowski:2017.strong.magnetized.acc.disk.bfield.decay}. Absent any source of new flux in the midplane, it seems plausible that escape of toroidal field from the midplane and/or turbulent resistivity could eventually weaken $B_{\phi}$, perhaps producing something more like the magnetically elevated disks in \citet{johansen.levin:2008.high.mdot.magnetized.disks} with a minimum in $\beta$ at the midplane. And we are not arguing that magnetic buoyancy is totally negligible here -- in contrast, in \S~\ref{sec:b:origin:turb}, \ref{sec:v:drive}, \&\ \ref{sec:vertical} we argued that buoyancy instabilities unique to low-$\beta$ disks might play an important role driving the turbulence in the disk and explaining the weak stratification we observe. However, we showed above in e.g.\ Fig.~\ref{fig:b.induction.eqn} and \S~\ref{sec:b:ov} that even if vertical buoyancy removed midplane toroidal field on the fastest possible timescale it can operate (a few $\Omega^{-1}$), this would be balanced by the growth of toroidal field just from new radial flux carried in with the midplane accretion flow (see also \citealt{shibata:1990.low.beta.disk.parker.suppressed.by.magnetic.tension}, who make a similar argument from both analytic considerations and idealized MHD simulations). In brief, the midplane mean toroidal field in the simulations is constantly sourced by advection of new radial and toroidal flux, ``closing the dynamo loop'' and replenishing/maintaining the magnetic-field strength. 

We stress that this possibility was indeed anticipated by \citet{salveson:2016.sims.mri.dominated.bfield.disks,fragile.sadowski:2017.strong.magnetized.acc.disk.bfield.decay}, as well as others such as \citet{shibata:1990.low.beta.disk.parker.suppressed.by.magnetic.tension,johansen.levin:2008.high.mdot.magnetized.disks,kudoh:2020.strong.b.field.agn.acc.disk.sims.compare}, all of whom emphasized the critical importance of physically-motivated boundary conditions for the accretion disk from larger (ISM) radii and therefore the source of flux (the motivation for our simulations in this paper). \citet{salveson:2016.sims.mri.dominated.bfield.disks} specifically stated that, given ``favorable conditions'' for the source of magnetic flux along with the accretion flow (which they noted could arise from the ``external'' field from the ISM/galactic scales), ``a strongly magnetized disc would necessarily follow.'' Though they envisioned primarily poloidal external flux, while we find a mix of radial and poloidal flux (see Fig.~\ref{fig:b.profile}), the overall effect will be similar, maintaining strong toroidal fields in the disk \citep{fragile.sadowski:2017.strong.magnetized.acc.disk.bfield.decay}. Therefore, taken together, the stable behavior of the strong toroidal fields here is expected, and does not contradict the results from the more idealized test-problem simulations considered in \citet{salveson:2016.sims.mri.dominated.bfield.disks,fragile.sadowski:2017.strong.magnetized.acc.disk.bfield.decay}.

\section{Angular Momentum of the Disk Relative to the Pre-Existing BH Spin}
\label{sec:angmom.alignment}

Our simulations do not have sufficient resolution to follow the spin of the SMBH in detail. However, even in the low-resolution ``progenitor'' simulation, we do dynamically follow the total angular momentum (AM) accreted by the SMBH ${\bf j}_{\rm BH}$ (from its formation as a seed, with ${\bf j}_{\rm BH}$  updated whenever some gas is accreted following the detailed numerical description in \citealt{hopkins:fire3.methods}). While imperfect, $\hat{\bf j}_{\rm BH}$ should serve as at least a plausible guess for the spin {\em direction} of the SMBH. Given the scenario occurring in our simulation (\S~\ref{sec:basic}) -- where a turbulent massive star-forming cloud complex in a highly chaotic, clumpy high-redshift massive galaxy is partially tidally disrupted on close passage to a pre-existing SMBH in the nucleus -- there is no reason to think that the accretion disk AM vector ${\bf j}_{\rm gas}$ should be preferentially aligned with this pre-existing BH AM vector (or the SMBH spin direction).

We can immediately verify this in the simulations. Comparing the vector directions $\hat{\bf j}_{\rm gas}$ and $\hat{\bf j}_{\rm BH}$, the accretion disk at the times we follow is approximately $\sim 140^{\circ}$ misaligned from $\hat{\bf j}_{\rm BH}$ interior to $\lesssim 0.1$\,pc (where the disk AM direction $\hat{\bf j}_{\rm gas}$ is quite stable within different annuli, to within a few degrees). This means it is likely retrograde and out-of-plane with the pre-existing SMBH spin. As the disk evolves, the AM direction will evolve too, tracing material further away from the SMBH at this time: at $\sim 1\,$pc (or $\sim 10\,$pc) the mean AM vector of the accreting gas is $\sim 100^{\circ}$ (or $\sim 130^{\circ}$) misaligned, so can vary by $\sim 40^{\circ}$. The corresponding enclosed gas masses and accretion timescales at the current accretion rate into $<80\,$au are: $\sim (3\times10^{4},\,6\times10^{5},\,10^{7})\,M_{\odot}$ and $\sim (10^{3},\,2\times10^{4},\,3\times10^{5})\,{\rm yr}$ for gas within $\sim (0.1,\,1,\,10)$\,pc. 

There is also no correlation between the direction $\hat{\bf j}_{\rm BH}$ and the secondary preferred direction of the accretion disk: namely the direction of the semi-major axis in the plane of the disk (since it is coherently eccentric). But this is expected even if $\hat{\bf j}_{\rm BH}$ and $\hat{\bf j}_{\rm gas}$ were aligned, since the semi-major axis direction precesses on timescales of order the orbital time at the outer disk radius. 

Our inner resolution scale ($\sim 80\,$au or $\sim 300\,R_{\rm schw}$) is much larger than any scale where BH spin directly influences the dynamics (e.g.\ via Lense-Thirring precession), so this is not directly important for the scales we resolve. However, simulations of smaller scales using these results as outer boundary conditions should include such effects. Moreover, this is consistent with the well-established observational result that the ``spin axis'' inferred from AGN jet directions does not appear to correlate with gas AM on any resolved macroscopic scales in galaxies \citep[see e.g.][and references therein]{schmitt:1997.radio.alignment.w.host,kinney:2000.bh.jet.directions,hopkins:torus,hopkins:agn.alignment,davies:2014.dusty.structures.misaligned.around.agn,reynolds:2021.bh.spin.review}. Indeed, the misalignment angles that we find, and their variation with scale and time on sub-pc scales, are consistent with those seen in previous simulations by \citet{daa:20.hyperrefinement.bh.growth}, who also emphasize the lack of correlation between the angular momentum vector at $\lesssim1\,$pc (at any given instant) and on much larger ($\gtrsim$\,kpc) scales. 

Together, this further suggests that extremely misaligned accretion -- which can itself even further help in promoting very rapid accretion on scales smaller than those we resolve here \citep[see e.g.][]{kaaz:2022.grmhd.sims.misaligned.acc.disks.spin} -- might be quite common in quasars whose large-scale fueling episodes resemble that here.

\begin{figure*}
	\centering\includegraphics[width=0.95\textwidth]{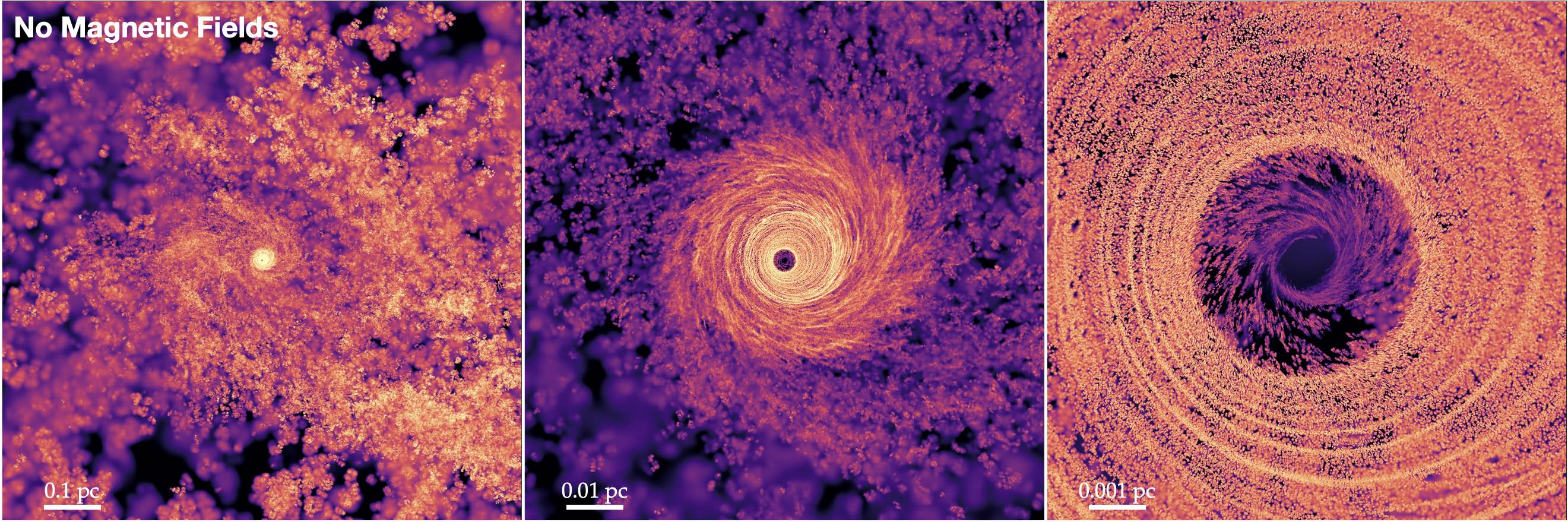}
	\caption{Face-on images showing gas surface density in a re-simulation of our fiducial simulation without magnetic fields (see \S~\ref{sec:no.mhd}), as in Figs.~\ref{fig:image.zoom}-\ref{fig:image.faceonedgeon.inner}, at the latest time to which we are able to run the simulation. We see much more vigorous clumping/fragmentation at radii $\sim 0.01-1\,$pc, a more compact disk that only emerges interior to $ \lesssim 0.01\,$pc, a series of tightly-wound $m=1$ modes with large amplitudes, giving a ``concentric ring'' appearance, and a central hole in the disk that is expanding (larger than our inner accretion boundary by a factor of several).
	\label{fig:nomhd.images.faceon}}
\end{figure*}

\begin{figure}
	\centering\includegraphics[width=0.99\columnwidth]{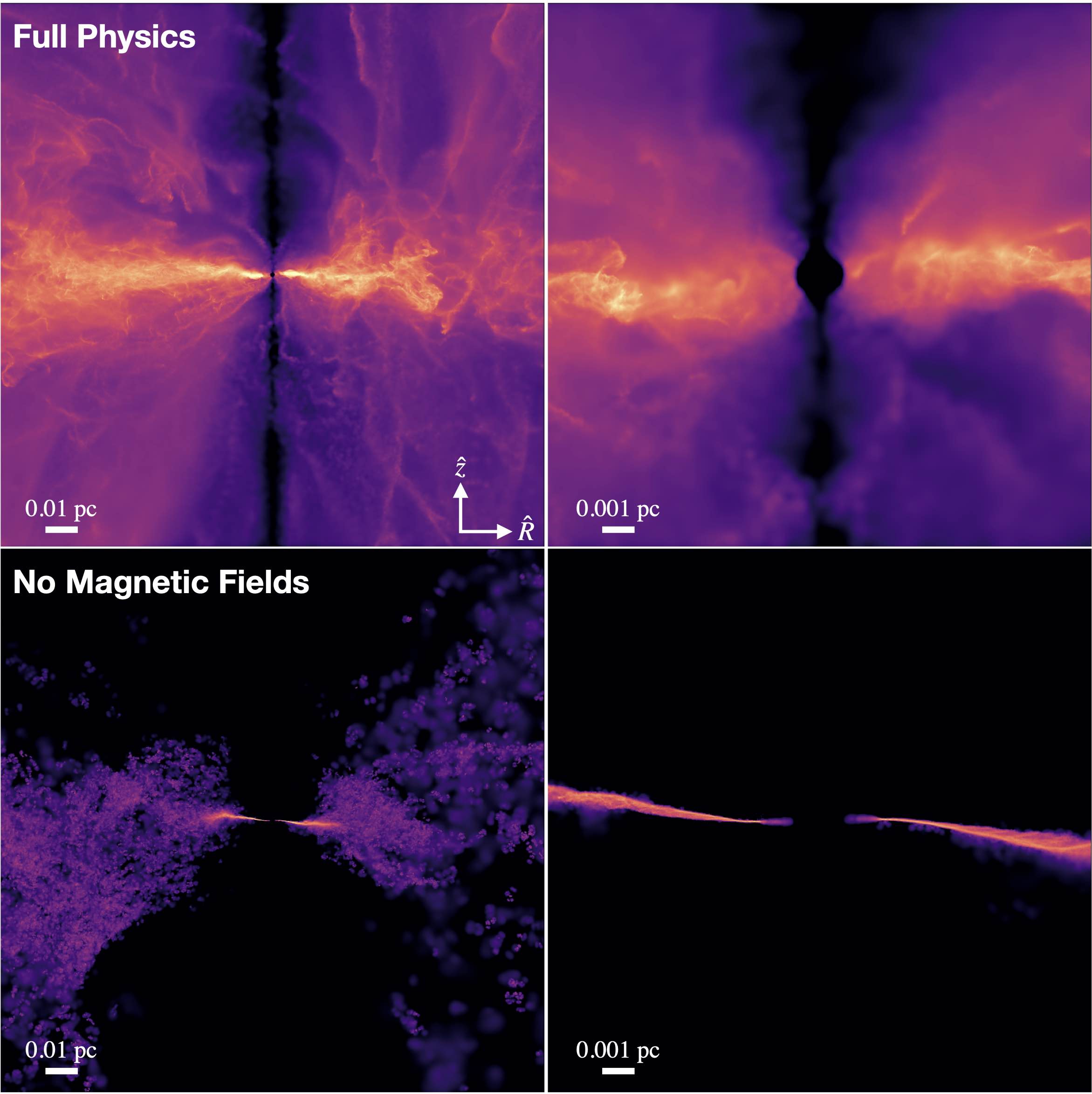}
	\caption{Edge-on images of gas surface density in our fiducial/``full-physics'' simulation ({\em top}) and re-simulation without MHD ({\em bottom}; as in Fig.~\ref{fig:nomhd.images.faceon}). Cylindrical $R$-$z$ coordinates are used to better see the disk structure. We clearly see that without magnetic fields, the disk is much more compact and razor thin, featuring almost no extended vertical atmosphere/halo/coronal structure. 
	\label{fig:nomhd.images.edgeon}}
\end{figure}

\begin{figure*}
	\centering\includegraphics[width=0.48\textwidth]{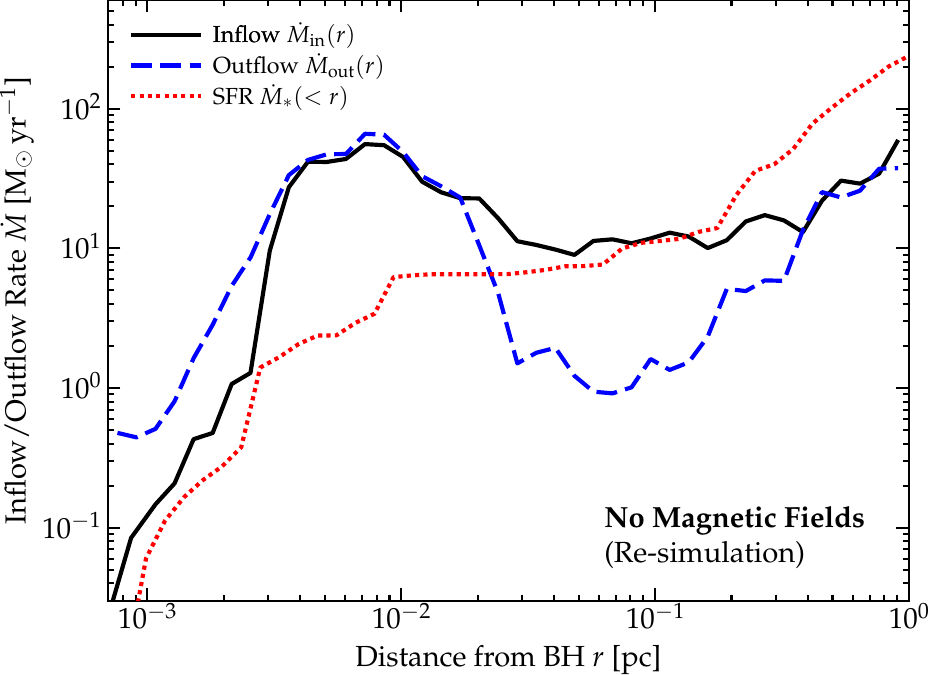}
	\centering\includegraphics[width=0.48\textwidth]{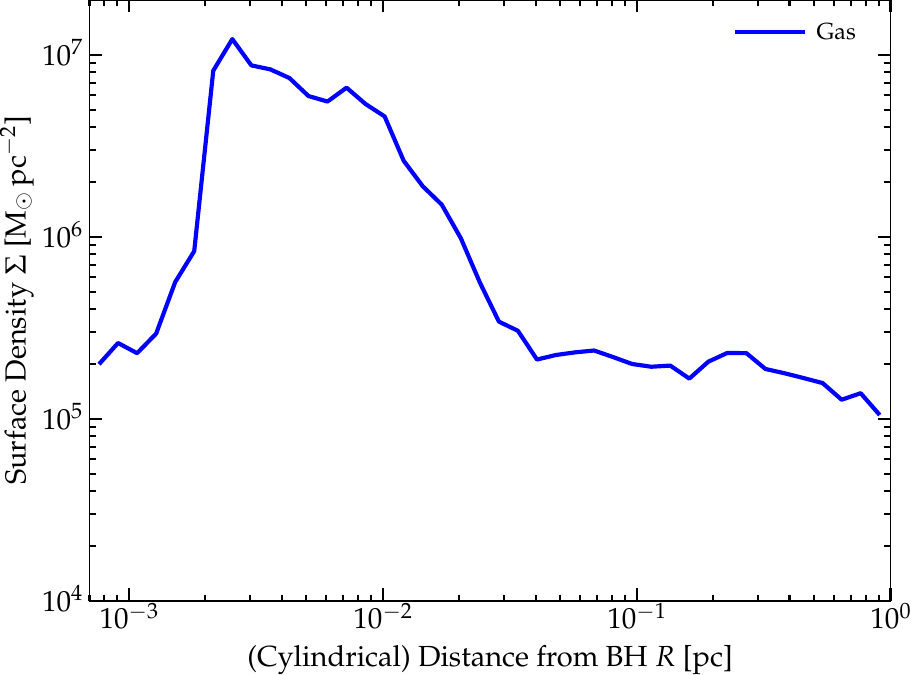}
	\centering\includegraphics[width=0.48\textwidth]{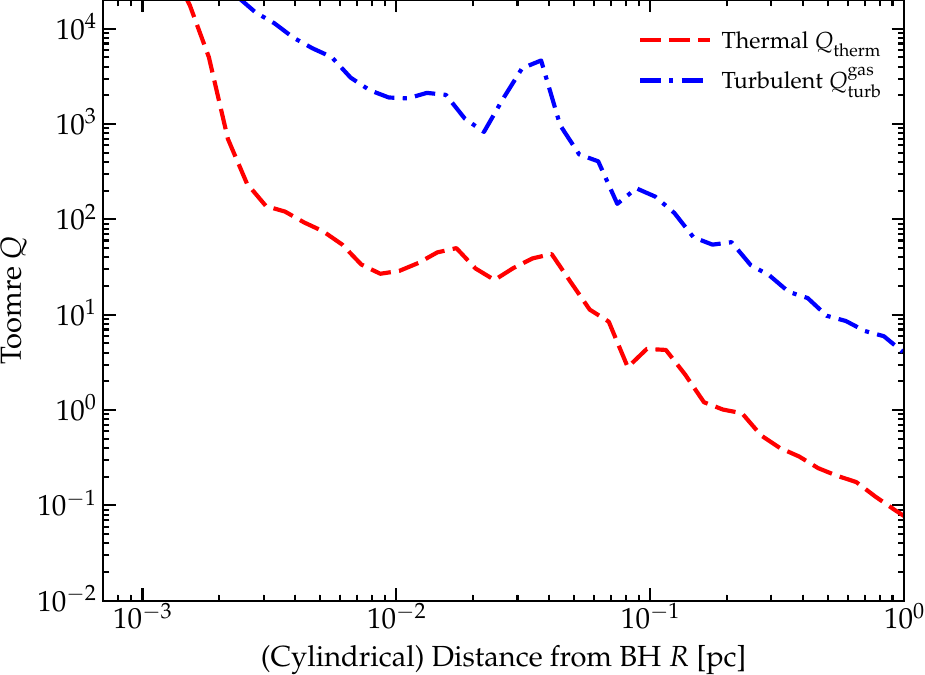}
	\centering\includegraphics[width=0.48\textwidth]{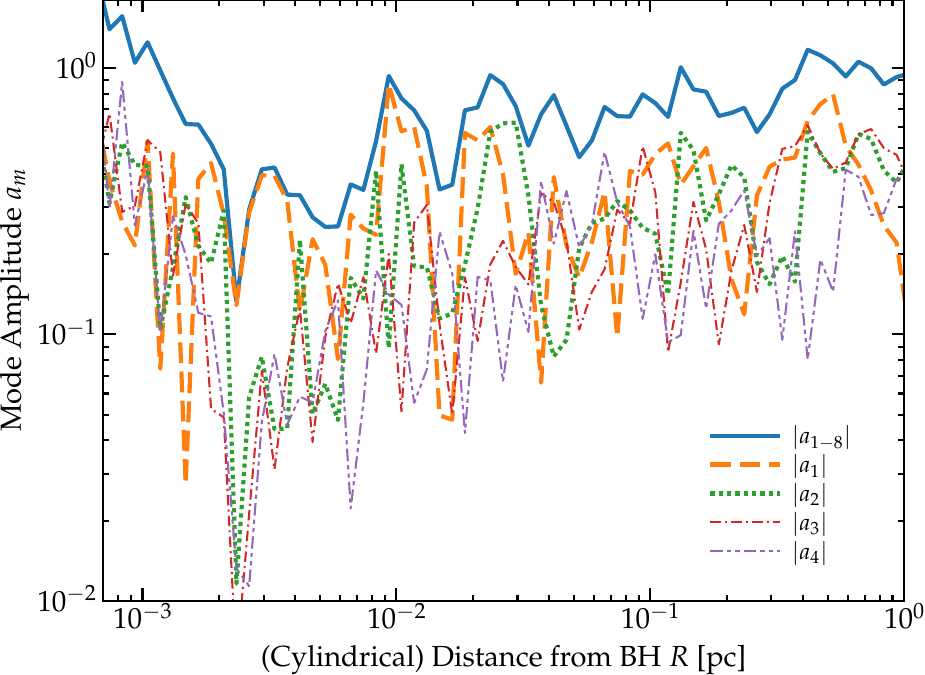}
	\centering\includegraphics[width=0.47\textwidth]{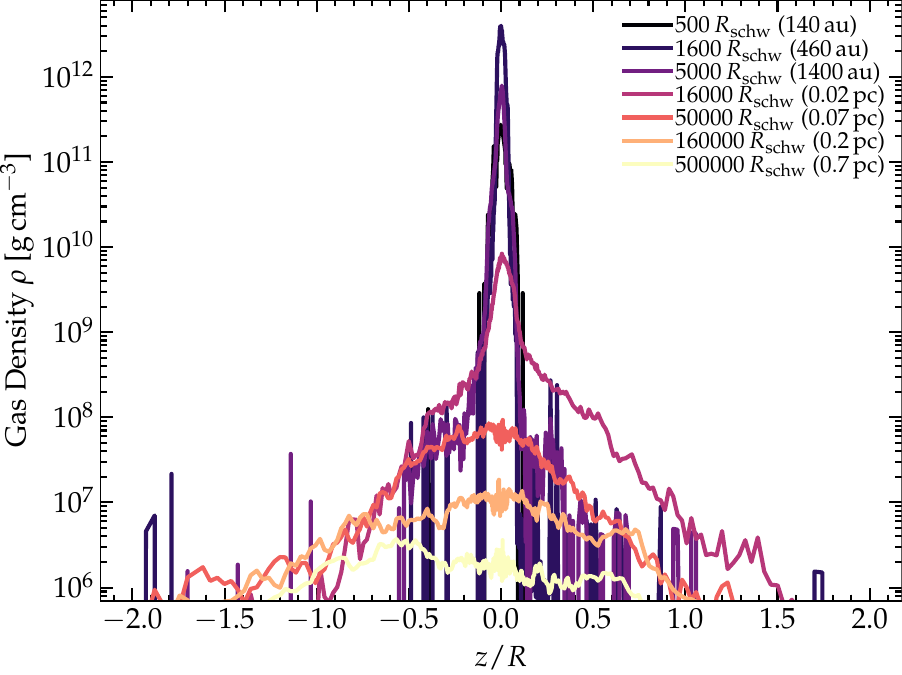}
	\centering\includegraphics[width=0.475\textwidth]{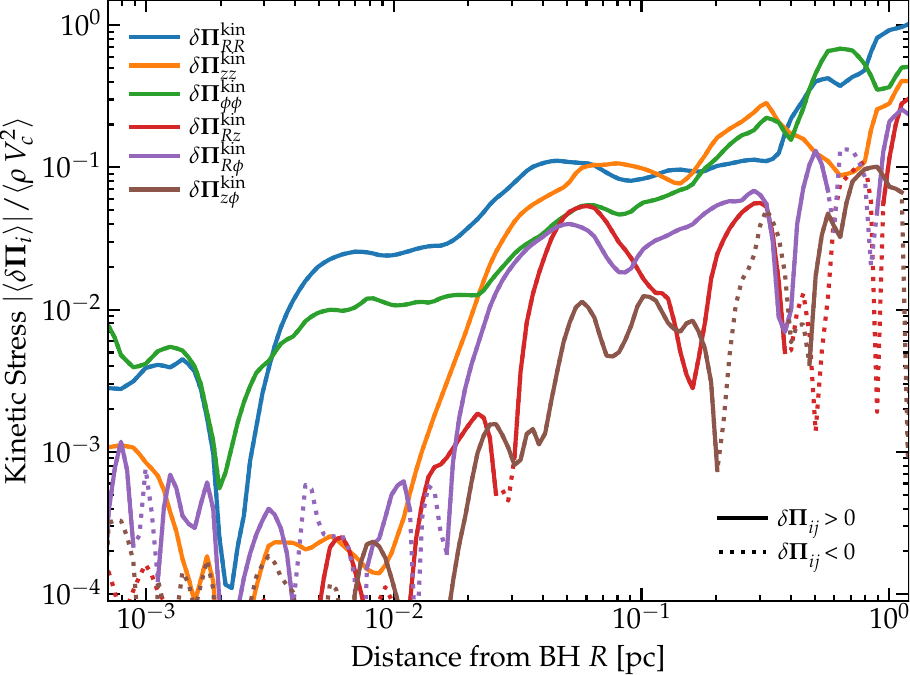}
	\caption{Quantitative properties of our re-simulation {\em without} magnetic fields (\S~\ref{sec:no.mhd}) after we further hyper-refine to $< 10^{-6}\,{\rm M_{\odot}}$ mass resolution in the innermost $<0.01\,$pc, at the latest time we evolve it.
	{\em Top Left:} Inflow $\dot{M}_{\rm in}$, outflow $\dot{M}_{\rm out}$, \&\ SFRs $\dot{M}_{\ast}$ (as Fig.~\ref{fig:radial.profile.general}). Without MHD, runaway fragmentation produces orders-of-magnitude larger SFRs ($\dot{M}_{\ast} > \dot{M}_{\rm in}$) and lower inflow rates, with $\dot{M}_{\rm in}$ dropping rapidly as $r\rightarrow0$ and outflow $\dot{M}_{\rm out} > \dot{M}_{\rm in}$ at this time.
	{\em Top Right:} Surface density profile. Absent magnetic support, fragmentation produces a sharper outer truncation of the disk at $\gtrsim 0.01\,$pc, seen in the steep $\Sigma_{\rm gas}(R)$, while the $\dot{M}_{\rm out} > \dot{M}_{\rm in}$ in the inner disk has produced the ``hole'' visible here as lower $\Sigma_{\rm gas}(R \lesssim 0.002\,{\rm pc})$ and in Fig.~\ref{fig:nomhd.images.faceon}.
	{\em Middle Left:} Thermal and turbulent Toomre $Q$ parameter (as in Fig.~\ref{fig:radial.profile.general}). The outer disk truncation corresponds to where the thermal $Q$ falls to $\lesssim 10$, where we would expect more catastrophic turbulent fragmentation (absent MHD), with strong gravito-turbulence at smaller radii (given that $t_{\rm cool} \ll t_{\rm dyn}$ on scales here). 
	{\em Middle Right:} Amplitude of the non-axisymmetric modes (as in Fig.~\ref{fig:eccentricity.profile}). Consistent with strong gravito-turbulence and the morphology in Fig.~\ref{fig:nomhd.images.faceon}, we see tightly-wound spiral modes with $\mathcal{O}(|a_{1}|)\sim 1$, much stronger than our full-physics case.
	{\em Bottom Left:} Vertical gas density profile (as in Fig.~\ref{fig:vertical}). Without magnetic support the disk is much thinner ($H/R \lesssim 0.01$ inside the disk region at $\lesssim 0.01\,$pc), but also we see orders-of-magnitude lower density ``above'' the disk (the extended, slowly-falling $\rho(|z|)$ atmosphere/corona from Fig.~\ref{fig:vertical} is entirely absent without MHD inside the disk radii, per Fig.~\ref{fig:nomhd.images.edgeon}). 
	{\em Bottom Right:} Reynolds (fluctuating kinetic) stress tensor components $\delta\boldsymbol{\Pi}^{\rm kin}_{ij} \equiv \rho \delta v_{i} \delta v_{j}$ (as in Fig.~\ref{fig:stress.terms}). 
	Within the disk, the $R\phi$ Reynolds stress responsible for angular momentum transfer is more than an order of magnitude weaker absent MHD, and has a fluctuating sign ($<0$ here indicating angular momentum {\em gain}, consistent with the net $\dot{M}_{\rm out} > \dot{M}_{\rm in}$ seen above). 
	Without magnetic fields, the turbulence is extremely anisotropic, evidenced by $\langle \delta v_{R}^{2} \rangle \sim \langle \delta v_{\phi}^{2} \rangle \sim 100\,\langle \delta v_{z}^{2} \rangle$, which is expected as gravito-turbulence alone cannot efficiently drive {\em vertical} motions.
	\label{fig:nomhd.profiles}}
\end{figure*}

\section{What Happens Without Magnetic Fields?}
\label{sec:no.mhd}

\paperone\ discusses several numerical experiments in detail, including an example where we begin our ``hyper-refinement'' stage of the simulation without magnetic fields (but from otherwise identical initial conditions). There we showed that this produces relatively weak effects on $\gg$\,pc scales (in the galactic ISM), consistent with the vast majority of previous simulation studies of magnetic fields in cosmological galaxy formation simulations \citep{su:2016.weak.mhd.cond.visc.turbdiff.fx,su:fire.feedback.alters.magnetic.amplification.morphology,su:2018.stellar.fb.fails.to.solve.cooling.flow,hopkins:cr.mhd.fire2,ji:fire.cr.cgm,steinwandel:2019.magnetic.bouyancy.galactic.scales,steinwandel:lmc.mass.galaxy.outflows.mfm.validation,martin.alvarez:2021.bfield.amplification,ponnada:fire.magnetic.fields.vs.obs,whitworth:no.mhd.fx.galform.sims}. But as expected from our arguments above, the effects on the disk at scales $\ll$\,pc can be dramatic. Here we explore the disks that form without MHD in more detail.

Briefly, we note the resolution and run-time of this simulation. We initially run the simulation with our ``default'' resolution of the full-physics simulation with MHD (mass resolution $\sim (1-5)\times10^{-3}\,M_{\odot}$ in the high-resolution region inside $\lesssim 10\,$pc), but for a more limited time (equivalent to $\sim 3000\,\Omega_{\rm inner}^{-1}$ or $\sim 100\,$yr at the very highest resolution level, and $\sim 10^{5}$\,yr at intermediate resolution following the infall during refinement; see \paperone\ for details of the refinement scheme). This reduced time is chosen both because it is a counterfactual numerical experiment and because it is notably more numerically expensive compared to our default simulation, owing to the much more rapid fragmentation into extremely dense sub-clumps, the formation of extremely massive stars from said fragmentation (see \paperfour), and the razor-thin disk. This razor-thin disk, discussed below, means that even at our extremely high resolution, it is challenging to resolve the vertical scale-height of the inner disk at $R \ll 0.01\,$pc. We therefore re-start this simulation with an even further layer of refinement continuing to a minimum mass resolution of $\Delta m_{\rm min} \approx 0.003\,{\rm M_{\odot}}\,(r / 0.1\,{\rm pc})^{2}$ at $r<0.1\,$pc, reaching a highest resolution of $\Delta m <10^{-7}\,{\rm M_{\odot}}$ at our innermost radii $<100\,$au. This is extremely expensive computationally, so we evolve it for a much shorter time at its highest resolution ($\sim 100\,\Omega_{\rm inner}^{-1}$), just to ensure that the properties at the innermost radii can be reliably modeled.

\subsection{Runaway Star Formation}
\label{sec:no.mhd:sf}

The major focus of our comparison in \paperone\ was to show that, absent magnetic fields, on sub-pc scales (where MHD torques take over as dominant from gravitational torques in our default simulation per Fig.~\ref{fig:torque.profile}) fragmentation and subsequent star formation run away catastrophically. Whereas in our default simulations (including magnetic fields) the fields stabilize the disk and lead to a sharp suppression of the SFR per unit area or volume at radii $\lesssim 0.1-1\,$pc, the volumetric SFR in a simulation without MHD continues to rise steeply as $r\rightarrow 0$. Per Fig.~\ref{fig:nomhd.profiles}, the SFR interior to $<1\,$pc (or $<0.1\,$pc), for example, rises from $\sim 10\,{\rm M_{\odot}\,yr^{-1}}$ ($\sim 0.1\,{\rm M_{\odot}\,yr^{-1}}$) with MHD to $\sim 250\,{\rm M_{\odot}\,yr^{-1}}$ ($\sim 5\,{\rm M_{\odot}\,yr^{-1}}$) without MHD. As discussed below, the SFR at $<1\,$pc significantly exceeds the total mass inflow rate into this annulus, strongly suppressing inflows to the BH accretion disk. 

The physical reasons for this -- again the main focus of \paperone\ -- are straightforward. Most importantly, the lack of magnetic fields means the disk is no longer stabilized at radii $\sim 0.01-1\,$pc against catastrophic gravito-turbulent fragmentation. On top of this, the lack of strong magnetic stresses/torques means that gas cannot inflow efficiently so ``piles up'' at large radii, further accelerating fragmentation. This directly enhances the ratio of SFR to inflow rate at $\sim 0.1-1\,$pc, therefore further suppressing inflow to even smaller radii. Thus, {\em without} magnetic fields, we indeed see the classic hydrodynamic problem -- reviewed in e.g.\ \citet{shlosman:inefficient.viscosities,shlosman:midscale.accretion,goodman:qso.disk.selfgrav} -- of runaway rapid fragmentation in the outer disk region when inflow rates from the ISM are large.

\subsection{Vastly Lower Accretion Rates}
\label{sec:no.mhd:mdot}

Partly as a result of the runaway star formation, but also because of inefficient MHD torques (discussed below), we see that the surviving inflow rates of gas into $\ll 1$\,pc are reduced dramatically without MHD. The time-averaged net gas inflow rate through our inner boundary at $<80\,$au is reduced from $\sim 20-30\,{\rm M_{\odot}\,yr^{-1}}$ in our default simulation, to $\lesssim 0.1\,{\rm M_{\odot}\,yr^{-1}}$ in the simulation without MHD (a factor of $\sim 200-500$ reduction). But even this accretion rate appears to be falling towards the end of the relatively short duration of the rerun without MHD, because the torques in the inner disk seem to be spinning up the disk and create a growing hole in the center of the disk. This can be seen by-eye in Figs.~\ref{fig:nomhd.images.faceon}-\ref{fig:nomhd.images.edgeon}. Indeed, from Fig.~\ref{fig:nomhd.profiles}, comparing either the total inflow+outflow rate within a given annulus ($\dot{M}_{\rm in} + \dot{M}_{\rm out} = 4\pi\,\langle v_{r}\,\rho\,r^{2} \rangle$), or from the total $R\phi$ kinetic stress ($\langle \boldsymbol{\Pi}^{\rm kin}_{R\phi} \rangle = \langle \rho\,v_{\phi}\,v_{R}\rangle \sim v_{\rm K}\,\langle \rho\,v_{R} \rangle$), we clearly see that the mean radial velocity in the disk midplane is {\em outward} -- i.e.\ we actually have a decretion disk at this time. 

Some gas still ``leaks'' through owing to non-equilibrium motions and vertical inflows joining onto the inner disk. This means the inflow rate is highly intermittent/bursty, dominated by occasional clumps that make it through the inner region. The vertical inflows contribute little in a time-averaged sense and would likely be ejected if we included some form of jets and/or harder radiation emitted by the unresolved disk interior to $<80\,$au, since they are primarily polar. Notably, if we evolved the disk longer, this net decretion, coupled to the larger SFR compared to gas inflow rate, would lead to even further depletion of the inner gas and therefore further suppression of the time-averaged gas accretion rate. Indeed, if we monitor the gas mass inside of $\lesssim 100\,$au, we see that after it initially rises in an inflow event, over the time period of the last $\sim 600\,\Omega_{\rm inner}^{-1}$ at $R \sim 80\,$au (just $\sim 20\,$yr) it drops precipitously by a factor of $\sim 3$ (then it increases slightly as the ring of gas that builds up goes unstable and a clump falls back in towards the central $<100\,$au). In contrast, we showed these properties were stable in our full-physics simulations (with MHD) for timescales at least $\gtrsim 10^{5}\,\Omega_{\rm inner}^{-1}$ ($\sim 10^{4}\,$yr, the duration of our default simulation at highest resolution). So again {\em without} magnetic fields, we see orders-of-magnitude lower accretion rates, and it appears difficult to sustain near-Eddington (let alone significantly super-Eddington) inflows into and through the disk.

Of course it is possible that on much longer timescales ($\gtrsim$\,Myr) the runaway star formation seen here could produce non-linearly different conditions that eventually lead to efficient accretion. But this does not change our generic conclusion that for a given set of initial/boundary conditions, magnetic fields play a critical role on these scales.

\subsection{A Razor-Thin, Compact, Residual Gravito-Turbulent Disk}
\label{sec:no.mhd:disk}

\subsubsection{Disk Size and Scale Height}
\label{sec:no.mhd:disk:size.height}

That said, there is still clearly in Figs.~\ref{fig:nomhd.images.faceon}-\ref{fig:nomhd.images.edgeon} some disk that forms. But it is visually obvious in Fig.~\ref{fig:nomhd.images.edgeon} that this disk is radically different from that in our full-physics simulations. First, consider the disk size. Per Fig.~\ref{fig:nomhd.profiles}, and by-eye in Fig.~\ref{fig:nomhd.images.edgeon}, we see that the surface density of the disk falls of rapidly, and the ratios $H/R$ and $\delta v_{\rm turb}/V_{c}$ all increase rapidly, outside of $R\gtrsim 0.01\,$pc -- so the ``outer'' disk radius here is at least an order of magnitude smaller than in the full-physics case. This corresponds to the radii where the Toomre $Q \sim c_{s}\,\kappa/(\pi\,G\,\Sigma) \sim 140\,(R/{\rm 0.01\,pc})^{-3/2}\,(\Sigma/10^{6}\,{\rm M_{\odot}\,pc^{-2}})^{-1}\,(c_{s}/8\,{\rm km\,s^{-1}})$ falls to $\lesssim 10$, where more catastrophic turbulent fragmentation should set in (as opposed to ``gravito-turbulent'' fragmentation which can at least maintain some semblance of disk structure; see e.g.\ \citealt{rice:2005.disk.frag.firstlook,hopkins:2013.turb.planet.direct.collapse} for examples) given the fact that the disk is quasi-isothermal over the limited dynamic range of radii resolved here at $\sim 8000\,$K. So it is not surprising that the disk can exist at these radii, given the (much lower) accretion rates seen here and its reasonably warm temperatures.

Second, we see that the disk is extremely thin, with $H/R \lesssim 0.01$ through most of its resolved extent. We see subtle warps, and a roughly concentric-ring morphology owing to tightly-wound modes discussed below. And not only is the disk scale-height quite small, but whereas in our full-physics simulations we saw a fairly extended, more slowly-falling (power-law-like) vertical atmosphere/corona above the disk ($|z|\gg H$; see Fig.~\ref{fig:vertical}), without MHD we see a much more stark (exponential or super-exponential) transition with extremely low densities outside the very narrow midplane (Fig.~\ref{fig:nomhd.profiles}). This, of course, is not surprising, given the lack of the dominant pressure support from magnetic fields seen in our full-physics simulations, but further accelerates fragmentation and gravito-turbulence in the disk. 

It is also worth noting that absent MHD, the disk is radiation-pressure dominated. This is expected without magnetic pressure, given the accretion rates and other properties (see Fig.~\ref{fig:radial.profile.general}, or comparison to SS73-like disks in \paperthree).

\subsubsection{Extremely Anisotropic Gravito-Turbulence and Tightly-Wound Modes}
\label{sec:no.mhd:disk:turb.stresses}

The disk without MHD is strongly gravito-turbulent and rapidly-cooling, consistent with its runaway fragmentation. Assuming black-body cooling from a thin disk balancing accretion, we would expect in the outer disk 
$T_{\rm midplane} \sim 2400\,{\rm K}\,(\dot{M}_{0.1} \Sigma_{6} \tilde{\kappa}/R_{0.01}^{3})^{1/4}$, 
and 
$t_{\rm cool}/t_{\rm dyn} \sim  0.003\,(R_{0.01}^{3} \Sigma_{6}^{5} \tilde{\kappa}/\dot{M}_{0.1}^{5})^{5/4}$ 
(where $\dot{M}_{0.1}\equiv \dot{M}_{\rm in} / 0.1\,{\rm M_{\odot}\,yr^{-1}}$; $\Sigma_{6} \equiv \Sigma_{\rm gas}/10^{6}\,{\rm M_{\odot}\,pc^{-2}}$; $R_{0.01}\equiv R/0.01\,{\rm pc}$; and $\tilde{\kappa} \equiv \kappa/\kappa_{\rm es}$), and therefore highly super-sonic turbulence in the disk plane with $\mathcal{M}_{s} \sim  (t_{\rm cool}/t_{\rm dyn})^{-1/2} \sim 10-50$. These expectations are broadly consistent with what we see in the simulation inside the disk region ($R\lesssim 0.01\,$pc). Although we do note that just as in our full-physics simulations, non-LTE effects and external heating become more important to the disk temperature at larger radii.

The $m=1$ modes reach high (order-unity) amplitudes $|a_{1}| \sim 0.1-1$, and are very tightly-wound with almost all the power at short characteristic wavelengths $k \sim 1/H$.  This produces a morphology that resembles a series of narrow concentric rings of alternating high and low density. This is all expected for strong gravito-turbulence in a very thin ($H\ll R$) differentially rotating disk \citep{gammie:2001.cooling.in.keplerian.disks,paardekooper:2012.stochastic.disk.frag,meru:2012.nonconvergence.disk.frag}. Consistent with this, the velocity fluctuations are highly anisotropic: while the {\em in-plane} motions driven by these modes have relatively large $|\delta v_{R}| \sim |\delta v_{\phi} | \sim |a_{1}|\,V_{c} \sim 0.1\,V_{c}$, the {\em vertical} motion is an order-of-magnitude smaller, $|\delta v_{z}| \sim 0.01\,V_{c}$. Again, this is consistent with previous idealized simulations in the regime without efficient magnetic fields and/or stellar feedback to isotropize the turbulence and maintain a thick disk \citep{hopkins:fb.ism.prop}: ``pure gravitoturbulent'' driving is unable to maintain appreciable vertical dispersion. 

Examining the kinetic stress tensor in Fig.~\ref{fig:nomhd.profiles} in more detail, it is worth noting that the majority of the $RR$ and $\phi\phi$ dispersion ($\langle \rho \delta v_{R} \delta v_{R} \rangle$ and $\langle \rho \delta v_{\phi} \delta v_{\phi} \rangle$) can be directly attributed to the $m=1$ modes: subtracting the best-fit $m=1$ component from e.g.\ $\delta v_{R}(R,\,\phi)$ at each $R$ leads to an order-of-magnitude reduction in $\delta \boldsymbol{\Pi}^{\rm kin}_{R R}$. The latter is the true ``turbulent'' component. But even this is order-of-magnitude larger than the much smaller vertical dispersion, as noted above. Turning to the $\langle \delta\boldsymbol{\Pi}^{\rm kin}_{R\phi} \rangle$ component of the Reynolds stress relevant for angular momentum transport, we see that not only is it order-of-magnitude smaller than the Reynolds stress in our full-physics simulations (Fig.~\ref{fig:stress.terms}), but it often has the opposite sign, i.e.\ $\langle \delta\boldsymbol{\Pi}^{\rm kin}_{R\phi} \rangle = \langle \rho \delta v_{R} \delta v_{\phi} \rangle < 0$, driving outflow rather than inflow. Both the large eccentric fluctuations in $v_{R}$, and the sign variations in the Reynolds stress, are directly reflected in the inflow/outflow rates in Fig.~\ref{fig:nomhd.profiles}. 

In short, while we see the expected strongly non-linear $m=1$ modes, in-plane supersonic turbulence, and efficient fragmentation, as expected for strong gravito-turbulence in a hydrodynamic disk with $t_{\rm cool} \ll t_{\rm dyn}$, the effective \citet{shakurasunyaev73}-equivalent $\alpha$ parameter is often smaller than $\sim \mathcal{M}_{s}^{2}$, and the Reynolds stresses can even produce net outflow. Taken together, this demonstrates a simple but important point: we cannot simply assume the properties of the disk without magnetic fields would be the same as that with magnetic fields, but with the fields simply ``removed'' (i.e.\ that we would have the same Reynolds stress) -- the disk is non-linearly different in crucial aspects.

\section{Conclusions}
\label{sec:conclusions}

\paperone\ presented the first numerical simulations to follow gas dynamics in a fully-cosmological setup from scales of $\gtrsim$\,Mpc down to $<80\,$au or $<300$ Schwarzschild radii from a SMBH accreting in a quasar phase (Fig.~\ref{fig:image.zoom}). The simulations capture a diverse range of key physical effects relevant to evolution on these scales including multi-band radiation transport; coupling to non-LTE and non-equilibrium atomic/molecular/ionized gas thermo-chemistry and radiative cooling; resolved individual (proto- and main-sequence) star formation and stellar evolution, with associated ``feedback'' in the form of jets, radiation, stellar mass-loss, and supernovae; and magneto-hydrodynamics with kinetic and non-ideal effects, and amplification only from trace cosmological fields in the inter-galactic medium. In \paperone\ we studied the large-scale properties of these simulations and how star formation is suppressed close to the BH and a true quasar accretion disk forms on sub-pc scales (Figs.~\ref{fig:image.faceonedgeon.inner}, \ref{fig:image.wedgeplot}). In this paper, we present a detailed study of the accretion physics, structure, and origins of the magnetic fields in these simulations. Our key conclusions include:

\begin{enumerate}[labelindent=0pt,labelwidth=10pt,labelsep*=0pt,leftmargin=!,align=parleft]

\item{\bf The accretion disk is {\em magnetically-dominated}}, with plasma $\beta \sim 10^{-6}-10^{-2}$ even in the midplane. In the inner disk, the field is dominated by the mean toroidal $\langle B_{\phi} \rangle$, but with non-negligible mean radial field $\langle B_{R} \rangle$ and fluctuating toroidal/radial/vertical components $|\delta B_{R,\,\phi,\,z}|$ (Figs.~\ref{fig:radial.profile.general}, \ref{fig:b.profile}).

\item{\bf The magnetic fields arise from simple flux-freezing}, viz.\ {\bf the dynamo is closed by advection of radial flux}. The field strengths are amplified smoothly from sub-nanoGauss intergalactic fields and typical few-microGauss interstellar fields at $\sim$\,kpc scales, without some sharp change at smaller radii (Fig.~\ref{fig:bfield.demo}). The accretion disk initially forms from capture of gas with tangled, quasi-isotropic fields in the galactic nucleus; these are stretched into radial fields as the gas initially falls through the BHROI, and then into toroidal fields as it circularizes (Figs.~\ref{fig:b.tests.goodmodels}, \ref{fig:b.tests.badmodels}), so it is a ``flux-fed'' and ``flux-frozen'' disk in a general sense. This leaves characteristic imprints such as sign flips in the toroidal field that are advected with the gas as it accretes (Figs.~\ref{fig:bfields.faceon.edgeon}, \ref{fig:bfields.faceon.zoom}, \ref{fig:b.time.flip}, \ref{fig:b.induction.eqn}). Given the rapid inflow rates and corresponding advection of flux, we expect and confirm that the toroidal fields do not damp, even over timescales $\gtrsim 10^{5}$ times the dynamical time at our innermost resolved radii (Figs.~\ref{fig:b.time.flip}, \ref{fig:b.induction.eqn}). 

\item{\bf The toroidal field is stronger than the often-quoted limit for linear growth of the ``traditional'' MRI} ($v_{A} > \sqrt{c_{s}\,v_{\rm K}}$; Figs.~\ref{fig:b.tests.badmodels}, \ref{fig:instability.map}). Instead, we see that the disk lies nominally in the regime of the related Type II/III or SSMI/SHMI magnetic instabilities from e.g.\ \citet{kim.ostriker:2000.mhd.instab.shearing.cold.winds,pessah.psaltis:2005.mri.extensions.stronger.fields,das:2018.pessah.psaltis.limit.mri}, when analyzed according to those previous analytic linear studies. The fluctuating field shows coherent structure with most of the power on wavelengths of roughly the scale height $H$ (Figs.~\ref{fig:bfields.edgeon.smallscale}, \ref{fig:bfields.edgeon.oplotrho}).

\item{\bf The disk is strongly turbulent}, with trans-\Alf{ic} and highly super-sonic, broadly isotropic velocity fluctuations (Figs.~\ref{fig:v.profile}, \ref{fig:mach}). The velocity and magnetic fluctuations are consistent with one another and the stresses. We see some coherent vertical infall into the disk (Figs.~\ref{fig:vfields.lines}, \ref{fig:vfields.lines.rho}) but this is negligible compared to inflow through the disk. 

\item{\bf The disk is coherently eccentric at large radii} with a coherent dimensionless eccentricity of $|a_{1}| \gtrsim 0.1$ throughout much of the disk (Fig.~\ref{fig:eccentricity.profile}). This is driven by the infall and non-Keplerian potential at larger radii outside the BHROI. To first order this dominates the deviations from perfectly circular orbits (Fig.~\ref{fig:vfields.lines}) and may produce shocks that drive some of the turbulence especially at the largest radii.  But these effects do not dominate the net torque/angular momentum transport of the gas, especially at smaller radii (Figs.~\ref{fig:torque.map}, \ref{fig:stress.terms}, \ref{fig:maxwell.reynolds.turb}; \S~\ref{sec:torque:stress} \&\ \ref{sec:no.mhd}).

\item{\bf The disk is weakly stratified}. This is due to a combination of the vigorous highly super-sonic turbulent transport and non-LTE effects on the thermo-chemistry in the more tenuous disk atmosphere and outer disk. As a result, the magnetic-field strength declines only weakly away from the midplane, the temperature is inversely stratified (as expected for a corona/tenuous atmosphere), and the plasma $\beta$ is weakly stratified but lowest in the disk midplane (Fig.~\ref{fig:vertical}).

\item{\bf Accretion is driven by a combination of Maxwell and Reynolds stresses}. Within the accretion disk, the torques on the gas are dominated by a quasi-turbulent  (Fig.~\ref{fig:torque.map}) MHD torque (as opposed to e.g.\ torques from gravitational or radiation pressure forces; Fig.~\ref{fig:torque.profile}). Within this, we see that the usual Maxwell+Reynolds stresses are the dominant sources of angular momentum transport (Fig.~\ref{fig:stress.terms}). In the outer disk, where the turbulence is mildly super-\Alf{ic}, Reynolds stresses dominate, while in the inner disk, the mean-field Maxwell stress dominates, followed by the fluctuating Maxwell and then Reynolds stress, but the three terms are always within an order-of-magnitude of one another (Fig.~\ref{fig:maxwell.reynolds.turb}). 

\item{\bf The disk is not rapidly fragmenting}. This is the main subject of \paperone\ so we refer to that study for details. But we confirm here (e.g.\ Figs.~\ref{fig:image.wedgeplot}, \ref{fig:radial.profile.general}) that the accretion disk is stable against catastrophic turbulent or gravito-turbulent fragmentation on all scales $\ll$\,pc, and that the inflow rates are much larger than the star formation rates. As discussed in more detail in \paperone\ and \paperfour, on all scales of interest here, the mass loss, injection, and turbulent driving contributions from star formation and/or stellar feedback are completely negligible compared to the other terms we study.

\item{\bf These effects can produce super-Eddington accretion}. The torques (Fig.~\ref{fig:torque.profile}), Maxwell \&\ Reynolds stresses (Fig.~\ref{fig:maxwell.reynolds.turb}), mean radial-flow velocities (Fig.~\ref{fig:v.profile}), time-steady structure of the disk (Fig.~\ref{fig:radial.profile.general}), and directly measured inflow/accretion rates (Fig.~\ref{fig:radial.profile.general}) are all consistent with a sustained gas inflow rate that is remarkably constant in both space and time:  $\dot{M}_{\rm in} \sim 20-30\,{\rm M_{\odot}\,yr^{-1}}$ is sustained in steady state from scales $\lesssim 80\,$au to $\sim 1\,$pc for the duration of our simulation ($\gtrsim 10^{5}$ inner dynamical times, or $\sim 10^{4}$\,yr at the highest refinement level). Given the SMBH mass of $\simeq 1.3\times10^{7}\,{\rm M_{\odot}}$, this is up to $\sim 100$ times the canonical optically thin electron-scattering Eddington mass accretion rate for a nominal radiative efficiency of $\epsilon_{r}=0.1$. Whether this can be sustained to horizon scales (and whether the accretion is radiatively efficient on those scales), and the effects of whatever radiation emerges from the inner disk externally illuminating the outer disk, are important subjects for future study.

\item{\bf \ The accretion disk is likely mis-aligned with the BH spin}. The quasar episode here is triggered by tidal interactions with passing, highly-turbulent giant molecular cloud complexes in a clumpy, turbulent, merging galaxy. As such, it is unsurprising that we find the inner accretion disk angular momentum has essentially no correlation with the angular momentum vector of previous generations of BH accretion at earlier cosmic times (which we use as a proxy for the BH spin direction). The disk here is mis-aligned by $\sim 140^{\circ}$ (\S~\ref{sec:angmom.alignment}), so is both retrograde and mis-aligned.

\end{enumerate}

We further validate these results by comparing an equivalent simulation without any magnetic fields, which we show produces completely different results (Figs.~\ref{fig:nomhd.images.faceon}-\ref{fig:nomhd.profiles}): it undergoes catastrophic gravito-turbulent fragmentation, with orders-of-magnitude higher star formation rates and orders-of-magnitude lower gas inflow rates; the gas inflow rate drops rapidly towards the center; the disk is razor-thin, with extremely anisotropic turbulence (strong in-plane gravito-turbulent modes but very weak vertical stirring/mixing); the Reynolds stresses are an order-of-magnitude weaker and often have the opposite sign (pushing the disk outwards); the $m=1$ modes reach much stronger amplitudes and are tightly wound up to short radial wavelengths of order the pressure scale-length, leading to a concentric-ring-like morphology; and the disk mass and outer extent are reduced by more than an order-of-magnitude (with no real disk outside of $\gtrsim 0.01\,$pc and the disk mass inside $\lesssim 0.001\,$pc that can be supported with $Q\gg 1$ reduced by orders-of-magnitude). So it is clear that magnetic fields play an absolutely fundamental role on these scales.

In a companion paper (\paperthree), we also construct a simple self-similar analytic model for the flux-frozen strongly-magnetized, super-sonically turbulent disks that form consistently in the simulations. We show there that this can, at least qualitatively, reproduce the most important features of the simulations described above, although it is certainly an over-simplification that cannot capture all of the subtleties observed in the simulations (including e.g.\ behaviors which are clearly not strictly scale-free/power-law-like; see e.g.\ Fig.~\ref{fig:radial.profile.general}). In addition to providing some additional consistency checks and aid in interpreting the simulations, these models at least suggest that there is no obvious barrier to extrapolating the behavior seen on our resolved scales down to even smaller scales where dedicated GRMHD simulations are required. 

There are many obvious ways in which to extend and improve on the simulations here in future work. In principle, one could imagine refining even further, to smaller radii. However, doing so while retaining all of the physics of star formation, molecular cooling, cosmological expansion (let alone the mass of all the gas, stars, and dark matter to $\gtrsim$\,Mpc scales) is both unnecessary and imposes a huge computational overhead. Moreover, at some scales additional physics (e.g.\ general relativistic effects) will become important. Our goal is therefore not to extend these simulations directly to the ISCO, but to provide new motivation for exploration of disks like that we see here in dedicated GRMHD accretion-disk simulations. Such simulations can reach from scales within the ISCO out to hundreds of gravitational radii -- directly overlapping the innermost resolved radii here. So it would be possible to either directly take our inner boundary conditions to set up such smaller-scale simulations, or to use a form like the analytic models in \paperthree\ to set up initial conditions for idealized disk simulations. 

These simulations can then be used to not only survey different parameter space within the broad category of magnetically-dominated disks, but also to make first-principles predictions for the radiation and jets/outflows that should emerge from the accretion disk on scales $\ll 80\,$au, but which we cannot resolve here. Because this is a first experiment and these kinds of flux-frozen accretion disks have not been explored on such radii, it remains deeply ambiguous whether the disk here should, for example, be radiatively efficient or not, let alone what properties (and orientation) a jet should have. For these reasons and because our goal was to predict the accretion rates and disk outer boundary conditions in the first place, we again caution that these simulations take a simple accretion inner boundary, even though it seems likely that given the combination of strong fields and high accretion rates, significant jets or outflows and radiation must emerge from that boundary. Ultimately, those ``feedback'' properties could be re-introduced to simulations like those here -- running, for example, with some injection/boundary conditions motivated by those smaller-scale simulations -- to follow ``back up'' to larger scales. 

Other important extensions of the work here include exploring the accretion disks that form in different galaxies and at different times. While all of our experiments here suggest that there is nothing ``special'' or ``pathological'' about the time chosen here for hyper-refinement (relative to any other time that features large inflow rates into the BHROI and therefore potential quasar-level activity), and as noted above the accretion rates and implied quasar luminosities correspond to quasars around the ``knee'' of the observed luminosity function at these redshifts ($z\sim 4-5$), it is important to validate this directly. More important still would be to explore how the accretion disks behave in qualitatively different regimes: for example, at lower accretion rates, spanning the vast range between the most luminous sources (like the simulation here) through to ``intermediate'' luminosity quasars, then Seyferts, low-luminosity AGN, and ultimately, extremely low-accretion-rate systems like those in M87 or our own Galaxy.

Further, there are many more properties to study in the simulations here, which could enable a unique exploration and prediction space. This includes the structure of the obscuring torus, the nature of the broad-line region and narrow-line region transition, the IMF of stars forming in the circum-quasar medium and quasar accretion disk, the consequences for transient and gravitational-wave sources from stars and stellar-mass black holes in the disk, predictions for observable signatures of the strongly magnetized accretion disk, and more. We hope to explore these areas and other phenomenology in future work.

\begin{acknowledgements}
We would like to thank Alexander Tchekhovskoy and Mitch Begelman for insightful discussions and a number of helpful suggestions regarding early versions of this manuscript. Support for PFH was provided by NSF Research Grants 1911233, 20009234, 2108318, NSF CAREER grant 1455342, NASA grants 80NSSC18K0562, HST-AR-15800. 
CAFG was supported by NSF through grants AST-2108230  and CAREER award AST-1652522; by NASA through grants 17-ATP17-0067 and 21-ATP21-0036; by STScI through grant HST-GO-16730.016-A; and by CXO through grant TM2-23005X. 
DAA acknowledges support by NSF grants AST-2009687 and AST-2108944, CXO grant TM2-23006X, Simons Foundation Award CCA-1018464, and Cottrell Scholar Award CS-CSA-2023-028 by the Research Corporation for Science Advancement.
Numerical calculations were run on the Caltech compute cluster ``Wheeler,'' allocations AST21010 and AST20016 supported by the NSF and TACC, and NASA HEC SMD-16-7592.
\end{acknowledgements}

\bibliographystyle{mn2e}
\bibliography{ms_extracted}

\begin{thebibliography}{}
\makeatletter
\relax
\def\mn@urlcharsother{\let\do\@makeother \do\$\do\&\do\#\do\^\do\_\do\%\do\~}
\def\mn@doi{\begingroup\mn@urlcharsother \@ifnextchar [ {\mn@doi@}
  {\mn@doi@[]}}
\def\mn@doi@[#1]#2{\def\@tempa{#1}\ifx\@tempa\@empty \href
  {http://dx.doi.org/#2} {doi:#2}\else \href {http://dx.doi.org/#2} {#1}\fi
  \endgroup}
\def\mn@eprint#1#2{\mn@eprint@#1:#2::\@nil}
\def\mn@eprint@arXiv#1{\href {http://arxiv.org/abs/#1} {{\tt arXiv:#1}}}
\def\mn@eprint@dblp#1{\href {http://dblp.uni-trier.de/rec/bibtex/#1.xml}
  {dblp:#1}}
\def\mn@eprint@#1:#2:#3:#4\@nil{\def\@tempa {#1}\def\@tempb {#2}\def\@tempc
  {#3}\ifx \@tempc \@empty \let \@tempc \@tempb \let \@tempb \@tempa \fi \ifx
  \@tempb \@empty \def\@tempb {arXiv}\fi \@ifundefined
  {mn@eprint@\@tempb}{\@tempb:\@tempc}{\expandafter \expandafter \csname
  mn@eprint@\@tempb\endcsname \expandafter{\@tempc}}}

\bibitem[\protect\citeauthoryear{{Abramowicz} \& {Fragile}}{{Abramowicz} \&
  {Fragile}}{2013}]{abramowicz:accretion.theory.review}
{Abramowicz} M.~A.,  {Fragile} P.~C.,  2013, \mn@doi [Living Reviews in
  Relativity] {10.12942/lrr-2013-1}, \href
  {https://ui.adsabs.harvard.edu/abs/2013LRR....16....1A} {16, 1}

\bibitem[\protect\citeauthoryear{{Abramowicz}, {Czerny}, {Lasota}  \&
  {Szuszkiewicz}}{{Abramowicz} et~al.}{1988}]{abramowicz:1988.slim.disks}
{Abramowicz} M.~A.,  {Czerny} B.,  {Lasota} J.~P.,   {Szuszkiewicz} E.,  1988,
  \mn@doi [\apj] {10.1086/166683}, \href
  {https://ui.adsabs.harvard.edu/abs/1988ApJ...332..646A} {332, 646}

\bibitem[\protect\citeauthoryear{{Aller} \& {Richstone}}{{Aller} \&
  {Richstone}}{2007}]{aller:mbh.esph}
{Aller} M.~C.,  {Richstone} D.~O.,  2007, \mn@doi [\apj] {10.1086/519298},
  \href {http://adsabs.harvard.edu/abs/2007ApJ...665..120A} {665, 120}

\bibitem[\protect\citeauthoryear{{Angl{\'e}s-Alc{\'a}zar}
  et~al.,}{{Angl{\'e}s-Alc{\'a}zar}
  et~al.}{2021}]{daa:20.hyperrefinement.bh.growth}
{Angl{\'e}s-Alc{\'a}zar} D.,  et~al., 2021, \mn@doi [\apj]
  {10.3847/1538-4357/ac09e8}, \href
  {https://ui.adsabs.harvard.edu/abs/2021ApJ...917...53A} {917, 53}

\bibitem[\protect\citeauthoryear{{Bacon}, {Emsellem}, {Combes}, {Copin},
  {Monnet}  \& {Martin}}{{Bacon} et~al.}{2001}]{bacon:m31.disk}
{Bacon} R.,  {Emsellem} E.,  {Combes} F.,  {Copin} Y.,  {Monnet} G.,   {Martin}
  P.,  2001, \mn@doi [\aap] {10.1051/0004-6361:20010317}, \href
  {http://adsabs.harvard.edu/abs/2001A%26A...371..409B} {371, 409}

\bibitem[\protect\citeauthoryear{{Balbus} \& {Hawley}}{{Balbus} \&
  {Hawley}}{1998}]{balbus.hawley.review.1998}
{Balbus} S.~A.,  {Hawley} J.~F.,  1998, \mn@doi [Reviews of Modern Physics]
  {10.1103/RevModPhys.70.1}, \href
  {http://adsabs.harvard.edu/abs/1998RvMP...70....1B} {70, 1}

\bibitem[\protect\citeauthoryear{{Barnes} \& {Hernquist}}{{Barnes} \&
  {Hernquist}}{1996}]{barneshernquist96}
{Barnes} J.~E.,  {Hernquist} L.,  1996, \mn@doi [\apj] {10.1086/177957}, \href
  {http://adsabs.harvard.edu/cgi-bin/nph-bib_query?bibcode=1996ApJ...471..115B&db_key=AST}
  {471, 115}

\bibitem[\protect\citeauthoryear{{Beattie}, {Federrath}  \& {Seta}}{{Beattie}
  et~al.}{2020}]{beattie:2020.sims.field.fluctuations.supersonic.subalfvenic.turb}
{Beattie} J.~R.,  {Federrath} C.,   {Seta} A.,  2020, \mn@doi [\mnras]
  {10.1093/mnras/staa2257}, \href
  {https://ui.adsabs.harvard.edu/abs/2020MNRAS.498.1593B} {498, 1593}

\bibitem[\protect\citeauthoryear{{Beattie}, {Krumholz}, {Skalidis},
  {Federrath}, {Seta}, {Crocker}, {Mocz}  \& {Kriel}}{{Beattie}
  et~al.}{2022}]{beattie:2022.energy.equipartition.structure.supersonic.subalfvenic.turbulence}
{Beattie} J.~R.,  {Krumholz} M.~R.,  {Skalidis} R.,  {Federrath} C.,  {Seta}
  A.,  {Crocker} R.~M.,  {Mocz} P.,   {Kriel} N.,  2022, \mn@doi [\mnras]
  {10.1093/mnras/stac2099}, \href
  {https://ui.adsabs.harvard.edu/abs/2022MNRAS.515.5267B} {515, 5267}

\bibitem[\protect\citeauthoryear{{Beck}}{{Beck}}{2015}]{beck:2015.b.field.review}
{Beck} R.,  2015, \mn@doi [\aapr] {10.1007/s00159-015-0084-4}, \href
  {https://ui.adsabs.harvard.edu/abs/2015A&ARv..24....4B} {24, 4}

\bibitem[\protect\citeauthoryear{{Begelman} \& {Armitage}}{{Begelman} \&
  {Armitage}}{2023}]{begelman:2023.mri.saturation.estimates}
{Begelman} M.~C.,  {Armitage} P.~J.,  2023, \mn@doi [\mnras]
  {10.1093/mnras/stad914}, \href
  {https://ui.adsabs.harvard.edu/abs/2023MNRAS.521.5952B} {521, 5952}

\bibitem[\protect\citeauthoryear{{Begelman} \& {Pringle}}{{Begelman} \&
  {Pringle}}{2007}]{begelman.pringle:2007.acc.disks.strong.toroidal.fields}
{Begelman} M.~C.,  {Pringle} J.~E.,  2007, \mn@doi [\mnras]
  {10.1111/j.1365-2966.2006.11372.x}, \href
  {https://ui.adsabs.harvard.edu/abs/2007MNRAS.375.1070B} {375, 1070}

\bibitem[\protect\citeauthoryear{{Begelman} \& {Silk}}{{Begelman} \&
  {Silk}}{2023}]{begelman.silk:2023.magnetically.boosted.accretion}
{Begelman} M.~C.,  {Silk} J.,  2023, \mn@doi [arXiv e-prints]
  {10.48550/arXiv.2305.19081}, \href
  {https://ui.adsabs.harvard.edu/abs/2023arXiv230519081B} {p. arXiv:2305.19081}

\bibitem[\protect\citeauthoryear{{Bisnovatyi-Kogan} \&
  {Ruzmaikin}}{{Bisnovatyi-Kogan} \&
  {Ruzmaikin}}{1976}]{bisnovatyi.kogan:1976.mad.disk}
{Bisnovatyi-Kogan} G.~S.,  {Ruzmaikin} A.~A.,  1976, \mn@doi [\apss]
  {10.1007/BF01225967}, \href
  {https://ui.adsabs.harvard.edu/abs/1976Ap&SS..42..401B} {42, 401}

\bibitem[\protect\citeauthoryear{{Bonnerot}, {Price}, {Lodato}  \&
  {Rossi}}{{Bonnerot} et~al.}{2017}]{bonnerot:2017.bfield.in.tdes}
{Bonnerot} C.,  {Price} D.~J.,  {Lodato} G.,   {Rossi} E.~M.,  2017, \mn@doi
  [\mnras] {10.1093/mnras/stx1210}, \href
  {https://ui.adsabs.harvard.edu/abs/2017MNRAS.469.4879B} {469, 4879}

\bibitem[\protect\citeauthoryear{{Brandenburg}, {Nordlund}, {Stein}  \&
  {Torkelsson}}{{Brandenburg}
  et~al.}{1995}]{brandenburg:mri.predicted.high.ratio.maxwell.to.reynolds}
{Brandenburg} A.,  {Nordlund} A.,  {Stein} R.~F.,   {Torkelsson} U.,  1995,
  \mn@doi [\apj] {10.1086/175831}, \href
  {https://ui.adsabs.harvard.edu/abs/1995ApJ...446..741B} {446, 741}

\bibitem[\protect\citeauthoryear{{Bullock} \& {Johnston}}{{Bullock} \&
  {Johnston}}{2005}]{bullock.johnston:2005:stellar.halos}
{Bullock} J.~S.,  {Johnston} K.~V.,  2005, \mn@doi [\apj] {10.1086/497422},
  \href {https://ui.adsabs.harvard.edu/abs/2005ApJ...635..931B} {635, 931}

\bibitem[\protect\citeauthoryear{{Choban}, {Kere{\v{s}}}, {Hopkins},
  {Sandstrom}, {Hayward}  \& {Faucher-Gigu{\`e}re}}{{Choban}
  et~al.}{2022}]{choban:2022.fire.dust.growth.destruction.chemistry}
{Choban} C.~R.,  {Kere{\v{s}}} D.,  {Hopkins} P.~F.,  {Sandstrom} K.~M.,
  {Hayward} C.~C.,   {Faucher-Gigu{\`e}re} C.-A.,  2022, \mn@doi [\mnras]
  {10.1093/mnras/stac1542}, \href
  {https://ui.adsabs.harvard.edu/abs/2022MNRAS.514.4506C} {514, 4506}

\bibitem[\protect\citeauthoryear{{Cochrane} et~al.,}{{Cochrane}
  et~al.}{2023}]{cochrane:2023.agn.winds.galaxy.size.effects}
{Cochrane} R.~K.,  et~al., 2023, \mn@doi [\mnras] {10.1093/mnras/stad1528},
  \href {https://ui.adsabs.harvard.edu/abs/2023MNRAS.523.2409C} {523, 2409}

\bibitem[\protect\citeauthoryear{{Crenshaw} et~al.}{{Crenshaw}
  et~al.}{2000}]{crenshaw:nlr}
{Crenshaw} D.~M.,  et~al., 2000, \mn@doi [\aj] {10.1086/301574}, \href
  {http://adsabs.harvard.edu/abs/2000AJ....120.1731C} {120, 1731}

\bibitem[\protect\citeauthoryear{{Croton} et~al.}{{Croton}
  et~al.}{2006}]{croton:sam}
{Croton} D.~J.,  et~al., 2006, \mn@doi [\mnras]
  {10.1111/j.1365-2966.2005.09675.x}, \href
  {http://adsabs.harvard.edu/cgi-bin/nph-bib_query?bibcode=2006MNRAS.365...11C&db_key=AST}
  {365, 11}

\bibitem[\protect\citeauthoryear{{Das}, {Begelman}  \& {Lesur}}{{Das}
  et~al.}{2018}]{das:2018.pessah.psaltis.limit.mri}
{Das} U.,  {Begelman} M.~C.,   {Lesur} G.,  2018, \mn@doi [\mnras]
  {10.1093/mnras/stx2518}, \href
  {https://ui.adsabs.harvard.edu/abs/2018MNRAS.473.2791D} {473, 2791}

\bibitem[\protect\citeauthoryear{{Datta}, {Mondal}  \& {Mukhopadhyay}}{{Datta}
  et~al.}{2022}]{datta:2022.am.transport.acc.disk.weak.bfields}
{Datta} S.~R.,  {Mondal} T.,   {Mukhopadhyay} B.,  2022, \mn@doi [\mnras]
  {10.1093/mnras/stac835}, \href
  {https://ui.adsabs.harvard.edu/abs/2022MNRAS.513..204D} {513, 204}

\bibitem[\protect\citeauthoryear{{Davies} et~al.,}{{Davies}
  et~al.}{2014}]{davies:2014.dusty.structures.misaligned.around.agn}
{Davies} R.~I.,  et~al., 2014, \mn@doi [\apj] {10.1088/0004-637X/792/2/101},
  \href {https://ui.adsabs.harvard.edu/abs/2014ApJ...792..101D} {792, 101}

\bibitem[\protect\citeauthoryear{{Di Matteo}, {Springel}  \& {Hernquist}}{{Di
  Matteo} et~al.}{2005}]{dimatteo:msigma}
{Di Matteo} T.,  {Springel} V.,   {Hernquist} L.,  2005, \mn@doi [\nat]
  {10.1038/nature03335}, \href
  {http://adsabs.harvard.edu/cgi-bin/nph-bib_query?bibcode=2005Natur.433..604D&db_key=AST}
  {433, 604}

\bibitem[\protect\citeauthoryear{{Dunn} et~al.}{{Dunn}
  et~al.}{2010}]{dunn:agn.fb.from.strong.outflows}
{Dunn} J.~P.,  et~al., 2010, \mn@doi [\apj] {10.1088/0004-637X/709/2/611},
  \href {http://adsabs.harvard.edu/abs/2010ApJ...709..611D} {709, 611}

\bibitem[\protect\citeauthoryear{{Faucher-Gigu{\`e}re} \&
  {Quataert}}{{Faucher-Gigu{\`e}re} \& {Quataert}}{2012}]{fgq2012}
{Faucher-Gigu{\`e}re} C.-A.,  {Quataert} E.,  2012, \mn@doi [\mnras]
  {10.1111/j.1365-2966.2012.21512.x}, \href
  {https://ui.adsabs.harvard.edu/abs/2012MNRAS.425..605F} {425, 605}

\bibitem[\protect\citeauthoryear{{Faucher-Gigu{\`e}re}, {Quataert}  \&
  {Murray}}{{Faucher-Gigu{\`e}re} et~al.}{2012}]{fgqm2012}
{Faucher-Gigu{\`e}re} C.-A.,  {Quataert} E.,   {Murray} N.,  2012, \mn@doi
  [\mnras] {10.1111/j.1365-2966.2011.20120.x}, \href
  {https://ui.adsabs.harvard.edu/abs/2012MNRAS.420.1347F} {420, 1347}

\bibitem[\protect\citeauthoryear{{Federrath}, {Schober}, {Bovino}  \&
  {Schleicher}}{{Federrath} et~al.}{2014}]{federrath:supersonic.turb.dynamo}
{Federrath} C.,  {Schober} J.,  {Bovino} S.,   {Schleicher} D.~R.~G.,  2014,
  \mn@doi [\apjl] {10.1088/2041-8205/797/2/L19}, \href
  {http://adsabs.harvard.edu/abs/2014ApJ...797L..19F} {797, L19}

\bibitem[\protect\citeauthoryear{{Ferrarese} \& {Merritt}}{{Ferrarese} \&
  {Merritt}}{2000}]{FM00}
{Ferrarese} L.,  {Merritt} D.,  2000, \mn@doi [\apjl] {10.1086/312838}, \href
  {http://adsabs.harvard.edu/cgi-bin/nph-bib_query?bibcode=2000ApJ...539L...9F&db_key=AST}
  {539, L9}

\bibitem[\protect\citeauthoryear{{Foglizzo} \& {Tagger}}{{Foglizzo} \&
  {Tagger}}{1994}]{foglizzo:1994.parker.instab.with.differential.rotation}
{Foglizzo} T.,  {Tagger} M.,  1994, \mn@doi [\aap]
  {10.48550/arXiv.astro-ph/9403019}, \href
  {https://ui.adsabs.harvard.edu/abs/1994A&A...287..297F} {287, 297}

\bibitem[\protect\citeauthoryear{{Forgan}, {Price}  \& {Bonnell}}{{Forgan}
  et~al.}{2017}]{forgan:2017.mhd.gravitoturb.sims}
{Forgan} D.,  {Price} D.~J.,   {Bonnell} I.,  2017, \mn@doi [\mnras]
  {10.1093/mnras/stw3314}, \href
  {https://ui.adsabs.harvard.edu/abs/2017MNRAS.466.3406F} {466, 3406}

\bibitem[\protect\citeauthoryear{{Fragile} \& {S{\k{a}}dowski}}{{Fragile} \&
  {S{\k{a}}dowski}}{2017}]{fragile.sadowski:2017.strong.magnetized.acc.disk.bfield.decay}
{Fragile} P.~C.,  {S{\k{a}}dowski} A.,  2017, \mn@doi [\mnras]
  {10.1093/mnras/stx274}, \href
  {https://ui.adsabs.harvard.edu/abs/2017MNRAS.467.1838F} {467, 1838}

\bibitem[\protect\citeauthoryear{{Frank}, {King}  \& {Raine}}{{Frank}
  et~al.}{2002}]{frank:2002.accretion.book}
{Frank} J.,  {King} A.,   {Raine} D.~J.,  2002, {Accretion Power in
  Astrophysics: Third Edition}, isbn 0521620538 edn.
Cambridge, UK: Cambridge University Press, Cambridge, UK

\bibitem[\protect\citeauthoryear{{Gaburov}, {Johansen}  \& {Levin}}{{Gaburov}
  et~al.}{2012}]{gaburov:2012.public.moving.mesh.code}
{Gaburov} E.,  {Johansen} A.,   {Levin} Y.,  2012, \mn@doi [\apj]
  {10.1088/0004-637X/758/2/103}, \href
  {http://adsabs.harvard.edu/abs/2012ApJ...758..103G} {758, 103}

\bibitem[\protect\citeauthoryear{{Gammie}}{{Gammie}}{2001}]{gammie:2001.cooling.in.keplerian.disks}
{Gammie} C.~F.,  2001, \mn@doi [\apj] {10.1086/320631}, \href
  {http://adsabs.harvard.edu/abs/2001ApJ...553..174G} {553, 174}

\bibitem[\protect\citeauthoryear{{Gandhi}, {Wetzel}, {Hopkins}, {Shappee},
  {Wheeler}  \& {Faucher-Gigu{\`e}re}}{{Gandhi}
  et~al.}{2022}]{gandhi:2022.sne.1a.comparisons}
{Gandhi} P.~J.,  {Wetzel} A.,  {Hopkins} P.~F.,  {Shappee} B.~J.,  {Wheeler}
  C.,   {Faucher-Gigu{\`e}re} C.-A.,  2022, \mn@doi [\mnras]
  {10.1093/mnras/stac2228}, \href
  {https://ui.adsabs.harvard.edu/abs/2022MNRAS.516.1941G} {516, 1941}

\bibitem[\protect\citeauthoryear{{Gebhardt} et~al.}{{Gebhardt}
  et~al.}{2000}]{Gebhardt00}
{Gebhardt} K.,  et~al., 2000, \mn@doi [\apjl] {10.1086/312840}, \href
  {http://adsabs.harvard.edu/cgi-bin/nph-bib_query?bibcode=2000ApJ...539L..13G&db_key=AST}
  {539, L13}

\bibitem[\protect\citeauthoryear{{Goldreich} \& {Sridhar}}{{Goldreich} \&
  {Sridhar}}{1995}]{GS95.turbulence}
{Goldreich} P.,  {Sridhar} S.,  1995, \mn@doi [\apj] {10.1086/175121}, \href
  {https://ui.adsabs.harvard.edu/abs/1995ApJ...438..763G} {438, 763}

\bibitem[\protect\citeauthoryear{{Goodman}}{{Goodman}}{2003}]{goodman:qso.disk.selfgrav}
{Goodman} J.,  2003, \mn@doi [\mnras] {10.1046/j.1365-8711.2003.06241.x}, \href
  {http://adsabs.harvard.edu/abs/2003MNRAS.339..937G} {339, 937}

\bibitem[\protect\citeauthoryear{{Grudi{\'c}}}{{Grudi{\'c}}}{2021}]{grudic:2021.accelerating.hydro.with.adaptive.force.updates}
{Grudi{\'c}} M.~Y.,  2021, \mn@doi [\mnras] {10.1093/mnras/stab2208}, \href
  {https://ui.adsabs.harvard.edu/abs/2021MNRAS.507.1064G} {507, 1064}

\bibitem[\protect\citeauthoryear{{Grudi{\'c}} \& {Hopkins}}{{Grudi{\'c}} \&
  {Hopkins}}{2020}]{grudic:2020.tidal.timestep.criterion}
{Grudi{\'c}} M.~Y.,  {Hopkins} P.~F.,  2020, \mn@doi [\mnras]
  {10.1093/mnras/staa1453}, \href
  {https://ui.adsabs.harvard.edu/abs/2020MNRAS.495.4306G} {495, 4306}

\bibitem[\protect\citeauthoryear{{Grudi{\'c}}, {Guszejnov}, {Hopkins}, {Offner}
   \& {Faucher-Gigu{\`e}re}}{{Grudi{\'c}}
  et~al.}{2021}]{grudic:starforge.methods}
{Grudi{\'c}} M.~Y.,  {Guszejnov} D.,  {Hopkins} P.~F.,  {Offner} S. S.~R.,
  {Faucher-Gigu{\`e}re} C.-A.,  2021, \mn@doi [\mnras]
  {10.1093/mnras/stab1347}, \href
  {https://ui.adsabs.harvard.edu/abs/2021MNRAS.506.2199G} {506, 2199}

\bibitem[\protect\citeauthoryear{{Grudi{\'c}}, {Guszejnov}, {Offner}, {Rosen},
  {Raju}, {Faucher-Gigu{\`e}re}  \& {Hopkins}}{{Grudi{\'c}}
  et~al.}{2022}]{grudic:2022.sf.fullstarforge.imf}
{Grudi{\'c}} M.~Y.,  {Guszejnov} D.,  {Offner} S. S.~R.,  {Rosen} A.~L.,
  {Raju} A.~N.,  {Faucher-Gigu{\`e}re} C.-A.,   {Hopkins} P.~F.,  2022, \mn@doi
  [\mnras] {10.1093/mnras/stac526}, \href
  {https://ui.adsabs.harvard.edu/abs/2022MNRAS.512..216G} {512, 216}

\bibitem[\protect\citeauthoryear{{Guillochon} \& {McCourt}}{{Guillochon} \&
  {McCourt}}{2017}]{2017ApJ...834L..19G}
{Guillochon} J.,  {McCourt} M.,  2017, \mn@doi [\apjl]
  {10.3847/2041-8213/834/2/L19}, \href
  {https://ui.adsabs.harvard.edu/abs/2017ApJ...834L..19G} {834, L19}

\bibitem[\protect\citeauthoryear{{Guszejnov}, {Grudi{\'c}}, {Offner},
  {Boylan-Kolchin}, {Faucher-Gigu{\`e}re}, {Wetzel}, {Benincasa}  \&
  {Loebman}}{{Guszejnov} et~al.}{2020}]{guszejnov:fire.gmc.props.vs.z}
{Guszejnov} D.,  {Grudi{\'c}} M.~Y.,  {Offner} S. S.~R.,  {Boylan-Kolchin} M.,
  {Faucher-Gigu{\`e}re} C.-A.,  {Wetzel} A.,  {Benincasa} S.~M.,   {Loebman}
  S.,  2020, \mn@doi [\mnras] {10.1093/mnras/stz3527}, \href
  {https://ui.adsabs.harvard.edu/abs/2020MNRAS.492..488G} {492, 488}

\bibitem[\protect\citeauthoryear{{Guszejnov}, {Grudi{\'c}}, {Hopkins}, {Offner}
   \& {Faucher-Gigu{\`e}re}}{{Guszejnov}
  et~al.}{2021}]{guszejnov:2020.starforge.jets}
{Guszejnov} D.,  {Grudi{\'c}} M.~Y.,  {Hopkins} P.~F.,  {Offner} S. S.~R.,
  {Faucher-Gigu{\`e}re} C.-A.,  2021, \mn@doi [\mnras] {10.1093/mnras/stab278},
  \href {https://ui.adsabs.harvard.edu/abs/2021MNRAS.502.3646G} {502, 3646}

\bibitem[\protect\citeauthoryear{{Guszejnov}, {Raju}, {Offner}, {Grudi{\'c}},
  {Faucher-Gigu{\`e}re}, {Hopkins}  \& {Rosen}}{{Guszejnov}
  et~al.}{2022a}]{guszejnov:starforge.environment.multiplicity}
{Guszejnov} D.,  {Raju} A.~N.,  {Offner} S. S.~R.,  {Grudi{\'c}} M.~Y.,
  {Faucher-Gigu{\`e}re} C.-A.,  {Hopkins} P.~F.,   {Rosen} A.~L.,  2022a,
  \mn@doi [\mnras] {10.1093/mnras/stac3268}, \href
  {https://ui.adsabs.harvard.edu/abs/2022MNRAS.tmp.3043G} {}

\bibitem[\protect\citeauthoryear{{Guszejnov}, {Markey}, {Offner}, {Grudi{\'c}},
  {Faucher-Gigu{\`e}re}, {Rosen}  \& {Hopkins}}{{Guszejnov}
  et~al.}{2022b}]{guszejnov:2022.starforge.cluster.assembly}
{Guszejnov} D.,  {Markey} C.,  {Offner} S. S.~R.,  {Grudi{\'c}} M.~Y.,
  {Faucher-Gigu{\`e}re} C.-A.,  {Rosen} A.~L.,   {Hopkins} P.~F.,  2022b,
  \mn@doi [\mnras] {10.1093/mnras/stac1737}, \href
  {https://ui.adsabs.harvard.edu/abs/2022MNRAS.515..167G} {515, 167}

\bibitem[\protect\citeauthoryear{{Guszejnov}, {Grudi{\'c}}, {Offner},
  {Faucher-Gigu{\`e}re}, {Hopkins}  \& {Rosen}}{{Guszejnov}
  et~al.}{2022c}]{guszejnov:environment.feedback.starforge.imf}
{Guszejnov} D.,  {Grudi{\'c}} M.~Y.,  {Offner} S. S.~R.,  {Faucher-Gigu{\`e}re}
  C.-A.,  {Hopkins} P.~F.,   {Rosen} A.~L.,  2022c, \mn@doi [\mnras]
  {10.1093/mnras/stac2060}, \href
  {https://ui.adsabs.harvard.edu/abs/2022MNRAS.515.4929G} {515, 4929}

\bibitem[\protect\citeauthoryear{{Habibi} \& {Abbassi}}{{Habibi} \&
  {Abbassi}}{2019}]{habibi:2019.thermal.instabilities.magnetically.dominated.disks}
{Habibi} A.,  {Abbassi} S.,  2019, \mn@doi [\apj] {10.3847/1538-4357/ab5793},
  \href {https://ui.adsabs.harvard.edu/abs/2019ApJ...887..256H} {887, 256}

\bibitem[\protect\citeauthoryear{{Hernquist} \& {Mihos}}{{Hernquist} \&
  {Mihos}}{1995}]{hernquist.mihos:minor.mergers}
{Hernquist} L.,  {Mihos} J.~C.,  1995, \mn@doi [\apj] {10.1086/175940}, \href
  {http://adsabs.harvard.edu/cgi-bin/nph-bib_query?bibcode=1995ApJ...448...41H&db_key=AST}
  {448, 41}

\bibitem[\protect\citeauthoryear{{Hirabayashi} \& {Hoshino}}{{Hirabayashi} \&
  {Hoshino}}{2016}]{hirabayashi:2016.toroidal.field.bouyancy.modes}
{Hirabayashi} K.,  {Hoshino} M.,  2016, \mn@doi [\apj]
  {10.3847/0004-637X/822/2/87}, \href
  {https://ui.adsabs.harvard.edu/abs/2016ApJ...822...87H} {822, 87}

\bibitem[\protect\citeauthoryear{{Hobbs} \& {Nayakshin}}{{Hobbs} \&
  {Nayakshin}}{2009}]{hobbs:2009.mw.nucdisk.sim}
{Hobbs} A.,  {Nayakshin} S.,  2009, \mn@doi [\mnras]
  {10.1111/j.1365-2966.2008.14359.x}, \href
  {http://adsabs.harvard.edu/abs/2009MNRAS.394..191H} {394, 191}

\bibitem[\protect\citeauthoryear{{Hogg} \& {Reynolds}}{{Hogg} \&
  {Reynolds}}{2018}]{hogg:2018.truncated.disks.sims}
{Hogg} J.~D.,  {Reynolds} C.~S.,  2018, \mn@doi [\apj]
  {10.3847/1538-4357/aaa6c6}, \href
  {https://ui.adsabs.harvard.edu/abs/2018ApJ...854....6H} {854, 6}

\bibitem[\protect\citeauthoryear{{Hopkins}}{{Hopkins}}{2010}]{hopkins:slow.modes}
{Hopkins} P.~F.,  2010, \mnras, in press, arXiv:1009.4702 [astro-ph], \href
  {http://adsabs.harvard.edu/abs/2010arXiv1009.4702H} {}

\bibitem[\protect\citeauthoryear{{Hopkins}}{{Hopkins}}{2013}]{hopkins:frag.theory}
{Hopkins} P.~F.,  2013, \mn@doi [\mnras] {10.1093/mnras/sts704}, \href
  {http://adsabs.harvard.edu/abs/2013MNRAS.430.1653H} {430, 1653}

\bibitem[\protect\citeauthoryear{{Hopkins}}{{Hopkins}}{2015}]{hopkins:gizmo}
{Hopkins} P.~F.,  2015, \mn@doi [\mnras] {10.1093/mnras/stv195}, \href
  {http://adsabs.harvard.edu/abs/2015MNRAS.450...53H} {450, 53}

\bibitem[\protect\citeauthoryear{{Hopkins}}{{Hopkins}}{2016}]{hopkins:cg.mhd.gizmo}
{Hopkins} P.~F.,  2016, \mn@doi [\mnras] {10.1093/mnras/stw1578}, \href
  {http://adsabs.harvard.edu/abs/2016MNRAS.462..576H} {462, 576}

\bibitem[\protect\citeauthoryear{{Hopkins}}{{Hopkins}}{2017}]{hopkins:gizmo.diffusion}
{Hopkins} P.~F.,  2017, \mn@doi [\mnras] {10.1093/mnras/stw3306}, \href
  {http://adsabs.harvard.edu/abs/2017MNRAS.466.3387H} {466, 3387}

\bibitem[\protect\citeauthoryear{{Hopkins}}{{Hopkins}}{2022}]{hopkins:cr.spectra.accurate.integration}
{Hopkins} P.~F.,  2022, arXiv:2202.05283, \href
  {https://ui.adsabs.harvard.edu/abs/2022arXiv220205283H} {p. arXiv:2202.05283}

\bibitem[\protect\citeauthoryear{{Hopkins}}{{Hopkins}}{2023b}]{hopkins:superzoom.imf}
{Hopkins} P.~F. e.~a.,  2023b, in prep.

\bibitem[\protect\citeauthoryear{{Hopkins}}{{Hopkins}}{2023a}]{hopkins:superzoom.analytic}
{Hopkins} P.~F. e.~a.,  2023a, in prep.

\bibitem[\protect\citeauthoryear{{Hopkins} \& {Christiansen}}{{Hopkins} \&
  {Christiansen}}{2013}]{hopkins:2013.turb.planet.direct.collapse}
{Hopkins} P.~F.,  {Christiansen} J.~L.,  2013, \mn@doi [\apj]
  {10.1088/0004-637X/776/1/48}, \href
  {http://adsabs.harvard.edu/abs/2013ApJ...776...48H} {776, 48}

\bibitem[\protect\citeauthoryear{{Hopkins} \& {Grudi{\'c}}}{{Hopkins} \&
  {Grudi{\'c}}}{2019}]{hopkins:2019.grudic.photon.momentum.rad.pressure.coupling}
{Hopkins} P.~F.,  {Grudi{\'c}} M.~Y.,  2019, \mn@doi [\mnras]
  {10.1093/mnras/sty3089}, \href
  {https://ui.adsabs.harvard.edu/abs/2019MNRAS.483.4187H} {483, 4187}

\bibitem[\protect\citeauthoryear{{Hopkins} \& {Quataert}}{{Hopkins} \&
  {Quataert}}{2010a}]{hopkins:m31.disk}
{Hopkins} P.~F.,  {Quataert} E.,  2010a, \mn@doi [\mnras]
  {10.1111/j.1745-3933.2010.00855.x}, \href
  {http://adsabs.harvard.edu/abs/2010arXiv1002.1079H} {405, L41}

\bibitem[\protect\citeauthoryear{{Hopkins} \& {Quataert}}{{Hopkins} \&
  {Quataert}}{2010b}]{hopkins:zoom.sims}
{Hopkins} P.~F.,  {Quataert} E.,  2010b, \mn@doi [\mnras]
  {10.1111/j.1365-2966.2010.17064.x}, \href
  {http://adsabs.harvard.edu/abs/2009arXiv0912.3257H} {407, 1529}

\bibitem[\protect\citeauthoryear{{Hopkins} \& {Quataert}}{{Hopkins} \&
  {Quataert}}{2011}]{hopkins:inflow.analytics}
{Hopkins} P.~F.,  {Quataert} E.,  2011, \mn@doi [\mnras]
  {10.1111/j.1365-2966.2011.18542.x}, \href
  {http://adsabs.harvard.edu/abs/2010arXiv1007.2647H} {415, 1027}

\bibitem[\protect\citeauthoryear{{Hopkins} \& {Raives}}{{Hopkins} \&
  {Raives}}{2016}]{hopkins:mhd.gizmo}
{Hopkins} P.~F.,  {Raives} M.~J.,  2016, \mn@doi [\mnras]
  {10.1093/mnras/stv2180}, \href
  {http://adsabs.harvard.edu/abs/2016MNRAS.455...51H} {455, 51}

\bibitem[\protect\citeauthoryear{{Hopkins}, {Hernquist}, {Cox}, {Di Matteo},
  {Martini}, {Robertson}  \& {Springel}}{{Hopkins}
  et~al.}{2005a}]{hopkins:lifetimes.methods}
{Hopkins} P.~F.,  {Hernquist} L.,  {Cox} T.~J.,  {Di Matteo} T.,  {Martini} P.,
   {Robertson} B.,   {Springel} V.,  2005a, \mn@doi [\apj] {10.1086/432438},
  \href
  {http://adsabs.harvard.edu/cgi-bin/nph-bib_query?bibcode=2005ApJ...630..705H&db_key=AST}
  {630, 705}

\bibitem[\protect\citeauthoryear{{Hopkins}, {Hernquist}, {Cox}, {Di Matteo},
  {Robertson}  \& {Springel}}{{Hopkins}
  et~al.}{2005b}]{hopkins:lifetimes.obscuration}
{Hopkins} P.~F.,  {Hernquist} L.,  {Cox} T.~J.,  {Di Matteo} T.,  {Robertson}
  B.,   {Springel} V.,  2005b, \mn@doi [\apj] {10.1086/432755}, \href
  {http://adsabs.harvard.edu/cgi-bin/nph-bib_query?bibcode=2005ApJ...632...81H&db_key=AST}
  {632, 81}

\bibitem[\protect\citeauthoryear{{Hopkins}, {Hernquist}, {Cox}, {Di Matteo},
  {Robertson}  \& {Springel}}{{Hopkins} et~al.}{2006}]{hopkins:qso.all}
{Hopkins} P.~F.,  {Hernquist} L.,  {Cox} T.~J.,  {Di Matteo} T.,  {Robertson}
  B.,   {Springel} V.,  2006, \mn@doi [\apjs] {10.1086/499298}, \href
  {http://adsabs.harvard.edu/cgi-bin/nph-bib_query?bibcode=2006ApJS..163....1H&db_key=AST}
  {163, 1}

\bibitem[\protect\citeauthoryear{{Hopkins}, {Richards}  \&
  {Hernquist}}{{Hopkins} et~al.}{2007a}]{hopkins:bol.qlf}
{Hopkins} P.~F.,  {Richards} G.~T.,   {Hernquist} L.,  2007a, \mn@doi [\apj]
  {10.1086/509629}, \href
  {http://adsabs.harvard.edu/cgi-bin/nph-bib_query?bibcode=2007ApJ...654..731H&db_key=AST}
  {654, 731}

\bibitem[\protect\citeauthoryear{{Hopkins}, {Hernquist}, {Cox}, {Robertson}  \&
  {Krause}}{{Hopkins} et~al.}{2007b}]{hopkins:bhfp.obs}
{Hopkins} P.~F.,  {Hernquist} L.,  {Cox} T.~J.,  {Robertson} B.,   {Krause} E.,
   2007b, \mn@doi [\apj] {10.1086/521601}, \href
  {http://adsabs.harvard.edu/abs/2007ApJ...669...67H} {669, 67}

\bibitem[\protect\citeauthoryear{{Hopkins}, {Cox}, {Kere{\v s}}  \&
  {Hernquist}}{{Hopkins} et~al.}{2008}]{hopkins:groups.ell}
{Hopkins} P.~F.,  {Cox} T.~J.,  {Kere{\v s}} D.,   {Hernquist} L.,  2008,
  \mn@doi [\apjs] {10.1086/524363}, \href
  {http://adsabs.harvard.edu/abs/2008ApJS..175..390H} {175, 390}

\bibitem[\protect\citeauthoryear{{Hopkins}, {Hayward}, {Narayanan}  \&
  {Hernquist}}{{Hopkins} et~al.}{2012a}]{hopkins:torus}
{Hopkins} P.~F.,  {Hayward} C.~C.,  {Narayanan} D.,   {Hernquist} L.,  2012a,
  \mn@doi [\mnras] {10.1111/j.1365-2966.2011.20035.x}, \href
  {http://adsabs.harvard.edu/abs/2011arXiv1108.3086H} {420, 320}

\bibitem[\protect\citeauthoryear{{Hopkins}, {Quataert}  \& {Murray}}{{Hopkins}
  et~al.}{2012b}]{hopkins:fb.ism.prop}
{Hopkins} P.~F.,  {Quataert} E.,   {Murray} N.,  2012b, \mn@doi [\mnras]
  {10.1111/j.1365-2966.2012.20578.x}, \href
  {http://adsabs.harvard.edu/abs/2012MNRAS.421.3488H} {421, 3488}

\bibitem[\protect\citeauthoryear{{Hopkins}, {Hernquist}, {Hayward}  \&
  {Narayanan}}{{Hopkins} et~al.}{2012c}]{hopkins:agn.alignment}
{Hopkins} P.~F.,  {Hernquist} L.,  {Hayward} C.~C.,   {Narayanan} D.,  2012c,
  \mn@doi [\mnras] {10.1111/j.1365-2966.2012.21449.x}, \href
  {http://adsabs.harvard.edu/abs/2012MNRAS.425.1121H} {425, 1121}

\bibitem[\protect\citeauthoryear{{Hopkins}, {Torrey}, {Faucher-Gigu{\`e}re},
  {Quataert}  \& {Murray}}{{Hopkins}
  et~al.}{2016}]{hopkins:qso.stellar.fb.together}
{Hopkins} P.~F.,  {Torrey} P.,  {Faucher-Gigu{\`e}re} C.-A.,  {Quataert} E.,
  {Murray} N.,  2016, \mn@doi [\mnras] {10.1093/mnras/stw289}, \href
  {http://adsabs.harvard.edu/abs/2016MNRAS.458..816H} {458, 816}

\bibitem[\protect\citeauthoryear{{Hopkins} et~al.,}{{Hopkins}
  et~al.}{2018}]{hopkins:fire2.methods}
{Hopkins} P.~F.,  et~al., 2018, \mn@doi [\mnras] {10.1093/mnras/sty1690}, \href
  {http://adsabs.harvard.edu/abs/2018MNRAS.480..800H} {480, 800}

\bibitem[\protect\citeauthoryear{{Hopkins}, {Grudi{\'c}}, {Wetzel},
  {Kere{\v{s}}}, {Faucher-Gigu{\`e}re}, {Ma}, {Murray}  \& {Butcher}}{{Hopkins}
  et~al.}{2020a}]{hopkins:radiation.methods}
{Hopkins} P.~F.,  {Grudi{\'c}} M.~Y.,  {Wetzel} A.,  {Kere{\v{s}}} D.,
  {Faucher-Gigu{\`e}re} C.-A.,  {Ma} X.,  {Murray} N.,   {Butcher} N.,  2020a,
  \mn@doi [\mnras] {10.1093/mnras/stz3129}, \href
  {https://ui.adsabs.harvard.edu/abs/2020MNRAS.491.3702H} {491, 3702}

\bibitem[\protect\citeauthoryear{{Hopkins} et~al.,}{{Hopkins}
  et~al.}{2020b}]{hopkins:cr.mhd.fire2}
{Hopkins} P.~F.,  et~al., 2020b, \mn@doi [\mnras] {10.1093/mnras/stz3321},
  \href {https://ui.adsabs.harvard.edu/abs/2020MNRAS.492.3465H} {492, 3465}

\bibitem[\protect\citeauthoryear{{Hopkins}, {Nadler}, {Grudic}, {Shen}, {Sands}
   \& {Jiang}}{{Hopkins} et~al.}{2022a}]{hopkins:tidal.softening}
{Hopkins} P.~F.,  {Nadler} E.~O.,  {Grudic} M.~Y.,  {Shen} X.,  {Sands} I.,
  {Jiang} F.,  2022a, arXiv e-prints, \href
  {https://ui.adsabs.harvard.edu/abs/2022arXiv221206851H} {p. arXiv:2212.06851}

\bibitem[\protect\citeauthoryear{{Hopkins}, {Squire}  \& {Butsky}}{{Hopkins}
  et~al.}{2022b}]{hopkins:m1.cr.closure}
{Hopkins} P.~F.,  {Squire} J.,   {Butsky} I.~S.,  2022b, \mn@doi [\mnras]
  {10.1093/mnras/stab2635}, \href
  {https://ui.adsabs.harvard.edu/abs/2022MNRAS.509.3779H} {509, 3779}

\bibitem[\protect\citeauthoryear{{Hopkins}, {Butsky}, {Panopoulou}, {Ji},
  {Quataert}, {Faucher-Gigu{\`e}re}  \& {Kere{\v{s}}}}{{Hopkins}
  et~al.}{2022c}]{hopkins:cr.multibin.mw.comparison}
{Hopkins} P.~F.,  {Butsky} I.~S.,  {Panopoulou} G.~V.,  {Ji} S.,  {Quataert}
  E.,  {Faucher-Gigu{\`e}re} C.-A.,   {Kere{\v{s}}} D.,  2022c, \mn@doi
  [\mnras] {10.1093/mnras/stac1791}, \href
  {https://ui.adsabs.harvard.edu/abs/2022MNRAS.516.3470H} {516, 3470}

\bibitem[\protect\citeauthoryear{{Hopkins}, {Squire}, {Butsky}  \&
  {Ji}}{{Hopkins} et~al.}{2022d}]{hopkins:2021.sc.et.models.incompatible.obs}
{Hopkins} P.~F.,  {Squire} J.,  {Butsky} I.~S.,   {Ji} S.,  2022d, \mn@doi
  [\mnras] {10.1093/mnras/stac2909}, \href
  {https://ui.adsabs.harvard.edu/abs/2022MNRAS.517.5413H} {517, 5413}

\bibitem[\protect\citeauthoryear{{Hopkins} et~al.,}{{Hopkins}
  et~al.}{2023a}]{hopkins:superzoom.overview}
{Hopkins} P.~F.,  et~al., 2023a, \mn@doi [arXiv e-prints]
  {10.48550/arXiv.2309.13115}, \href
  {https://ui.adsabs.harvard.edu/abs/2023arXiv230913115H} {p. arXiv:2309.13115}

\bibitem[\protect\citeauthoryear{{Hopkins} et~al.,}{{Hopkins}
  et~al.}{2023b}]{hopkins:fire3.methods}
{Hopkins} P.~F.,  et~al., 2023b, \mn@doi [\mnras] {10.1093/mnras/stac3489},
  \href {https://ui.adsabs.harvard.edu/abs/2023MNRAS.519.3154H} {519, 3154}

\bibitem[\protect\citeauthoryear{{Horiuchi}, {Matsumoto}, {Hanawa}  \&
  {Shibata}}{{Horiuchi}
  et~al.}{1988}]{horiuchi:1988.parker.growth.rate.vs.beta}
{Horiuchi} T.,  {Matsumoto} R.,  {Hanawa} T.,   {Shibata} K.,  1988, \pasj,
  \href {https://ui.adsabs.harvard.edu/abs/1988PASJ...40..147H} {40, 147}

\bibitem[\protect\citeauthoryear{{Indriolo} et~al.,}{{Indriolo}
  et~al.}{2015}]{indriolo:2015.cr.ionization.rate.vs.galactic.radius}
{Indriolo} N.,  et~al., 2015, \mn@doi [\apj] {10.1088/0004-637X/800/1/40},
  \href {https://ui.adsabs.harvard.edu/abs/2015ApJ...800...40I} {800, 40}

\bibitem[\protect\citeauthoryear{{Jafari}}{{Jafari}}{2019}]{jafari:2019.mhd.ppd.review}
{Jafari} A.,  2019, arXiv e-prints, \href
  {https://ui.adsabs.harvard.edu/abs/2019arXiv190409677J} {p. arXiv:1904.09677}

\bibitem[\protect\citeauthoryear{{Jaffe}}{{Jaffe}}{2019}]{jaffe:2019.magnetic.field.modeling.review}
{Jaffe} T.~R.,  2019, \mn@doi [Galaxies] {10.3390/galaxies7020052}, \href
  {https://ui.adsabs.harvard.edu/abs/2019Galax...7...52J} {7, 52}

\bibitem[\protect\citeauthoryear{{Ji} et~al.,}{{Ji}
  et~al.}{2020}]{ji:fire.cr.cgm}
{Ji} S.,  et~al., 2020, \mn@doi [\mnras] {10.1093/mnras/staa1849}, \href
  {https://ui.adsabs.harvard.edu/abs/2020MNRAS.496.4221J} {496, 4221}

\bibitem[\protect\citeauthoryear{{Jiang}, {Stone}  \& {Davis}}{{Jiang}
  et~al.}{2019}]{jiang:2019.supereddington.sims.low.outflow.efficiency.low.rad.eff}
{Jiang} Y.-F.,  {Stone} J.~M.,   {Davis} S.~W.,  2019, \mn@doi [\apj]
  {10.3847/1538-4357/ab29ff}, \href
  {https://ui.adsabs.harvard.edu/abs/2019ApJ...880...67J} {880, 67}

\bibitem[\protect\citeauthoryear{{Johansen} \& {Levin}}{{Johansen} \&
  {Levin}}{2008}]{johansen.levin:2008.high.mdot.magnetized.disks}
{Johansen} A.,  {Levin} Y.,  2008, \mn@doi [\aap]
  {10.1051/0004-6361:200810385}, \href
  {https://ui.adsabs.harvard.edu/abs/2008A&A...490..501J} {490, 501}

\bibitem[\protect\citeauthoryear{{Ju}, {Stone}  \& {Zhu}}{{Ju}
  et~al.}{2017}]{ju:2017.mri.acc.disks.with.spiral.modes}
{Ju} W.,  {Stone} J.~M.,   {Zhu} Z.,  2017, \mn@doi [\apj]
  {10.3847/1538-4357/aa705d}, \href
  {https://ui.adsabs.harvard.edu/abs/2017ApJ...841...29J} {841, 29}

\bibitem[\protect\citeauthoryear{{Kaaz}, {Liska}, {Jacquemin-Ide}, {Andalman},
  {Musoke}, {Tchekhovskoy}  \& {Porth}}{{Kaaz}
  et~al.}{2022}]{kaaz:2022.grmhd.sims.misaligned.acc.disks.spin}
{Kaaz} N.,  {Liska} M. T.~P.,  {Jacquemin-Ide} J.,  {Andalman} Z.~L.,  {Musoke}
  G.,  {Tchekhovskoy} A.,   {Porth} O.,  2022, \mn@doi [arXiv e-prints]
  {10.48550/arXiv.2210.10053}, \href
  {https://ui.adsabs.harvard.edu/abs/2022arXiv221010053K} {p. arXiv:2210.10053}

\bibitem[\protect\citeauthoryear{{Kalnajs}}{{Kalnajs}}{1971}]{kalnajs:1971}
{Kalnajs} A.~J.,  1971, \mn@doi [\apj] {10.1086/150957}, \href
  {http://adsabs.harvard.edu/abs/1971ApJ...166..275K} {166, 275}

\bibitem[\protect\citeauthoryear{{Kim} \& {Ostriker}}{{Kim} \&
  {Ostriker}}{2000}]{kim.ostriker:2000.mhd.instab.shearing.cold.winds}
{Kim} W.-T.,  {Ostriker} E.~C.,  2000, \mn@doi [\apj] {10.1086/309293}, \href
  {https://ui.adsabs.harvard.edu/abs/2000ApJ...540..372K} {540, 372}

\bibitem[\protect\citeauthoryear{{Kim} \& {Ostriker}}{{Kim} \&
  {Ostriker}}{2001}]{kim.ostriker:2001.gravitoturb.galactic.disks.mhd.conditions}
{Kim} W.-T.,  {Ostriker} E.~C.,  2001, \mn@doi [\apj] {10.1086/322330}, \href
  {https://ui.adsabs.harvard.edu/abs/2001ApJ...559...70K} {559, 70}

\bibitem[\protect\citeauthoryear{{Kim} \& {Ryu}}{{Kim} \&
  {Ryu}}{2001}]{kim:2001.turb.strongly.suppresses.parker.instability}
{Kim} J.,  {Ryu} D.,  2001, \mn@doi [\apjl] {10.1086/324422}, \href
  {https://ui.adsabs.harvard.edu/abs/2001ApJ...561L.135K} {561, L135}

\bibitem[\protect\citeauthoryear{{Kim}, {Hong}  \& {Ryu}}{{Kim}
  et~al.}{1997}]{kim:1997.parker.instab.linear.gravity.relevant.for.acc.disks}
{Kim} J.,  {Hong} S.~S.,   {Ryu} D.,  1997, \mn@doi [\apj] {10.1086/304410},
  \href {https://ui.adsabs.harvard.edu/abs/1997ApJ...485..228K} {485, 228}

\bibitem[\protect\citeauthoryear{{Kim}, {Ostriker}  \& {Stone}}{{Kim}
  et~al.}{2002}]{kim:2002.mhd.disk.instabilities}
{Kim} W.-T.,  {Ostriker} E.~C.,   {Stone} J.~M.,  2002, \mn@doi [\apj]
  {10.1086/344367}, \href {http://adsabs.harvard.edu/abs/2002ApJ...581.1080K}
  {581, 1080}

\bibitem[\protect\citeauthoryear{{King}}{{King}}{2003}]{king:msigma.superfb.1}
{King} A.,  2003, \mn@doi [\apjl] {10.1086/379143}, \href
  {http://adsabs.harvard.edu/abs/2003ApJ...596L..27K} {596, L27}

\bibitem[\protect\citeauthoryear{{Kinney}, {Schmitt}, {Clarke}, {Pringle},
  {Ulvestad}  \& {Antonucci}}{{Kinney}
  et~al.}{2000}]{kinney:2000.bh.jet.directions}
{Kinney} A.~L.,  {Schmitt} H.~R.,  {Clarke} C.~J.,  {Pringle} J.~E.,
  {Ulvestad} J.~S.,   {Antonucci} R.~R.~J.,  2000, \mn@doi [\apj]
  {10.1086/309016}, \href {http://adsabs.harvard.edu/abs/2000ApJ...537..152K}
  {537, 152}

\bibitem[\protect\citeauthoryear{{Kormendy}, {Bender}  \& {Cornell}}{{Kormendy}
  et~al.}{2011}]{kormendy:2011.bh.nodisk.corr}
{Kormendy} J.,  {Bender} R.,   {Cornell} M.~E.,  2011, \mn@doi [\nat]
  {10.1038/nature09694}, \href
  {http://adsabs.harvard.edu/abs/2011Natur.469..374K} {469, 374}

\bibitem[\protect\citeauthoryear{{Krause} et~al.,}{{Krause}
  et~al.}{2020}]{krause:2020.spiral.galaxy.halo.magnetic.geometries.and.coherence}
{Krause} M.,  et~al., 2020, \mn@doi [\aap] {10.1051/0004-6361/202037780}, \href
  {https://ui.adsabs.harvard.edu/abs/2020A&A...639A.112K} {639, A112}

\bibitem[\protect\citeauthoryear{{Kudoh}, {Wada}  \& {Norman}}{{Kudoh}
  et~al.}{2020}]{kudoh:2020.strong.b.field.agn.acc.disk.sims.compare}
{Kudoh} Y.,  {Wada} K.,   {Norman} C.,  2020, \mn@doi [\apj]
  {10.3847/1538-4357/abba39}, \href
  {https://ui.adsabs.harvard.edu/abs/2020ApJ...904....9K} {904, 9}

\bibitem[\protect\citeauthoryear{{Laor}, {Fiore}, {Elvis}, {Wilkes}  \&
  {McDowell}}{{Laor} et~al.}{1997}]{laor:warm.absorber}
{Laor} A.,  {Fiore} F.,  {Elvis} M.,  {Wilkes} B.~J.,   {McDowell} J.~C.,
  1997, \mn@doi [\apj] {10.1086/303696}, \href
  {http://adsabs.harvard.edu/cgi-bin/nph-bib_query?bibcode=1997ApJ...477...93L&db_key=AST}
  {477, 93}

\bibitem[\protect\citeauthoryear{{Lee} \& {Hong}}{{Lee} \&
  {Hong}}{2007}]{lee:2007.parker.instab.analysis.char.modes.wavelengths}
{Lee} S.~M.,  {Hong} S.~S.,  2007, \mn@doi [\apjs] {10.1086/509761}, \href
  {https://ui.adsabs.harvard.edu/abs/2007ApJS..169..269L} {169, 269}

\bibitem[\protect\citeauthoryear{{Lin}}{{Lin}}{2014}]{lin:2014.linear.stability.magnetized.selfgrav.ppds}
{Lin} M.-K.,  2014, \mn@doi [\apj] {10.1088/0004-637X/790/1/13}, \href
  {https://ui.adsabs.harvard.edu/abs/2014ApJ...790...13L} {790, 13}

\bibitem[\protect\citeauthoryear{{Lizano}, {Galli}, {Cai}  \& {Adams}}{{Lizano}
  et~al.}{2010}]{lizano:2010.stab.grav.instab.ppds.via.bfields}
{Lizano} S.,  {Galli} D.,  {Cai} M.~J.,   {Adams} F.~C.,  2010, \mn@doi [\apj]
  {10.1088/0004-637X/724/2/1561}, \href
  {https://ui.adsabs.harvard.edu/abs/2010ApJ...724.1561L} {724, 1561}

\bibitem[\protect\citeauthoryear{{Lubow}, {Papaloizou}  \& {Pringle}}{{Lubow}
  et~al.}{1994}]{lubow:1994.bfield.advection.acc.disks.vs.resistivity}
{Lubow} S.~H.,  {Papaloizou} J.~C.~B.,   {Pringle} J.~E.,  1994, \mn@doi
  [\mnras] {10.1093/mnras/267.2.235}, \href
  {https://ui.adsabs.harvard.edu/abs/1994MNRAS.267..235L} {267, 235}

\bibitem[\protect\citeauthoryear{{Ma}, {Hopkins}, {Wetzel}, {Kirby},
  {Angl{\'e}s-Alc{\'a}zar}, {Faucher-Gigu{\`e}re}, {Kere{\v s}}  \&
  {Quataert}}{{Ma} et~al.}{2017}]{ma:2016.disk.structure}
{Ma} X.,  {Hopkins} P.~F.,  {Wetzel} A.~R.,  {Kirby} E.~N.,
  {Angl{\'e}s-Alc{\'a}zar} D.,  {Faucher-Gigu{\`e}re} C.-A.,  {Kere{\v s}} D.,
   {Quataert} E.,  2017, \mn@doi [\mnras] {10.1093/mnras/stx273}, \href
  {http://adsabs.harvard.edu/abs/2017MNRAS.467.2430M} {467, 2430}

\bibitem[\protect\citeauthoryear{{Ma}, {Mao}, {Ordog}  \& {Brown}}{{Ma}
  et~al.}{2020}]{mao:galactic.bfields.revisions}
{Ma} Y.~K.,  {Mao} S.~A.,  {Ordog} A.,   {Brown} J.~C.,  2020, \mn@doi [\mnras]
  {10.1093/mnras/staa2105}, \href
  {https://ui.adsabs.harvard.edu/abs/2020MNRAS.497.3097M} {497, 3097}

\bibitem[\protect\citeauthoryear{{Machida}, {Nakamura}  \&
  {Matsumoto}}{{Machida}
  et~al.}{2006}]{machida:2006.mag.supported.disks.from.cooling.around.adafs}
{Machida} M.,  {Nakamura} K.~E.,   {Matsumoto} R.,  2006, \mn@doi [\pasj]
  {10.1093/pasj/58.1.193}, \href
  {https://ui.adsabs.harvard.edu/abs/2006PASJ...58..193M} {58, 193}

\bibitem[\protect\citeauthoryear{{Magorrian} et~al.}{{Magorrian}
  et~al.}{1998}]{magorrian}
{Magorrian} J.,  et~al., 1998, \mn@doi [\aj] {10.1086/300353}, \href
  {http://adsabs.harvard.edu/cgi-bin/nph-bib_query?bibcode=1998AJ....115.2285M&db_key=AST}
  {115, 2285}

\bibitem[\protect\citeauthoryear{{Mao}, {Gaensler}, {Haverkorn}, {Zweibel},
  {Madsen}, {McClure-Griffiths}, {Shukurov}  \& {Kronberg}}{{Mao}
  et~al.}{2010}]{mao:2010.no.coherent.vertical.Bfield.through.solar.circle}
{Mao} S.~A.,  {Gaensler} B.~M.,  {Haverkorn} M.,  {Zweibel} E.~G.,  {Madsen}
  G.~J.,  {McClure-Griffiths} N.~M.,  {Shukurov} A.,   {Kronberg} P.~P.,  2010,
  \mn@doi [\apj] {10.1088/0004-637X/714/2/1170}, \href
  {https://ui.adsabs.harvard.edu/abs/2010ApJ...714.1170M} {714, 1170}

\bibitem[\protect\citeauthoryear{{Martin-Alvarez}, {Katz}, {Sijacki},
  {Devriendt}  \& {Slyz}}{{Martin-Alvarez}
  et~al.}{2021}]{martin.alvarez:2021.bfield.amplification}
{Martin-Alvarez} S.,  {Katz} H.,  {Sijacki} D.,  {Devriendt} J.,   {Slyz} A.,
  2021, \mn@doi [\mnras] {10.1093/mnras/stab968}, \href
  {https://ui.adsabs.harvard.edu/abs/2021MNRAS.504.2517M} {504, 2517}

\bibitem[\protect\citeauthoryear{{Martin-Alvarez}, {Devriendt}, {Slyz},
  {Sijacki}, {Richardson}  \& {Katz}}{{Martin-Alvarez}
  et~al.}{2022}]{martin.alvarez:2022.cosmological.turb.dynamo}
{Martin-Alvarez} S.,  {Devriendt} J.,  {Slyz} A.,  {Sijacki} D.,  {Richardson}
  M. L.~A.,   {Katz} H.,  2022, \mn@doi [\mnras] {10.1093/mnras/stac1099},
  \href {https://ui.adsabs.harvard.edu/abs/2022MNRAS.513.3326M} {513, 3326}

\bibitem[\protect\citeauthoryear{{Mercedes-Feliz} et~al.,}{{Mercedes-Feliz}
  et~al.}{2023}]{mercedes.feliz:2023.agn.feedback.positive.negative}
{Mercedes-Feliz} J.,  et~al., 2023, \mn@doi [\mnras] {10.1093/mnras/stad2079},
  \href {https://ui.adsabs.harvard.edu/abs/2023MNRAS.tmp.2027M} {}

\bibitem[\protect\citeauthoryear{{Meru} \& {Bate}}{{Meru} \&
  {Bate}}{2012}]{meru:2012.nonconvergence.disk.frag}
{Meru} F.,  {Bate} M.~R.,  2012, \mnras, in press, arXiv:1209.1107, \href
  {http://adsabs.harvard.edu/abs/2012arXiv1209.1107M} {}

\bibitem[\protect\citeauthoryear{{Miller} \& {Stone}}{{Miller} \&
  {Stone}}{2000}]{miller.stone:2000.magnetically.elevated.disk}
{Miller} K.~A.,  {Stone} J.~M.,  2000, \mn@doi [\apj] {10.1086/308736}, \href
  {https://ui.adsabs.harvard.edu/abs/2000ApJ...534..398M} {534, 398}

\bibitem[\protect\citeauthoryear{{Mishra}, {Begelman}, {Armitage}  \&
  {Simon}}{{Mishra}
  et~al.}{2020}]{mishra:2020.elevated.disks.sims.strongly.sensitive.initial.beta}
{Mishra} B.,  {Begelman} M.~C.,  {Armitage} P.~J.,   {Simon} J.~B.,  2020,
  \mn@doi [\mnras] {10.1093/mnras/stz3572}, \href
  {https://ui.adsabs.harvard.edu/abs/2020MNRAS.492.1855M} {492, 1855}

\bibitem[\protect\citeauthoryear{{Mocz}, {Burkhart}, {Hernquist}, {McKee}  \&
  {Springel}}{{Mocz} et~al.}{2017}]{2017ApJ...838...40M}
{Mocz} P.,  {Burkhart} B.,  {Hernquist} L.,  {McKee} C.~F.,   {Springel} V.,
  2017, \mn@doi [\apj] {10.3847/1538-4357/aa6475}, \href
  {https://ui.adsabs.harvard.edu/abs/2017ApJ...838...40M} {838, 40}

\bibitem[\protect\citeauthoryear{{Moster}, {Macci{\`o}}, {Somerville},
  {Johansson}  \& {Naab}}{{Moster}
  et~al.}{2010}]{moster:2010.thin.disk.heating.vs.gas}
{Moster} B.~P.,  {Macci{\`o}} A.~V.,  {Somerville} R.~S.,  {Johansson} P.~H.,
  {Naab} T.,  2010, \mn@doi [\mnras] {10.1111/j.1365-2966.2009.16190.x}, \href
  {http://adsabs.harvard.edu/abs/2010MNRAS.403.1009M} {403, 1009}

\bibitem[\protect\citeauthoryear{{Murray}, {Quataert}  \& {Thompson}}{{Murray}
  et~al.}{2005}]{murray:momentum.winds}
{Murray} N.,  {Quataert} E.,   {Thompson} T.~A.,  2005, \mn@doi [\apj]
  {10.1086/426067}, \href
  {http://adsabs.harvard.edu/cgi-bin/nph-bib_query?bibcode=2005ApJ...618..569M&db_key=AST}
  {618, 569}

\bibitem[\protect\citeauthoryear{{Narayan} \& {Yi}}{{Narayan} \&
  {Yi}}{1995}]{narayan.yi.95:adaf.lowmass.bhs}
{Narayan} R.,  {Yi} I.,  1995, \mn@doi [\apj] {10.1086/176343}, \href
  {http://adsabs.harvard.edu/abs/1995ApJ...452..710N} {452, 710}

\bibitem[\protect\citeauthoryear{{Narayan}, {Igumenshchev}  \&
  {Abramowicz}}{{Narayan} et~al.}{2003}]{narayan:2003.mad.disk}
{Narayan} R.,  {Igumenshchev} I.~V.,   {Abramowicz} M.~A.,  2003, \mn@doi
  [\pasj] {10.1093/pasj/55.6.L69}, \href
  {https://ui.adsabs.harvard.edu/abs/2003PASJ...55L..69N} {55, L69}

\bibitem[\protect\citeauthoryear{{Nayakshin}}{{Nayakshin}}{2005}]{nayakshin:2005.warped.disk.obsc}
{Nayakshin} S.,  2005, \mn@doi [\mnras] {10.1111/j.1365-2966.2005.08913.x},
  \href {http://adsabs.harvard.edu/abs/2005MNRAS.359..545N} {359, 545}

\bibitem[\protect\citeauthoryear{{Noguchi}}{{Noguchi}}{1988}]{noguchi:merger.induced.bars.gas.forcing}
{Noguchi} M.,  1988, \aap, \href
  {http://adsabs.harvard.edu/abs/1988A%26A...203..259N} {203, 259}

\bibitem[\protect\citeauthoryear{{Novikov} \& {Thorne}}{{Novikov} \&
  {Thorne}}{1973}]{novikov.thorne:1973.astro.of.bhs}
{Novikov} I.~D.,  {Thorne} K.~S.,  1973, in Black Holes (Les Astres Occlus). pp
  343--450

\bibitem[\protect\citeauthoryear{{Oda}, {Machida}, {Nakamura}  \&
  {Matsumoto}}{{Oda}
  et~al.}{2009}]{oda:2009.analytic.mag.disk.structure.models}
{Oda} H.,  {Machida} M.,  {Nakamura} K.~E.,   {Matsumoto} R.,  2009, \mn@doi
  [\apj] {10.1088/0004-637X/697/1/16}, \href
  {https://ui.adsabs.harvard.edu/abs/2009ApJ...697...16O} {697, 16}

\bibitem[\protect\citeauthoryear{{Paardekooper}}{{Paardekooper}}{2012}]{paardekooper:2012.stochastic.disk.frag}
{Paardekooper} S.-J.,  2012, \mn@doi [\mnras]
  {10.1111/j.1365-2966.2012.20553.x}, \href
  {http://adsabs.harvard.edu/abs/2012MNRAS.421.3286P} {421, 3286}

\bibitem[\protect\citeauthoryear{{Paczy{\'n}sky} \& {Wiita}}{{Paczy{\'n}sky} \&
  {Wiita}}{1980}]{paczynsky.wiita:1980.slim.disk}
{Paczy{\'n}sky} B.,  {Wiita} P.~J.,  1980, \aap, \href
  {https://ui.adsabs.harvard.edu/abs/1980A&A....88...23P} {88, 23}

\bibitem[\protect\citeauthoryear{{Pan} \& {Scannapieco}}{{Pan} \&
  {Scannapieco}}{2010}]{pan:2010.turbulent.mixing.times}
{Pan} L.,  {Scannapieco} E.,  2010, \mn@doi [\apj]
  {10.1088/0004-637X/721/2/1765}, \href
  {http://adsabs.harvard.edu/abs/2010ApJ...721.1765P} {721, 1765}

\bibitem[\protect\citeauthoryear{{Parker}}{{Parker}}{1966}]{parker:1966.magnetic.instability.disks}
{Parker} E.~N.,  1966, \mn@doi [\apj] {10.1086/148828}, \href
  {https://ui.adsabs.harvard.edu/abs/1966ApJ...145..811P} {145, 811}

\bibitem[\protect\citeauthoryear{{Pessah} \& {Psaltis}}{{Pessah} \&
  {Psaltis}}{2005}]{pessah.psaltis:2005.mri.extensions.stronger.fields}
{Pessah} M.~E.,  {Psaltis} D.,  2005, \mn@doi [\apj] {10.1086/430940}, \href
  {https://ui.adsabs.harvard.edu/abs/2005ApJ...628..879P} {628, 879}

\bibitem[\protect\citeauthoryear{{Pessah}, {Chan}  \& {Psaltis}}{{Pessah}
  et~al.}{2006}]{pessah:2006.mri.signature.ratio.maxwell.reynolds}
{Pessah} M.~E.,  {Chan} C.-K.,   {Psaltis} D.,  2006, \mn@doi [\mnras]
  {10.1111/j.1365-2966.2006.10824.x}, \href
  {https://ui.adsabs.harvard.edu/abs/2006MNRAS.372..183P} {372, 183}

\bibitem[\protect\citeauthoryear{{Pitts} \& {Tayler}}{{Pitts} \&
  {Tayler}}{1985}]{pitts.tayler:1985.toroidal.buoyancy.modes.rotating.stars}
{Pitts} E.,  {Tayler} R.~J.,  1985, \mn@doi [\mnras] {10.1093/mnras/216.2.139},
  \href {https://ui.adsabs.harvard.edu/abs/1985MNRAS.216..139P} {216, 139}

\bibitem[\protect\citeauthoryear{{Ponnada} et~al.,}{{Ponnada}
  et~al.}{2022}]{ponnada:fire.magnetic.fields.vs.obs}
{Ponnada} S.~B.,  et~al., 2022, \mn@doi [\mnras] {10.1093/mnras/stac2448},
  \href {https://ui.adsabs.harvard.edu/abs/2022MNRAS.516.4417P} {516, 4417}

\bibitem[\protect\citeauthoryear{{Pringle}}{{Pringle}}{1981}]{pringle:accretion.review}
{Pringle} J.~E.,  1981, \mn@doi [\araa] {10.1146/annurev.aa.19.090181.001033},
  \href {http://adsabs.harvard.edu/abs/1981ARA%26A..19..137P} {19, 137}

\bibitem[\protect\citeauthoryear{{Reynolds}}{{Reynolds}}{2021}]{reynolds:2021.bh.spin.review}
{Reynolds} C.~S.,  2021, \mn@doi [\araa] {10.1146/annurev-astro-112420-035022},
  \href {https://ui.adsabs.harvard.edu/abs/2021ARA&A..59..117R} {59, 117}

\bibitem[\protect\citeauthoryear{{Rice}, {Lodato}  \& {Armitage}}{{Rice}
  et~al.}{2005}]{rice:2005.disk.frag.firstlook}
{Rice} W.~K.~M.,  {Lodato} G.,   {Armitage} P.~J.,  2005, \mn@doi [\mnras]
  {10.1111/j.1745-3933.2005.00105.x}, \href
  {http://adsabs.harvard.edu/abs/2005MNRAS.364L..56R} {364, L56}

\bibitem[\protect\citeauthoryear{{Rieder} \& {Teyssier}}{{Rieder} \&
  {Teyssier}}{2017}]{rieder.teyssier:2017.turb.dynamo.saturation.few.percent.fraction.b}
{Rieder} M.,  {Teyssier} R.,  2017, \mn@doi [\mnras] {10.1093/mnras/stx1670},
  \href {http://adsabs.harvard.edu/abs/2017MNRAS.471.2674R} {471, 2674}

\bibitem[\protect\citeauthoryear{{Riols} \& {Latter}}{{Riols} \&
  {Latter}}{2016}]{riols:2016.mhd.ppd.gravitoturb}
{Riols} A.,  {Latter} H.,  2016, \mn@doi [\mnras] {10.1093/mnras/stw1112},
  \href {https://ui.adsabs.harvard.edu/abs/2016MNRAS.460.2223R} {460, 2223}

\bibitem[\protect\citeauthoryear{{Robertson} \& {Goldreich}}{{Robertson} \&
  {Goldreich}}{2012}]{robertson.goldreich:2012.adiabatic.turb.heating}
{Robertson} B.,  {Goldreich} P.,  2012, \mn@doi [\apjl]
  {10.1088/2041-8205/750/2/L31}, \href
  {https://ui.adsabs.harvard.edu/abs/2012ApJ...750L..31R} {750, L31}

\bibitem[\protect\citeauthoryear{{Rodrigues}, {Sarson}, {Shukurov}, {Bushby}
  \& {Fletcher}}{{Rodrigues}
  et~al.}{2016}]{rodrigues:2016.parker.instab.disk.galaxies}
{Rodrigues} L.~F.~S.,  {Sarson} G.~R.,  {Shukurov} A.,  {Bushby} P.~J.,
  {Fletcher} A.,  2016, \mn@doi [\apj] {10.3847/0004-637X/816/1/2}, \href
  {https://ui.adsabs.harvard.edu/abs/2016ApJ...816....2R} {816, 2}

\bibitem[\protect\citeauthoryear{{Salpeter}}{{Salpeter}}{1964}]{salpeter64}
{Salpeter} E.~E.,  1964, \mn@doi [\apj] {10.1086/147973}, \href
  {http://adsabs.harvard.edu/cgi-bin/nph-bib_query?bibcode=1964ApJ...140..796S&db_key=AST}
  {140, 796}

\bibitem[\protect\citeauthoryear{{Salvesen}, {Simon}, {Armitage}  \&
  {Begelman}}{{Salvesen}
  et~al.}{2016a}]{salveson:2016.sims.mri.dominated.bfield.disks}
{Salvesen} G.,  {Simon} J.~B.,  {Armitage} P.~J.,   {Begelman} M.~C.,  2016a,
  \mn@doi [\mnras] {10.1093/mnras/stw029}, \href
  {https://ui.adsabs.harvard.edu/abs/2016MNRAS.457..857S} {457, 857}

\bibitem[\protect\citeauthoryear{{Salvesen}, {Armitage}, {Simon}  \&
  {Begelman}}{{Salvesen}
  et~al.}{2016b}]{salveson:2016.decaying.fields.poloidal.flux}
{Salvesen} G.,  {Armitage} P.~J.,  {Simon} J.~B.,   {Begelman} M.~C.,  2016b,
  \mn@doi [\mnras] {10.1093/mnras/stw1231}, \href
  {https://ui.adsabs.harvard.edu/abs/2016MNRAS.460.3488S} {460, 3488}

\bibitem[\protect\citeauthoryear{{Scheuer} \& {Feiler}}{{Scheuer} \&
  {Feiler}}{1996}]{scheuer:1996.bh.acc.disk.alignment}
{Scheuer} P.~A.~G.,  {Feiler} R.,  1996, \mnras, \href
  {http://adsabs.harvard.edu/abs/1996MNRAS.282..291S} {282, 291}

\bibitem[\protect\citeauthoryear{{Schmidt}}{{Schmidt}}{1963}]{schmidt:1963.qso.redshift}
{Schmidt} M.,  1963, \mn@doi [\nat] {10.1038/1971040a0}, \href
  {https://ui.adsabs.harvard.edu/abs/1963Natur.197.1040S} {197, 1040}

\bibitem[\protect\citeauthoryear{{Schmitt}, {Kinney}, {Storchi-Bergmann}  \&
  {Antonucci}}{{Schmitt} et~al.}{1997}]{schmitt:1997.radio.alignment.w.host}
{Schmitt} H.~R.,  {Kinney} A.~L.,  {Storchi-Bergmann} T.,   {Antonucci} R.,
  1997, \mn@doi [\apj] {10.1086/303744}, \href
  {http://adsabs.harvard.edu/abs/1997ApJ...477..623S} {477, 623}

\bibitem[\protect\citeauthoryear{{Seta} \& {Federrath}}{{Seta} \&
  {Federrath}}{2022}]{seta.federrath:2022.turb.dynamo.twophase.medium}
{Seta} A.,  {Federrath} C.,  2022, \mn@doi [\mnras] {10.1093/mnras/stac1400},
  \href {https://ui.adsabs.harvard.edu/abs/2022MNRAS.514..957S} {514, 957}

\bibitem[\protect\citeauthoryear{{Shakura} \& {Sunyaev}}{{Shakura} \&
  {Sunyaev}}{1973}]{shakurasunyaev73}
{Shakura} N.~I.,  {Sunyaev} R.~A.,  1973, \aap, \href
  {http://adsabs.harvard.edu/cgi-bin/nph-bib_query?bibcode=1973A%26A....24..337S&db_key=AST}
  {24, 337}

\bibitem[\protect\citeauthoryear{{Shen}, {Hopkins}, {Faucher-Gigu{\`e}re},
  {Alexander}, {Richards}, {Ross}  \& {Hickox}}{{Shen}
  et~al.}{2020}]{shen:bolometric.qlf.update}
{Shen} X.,  {Hopkins} P.~F.,  {Faucher-Gigu{\`e}re} C.-A.,  {Alexander} D.~M.,
  {Richards} G.~T.,  {Ross} N.~P.,   {Hickox} R.~C.,  2020, \mn@doi [\mnras]
  {10.1093/mnras/staa1381}, \href
  {https://ui.adsabs.harvard.edu/abs/2020MNRAS.495.3252S} {495, 3252}

\bibitem[\protect\citeauthoryear{{Shi}, {Kremer}, {Grudi{\'c}},
  {Gerling-Dunsmore}  \& {Hopkins}}{{Shi}
  et~al.}{2022}]{shi:2022.hyper.eddington.no.bhfb}
{Shi} Y.,  {Kremer} K.,  {Grudi{\'c}} M.~Y.,  {Gerling-Dunsmore} H.~J.,
  {Hopkins} P.~F.,  2022, \mnras, submitted, arXiv e-prints, \href
  {https://ui.adsabs.harvard.edu/abs/2022arXiv220805025S} {p. arXiv:2208.05025}

\bibitem[\protect\citeauthoryear{{Shibata}, {Tajima}  \& {Matsumoto}}{{Shibata}
  et~al.}{1990}]{shibata:1990.low.beta.disk.parker.suppressed.by.magnetic.tension}
{Shibata} K.,  {Tajima} T.,   {Matsumoto} R.,  1990, \mn@doi [\apj]
  {10.1086/168382}, \href
  {https://ui.adsabs.harvard.edu/abs/1990ApJ...350..295S} {350, 295}

\bibitem[\protect\citeauthoryear{{Shlosman} \& {Begelman}}{{Shlosman} \&
  {Begelman}}{1989}]{shlosman:inefficient.viscosities}
{Shlosman} I.,  {Begelman} M.~C.,  1989, \mn@doi [\apj] {10.1086/167526}, \href
  {http://adsabs.harvard.edu/abs/1989ApJ...341..685S} {341, 685}

\bibitem[\protect\citeauthoryear{{Shlosman}, {Begelman}  \& {Frank}}{{Shlosman}
  et~al.}{1990}]{shlosman:midscale.accretion}
{Shlosman} I.,  {Begelman} M.~C.,   {Frank} J.,  1990, \mn@doi [\nat]
  {10.1038/345679a0}, \href {http://adsabs.harvard.edu/abs/1990Natur.345..679S}
  {345, 679}

\bibitem[\protect\citeauthoryear{{Silk} \& {Rees}}{{Silk} \&
  {Rees}}{1998}]{silkrees:msigma}
{Silk} J.,  {Rees} M.~J.,  1998, \aap, \href
  {http://adsabs.harvard.edu/cgi-bin/nph-bib_query?bibcode=1998A%26A...331L...1S&db_key=AST}
  {331, L1}

\bibitem[\protect\citeauthoryear{{S{\k{a}}dowski}}{{S{\k{a}}dowski}}{2016}]{sadowski:2016.mag.elevated.disk.sims.radpressure.midplane.thermal.instability.suppressed}
{S{\k{a}}dowski} A.,  2016, \mn@doi [\mnras] {10.1093/mnras/stw913}, \href
  {https://ui.adsabs.harvard.edu/abs/2016MNRAS.459.4397S} {459, 4397}

\bibitem[\protect\citeauthoryear{{Skalidis}, {Sternberg}, {Beattie}, {Pavlidou}
   \& {Tassis}}{{Skalidis}
  et~al.}{2021}]{skalidis:2021.sub.alfvenic.turb.cdf.method.revision}
{Skalidis} R.,  {Sternberg} J.,  {Beattie} J.~R.,  {Pavlidou} V.,   {Tassis}
  K.,  2021, \mn@doi [\aap] {10.1051/0004-6361/202142045}, \href
  {https://ui.adsabs.harvard.edu/abs/2021A&A...656A.118S} {656, A118}

\bibitem[\protect\citeauthoryear{{Skalidis}, {Tassis}  \&
  {Pavlidou}}{{Skalidis}
  et~al.}{2022}]{skalidis:2022.lagrangian.models.sub.alfvenic.supersonic.turb}
{Skalidis} R.,  {Tassis} K.,   {Pavlidou} V.,  2022, \mn@doi [arXiv e-prints,
  arXiv:2209.14143] {10.48550/arXiv.2209.14143}, \href
  {https://ui.adsabs.harvard.edu/abs/2022arXiv220914143S} {p. arXiv:2209.14143}

\bibitem[\protect\citeauthoryear{{Soltan}}{{Soltan}}{1982}]{soltan82}
{Soltan} A.,  1982, \mnras, \href
  {http://adsabs.harvard.edu/cgi-bin/nph-bib_query?bibcode=1982MNRAS.200..115S&db_key=AST}
  {200, 115}

\bibitem[\protect\citeauthoryear{{Steinwandel}, {Beck}, {Arth}, {Dolag},
  {Moster}  \& {Nielaba}}{{Steinwandel}
  et~al.}{2019}]{steinwandel:2019.magnetic.bouyancy.galactic.scales}
{Steinwandel} U.~P.,  {Beck} M.~C.,  {Arth} A.,  {Dolag} K.,  {Moster} B.~P.,
  {Nielaba} P.,  2019, \mn@doi [\mnras] {10.1093/mnras/sty3083}, \href
  {https://ui.adsabs.harvard.edu/abs/2019MNRAS.483.1008S} {483, 1008}

\bibitem[\protect\citeauthoryear{{Steinwandel}, {Kim}, {Bryan}, {Ostriker},
  {Somerville}  \& {Fielding}}{{Steinwandel}
  et~al.}{2022}]{steinwandel:lmc.mass.galaxy.outflows.mfm.validation}
{Steinwandel} U.~P.,  {Kim} C.-G.,  {Bryan} G.~L.,  {Ostriker} E.~C.,
  {Somerville} R.~S.,   {Fielding} D.~B.,  2022, arXiv e-prints, \href
  {https://ui.adsabs.harvard.edu/abs/2022arXiv221203898S} {p. arXiv:2212.03898}

\bibitem[\protect\citeauthoryear{{Sturm} et~al.}{{Sturm}
  et~al.}{2011}]{sturm:2011.ulirg.herschel.outflows}
{Sturm} E.,  et~al., 2011, \mn@doi [\apjl] {10.1088/2041-8205/733/1/L16}, \href
  {http://adsabs.harvard.edu/abs/2011ApJ...733L..16S} {733, L16+}

\bibitem[\protect\citeauthoryear{{Su}, {Hopkins}, {Hayward},
  {Faucher-Gigu{\`e}re}, {Kere{\v s}}, {Ma}  \& {Robles}}{{Su}
  et~al.}{2017}]{su:2016.weak.mhd.cond.visc.turbdiff.fx}
{Su} K.-Y.,  {Hopkins} P.~F.,  {Hayward} C.~C.,  {Faucher-Gigu{\`e}re} C.-A.,
  {Kere{\v s}} D.,  {Ma} X.,   {Robles} V.~H.,  2017, \mn@doi [\mnras]
  {10.1093/mnras/stx1463}, \href
  {http://adsabs.harvard.edu/abs/2017MNRAS.471..144S} {471, 144}

\bibitem[\protect\citeauthoryear{{Su}, {Hayward}, {Hopkins}, {Quataert},
  {Faucher-Gigu{\`e}re}  \& {Kere{\v s}}}{{Su}
  et~al.}{2018}]{su:fire.feedback.alters.magnetic.amplification.morphology}
{Su} K.-Y.,  {Hayward} C.~C.,  {Hopkins} P.~F.,  {Quataert} E.,
  {Faucher-Gigu{\`e}re} C.-A.,   {Kere{\v s}} D.,  2018, \mn@doi [\mnras]
  {10.1093/mnrasl/slx172}, \href
  {http://adsabs.harvard.edu/abs/2018MNRAS.473L.111S} {473, L111}

\bibitem[\protect\citeauthoryear{{Su} et~al.,}{{Su}
  et~al.}{2019}]{su:2018.stellar.fb.fails.to.solve.cooling.flow}
{Su} K.-Y.,  et~al., 2019, \mn@doi [\mnras] {10.1093/mnras/stz1494}, \href
  {https://ui.adsabs.harvard.edu/abs/2019MNRAS.487.4393S} {487, 4393}

\bibitem[\protect\citeauthoryear{{Tayler}}{{Tayler}}{1973}]{tayler:1973.magnetic.instabilities.interchange}
{Tayler} R.~J.,  1973, \mn@doi [\mnras] {10.1093/mnras/161.4.365}, \href
  {https://ui.adsabs.harvard.edu/abs/1973MNRAS.161..365T} {161, 365}

\bibitem[\protect\citeauthoryear{{Tchekhovskoy}, {Narayan}  \&
  {McKinney}}{{Tchekhovskoy} et~al.}{2010}]{tchekhovskoy:bh.spin.vs.radio.pwr}
{Tchekhovskoy} A.,  {Narayan} R.,   {McKinney} J.~C.,  2010, \mn@doi [\apj]
  {10.1088/0004-637X/711/1/50}, \href
  {http://adsabs.harvard.edu/abs/2009arXiv0911.2228T} {711, 50}

\bibitem[\protect\citeauthoryear{{Tchekhovskoy}, {Narayan}  \&
  {McKinney}}{{Tchekhovskoy} et~al.}{2011}]{tchekhovskoy:2011.mad.disk.jets}
{Tchekhovskoy} A.,  {Narayan} R.,   {McKinney} J.~C.,  2011, \mn@doi [\mnras]
  {10.1111/j.1745-3933.2011.01147.x}, \href
  {https://ui.adsabs.harvard.edu/abs/2011MNRAS.418L..79T} {418, L79}

\bibitem[\protect\citeauthoryear{{Terquem} \& {Papaloizou}}{{Terquem} \&
  {Papaloizou}}{1996}]{terquem:1996.toroidal.field.bouyancy.instability}
{Terquem} C.,  {Papaloizou} J. C.~B.,  1996, \mn@doi [\mnras]
  {10.1093/mnras/279.3.767}, \href
  {https://ui.adsabs.harvard.edu/abs/1996MNRAS.279..767T} {279, 767}

\bibitem[\protect\citeauthoryear{{Thompson}, {Quataert}  \&
  {Murray}}{{Thompson} et~al.}{2005}]{thompson:rad.pressure}
{Thompson} T.~A.,  {Quataert} E.,   {Murray} N.,  2005, \mn@doi [\apj]
  {10.1086/431923}, \href {http://adsabs.harvard.edu/abs/2005ApJ...630..167T}
  {630, 167}

\bibitem[\protect\citeauthoryear{{Torrey} et~al.,}{{Torrey}
  et~al.}{2020}]{torrey:2020.agn.wind.bal.gal.fx.fire}
{Torrey} P.,  et~al., 2020, \mn@doi [\mnras] {10.1093/mnras/staa2222}, \href
  {https://ui.adsabs.harvard.edu/abs/2020MNRAS.497.5292T} {497, 5292}

\bibitem[\protect\citeauthoryear{{Tremaine}}{{Tremaine}}{2001}]{tremaine:slow.keplerian.modes}
{Tremaine} S.,  2001, \mn@doi [\aj] {10.1086/319398}, \href
  {http://adsabs.harvard.edu/abs/2001AJ....121.1776T} {121, 1776}

\bibitem[\protect\citeauthoryear{{Van der Kruit}}{{Van der
  Kruit}}{1988}]{vanderkruit:1988.vertical.light.profile.fitting.functions}
{Van der Kruit} P.~C.,  1988, \aap, \href
  {https://ui.adsabs.harvard.edu/abs/1988A&A...192..117V} {192, 117}

\bibitem[\protect\citeauthoryear{{Vazquez-Semadeni} \&
  {Gazol}}{{Vazquez-Semadeni} \&
  {Gazol}}{1995}]{vazquez-semadeni:1995.turb.jeans.instab}
{Vazquez-Semadeni} E.,  {Gazol} A.,  1995, \aap, \href
  {http://adsabs.harvard.edu/abs/1995A%26A...303..204V} {303, 204}

\bibitem[\protect\citeauthoryear{{Vishniac}}{{Vishniac}}{1995}]{vishniac:1995.mhd.turb.acc.disks}
{Vishniac} E.~T.,  1995, in {Pena} M.,  {Kurtz} S.,  eds,  Revista Mexicana de
  Astronomia y Astrofisica Conference Series Vol. 3, Revista Mexicana de
  Astronomia y Astrofisica Conference Series. p.~69 (\mn@eprint {arXiv}
  {astro-ph/9508085}), \mn@doi{10.48550/arXiv.astro-ph/9508085}

\bibitem[\protect\citeauthoryear{{V{\"o}lk} \& {Aplers}}{{V{\"o}lk} \&
  {Aplers}}{1973}]{voelk.aplers:1973.alfven.packet.eos}
{V{\"o}lk} H.~J.,  {Aplers} W.,  1973, \mn@doi [\apss] {10.1007/BF00642204},
  \href {https://ui.adsabs.harvard.edu/abs/1973Ap&SS..20..267V} {20, 267}

\bibitem[\protect\citeauthoryear{{Wellons} et~al.,}{{Wellons}
  et~al.}{2022}]{wellons:2022.smbh.growth}
{Wellons} S.,  et~al., 2022, arXiv e-prints, \href
  {https://ui.adsabs.harvard.edu/abs/2022arXiv220306201W} {p. arXiv:2203.06201}

\bibitem[\protect\citeauthoryear{{White}, {Stone}  \& {Quataert}}{{White}
  et~al.}{2019}]{white:2019.mad.disk.sims}
{White} C.~J.,  {Stone} J.~M.,   {Quataert} E.,  2019, \mn@doi [\apj]
  {10.3847/1538-4357/ab0c0c}, \href
  {https://ui.adsabs.harvard.edu/abs/2019ApJ...874..168W} {874, 168}

\bibitem[\protect\citeauthoryear{{Whitworth}, {Smith}, {Klessen}, {Mac Low},
  {Glover}, {Tress}, {Pakmor}  \& {Soler}}{{Whitworth}
  et~al.}{2022}]{whitworth:no.mhd.fx.galform.sims}
{Whitworth} D.~J.,  {Smith} R.~J.,  {Klessen} R.~S.,  {Mac Low} M.-M.,
  {Glover} S. C.~O.,  {Tress} R.,  {Pakmor} R.,   {Soler} J.~D.,  2022, arXiv
  e-prints, \href {https://ui.adsabs.harvard.edu/abs/2022arXiv221004922W} {p.
  arXiv:2210.04922}

\bibitem[\protect\citeauthoryear{{Williams}, {Maiolino}, {Krongold},
  {Carniani}, {Cresci}, {Mannucci}  \& {Marconi}}{{Williams}
  et~al.}{2017}]{Williams2017}
{Williams} R.~J.,  {Maiolino} R.,  {Krongold} Y.,  {Carniani} S.,  {Cresci} G.,
   {Mannucci} F.,   {Marconi} A.,  2017, \mn@doi [\mnras]
  {10.1093/mnras/stx311}, \href
  {http://adsabs.harvard.edu/abs/2017MNRAS.467.3399W} {467, 3399}

\bibitem[\protect\citeauthoryear{{Xie} \& {Zdziarski}}{{Xie} \&
  {Zdziarski}}{2019}]{fuguo:2019.mad.disk.sims.rad.properties}
{Xie} F.-G.,  {Zdziarski} A.~A.,  2019, \mn@doi [\apj]
  {10.3847/1538-4357/ab5848}, \href
  {https://ui.adsabs.harvard.edu/abs/2019ApJ...887..167X} {887, 167}

\bibitem[\protect\citeauthoryear{{Younger}, {Hopkins}, {Cox}  \&
  {Hernquist}}{{Younger} et~al.}{2008}]{younger:minor.mergers}
{Younger} J.~D.,  {Hopkins} P.~F.,  {Cox} T.~J.,   {Hernquist} L.,  2008,
  \mn@doi [\apj] {10.1086/591639}, \href
  {http://adsabs.harvard.edu/abs/2008ApJ...686..815Y} {686, 815}

\bibitem[\protect\citeauthoryear{{Zakamska} et~al.,}{{Zakamska}
  et~al.}{2016}]{Zakamska2016}
{Zakamska} N.~L.,  et~al., 2016, \mn@doi [\mnras] {10.1093/mnras/stw718}, \href
  {http://adsabs.harvard.edu/abs/2016MNRAS.459.3144Z} {459, 3144}

\makeatother
\end{thebibliography}

\end{document}